%% file: BASS_AGNs_Milli_Prop_arXiv.tex
\documentclass[twocolumn,tighten]{aastex631}

\usepackage{graphicx}
\usepackage{bm}
\usepackage{url}
\usepackage{threeparttable}
\usepackage{amsmath,amssymb}
\usepackage{xspace}
\usepackage{ulem}
\usepackage{subfigure}
\usepackage{longtable}
\usepackage{rotating}

\shorttitle{Millimeter-wave Emission Catalog}
\shortauthors{Kawamuro et al.}

\begin{document}
\title{
    BASS XXXIV:
    A Catalog of the Nuclear Mm-wave Continuum Emission Properties of AGNs Constrained on Scales $\lesssim$ 100--200\, pc
} 

\correspondingauthor{Taiki Kawamuro}
\email{taiki.kawamuro@mail.udp.cl}

\author[0000-0002-6808-2052]{Taiki Kawamuro}
\altaffiliation{FONDECYT postdoctoral fellow}
\affil{N\'{u}cleo de Astronom\'{i}a de la Facultad de Ingenier\'{i}a, Universidad Diego Portales, Av. Ej\'{e}ercito Libertador 441, Santiago, Chile}
\affil{National Astronomical Observatory of Japan, Osawa, Mitaka, Tokyo 181-8588, Japan}
\affil{RIKEN Cluster for Pioneering Research, 2-1 Hirosawa, Wako, Saitama 351-0198, Japan}

\author[0000-0001-5231-2645]{Claudio Ricci} 
\affil{N\'{u}cleo de Astronom\'{i}a de la Facultad de Ingenier\'{i}a, Universidad Diego Portales, Av. Ej\'{e}ercito Libertador 441, Santiago, Chile}
\affil{Kavli Institute for Astronomy and Astrophysics, Peking University, Beijing 100871, People’s Republic of China}

\author[0000-0002-7962-5446]{Richard F. Mushotzky}
\affil{Department of Astronomy, University of Maryland, College Park, MD 20742, USA}

\author[0000-0001-6186-8792]{Masatoshi Imanishi} 
\affil{National Astronomical Observatory of Japan, Osawa, Mitaka, Tokyo 181-8588, Japan}
\affil{Department of Astronomy, School of Science, Graduate University for Advanced Studies (SOKENDAI), 2-21-1 Osawa, Mitaka, Tokyo 181-8588, Japan}

\author[0000-0002-8686-8737]{Franz E. Bauer}
\affiliation{Instituto de Astrof\'{\i}sica  and Centro de Astroingenier{\'{\i}}a, Facultad de F\'{i}sica, Pontificia Universidad Cat\'{o}lica de Chile, Casilla 306, Santiago 22, Chile}
\affiliation{Millennium Institute of Astrophysics (MAS), Nuncio Monse{\~{n}}or S{\'{o}}tero Sanz 100, Providencia, Santiago, Chile}
\affiliation{Space Science Institute, 4750 Walnut Street, Suite 205, Boulder, Colorado 80301, USA}

\author[0000-0001-5742-5980]{Federica Ricci}
\affil{Dipartimento di Fisica e Astronomia, Università di Bologna, via Gobetti 93/2, 40129 Bologna, Italy}
\affil{INAF Osservatorio Astronomico di Bologna, via Gobetti 93/3, 40129 Bologna, Italy}
\affil{Dipartimento di Matematica e Fisica, Universitá degli Studi Roma Tre, Via della Vasca Navale 84, I-00146, Roma, Italy}

\author[0000-0002-7998-9581]{Michael J. Koss} 
\affil{Eureka Scientific, 2452 Delmer Street Suite 100, Oakland, CA 94602-3017, USA}

\author[0000-0003-3474-1125]{George C. Privon}
\affil{National Radio Astronomy Observatory, 520 Edgemont Road, Charlottesville, VA 22903, USA}
\affil{Department of Astronomy, University of Florida, P.O. Box 112055, Gainesville, FL 32611, USA}

\author[0000-0002-3683-7297]{Benny Trakhtenbrot}
\affil{School of Physics and Astronomy, Tel Aviv University, Tel Aviv 69978, Israel}

\author[0000-0001-9452-0813]{Takuma Izumi}
\altaffiliation{NAOJ fellow} 
\affil{National Astronomical Observatory of Japan, Osawa, Mitaka, Tokyo 181-8588, Japan}
\affil{Department of Astronomy, School of Science, Graduate University for Advanced Studies (SOKENDAI), 2-21-1 Osawa, Mitaka, Tokyo 181-8588, Japan}

\author[0000-0002-4377-903X]{Kohei Ichikawa}
\affil{Frontier Research Institute for Interdisciplinary Sciences, Tohoku University, Sendai 980-8578, Japan}
\affil{Astronomical Institute, Tohoku University, Aramaki, Aoba-ku, Sendai, Miyagi 980-8578, Japan}
\affil{Max-Planck-Institut f{\" u}r extraterrestrische Physik (MPE), Giessenbachstrasse 1, D-85748 Garching bei M{\" u}nchen, Germanyn}

\author[0000-0003-0006-8681]{Alejandra F. Rojas}
\altaffiliation{FONDECYT postdoctoral fellow}
\affil{Centro de Astronom\'{i}a (CITEVA), Universidad de Antofagasta, Avenida Angamos 601, Antofagasta, Chile}
\affil{N\'{u}cleo de Astronom\'{i}a de la Facultad de Ingenier\'{i}a, Universidad Diego Portales, Av. Ej\'{e}ercito Libertador 441, Santiago, Chile}

\author[0000-0001-5785-7038]{Krista Lynne Smith}
\altaffiliation{Einstein fellow}
\affil{KIPAC at SLAC, Stanford University, Menlo Park, CA 94025, USA}
\affil{Southern Methodist University, Department of Physics, Dallas, TX 75205, USA}

\author[0000-0002-2125-4670]{Taro Shimizu}
\affil{Department of Physics and Astronomy, University College London, Gower Street, London WC1E 6BT, UK}

\author[0000-0002-5037-951X]{Kyuseok Oh}
\affil{Korea Astronomy and Space Science Institute, Daedeokdae-ro 776, Yuseong-gu, Daejeon 34055, Republic of Korea}

\author[0000-0002-8760-6157]{Jakob S. den Brok}
\affil{Institute for Particle Physics and Astrophysics, ETH Z\"{u}rich, Wolfgang-Pauli-Strasse 27, CH-8093 Z\"{u}rich, Switzerland}
\affil{Argelander Institute for Astronomy, Auf dem H\"{u}gel 71, 53231, Bonn, Germany}

\author[0000-0002-9850-6290]{Shunsuke Baba}
\affil{Graduate School of Science and Engineering, Kagoshima University, Korimoto, Kagoshima, Kagoshima 890-0065, Japan}

\author[0000-0003-0476-6647]{Mislav Balokovi\'{c}}
\affil{Yale Center for Astronomy \& Astrophysics, 52 Hillhouse Avenue, New Haven, CT 06511, USA}
\affil{Department of Physics, Yale University, P.O. Box 208120, New Haven, CT 06520, USA}

\author[0000-0001-9910-3234]{Chin-Shin Chang}
\affil{Joint ALMA Observatory, Alonso de Cordova 3107, Vitacura, Santiago, Chile}

\author[0000-0002-2603-2639]{Darshan Kakkad}
\affil{European Southern Observatory, Alonso de Cordova 3107, Vitacura, Casilla 19001, Santiago de Chile, Chile}
\affil{European Southern Observatory, Karl-Schwarzschild-Strasse 2, Garching bei M{\"u}nchen, Germany}

\author[0000-0001-8640-8522]{Ryan W. Pfeifle}
\affil{Department of Physics \& Astronomy, George Mason University, 4400 University Drive, MSN 3F3, Fairfax, VA 22030, USA}

\author[0000-0001-8433-550X]{Matthew J. Temple}
\altaffiliation{FONDECYT postdoctoral fellow}
\affil{N\'{u}cleo de Astronom\'{i}a de la Facultad de Ingenier\'{i}a, Universidad Diego Portales, Av. Ej\'{e}ercito Libertador 441, Santiago, Chile}

\author[0000-0001-7821-6715]{Yoshihiro Ueda}
\affil{Department of Astronomy, Kyoto University, Kyoto 606-8502, Japan}

\author[0000-0003-2992-8024]{Fiona Harrison}
\affil{Cahill Center for Astronomy and Astrophysics, California Institute of Technology, Pasadena, CA 91125, USA}

\author[0000-0003-2284-8603]{Meredith C. Powell}
\affil{Institute of Particle Astrophysics and Cosmology, Stanford University, 452 Lomita Mall, Stanford, CA 94305, USA}

\author[0000-0003-2686-9241]{Daniel Stern}
\affil{Jet Propulsion Laboratory, California Institute of Technology, 4800 Oak Grove Drive, MS 169-224, Pasadena, CA 91109, USA}

\author[0000-0002-0745-9792]{Meg Urry}
\affil{Yale Center for Astronomy \& Astrophysics, Physics Department, New Haven, CT 06520, USA}

\author[0000-0002-1233-9998]{David B. Sanders}
\affil{Institute for Astronomy, 2680 Woodlawn Drive, University of Hawaii, Honolulu, HI 96822, USA}

\begin{abstract}

We present a catalog of the millimeter-wave (mm-wave) continuum properties of 98 nearby ($z <$ 0.05) active galactic nuclei (AGNs) selected from the 70-month Swift/BAT hard X-ray catalog that have precisely determined X-ray spectral properties and subarcsec-resolution ALMA Band-6 (211--275\,GHz) observations as of 2021 April. Due to the hard-X-ray ($>$ 10\,keV) selection, the sample is nearly unbiased for obscured systems at least up to Compton-thick-level obscuration, 
and provides the largest number of AGNs with high physical resolution mm-wave data ($\lesssim$ 100--200\,pc). 
Our catalog reports emission peak coordinates, spectral indices, and peak fluxes and luminosities at 1.3\,mm (230\,GHz). 
Additionally, high-resolution mm-wave images are provided. 
Using the images and creating radial surface brightness profiles of mm-wave emission, we identify emission extending from the central source and isolated blob-like emission. Flags indicating the presence of these emission features are tabulated. Among 90 AGNs with significant detections of nuclear emission, 37 AGNs ($\approx$ 41\%) appear to have both or one of extended or blob-like components. 
We, in particular, investigate AGNs that show well-resolved mm-wave components and find that these seem to have a variety of origins (i.e., a jet, radio lobes, a secondary AGN, stellar clusters, a narrow line region, galaxy disk, active star-formation regions, and AGN-driven outflows), and some components have currently unclear origins.

\end{abstract}


\section{Introduction}\label{sec:int} 

Emission from active galactic nuclei (AGNs) has been identified  over a wide range of wavelengths, from radio to $\gamma$-rays.
However, the millimeter-wave (mm-wave, hereafter) emission of the AGN is still poorly constrained, as indicated by the fact that mm-wave data points were rarely presented in typical AGN spectral energy distributions \citep[SEDs; e.g.,][]{Elv94,Ho08,Hic18}.
Previous studies have suggested that the mm-wave emission cannot be explained simply by the extrapolation of a centimeter-wave (cm-wave) radio component (hereafter, ``radio" indicates cm-wave, or frequency below 30\,GHz) or that of an infrared (IR) one  \citep[e.g.,][]{Beh15,Ino18,Ino20}. 
It has been then proposed that the mm-wave component could be due to self-absorbed synchrotron emission from a compact region on a scale of $\lesssim$ $10^{-3}$ pc \citep[e.g.,][]{Lao08,Ino18,Ino20}. 
Thus, the study of mm-wave AGN emission could be essential to obtain a complete picture of the AGN process. 

To identify and better understand AGN-related mm-wave continuum emission,
we systematically analyzed high angular resolution ($<$ 1\arcsec) Atacama Large Millimeter/submillimeter Array (ALMA) Band-6 (211--275\,GHz) data of 98 nearby AGNs \citep[$z <$ 0.05;][hereafter Paper I]{Kaw22}.
The sample was created based on AGNs detected in the 70-month Swift/Burst Alert Telescope (BAT) hard X-ray catalog \citep{bau13}, which is nearly unbiased for obscured AGNs, at least up to the Compton-thick level of $N_{\rm H} \approx 10^{24}\,{\rm cm}^{-2}$ \citep[e.g.,][]{Bur11,ric15} and has been used in various types of research \citep[e.g.,][]{Kaw13,Kaw16b,Kaw16c,Ric17nat,Tan18,Kos18,Chi20,Bal20,Fis21}. 
Our sample provides improvement with respect to previous samples used to study mm-wave emission (e.g., \citealp{Beh15,Beh18}) in two different aspects.
i) The angular resolutions of our data are basically less than
0\farcs6, corresponding to $\lesssim$ 100--200\,pc for our targets, and are at least a few times better than those of the previous studies relying on $\sim$ 1\arcsec--2\arcsec\ resolutions. 
The higher resolutions reduce contaminating light from star formation (SF) processes in the host galaxies \cite[e.g., thermal emission from dust heated by stellar radiation, free-free emission from H\,{\sc ii} regions, and synchrotron emission from supernova remnants and other stellar processes;][]{Con92,Pan19}.
ii) The sample size is more than three times larger than the previous ones \citep{Beh15,Beh18}, which allows us to study the dependence of mm-wave emission on various parameters. 

In Paper I, we discovered tight correlations of nuclear peak mm-wave luminosity with AGN luminosities (bolometric, AGN mid-IR, 2--10\,keV, and 14--150\,keV luminosities).
Among these correlations, that with the 14--150\,keV luminosity has the smallest scatter (0.36\,dex), comparable with that observed between 12\,$\mu$m and X-ray luminosities \citep[e.g.,][]{Gan09}.
Focusing only on AGNs with the least SF contamination, selected based on mm-wave spectral indices inconsistent with SF processes, we found a similar correlation to that obtained for the whole sample.
This could indicate that, basically, a significant fraction of the observed mm-wave emission is related to the AGN activity traced by the X-ray emission. 

Paper~I focused on discussing the nuclear peak emission and its relations with AGN and host galaxy properties, and this paper (Paper~II, hereafter) presents a catalog of the mm-wave continuum properties of the 98 AGNs. 
For each object with significant detection of peak mm-wave emission, we provide its right ascension and declination (R.A. and decl.), spectral index, and flux and luminosity at 1.3 mm (230\,GHz). 
Furthermore, we show high-resolution mm-wave images of the sample. The images allowed us to identify objects that had resolved emission. 
These objects were classified, depending on the morphologies of resolved mm-wave emission, and we present brief discussions on some of them in the context of the existing literature. 

This paper is structured as follows. 
In Section~\ref{sec:sample}, we introduce our AGNs, their basic properties, 
and their ALMA data. These pieces of information are summarized in Tables~\ref{tab_app:sample} and \ref{tab_app:alma_data} in Appendix. 
In Section~\ref{sec:alma_data}, we describe our analysis of the ALMA Band-6 data and the identification of peak emission most likely related to an AGN. 
More details on the analysis were reported in Paper~I. 
The derived mm-wave properties of the peak emission are available from Table~\ref{tab_app:mm_prop}. 
In addition, we describe our quantitative and systematic identification of resolved mm-wave components based on high-resolution images and radial surface brightness profiles, which are displayed in Figures~\ref{fig_app:images} and \ref{fig_app:rads}, respectively. 
In Section~\ref{sec:ext}, we then summarize the possible origins inferred for particularly well-resolved components. 
The summary is based on discussions for individual objects, presented in Section~\ref{sec_app:notes} of Appendix. Finally, in Section~\ref{sec:sum}, we give our conclusions. 

Throughout the article, we adopt standard cosmological parameters ($H_0$ = 70 km\,s$^{-1}$\,Mpc$^{-1}$, $\Omega_{\rm m} = 0.3$, $\Omega_\Lambda = 0.7$). 

\section{Sample and Data}\label{sec:sample}

Our sample includes all nearby ($z <$ 0.05) 70-month BAT AGNs in \cite{bau13} with X-ray spectral properties precisely determined by \cite{ric17c} and with sub-arcsec resolution Band-6 (211--275 GHz) ALMA data, publicly available as of 2021 April.  
We focused on ALMA Band 6 because self-absorbed synchrotron emission from an AGN was expected to be more prominent at frequencies $\gtrsim$ 100--300\,GHz \citep{Ino18,Ino20}, and the band provided the largest sample of sources with $\gtrsim$ 100\,GHz data between Band 3 and Band 10. 
When searching the ALMA archive database, we considered a radius of 5\arcsec\ around BAT AGNs at $z < 0.05$. The radius corresponds to half of the radius of a typical size of the primary beam of the ALMA 12-m array in Band 6 ($\sim$ 20\arcsec). 
We note that some BAT AGNs were identified by \cite{Kos22_catalog} as Blazar-like objects, i.e., those with spectral energy distributions (SEDs) being dominated by non-thermal emission from radio to $\gamma$-rays and with radio properties being consistent with relativistic beaming. 
Specifically, the authors referred to the Roma Blazar Catalog \cite[BZCAT;][]{Mas09} and the follow-up work by \cite{pal19}. 
Although four are located at $z < 0.05$ and decl. $< 40^\circ$, which is an upper declination limit for ALMA observations, they had no publicly available Band-6 data and therefore are not included in our sample. 
In the BASS DR2 \citep{Kos22_catalog}, there are 364 AGNs at $z < 0.05$ which are not Blazar-like objects and can be observed by ALMA (i.e., decl. $< 40^\circ$), and among them, 98 AGNs (i.e., $\approx$ 27\% of the parent sample) had available ALMA data taken at sub-arcsec resolutions and were selected for our study. 

Table~\ref{tab_app:sample} in Appendix~\ref{sec_app:tab} summarizes the basic information of our objects: the identification number assigned to each AGN in our works (Papers~I and II), the BAT index in the 70-month catalog, Swift/BAT and counterpart names, the position of the AGN (R.A. and decl. in the ICRS frame), the reference for the position, the redshift ($z$), the distance ($D$), and the Seyfert type. The AGN positions are detailed in Section~\ref{sec:mm_det}. 
The distance measures (i.e., $z$ and $D$) and Seyfert types were taken from the BASS DR2 database \citep{Kos22_catalog,Kos22_overview}.  
We note that redshift-independent distances are adopted  for objects at distances below 50\,Mpc 
as the effect of peculiar velocity is expected to be significant compared to the recession velocity. 
For the distances, Koss et al. referred to the Extragalactic Distance Database of \cite{Tul09}, a catalog of the Cosmicflows project \citep{Cou17}, and the NASA/IPAC extragalactic database\footnote{The NASA/IPAC Extragalactic Database (NED) is funded by the National Aeronautics and Space Administration and operated by the California Institute of Technology (10.26132/NED1).}.

Our sample covers wide ranges in luminosity ($40 \lesssim \log(L_{\rm 14-150}/({\rm erg\,s^{-1}})) \lesssim 45$), black hole mass ($5 \lesssim \log(M_{\rm BH}/M_\odot) \lesssim 10$), and Eddington ratio ($-4 \lesssim \log \lambda_{\rm Edd} \lesssim 2$).
In spite of our limit of $z = 0.05$, or $D \sim$ 200\,Mpc, 
the sample consists of 81 objects at distances below 100\,Mpc. Also, according to the HyperLeda database \citep{Mak14}\footnote{http://leda.univ-lyon1.fr/}, 
55 and 36 objects are hosted by late-type and early-type galaxies, respectively.  
The host galaxies of the other seven objects were uncertain.
Regarding the nuclear type, 33, 23, and 42 objects 
were classified by \cite{Kos22_catalog} as type-1, type-1.9, and type-2 AGNs, respectively. As inferred from this optical classification, 
more than half, or 59 objects, are obscured with line-of-sight column densities above $10^{22}$ cm$^{-2}$. 
Finally, we remark that our sample includes some well-studied nearby AGNs at different wavelengths, such as NGC\,1068,
NGC\,1365, NGC\,3783, NGC\,4395, NGC\,4945, IC\,4329A,
the Circinus Galaxy, NGC\,5643, NGC\,6240, and IC\,5063. 
More details on our sample are described in Paper\,I.

In order to investigate AGN-related emission in Paper I by minimizing the impact of host galaxy contamination, we carefully selected the high-resolution ALMA data. Initially, we compiled a list of publicly available subarcsec-resolution data as of April 2021. Then, for objects observed in multiple projects, we selected only the data from the project that provided the highest angular resolution. 
Moreover, as some projects comprised different group observation unit set (OUS) IDs, each indicating the need to merge the data, in these cases, we exclusively utilized the data with the group OUS ID that offered the highest resolution.
Exceptionally, for NGC\,424, we selected the second highest resolution data, because the highest one shows many spurious lines in its spectra. 

The information on the ALMA data we analyzed is summarized in Table~\ref{tab_app:alma_data} in Appendix~\ref{sec_app:tab}. 
The table lists the ALMA project code, member OUS ID, observing mode (mosaic or not), field of view (FoV), exposure time, and observation date, which were taken from the ALMA archive database. 
Maximum recoverable scales (MRSs), listed as well, were calculated from calibrated data by considering the five percentile of the projected baseline lengths of antennas. In addition, we list the beam sizes along the minor and major axes, achieved with a weighting method in clean processes (see Section~\ref{sec:img_gen}) and their physical scales for the object distances.

\section{ALMA Data Analysis}\label{sec:alma_data}

\subsection{Continuum Image Generation}\label{sec:img_gen}

We reconstructed mm-wave continuum emission images from the ALMA data using the Common Astronomy Software Applications package \citep[CASA; ][]{McM07}. The raw data were reduced and calibrated by running scripts provided by the ARC in CASA with the versions adopted to create the scripts. 
For the rest of the analysis, we used CASA v.6.1.0.118.
By considering only channels devoid of any strong emission lines, continuum emission was imaged not only for each spectral window, but also by combining all available spectral windows. 
In these imaging processes, 
we utilized the task \texttt{tclean} along with the deconvolver \texttt{clark} and the Briggs weighting method employing \texttt{robust} = 0.5.
\texttt{clark} was adopted rather than the often-used alternative one \texttt{hogbom}, because the former properly subtracts clean components in the visibility space and also is faster, having enabled us to reconstruct images efficiently. 
The \texttt{robust} value does not offer the highest resolution or the best sensitivity, but we opted for it. 
This is because we were able to preserve nearly the highest resolution while retaining a sensitivity close to the case of \texttt{robust} = 2.0, which provides the lowest noise level. However, if no significant emission was detected from a target, we re-analyzed its data adopting \texttt{robust} = 2.0.
The detail is described later.
If data were obtained in the mosaic observation mode, we set \texttt{gridder} to mosaic. The \texttt{cell} size was set so that \texttt{cell} $< 1/3\times \theta^{\rm min}_{\rm beam}$, where $\theta^{\min}_{\rm beam}$ is the full width at half maximum (FWHM) of the beam size along the minor axis.
The \texttt{threshold} parameter was also set so that 
$3 \sigma_{\rm mm} <$ \texttt{threshold} $< 4 \sigma_{\rm mm}$, where $\sigma_{\rm mm}$ is a noise level derived from a region free from mm-wave emission. 
Finally, a primary beam correction was applied to all reconstructed images. 
Here, we mention that we exceptionally performed the self-calibration process for NGC 1068 and Circinus galaxy, as fringe-like features were clearly left in the images without the calibration.
According to the ALMA technical handbook\footnote{http://almascience.org/documents-and-tools/cycle8/alma-technical-handbook}, flux density measured in Band 6 can have a systematic uncertainty of 10\%, and this is taken into consideration throughout this paper.

\begin{figure*}
    \centering
    \includegraphics[width=6.9cm]{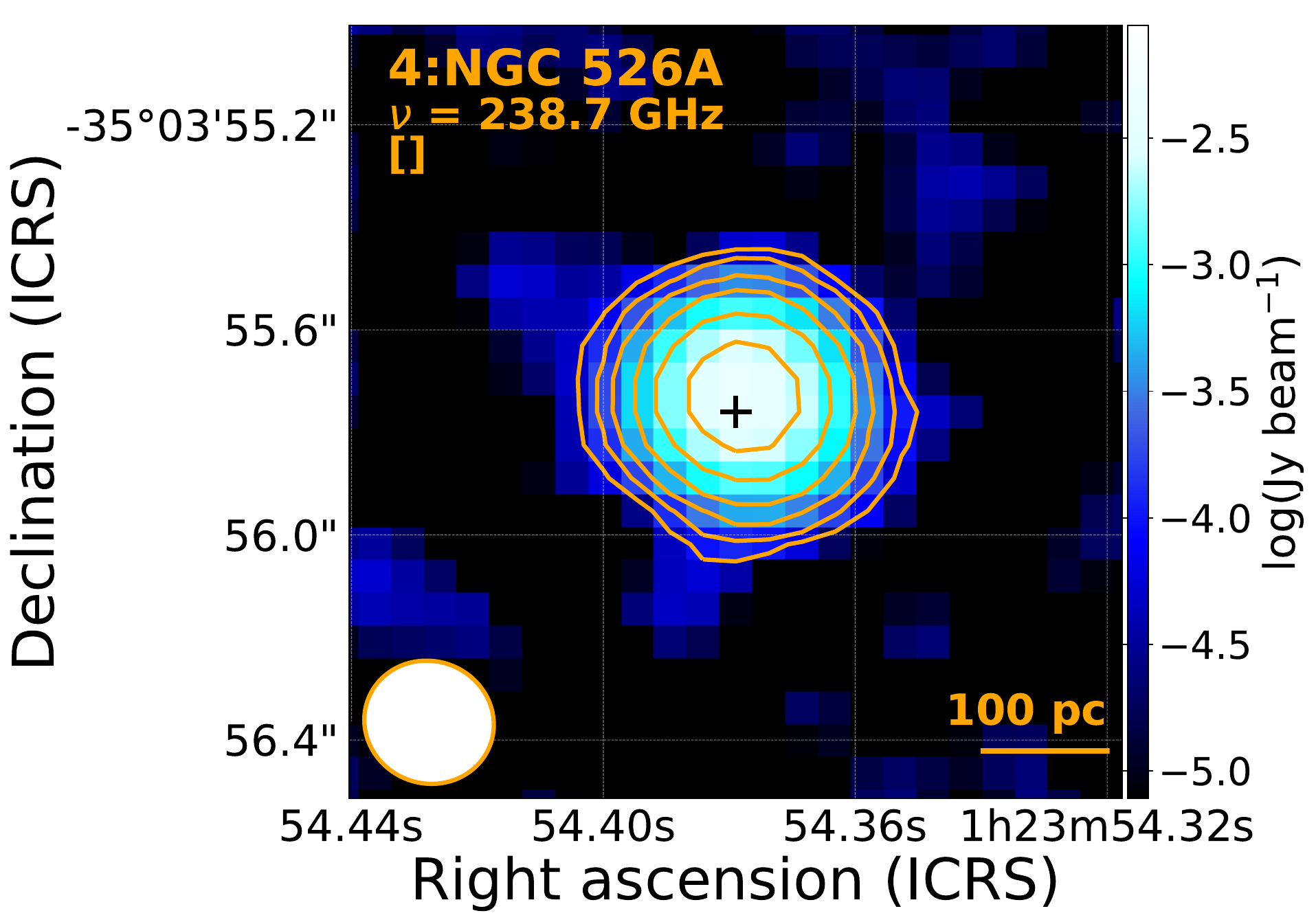} 
    \includegraphics[width=6.9cm]{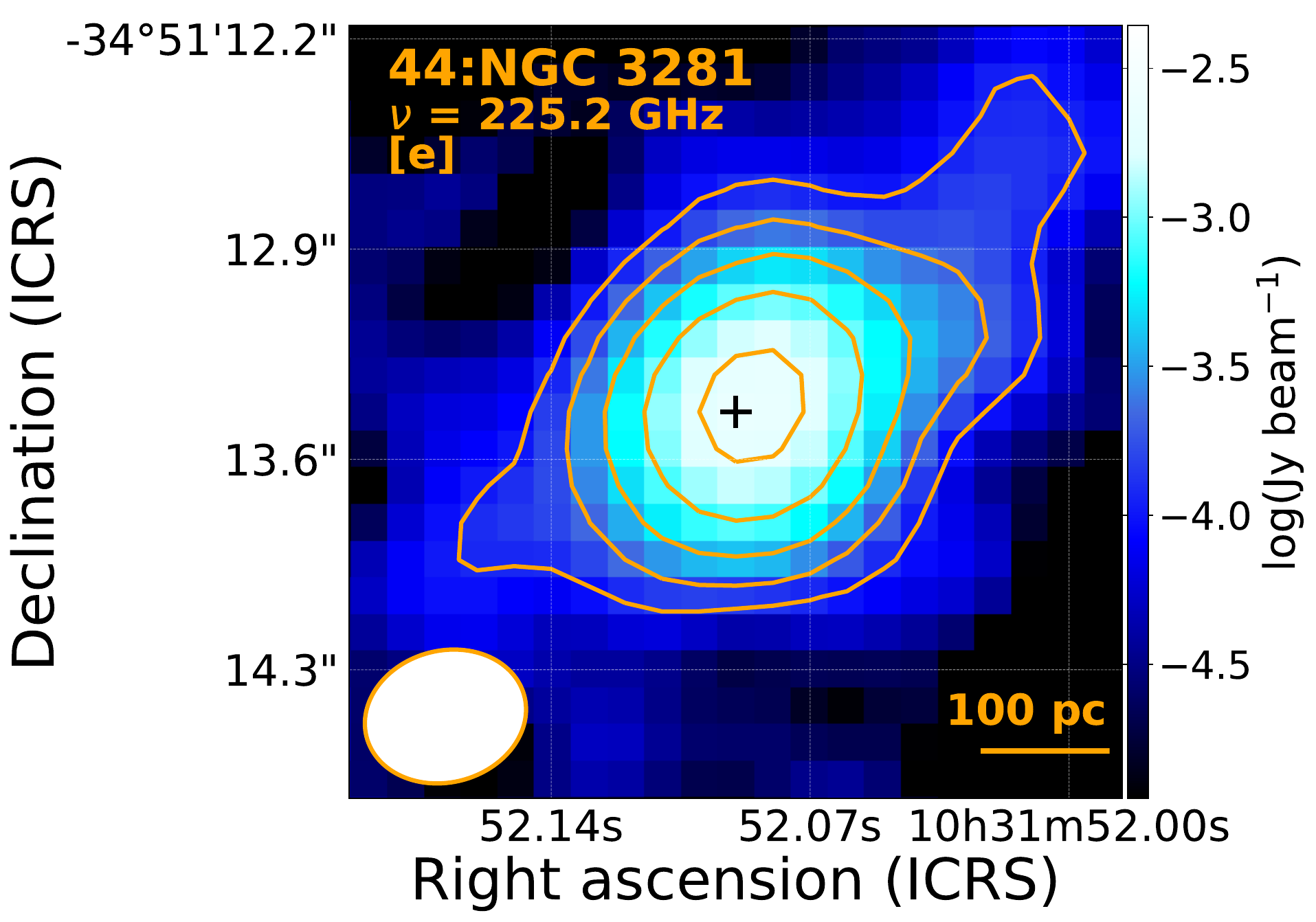} \\ 
    \includegraphics[width=6.9cm]{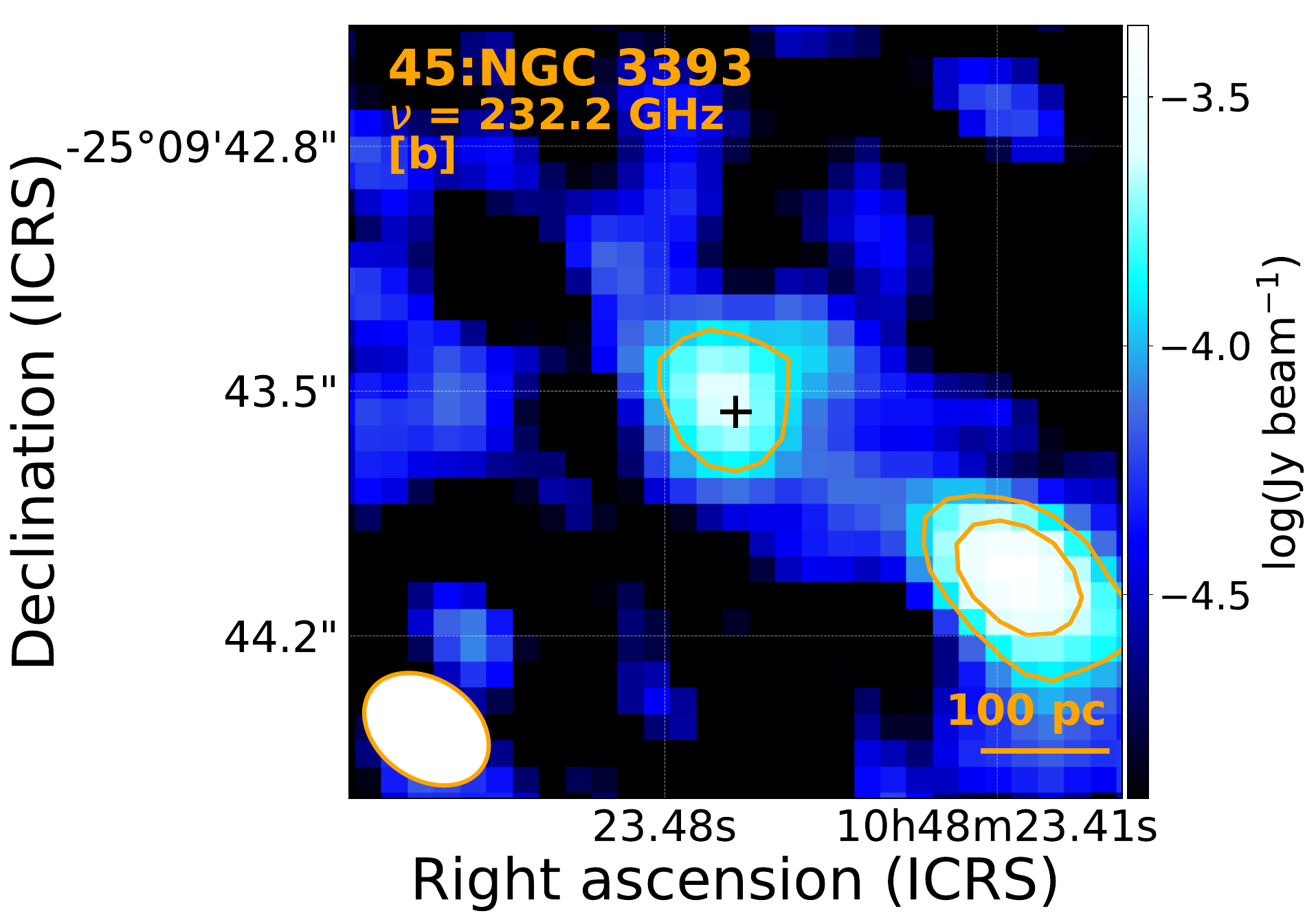}
    \includegraphics[width=6.9cm]{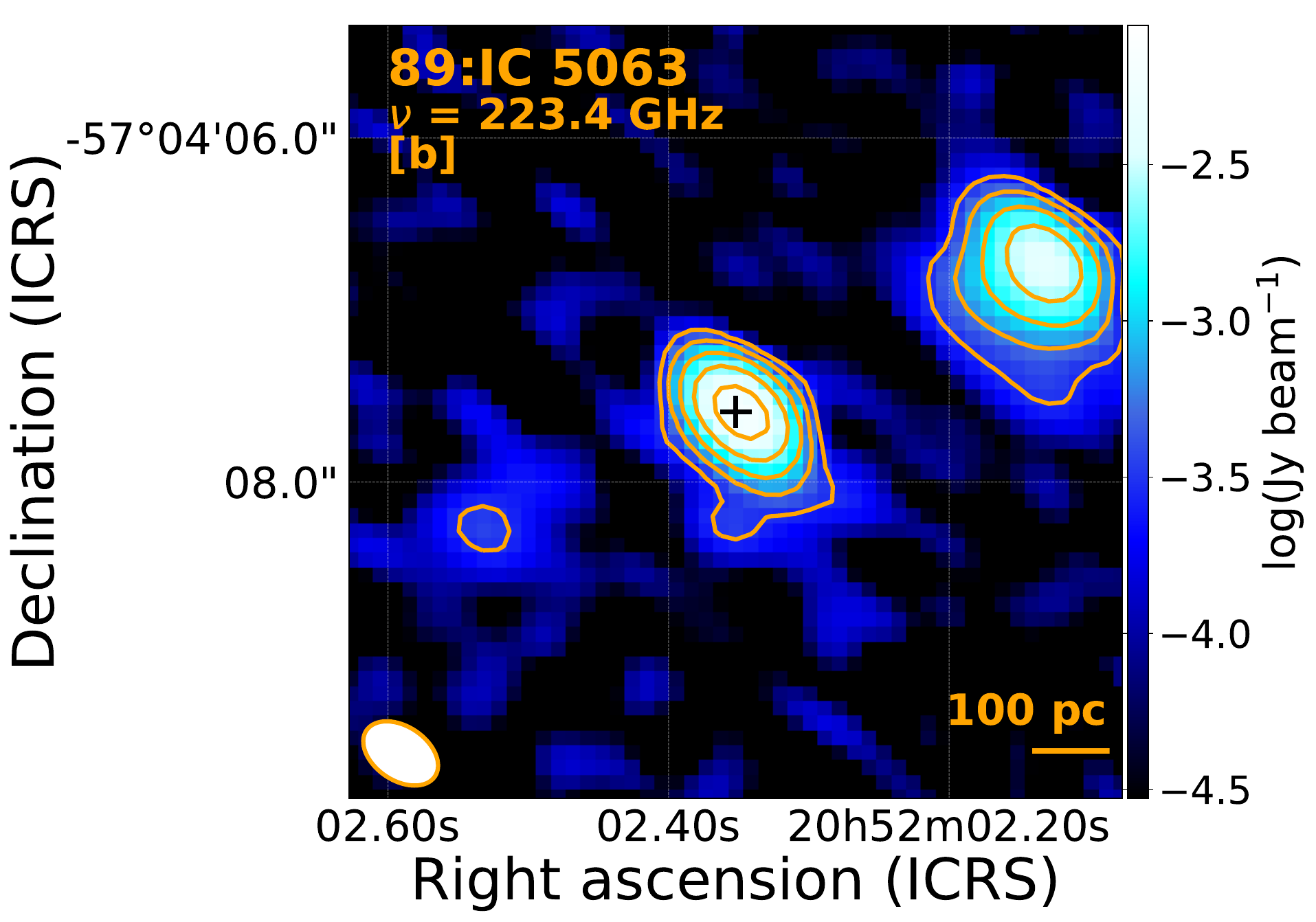}
    \caption{
    ALMA Band-6 images for NGC\,526A, NGC\,3281, NGC\,3393, and IC\,5063. 
    The beam sizes are indicated by the white ellipses in the bottom-left corners.
    In each panel, the identification number, assigned in this study (Table~\ref{tab_app:sample}) and Paper~I, is presented along with the object name. 
    The frequency is the rest-frame central frequency of the collapsed spectral window adopted, after removing any emission line flux. 
    The central black cross indicates the peak of the nuclear mm-wave emission, expected to be the AGN position. 
    A morphological parameter, which can consist of {e} and {b}, is also presented below the frequency. As detailed in Section~\ref{sec:ext}, {e} and {b} indicate the presence of extended emission connected with the central un-resolved component and the presence of isolated blob-like emission, respectively.
    Colors are assigned according to flux density per beam following the color bar on the right side. The orange contours indicate where flux densities per beam are $5\sigma_{\rm mm}$, $10\sigma_{\rm mm}$, $20\sigma_{\rm mm}$, $40\sigma_{\rm mm}$, $80\sigma_{\rm mm}$ and $160\sigma_{\rm mm}$.
    For NGC\,526A, NGC\,3281, NGC\,3393, and IC 5063, $\sigma_{\rm mm}$ = 0.016 mJy beam$^{-1}$, $\sigma_{\rm mm}$ = 0.022 mJy beam$^{-1}$, $\sigma_{\rm mm}$ = 0.025 mJy beam$^{-1}$, $\sigma_{\rm mm}$ = 0.059 mJy beam$^{-1}$, respectively. 
    ALMA images for all 98 objects are 
    summarized in Section~\ref{sec_app:fig} of Appendix. 
    }\label{fig:rep_objs}
\end{figure*}    

\subsection{Nuclear Mm-wave Source Detection}\label{sec:mm_det}

To identify mm-wave emission possibly associated with an AGN in each constructed ALMA image, we first created a list of AGN positions (R.A. and decl.) using data and catalogs at different wavelengths. As X-ray emission is one of the best probes for AGN activity, we primarily relied on 3--7\,keV images created by using Chandra data, which provide the highest resolution in the X-ray band. 
For 56 objects, we found on-axis Chandra data where the separation angle between a target and the focal plane is less than 1\arcmin. According to the Chandra X-Ray Center\footnote{https://cxc.harvard.edu/cal/ASPECT/celmon/},
the positional accuracy is 1\farcs4 at the 99\% level for such on-axis observations. Therefore,  for each target, we searched for a mm-wave peak within the 1\farcs4 radius around a Chandra-based AGN position.
As thresholds of 3--4$\sigma_{\rm mm}$ were adopted for the clean processes, only emission whose flux density per beam was greater than 5$\sigma_{\rm mm}$ was considered a detection. 

For the 42 objects that did not have Chandra data, we considered the IR ($\approx$ 3--22\,$\mu$m) ALLWISE catalog \citep{Wri00,Mai11}, which includes all our targets.
In the IR band, the AGN position can be inferred from the observation of dust emission due to an AGN \cite[e.g.,][]{Gan09,Wri00}.  
The positional accuracy of the catalog was estimated to be $\approx$ 0\farcs3 
by cross-matching with the 2MASS catalog\footnote{https://wise2.ipac.caltech.edu/docs/release/allsky/expsup/sec6\_4.html}, 
but we adopted 1\farcs5 as a search radius for WISE positions. This is because 
the 2MASS accuracy for a peak, or the nucleus, of a spatially resolved galaxy at the 99\% limit was estimated to be $\sim$ 1\farcs5 \citep{She17}. 

We adopted the mm-wave peaks identified by Chandra and WISE, basically in this order of priority, as candidates that are possibly associated with the AGNs. Further details can be found in Paper I. As a result, we identified significant ($> 5\sigma_{\rm mm}$) nuclear peaks for 89 AGNs, corresponding to a high detection rate of $\approx$ 91\%. 

For the nine objects without significant mm-wave peak detections, we performed the clean process with \texttt{robust} = 2. As a result, only for IC 4709, significant emission with $S^{\rm peak}_{\nu, \rm mm}/\sigma_{\rm mm} = 7.2$ was identified ($S^{\rm peak}_{\nu, \rm mm}/\sigma_{\rm mm} = 4.9$ if \texttt{robust} = 0.5). 
By considering this detection, mm-wave emission was detected from 90 AGNs (i.e., the detection fraction is 92\%).
As the last attempt to detect emission from the eight remaining objects, we proceeded to the analysis of alternative subarcsec-resolution data. Eventually, even with the newly adopted data, no significant signal was found from the eight objects.

In Table~\ref{tab_app:sample}, we list 
the positions of the identified mm-wave peaks as AGN positions. For the objects without significant mm-wave detections, we list AGN positions at different wavelengths. 
Figure~\ref{fig_app:images} of Appendix~\ref{sec_app:fig} summarizes the 98 images centered at these AGN positions.

\subsection{Non-nuclear Mm-wave Components}

The high-resolution ALMA observations ($\lesssim$ 0\farcs6, or $\lesssim$ 100--200\,pc) can reveal 
mm-wave emission that cannot be explained by a 
central un-resolved source. 
The capability can be seen in Figure~\ref{fig:rep_objs}, where NGC\,526A appears to show only an un-resolved component, while NGC\,3281, NGC\,3393, and IC\,5063 have other components. 
In the following, we refer to such resolved emission as non-nuclear emission, and in the next subsection, we describe our analyses for identifying non-nuclear components. 

\begin{figure*}
    \centering
    \includegraphics[width=8.9cm]{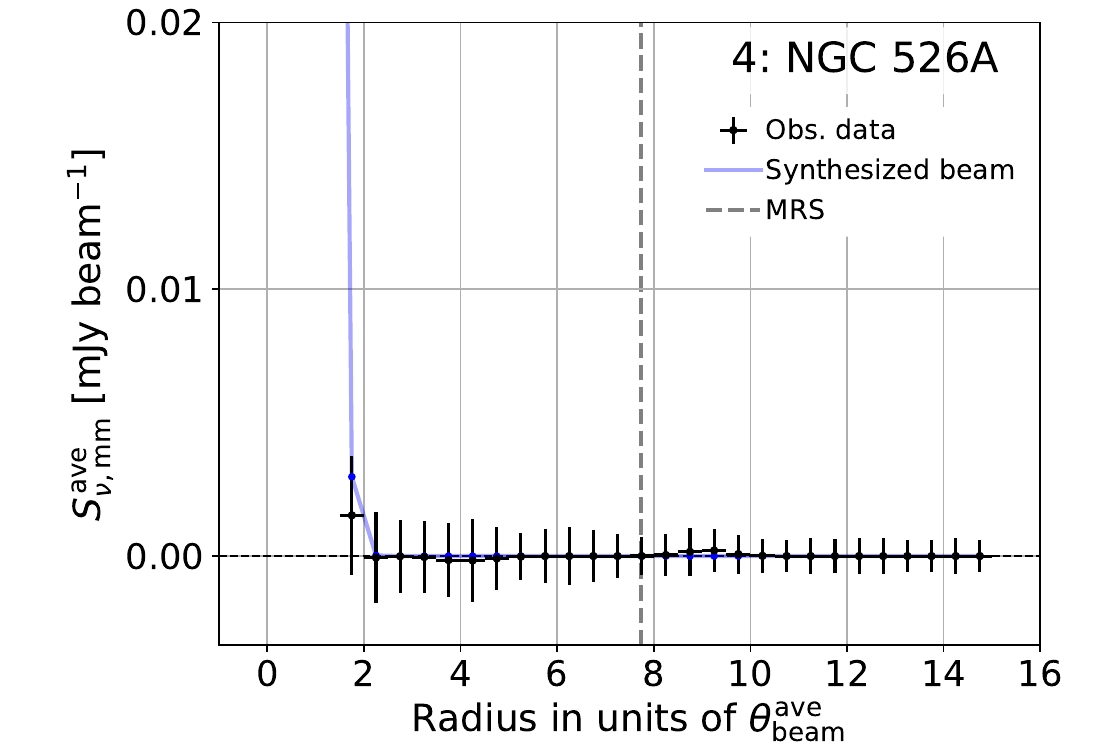} 
    \includegraphics[width=8.9cm]{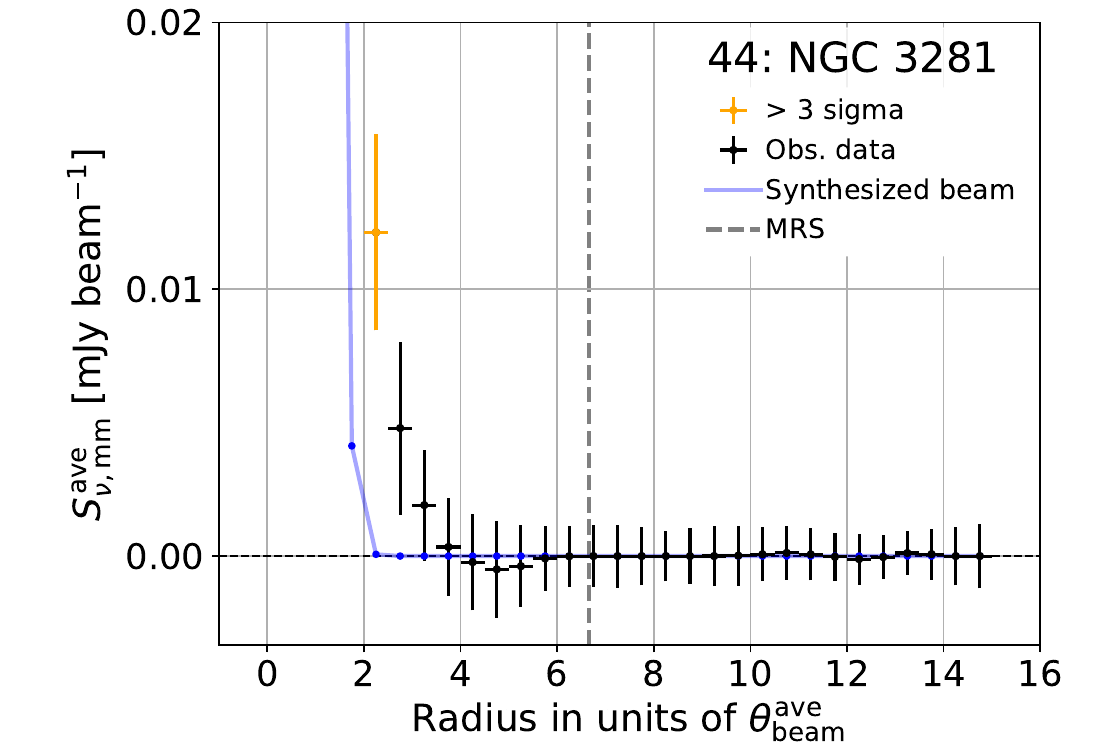} \\ 
    \includegraphics[width=8.9cm]{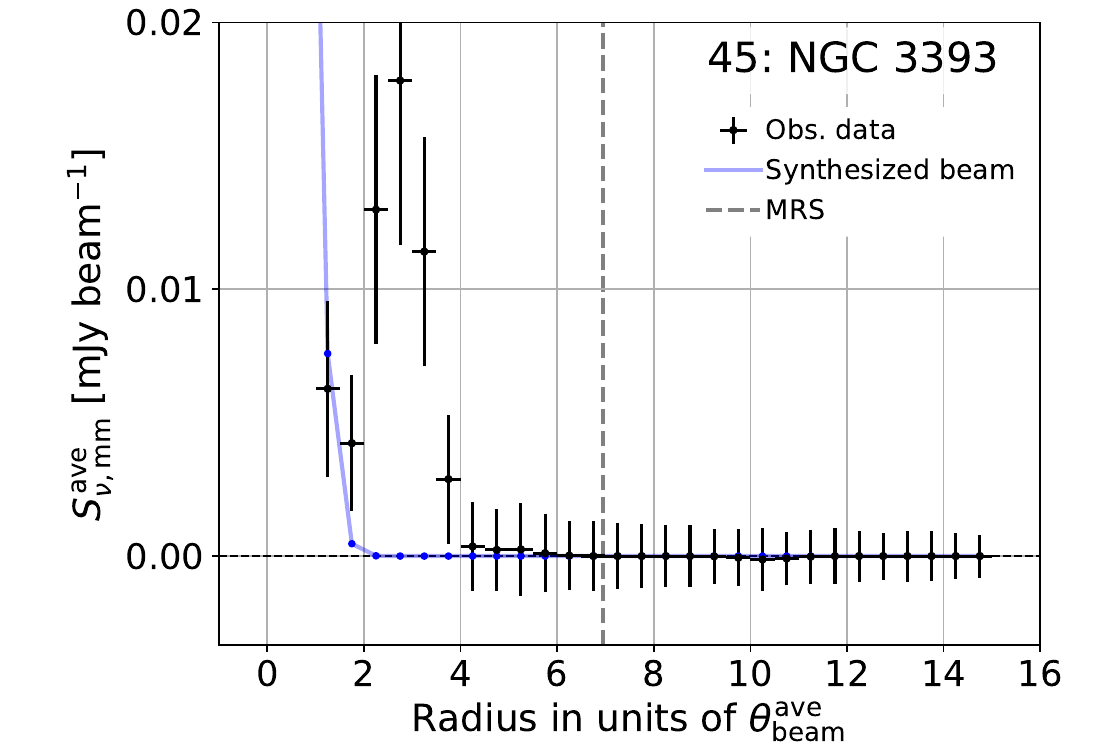}
    \includegraphics[width=8.9cm]{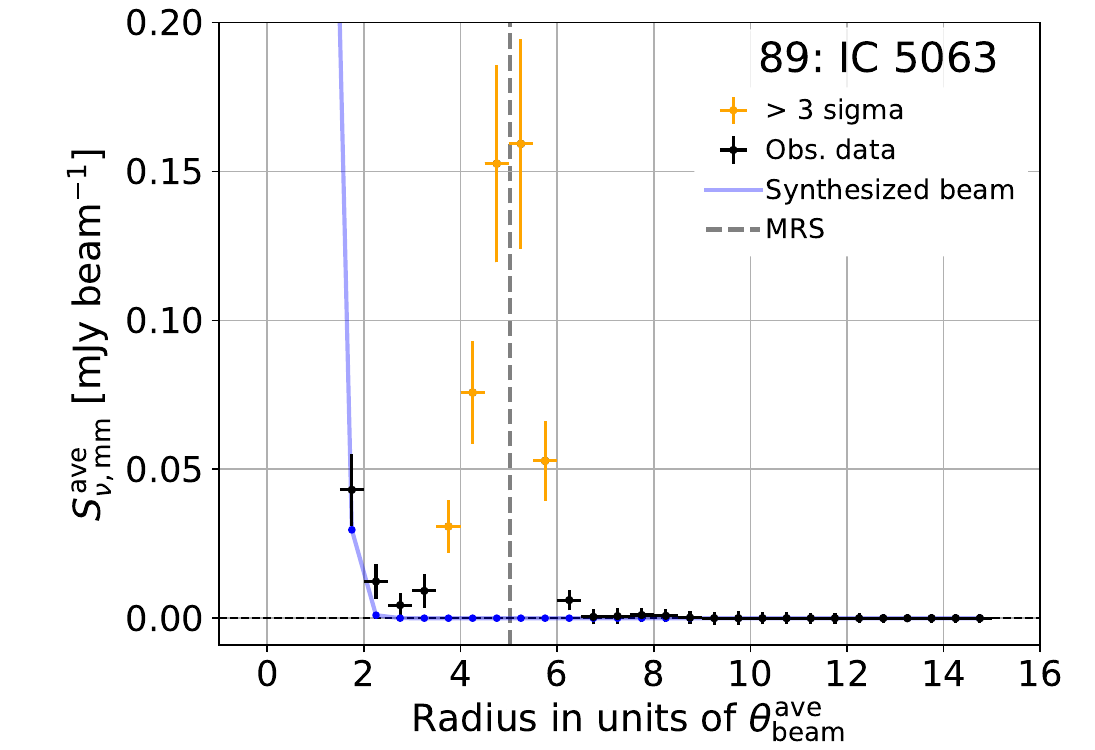} 
    \caption{
    Radial surface brightness profiles
    for NGC 526A, NGC 3281, NGC 3393, and IC 5063. 
    The radius is in units of the averaged beam size ($\theta^{\rm ave}_{\rm beam}$). 
    In each panel, 
    observed values are indicated with black error bars, 
    but particularly those detected at significances more than three times the SEs with respect to an un-resolved component (blue dots and lines) are shown in orange.  
    The dashed gray line indicates the MRS. 
    The ALMA images of the four objects are presented in Figure~\ref{fig:rep_objs}. 
    The radial profiles for all 98 AGNs are 
    summarized in Appendix~\ref{fig_app:rads}. 
    }\label{fig:ext_rad}
\end{figure*}    

\subsubsection{Quantitative and Systematic Identification of Non-nuclear Mm-wave Components}\label{sec:quant_id}

\begin{figure*}
    \centering  
    \includegraphics[width=8cm]{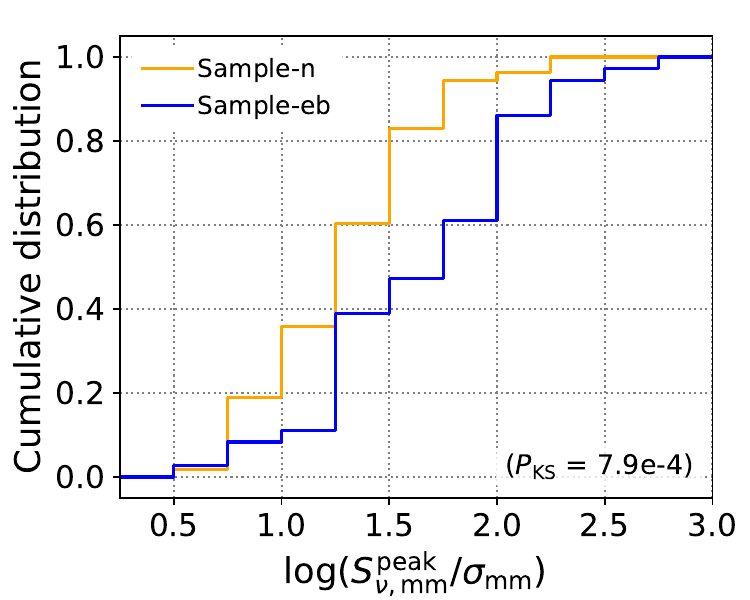}
    \includegraphics[width=8cm]{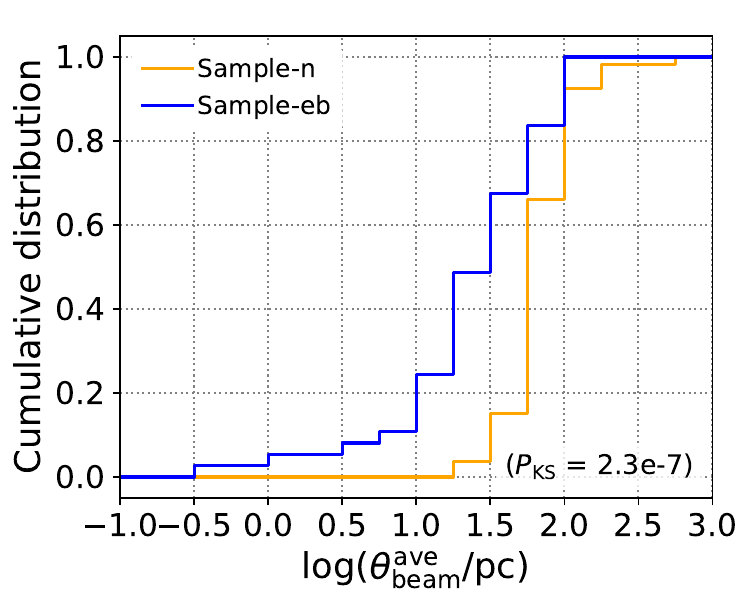}
    \caption{
    Left: cumulative detection-significance distribution of the objects showing both or one of extended (e) or blob-like (b) components (blue) and the others without such non-nuclear components (yellow). 
    The $p$-value obtained from the KS-test is indicated in the parenthesis. The small $p$-value suggests the significant difference in the distributions and the trend that the objects without non-nuclear components were detected at lower significances. 
    Right: same as the left, but for the beam size in units of pc. 
    The distributions and $p$-value suggest the trend that 
    the objects without non-nuclear components were observed at coarser physical resolutions. 
    }\label{fig:cd}
\end{figure*}    

We quantitatively identified non-nuclear components that extended from central un-resolved sources by deriving radial surface brightness profiles. 
Specifically, we defined concentric annuli centered on the nuclear emission peak. The radial width of each annulus was set to half of the average beam size of $\theta^{\rm ave}_{\rm beam}$ = $(\theta^{\rm min}_{\rm beam} \times \theta^{\rm maj}_{\rm beam})^{1/2}$, and the outer radius was set to $15\times\theta^{\rm ave}_{\rm beam}$. 
We then derived the averaged surface brightnesses ($S^{\rm ave}_{\nu, \rm mm}$ in units of Jy per beam) for all annuli using the \texttt{imstat} task. To confirm whether mm-wave emission is an excess over a central un-resolved source that extends by the point spread function (PSF), we also calculated the surface brightnesses of the achieved synthesized beam for the same annuli. Then, for comparison, the brightness in the innermost circle was adjusted to the observed one. 
As examples, Figure~\ref{fig:ext_rad} shows observed radial profiles and adjusted synthesized-beam ones for the four objects, mentioned previously (i.e., NGC\,526A, NGC\,3281, NGC\,3393, and IC\,5063). The profiles for all 98 AGNs are displayed in Figure~\ref{fig_app:rads} in Appendix. 
For NGC\,526A, no significant excess is found, consistent with the visual inspection; non-nuclear components do not exist. In contrast, the profile for NGC 3281 infers the presence of significant extended emission with respect to the un-resolved component above three times the standard errors (SEs). We used the obtained radial profiles for identifying such emission extending from the central un-resolved component, and labeled AGNs for which significant extended components were confirmed with ``{e}". 
Besides the extended emission, blob-like emission, which we define as emission separated from the central emission, can be seen in some objects. Clear examples are NGC 3393 and IC 5063, as seen in Figure~\ref{fig:rep_objs}. 
The blob-like features of IC 5063 are significantly 
detected also in its radial profile around the range $\theta^{\rm ave}_{\rm beam} \sim 4$--6 (Figure~\ref{fig:ext_rad}), while the blob-like component of NGC\,3393 becomes ambiguous in the radial profile around $\theta^{\rm ave}_{\rm beam} = 2$--4. This is likely due to noise in the same annulus, and indicates that, in the radial profile, we can easily miss significant blob-like components. 
We thus identified blob-like components in an alternative simple way; an image with 5$\sigma$ contours was created, and then any components that were isolated from the central emission enclosed by a $5\sigma_{\rm mm}$ contour and were significant above 5$\sigma_{\rm mm}$ were systematically regarded as blob-like components. We labeled AGNs for which blob-like components were found with ``{b}".

With the two flags, {e} and {b}, NGC 3281, NGC 3393, and IC 5063, shown in Figure~\ref{fig:rep_objs}, were labeled with {c}, {b}, and {b}, while no flag was raised for NGC\,526A. 
Of the 90 AGNs, 37 AGNs were quantitatively and systematically found to have non-nuclear emission, corresponding to $\approx$ 41\%. 
The flags are tabulated in Table~\ref{tab_app:mm_prop}.

We warn readers that structures larger than an MRS are resolved out and therefore there should be non-nuclear emission that we missed. Indeed, we confirmed that the size of each identified non-nuclear component is less than the MRS of the data used for the identification.

While the identification of the non-nuclear components depends on whether there are actually non-nuclear components and also on the MRS as described above, some other observational properties can be factors that affect the identifiability. 
As possible properties, we investigated two parameters: the detection significance for peak emission and the physical resolution. 
The former was considered because we have defined peak emission as the nuclear component, and therefore non-nuclear emission around that (if any) is basically fainter and would not be detected for sources detected at low significances. The latter was based on the simple fact that a coarse resolution makes it difficult to resolve emission components. 
To assess these hypotheses, we created two samples; one consists of the 53 AGNs without any morphological flags and the other includes the other 37 AGNs with significant
non-nuclear components. We refer to these samples as Sample-n and Sample-eb, respectively. 
Using the Kolmogorov-Smirnov-test (KS-test), we then compared their detection-significance and beam-size distributions, as shown in Figure~\ref{fig:cd}. 
The KS-test rejects their similarities with $p$-values of $P_{\rm KS} \approx 7.9\times10^{-4}$ and $P_{\rm KS} \approx 2.3\times10^{-7}$, respectively. Also, judging from the distributions in Figure~\ref{fig:cd}, we confirmed that the Sample-n objects were detected at lower significances and were observed at coarser physical resolutions, as expected.

\subsection{Mm-wave spectral indices}

\subsubsection{Caution on spectral indices}

Based on the ALMA data used for the imaging, we estimated spectral indices ($\alpha_{\rm mm}$, defined as $S^{\rm peak}_{\nu, \rm mm} \propto \nu^{-\alpha_{\rm mm}}$). The data are optimized to produce high-resolution images, but are not well suited for estimating the indices, for example, due to the narrow aggregate bandwidth ($<$ 4\,GHz). Therefore, we emphasize that there is a large uncertainty in each index, especially in the case where the signal-to-noise ratio of the nuclear emission is low ($\lesssim$ 50). 

\subsubsection{Actual Derivation}

The spectral index is necessary to derive the fundamental parameters of luminosity and flux.
The $k$-corrected luminosity can be derived using the equation below \citep[e.g.,][]{Nov17}:
\begin{equation}
    \nu L^{\rm peak}_{\nu, \rm mm} = \nu \frac{4\pi D^2}{(1+z)^{1-\alpha_{\rm mm}}}  \left(\frac{\nu}{\nu'}\right)^{-\alpha_{\rm mm}} S^{\rm peak}_{\nu', \rm mm},
\end{equation}
where $\nu$ is set to 230\,GHz, i.e., the representative frequency considered in this study, and $S^{\rm peak}_{\nu', \rm mm}$ is the peak flux density per beam at the observed frequency of $\nu'$. Each flux density per beam was derived from an image consisting of all available spectral windows. 
By following the definition of the luminosity, the flux is defined as follows:
\begin{equation}
    \nu F^{\rm peak}_{\nu, \rm mm} = \nu \frac{1}{(1+z)^{1-\alpha_{\rm mm}}}  \left(\frac{\nu}{\nu'}\right)^{-\alpha_{\rm mm}} S^{\rm peak}_{\nu', \rm mm}.
\end{equation}
While the effect of the $k$-correction 
is $\approx$ 0.8\% as the median value for the sample, 
that can be $\approx$ 5--10\% for six distant objects. This is comparable to the systematic uncertainty of $\sim$ 10\%, and therefore the $k$-correction was applied uniformly.

To obtain the spectral index of $\alpha_{\rm mm}$ and its error of $\Delta\alpha_{\rm mm}$ necessary to calculate the luminosity and flux and their errors, we fitted a power-law function to the flux measurements from individual spectral windows based on the chi-square method. 
For 70 objects that were observed with more than two spectral windows and for which peak emission was detected in all windows, spectral indices and statistical errors were estimated. 
To the errors, we added a systematic uncertainty of 0.2, which takes into account a possible flux calibration uncertainty between spectral windows at $\sim$ 230\,GHz \citep{Fra20}.
The mean and standard dispersion of the indices of 68 sources 
whose indices were individually calculated are $\alpha_{\rm mm} = 0.5\pm1.2$. 
Two objects, NGC\,1068 and NGC\,1194, were excluded from this calculation 
because their constraints could be unreliable, which is described in the paragraph after the next two ones; the actual indices assigned for the two objects to calculate $L^{\rm peak}_{\nu, \rm mm}$ and $F^{\rm peak}_{\nu, \rm mm}$ are also detailed therein. 
For the remaining 28 objects, the representative value of $\alpha_{\rm mm} = 0.5\pm1.2$ was adopted to calculate their luminosities and fluxes. 

\begin{figure}
    \centering  
    \includegraphics[width=8.6cm]{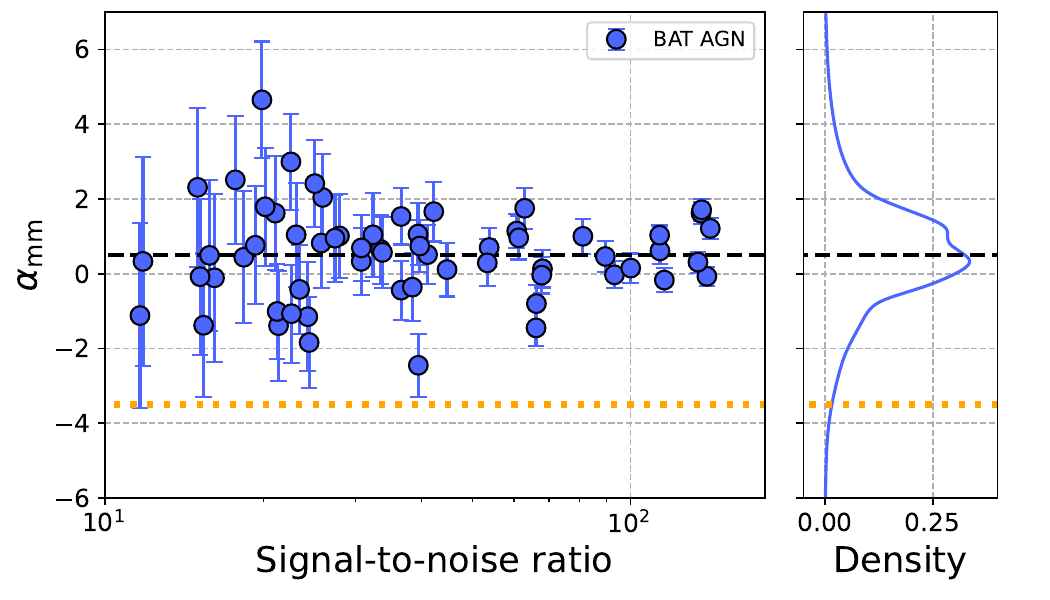}
    \caption{
    Spectral index as a function of signal-to-noise ratio of the nuclear emission (left panel) and the density derived by the kernel density estimation (right panel). 
    The black dashed lines correspond to 0.5, the mean for the BAT AGNs, while the yellow dotted lines indicate $\alpha_{\rm mm} = -3.5$, expected for the dust thermal emission. 
    }\label{fig:index_vs_sig}
\end{figure}

We emphasize here that many objects, especially those with low detection significance, have large uncertainties of $\geq$ 1 in their indices; the corresponding fraction is $\sim$ 40\%. 
This is seen in Figure~\ref{fig:index_vs_sig}, where the indices are plotted as a function of signal-to-noise ratio of the nuclear mm-wave emission. 
Therefore, we caution that many of the estimates are not so reliable, and that additional data are important for more reliable estimates.

In addition to showing the above fact, Figure~\ref{fig:index_vs_sig} suggests an interesting result that the indices are well distributed around the mean value of $\approx$ 0.5.
This suggests that the dust emission, for which a steep slope is expected (e.g., $\sim -3\sim-4$), is 
unlikely to be the dominant source, and flatter components, such as synchrotron and free-free emission, 
are likely to be more important.

Returning to NGC 1068, we find that its spectral index ($\alpha_{\rm mm} = 1.5\pm0.3$ around 270\,GHz) is largely different from an index of $\alpha_{\rm mm} \sim -1.3$ expected from a mm-wave band SED constructed by \cite{Ino20}, and also 
is a few sigma higher than the index more carefully derived by \cite{Michi23} (i.e., $\alpha_{\rm mm} \sim$ 0.5--0.8). The new study constructed an SED denser than that of \cite{Ino20} by using 
high-resolution ($\approx$ 0\farcs05) ALMA data. 
First, as a possible cause, we considered that this discrepancy might be ascribed to our data with a narrower spacing between the lowest and highest frequency spectral channels than those for the others (i.e., $\approx$ 5.4\,GHz with respect to $\sim$ 14--20\,GHz), except for NGC 1194. The object was also observed with a similar narrow spacing ($\approx$ 4.9\,GHz). To estimate the index better, we analyzed the second highest resolution data with $\theta^{\rm ave}_{\rm beam} \approx$ 0\farcs018 and a wider spacing of $\approx$ 19\,GHz. Then, 
a spectral index was constrained to be $\alpha_{\rm mm} =$ 0.4$\pm$0.2 around 260\,GHz. 
We additionally analyzed a Band-7 data with a comparable resolution ($\approx$ 0\farcs018) at 331\,GHz (group OUS ID = A001/X1465/X212e in 2019.1.00854.S). 
We then found that the inter-band spectral index between the Band-6 and Band-7 frequencies (256\,GHz and 331\,GHz) also inferred $\alpha_{\rm mm} \approx$ 0.4. 
The two measurements disfavor the result of \cite{Ino20} but are virtually consistent with the updated one of \cite{Michi23}. 
Thus, the true spectral index is possibly $\sim 0.4$, and it is suggested that the spectral index derived from the narrow spacing has a large systematic uncertainty.
Eventually, for NGC 1068, we conservatively adopted the representative value of $\alpha_{\rm mm} = 0.5\pm1.2$, consistent with the possible value of 0.4. 
For the remaining object NGC\,1194, as this did not have other sub-arcsec ALMA data, we also adopted the same representative value.

In addition to the narrow spectral coverage suspected as a cause of the unreliable constraint on the spectral index, we checked whether there were objects whose estimated indices could be unreliable due to observations with smaller aggregate bandwidths of $<$ 3.75\,GHz. As a result, it was confirmed that no objects were observed at such small aggregate bandwidths and have estimated indices.

Lastly, we attempted to improve the constraints on the spectral indices with large uncertainties of $\Delta \alpha_{\rm mm} \geq 1$.
For the 27 objects with $\Delta \alpha_{\rm mm} \geq 1$, we searched for different subarcsec-resolution Band-6 data, and found that only NGC\,4388 had available data. 
With the data, the spectral index was constrained to be $\alpha_{\rm mm} = -0.5\pm0.8$ around 220\,GHz, while that obtained with the highest resolution data was $\alpha_{\rm mm} = 1.0\pm1.0$ at $\sim$260\,GHz. 
This apparently large difference could be ascribed, in part, to a synchrotron self-absorption component having a peak around the frequencies, as suggested for nearby AGNs \citep[e.g.,][]{Ino18,Ino20}. Anyway, the derived luminosities and fluxes from the two different data 
differ by only $\approx$ 0.1 dex and are consistent within 1$\sigma$. 

With the indices estimated from the highest-resolution data, we calculated the luminosities and fluxes. To their errors, we added a systematic uncertainty of 10\%, as recommended by the ALMA technical handbook. Table~\ref{tab_app:mm_prop} in Appendix~\ref{sec_app:tab} provides the mm-wave properties obtained for all objects in our sample (e.g., spectral indices, luminosities, and fluxes).

\section{Non-nuclear Emission}\label{sec:ext}

Although the main purpose of this paper is to provide a catalog of the mm-wave properties derived in Paper\,I, we also discuss the origins of non-nuclear components found for some objects to obtain insights into the non-nuclear emission. This was omitted in Paper\,I to focus on discussing un-resolved nuclear mm-wave components. 
Of the 37 objects having morphology flags (i.e., e and b), for further discussions, we select 28 objects that showed largely extended non-nuclear components, or quantitatively those with the number of annuli where  
emission was detected being more than two and/or with blob-like components. Here, we exclude three objects with the b flag (NGC 424, Ark 120, and H 0557$-$385), as their blob-like components were significant in only a few pixels and it is ambiguous whether they are real emission or not. 

In Appendix~\ref{sec_app:notes}, we present our discussions on all the selected objects, and here only summarize our findings to avoid lengthy descriptions in the main text. Eventually, we have found that non-nuclear components in the investigated objects are related to various mechanisms and regions, such as a jet and radio lobes (e.g., NGC 3393 and IC 5063), a secondary AGN (NGC~6240), stellar clusters (NGC 1365), a narrow line region (e.g., IC 4518A), a galaxy disk (e.g., NGC 3281, NGC 4945 and NGC 5506), active SF (e.g., NGC 7130 and NGC 7469), and AGN-driven outflows (NGC 7582). Also, we have found that some objects (e.g., NGC 612, NGC 1566, IRAS 05189-2524, and NGC 3281) show components whose physical origins are unclear. In any case, further high-resolution observations are desired, perhaps unveiling  unknown mm-wave emitting mechanisms. 

\section{Conclusions}\label{sec:sum}

This paper provides a catalog of the properties of mm-wave continuum emission for the 98 nearby AGNs ($z <$ 0.05) selected from the 70-month Swift/BAT hard X-ray catalog \citep{bau13,ric17c}. 
Thanks to the hard-X-ray selection, the sample is almost unbiased for obscured systems up to the Compton-thick level absorption ($N_{\rm H} \sim 10^{24}$ cm$^{-2}$). 
The mm-wave properties were constrained by a systematic analysis of the ALMA band-6 (211--275\,GHz) data taken at high resolutions of $\lesssim$ 0\farcs6, corresponding to $\lesssim$ 100--200\,pc for almost all targets. 
Tables~\ref{tab_app:sample} and \ref{tab_app:alma_data} summarize the basic properties of the objects (e.g., names, distances, and Seyfert types) and the ALMA data used (e.g., achieved resolutions), respectively. Then, Table~\ref{tab_app:mm_prop} lists the derived mm-wave properties.
As Paper~I found that the peak mm-wave emission may generally be produced by an AGN, this catalog could be used for various AGN studies, such as a study of AGN SEDs, including mm-wave emission, which has often been ignored in past studies. Furthermore, high-resolution mm-wave images are presented in Figure~\ref{fig_app:images}. 
Also, we derived the radial profiles of observed surface brightnesses from the images, as summarized in Figure~\ref{fig_app:rads}, and we systematically identified significant non-nuclear emission extending from the central source in Section~\ref{sec:quant_id}. 
In addition, we systematically searched the original images for isolated blob-like mm-wave components. 
We then labeled AGNs that showed extended and blob-like components with ``e" and ``b", respectively. 
Among the 90 AGNs with significant detections of nuclear mm-wave emission, both or one of extended (e) or blob-like (b) components were identified for 37 AGNs (i.e., $\approx$ 41\%). 
In particular, for 28 AGNs that showed largely extended mm-wave components and/or blob-like components, we have briefly discussed them, while referring to the existing literature in Section~\ref{sec_app:notes}. 
Various mechanisms and regions (e.g., a jet and radio lobes, a secondary AGN, stellar clusters, a narrow-line region, a galaxy disk, active SF, and AGN-driven outflows) were eventually inferred, while some mm-wave components had unknown origins. To better understand the origin of these components, further investigations of the individual objects are encouraged.

\begin{acknowledgments} 
We thank the reviewer for the useful comments, which helped us improve the quality of the manuscript. 
We acknowledge support from FONDECYT Postdoctral Fellowships 3200470 (T.K.), 3210157 (A.R.) and 3220516 (M.J.T.), FONDECYT Iniciacion grant 11190831 (C.R.), FONDECYT Regular 1200495 and 1190818 (F.E.B.), ANID BASAL project FB210003 (C.R., F.E.B) and Millennium Science Initiative Program  – ICN12\_009 (F.E.B). T.K., T.I. and M.I. are supported by JSPS KAKENHI grant numbers JP20K14529/JP23K13153, JP20K14531, and JP21K03632, respectively.
B.T. acknowledges support from the European Research Council (ERC) under the European Union's Horizon 2020 research and innovation program (grant agreement 950533) and from the Israel Science Foundation (grant 1849/19).
K.O. acknowledges support from the National Research Foundation of Korea (NRF-2020R1C1C1005462).
M.B. acknowledges support from the YCAA Prize Postdoctoral Fellowship.
The scientific results reported in this article are based on data obtained from the Chandra Data Archive. This research has used software provided by the Chandra X-ray Center (CXC) in the application packages CIAO. 
This paper uses the following ALMA data: 
ADS/JAO.ALMA
\#2012.1.00139.S,
\#2012.1.00474.S,
\#2013.1.00525.S,
\#2013.1.00623.S,
\#2013.1.01161.S,
\#2015.1.00086.S,
\#2015.1.00370.S,
\#2015.1.00597.S,
\#2015.1.00872.S,
\#2016.1.00254.S,
\#2016.1.00839.S,
\#2016.1.01140.S,
\#2016.1.01279.S,
\#2016.1.01385.S,
\#2016.1.01553.S,
\#2017.1.00236.S,
\#2017.1.00255.S,
\#2017.1.00395.S,
\#2017.1.00598.S,
\#2017.1.00886.L,
\#2017.1.00904.S,
\#2017.1.01439.S,
\#2018.1.00006.S,
\#2018.1.00037.S,
\#2018.1.00211.S,
\#2018.1.00248.S,
\#2018.1.00538.S,
\#2018.1.00576.S,
\#2018.1.00581.S,
\#2018.1.00657.S,
\#2018.1.00978.S, and  
\#2019.1.01742.S. 
ALMA is a partnership of ESO (representing its member states), NSF (USA) and NINS (Japan), together with NRC (Canada), MOST and ASIAA (Taiwan), and KASI (Republic of Korea), in cooperation with the Republic of Chile. The Joint ALMA Observatory is operated by ESO, AUI/NRAO and NAOJ.
Data analysis was in part carried out on the Multi-wavelength Data Analysis System operated by the Astronomy Data Center (ADC), National Astronomical Observatory of Japan.
This publication makes use of data products from the Wide-field Infrared Survey Explorer, which is a joint project of the University of California, Los Angeles, and the Jet Propulsion Laboratory/California Institute of Technology, funded by the National Aeronautics and Space Administration.
This research has made use of the NASA/IPAC Extragalactic Database, which is funded by the National Aeronautics and Space Administration and operated by the California Institute of Technology.
This research has made use of the SIMBAD database, operated at CDS, Strasbourg, France. 
We acknowledge the usage of the HyperLeda database (http://leda.univ-lyon1.fr).

\end{acknowledgments}

\bibliography{ref.bib}

\appendix
\setcounter{table}{0}
\setcounter{section}{0}
\setcounter{figure}{0}
\renewcommand{\thetable}{\Alph{table}}
\renewcommand{\thefigure}{\Alph{figure}}

\section{Complete Tables}\label{sec_app:tab}

In this section, we present the complete tables of the basic properties of our 98 AGNs (Table~\ref{tab_app:sample}), the ALMA data we used (Table~\ref{tab_app:alma_data}), 
and the mm-wave properties (Table~\ref{tab_app:mm_prop}). 

\begin{longtable*}[c]{ccccccccccc}
\caption{Basic Properties of Our Targets\label{tab_app:sample}}\\
\multicolumn{11}{c}{\parbox{18cm}{
Notes. 
(1) Identification number in this paper (Paper~II) and Paper~I. 
(2) BAT index. 
(3,4) BAT and counterpart names, taken from \cite{ric17c}.  
(5,6) Right ascension and declination in units of degrees in the ICRS frame.
(7) Reference for the position in (5,6): 
A = ALMA (this work), W = ALLWISE \citep{Wri00,Mai11}, and
CXO = Chandra (this work). 
(8) Redshift.
(9,10) Distance and its error in units of Mpc. 
Particularly for objects at distances below 50\,Mpc, redshift-independent distances are adopted by referring to the Extragalactic Distance Database of 
\cite{Tul09}, a catalog of the Cosmicflows project \citep{Cou17}, and NASA/IPAC Extragalactic Database in this order \citep[see][for details]{Kos22_catalog}. 
Also, we note that (a) indicates that a small redshift error of the order of 10 km s$^{-1}$, used in computing the distance error, results in a very small error of $\ll$ 0.1\,Mpc and thus 0.1\,Mpc is denoted as a substitute. 
(11) Seyfert type. 
Information in the last four columns is taken from the BASS DR2 paper of \cite{Kos22_catalog} or the authors. 
This table is published in its entirety in machine-readable format. A portion of the table is shown here for guidance on its form and content.
}}\vspace{0.2cm}\\
\hline
(1) & (2) & (3) & (4) & (5) & (6)  & (7) & (8) & (9) & (10) & (11)\\ 
N & Index & BAT Name & Counterpart Name & R.A. & Decl. & Ref. & $z$ & $D$ & $\Delta D$  & Type \\
  &       &          &                  & (\arcdeg) & (\arcdeg) & & & (Mpc) & (Mpc) & \\
\hline \endfirsthead \hline
(1) & (2) & (3) & (4) & (5) & (6)  & (7) & (8) & (9) & (10) & (11)\\ 
N & Index & BAT name & Counterpart Name & R.A. & Decl. & Ref. & $z$ & $D$ & $\Delta D$  & Type \\
&       &          &                  & (\arcdeg) & (\arcdeg) & & & (Mpc) & (Mpc) & \\
\hline \endhead \hline \endfoot \hline\hline \endlastfoot
\input{table4sample} 
\end{longtable*}

\begin{longtable*}{cccccccccccc}
\caption{Basic Information of ALMA Data\label{tab_app:alma_data}}\\
\multicolumn{12}{c}{\parbox{18cm}{
Notes. 
(1) Identification number in this paper (Paper~II) and Paper~I. 
(2) ALMA project code.
(3) Member OUS ID. 
(4) Observation mode (mosaic or not). 
(5) Maximum recoverable scale in units of arcsec, calculated from the calibrated data. 
(6) Field of view in units of arcsec. 
(7) Exposure time in units of sec. 
(8) Observation date. 
Information in (6,7,8) is taken from the ALMA archive database. 
(9,10) Beam sizes (FWHM) in units of arcsec along the major and minor axes. These are estimated from a reconstructed ALMA image for each target. 
(11,12) Physical resolutions achieved in units of pc. 
This table is published in its entirety in machine-readable format. A portion of the table is shown here for guidance on its form and content.
}}\vspace{0.2cm}\\
\hline
(1) & (2) & (3) & (4) & (5) & (6)  & (7) & (8) & (9) & (10) & (11) & (12) \\ 
N & Project code & OUS ID & Mosaic & MRS & FoV & Exp. & Date & 
$\theta^{\rm maj}_{\rm beam}$  &
$\theta^{\rm min}_{\rm beam}$  & 
$\theta^{\rm maj}_{\rm beam,pc}$  & 
$\theta^{\rm min}_{\rm beam,pc}$  \\
 & & & & (\arcsec) & (\arcsec) & (sec)  &  & (\arcsec) & (\arcsec) & (pc) & (pc) \\ 
\hline \endfirsthead \hline
(1) & (2) & (3) & (4) & (5) & (6)  & (7) & (8) & (9) & (10) & (11) & (12) \\ 
N & Project code & OUS ID & Mosaic & MRS &  FoV  & Exp. & Date & 
$\theta^{\rm maj}_{\rm beam}$  &
$\theta^{\rm min}_{\rm beam}$  & 
$\theta^{\rm maj}_{\rm beam,pc}$  & 
$\theta^{\rm min}_{\rm beam,pc}$  \\
& & & & (\arcsec) & (\arcsec) & (sec)  &  & (\arcsec) & (\arcsec) & (pc) & (pc) \\ 
\hline \endhead \hline \endfoot \hline\hline \endlastfoot
\input{table4almadata} 
\end{longtable*}

\begin{longrotatetable}
\movetabledown=12mm
\begin{deluxetable*}{ccccccccccccccc}
\tablecaption{Complete List of Mm-wave Properties\label{tab_app:mm_prop} \\} 
\tablehead{
(1) & (2) & (3) & (4) & (5) & (6)  & (7) & (8) & (9) & (10)
& (11)  & (12) & (13) & (14) & (15)   \\ 
N & Det./SpN & $\nu_{\rm mm}/\nu^{\rm ent}_{\rm mm}$ & 
$W_{\rm mm}/W^{\rm ent}_{\rm mm}$ & 
$S^{\rm peak}_{\nu,\rm mm}$ & 
$\sigma_{\rm mm}$ & 
$S^{\rm peak}_{\nu,\rm mm}/\sigma_{\rm mm}$ & 
$f_{\rm F}$ &
$\log \nu F^{\rm peak}_{\nu,{\rm mm}}$ & 
$\log \nu L^{\rm peak}_{\nu,{\rm mm}}$ & 
$\Delta\log \nu F^{\rm peak}_{\nu,{\rm mm}}$ &  
$\alpha_{\rm mm}$ & 
$\Delta\alpha_{\rm mm}$ & 
$f_{\rm M}$ & $N_{\rm M}$  \\
& & & & & & & $f_{\rm L}$ & & & 
$\Delta\log \nu L^{\rm peak}_{\nu,{\rm mm}}$ &  
& &  \\
& & ({\scriptsize GHz}) &  ({\scriptsize GHz}) & ({\scriptsize mJy beam$^{-1}$}) & ({\scriptsize mJy beam$^{-1}$}) & & & ({\scriptsize erg cm$^{-2}$ s$^{-1}$}) & ({\scriptsize erg s$^{-1}$}) 
& ({\scriptsize erg cm$^{-2}$ s$^{-1}$}) & & &  \\ 
& & & & & & & & & & ({\scriptsize erg s$^{-1}$}) & & & & 
}
\startdata
\input{table4alma} 
\enddata
\tablecomments{
(1) Identification number in this paper (Paper~II) and Paper~I. 
(2) The number of spectral windows where mm-wave emission was significantly detected and that of spectral windows used for continuum detections. 
(3) Observed central frequency of the collapsed spectral window after removing any emission line flux and central frequency for the 
entire data. 
(4) Aggregate bandwidth used for continuum detections and bandwidth for the entire data. 
(5) Peak flux density per beam in units of mJy beam$^{-1}$. 
(6) Noise level in units of mJy beam$^{-1}$. We note that this does not consider systematic uncertainty. 
We also note that the quantities in (5) and (6) were derived in the images after the primary-beam correction. 
(7) Significance of peak emission. 
(8) Flag for flux and luminosity. The $<$ flag indicates that flux and luminosity must be regarded as upper limits at the $5\sigma_{\rm mm}$ level.  
(9,10) Flux in units of erg cm$^{-2}$ s$^{-1}$ and luminosity in units of erg s$^{-1}$ on a log scale. 
(11) Uncertainties in the flux and luminosity. 
(12) Spectral index defined as $S^{\rm peak}_{\nu, \rm mm} \propto \nu^{-\alpha_{\rm mm}}$. 
(13) Error in the spectral index. 
(14) Flag indicating what morphological components were
identified in Section~\ref{sec:ext}: e = extended emission  and b = blob-like emission.
(15) Number of annuli where significant non-nuclear emission was detected. This table is published in its entirety in machine-readable format. A portion of the table is shown here for guidance on its form and content.
}
\end{deluxetable*}
\end{longrotatetable}

\clearpage 
\appendix

\section{ALMA Images and Radial Profiles}\label{sec_app:fig}

This section provides Band-6 ALMA images 
and radial profiles of mm-wave emission in Figures~\ref{fig_app:images} and \ref{fig_app:rads}, respectively. 


\begin{figure*}
    \centering
\includegraphics[width=5.9cm]{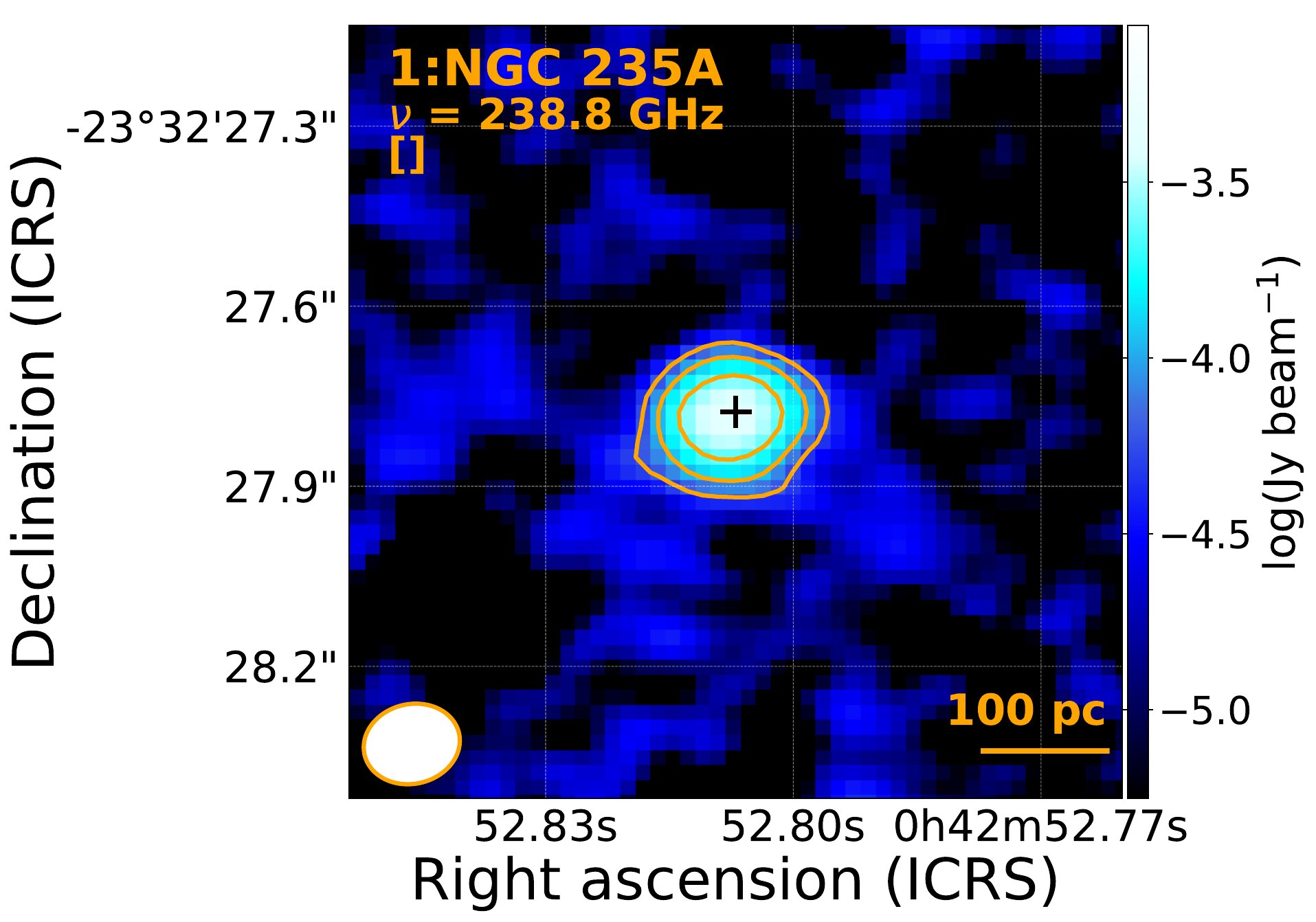}
\includegraphics[width=5.9cm]{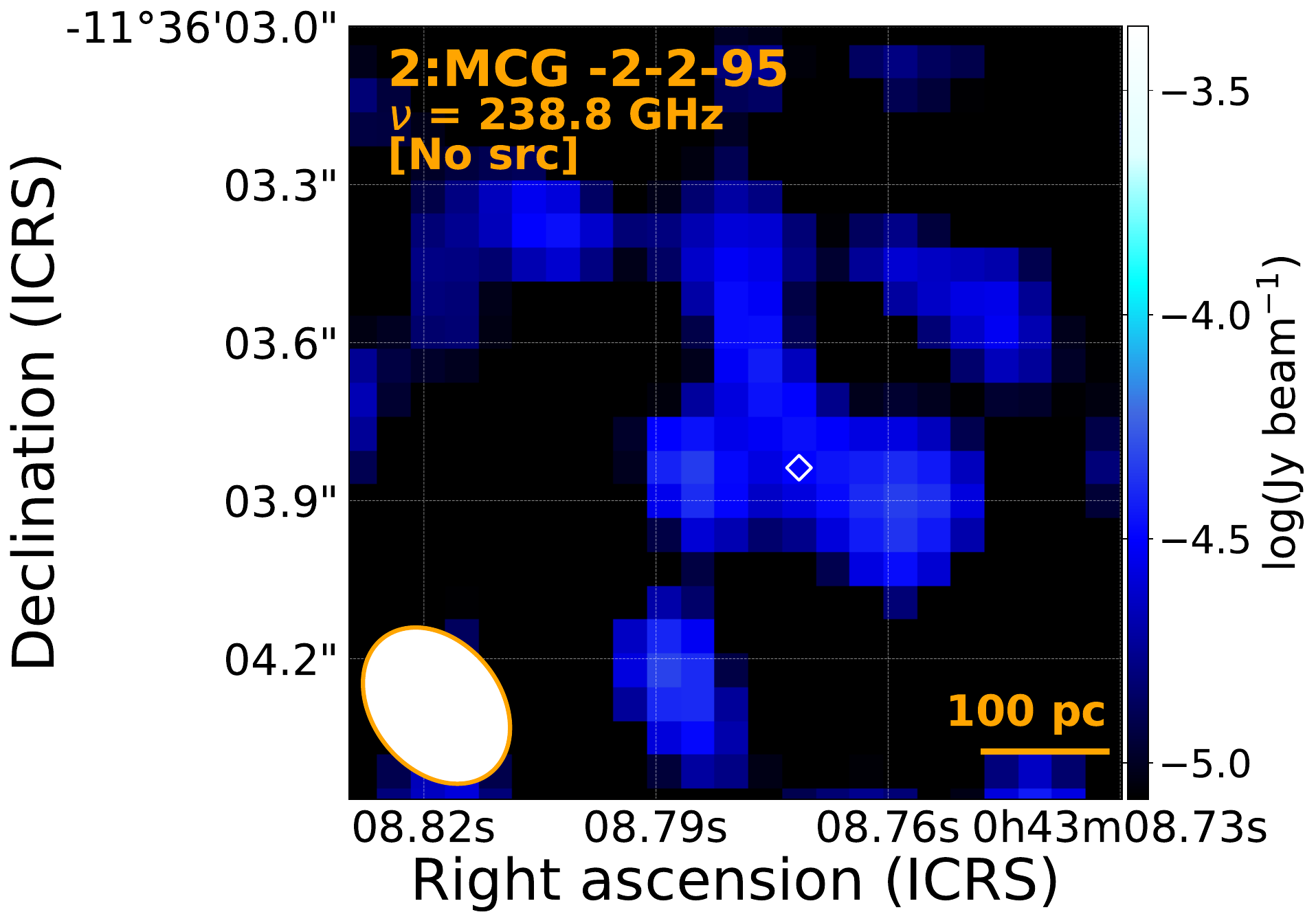}
\includegraphics[width=5.9cm]{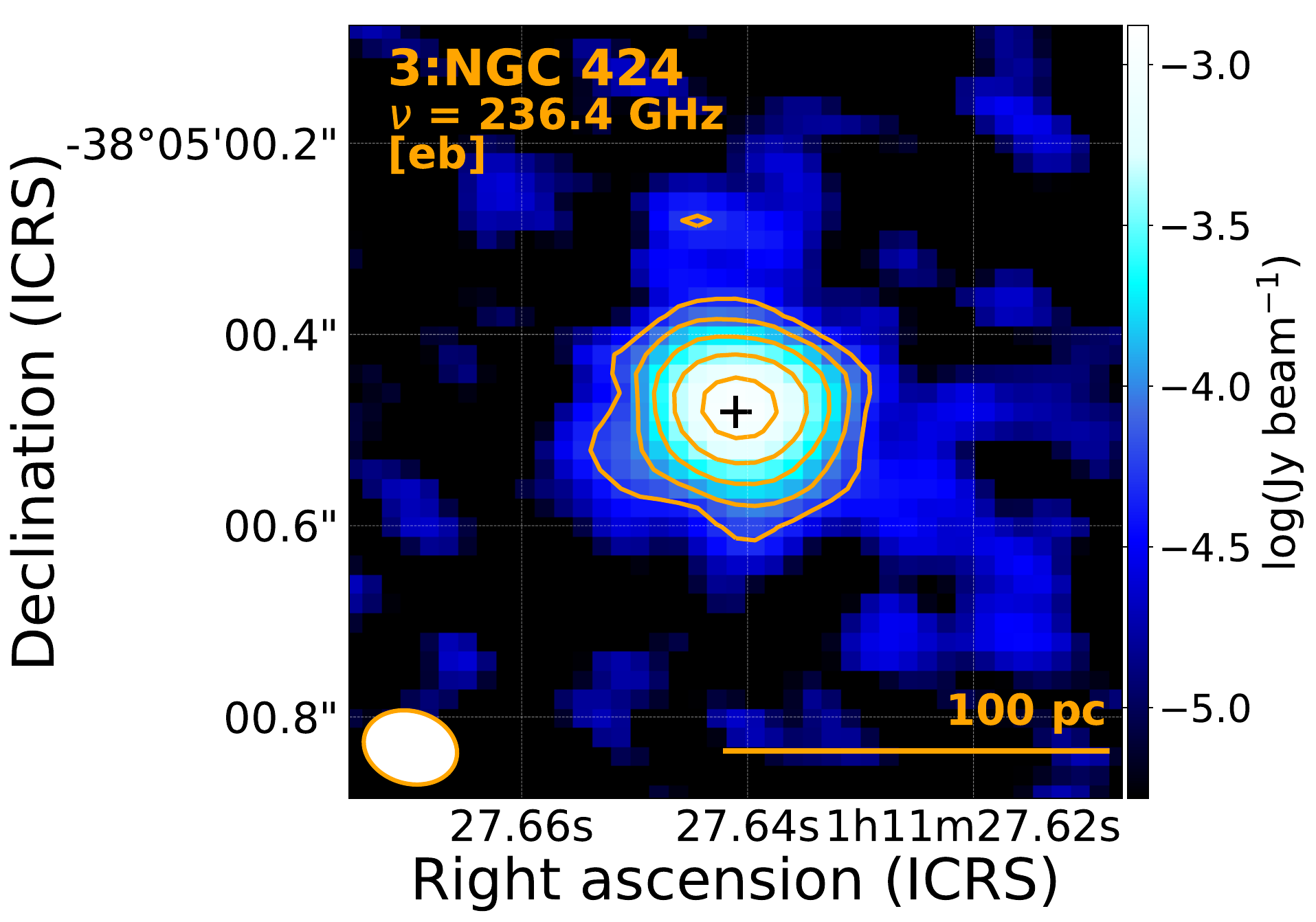}
\includegraphics[width=5.9cm]{004_NGC_526A_spwall_c_300pc.pdf}
\includegraphics[width=5.9cm]{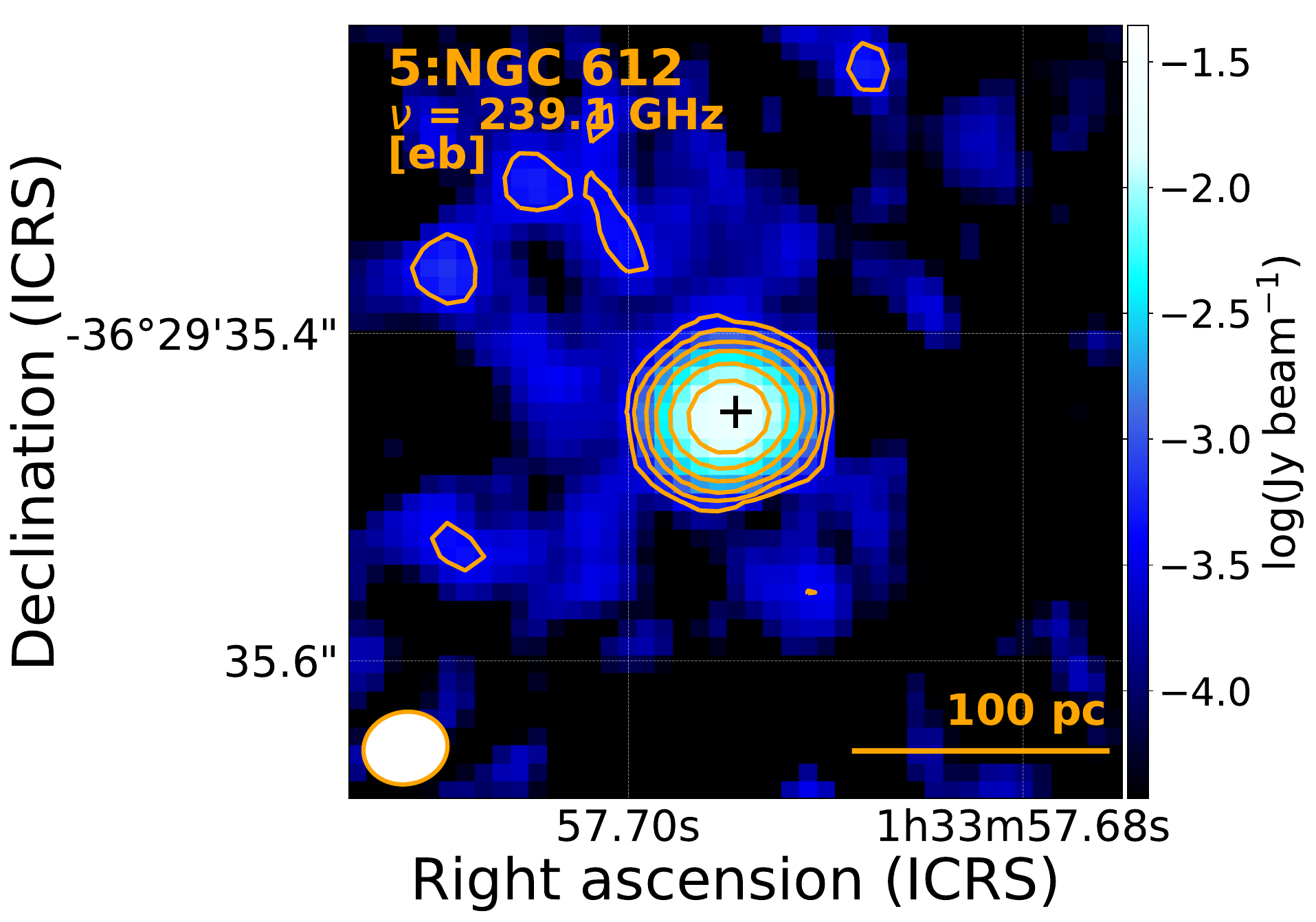}
\includegraphics[width=5.9cm]{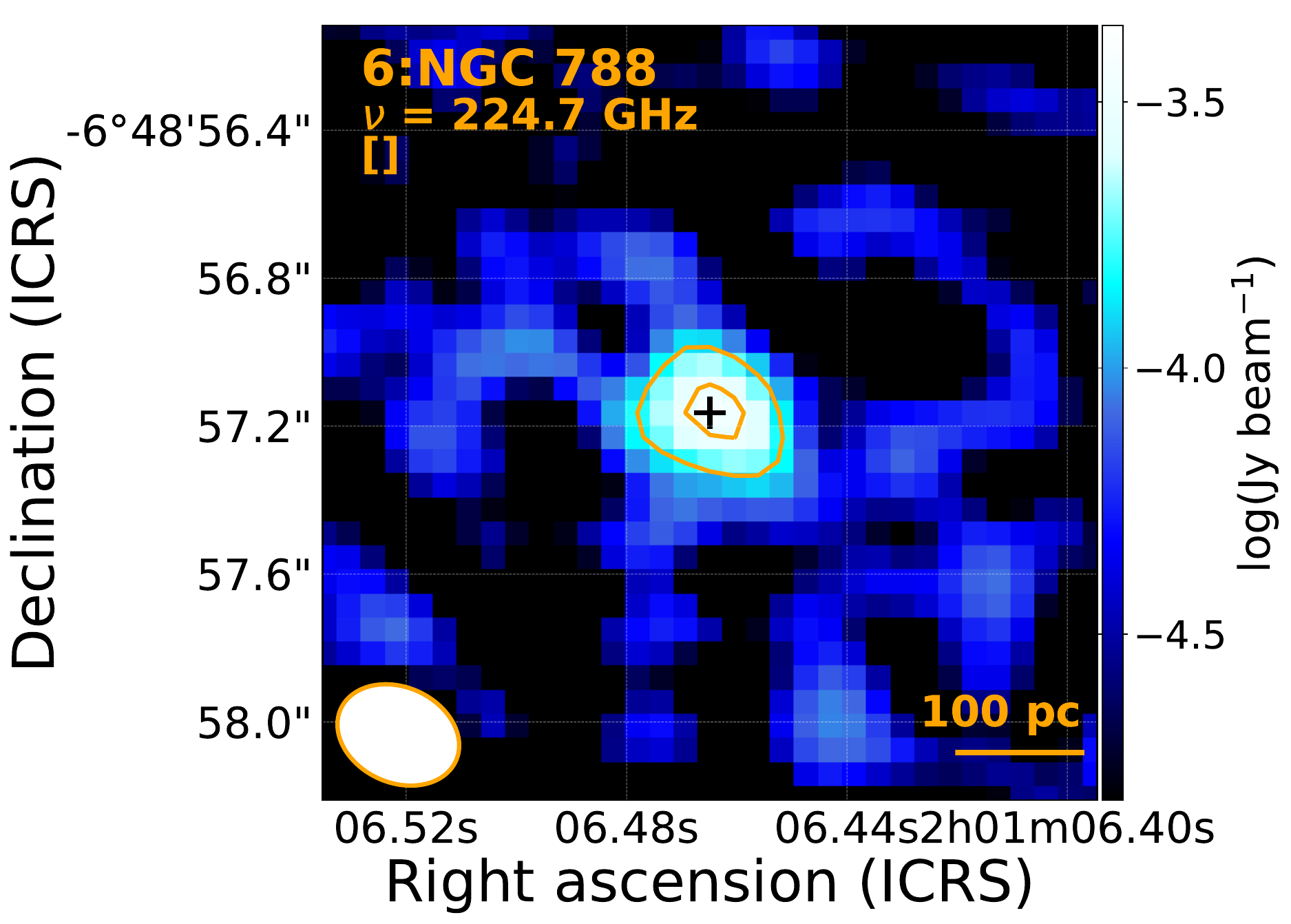}
\includegraphics[width=5.9cm]{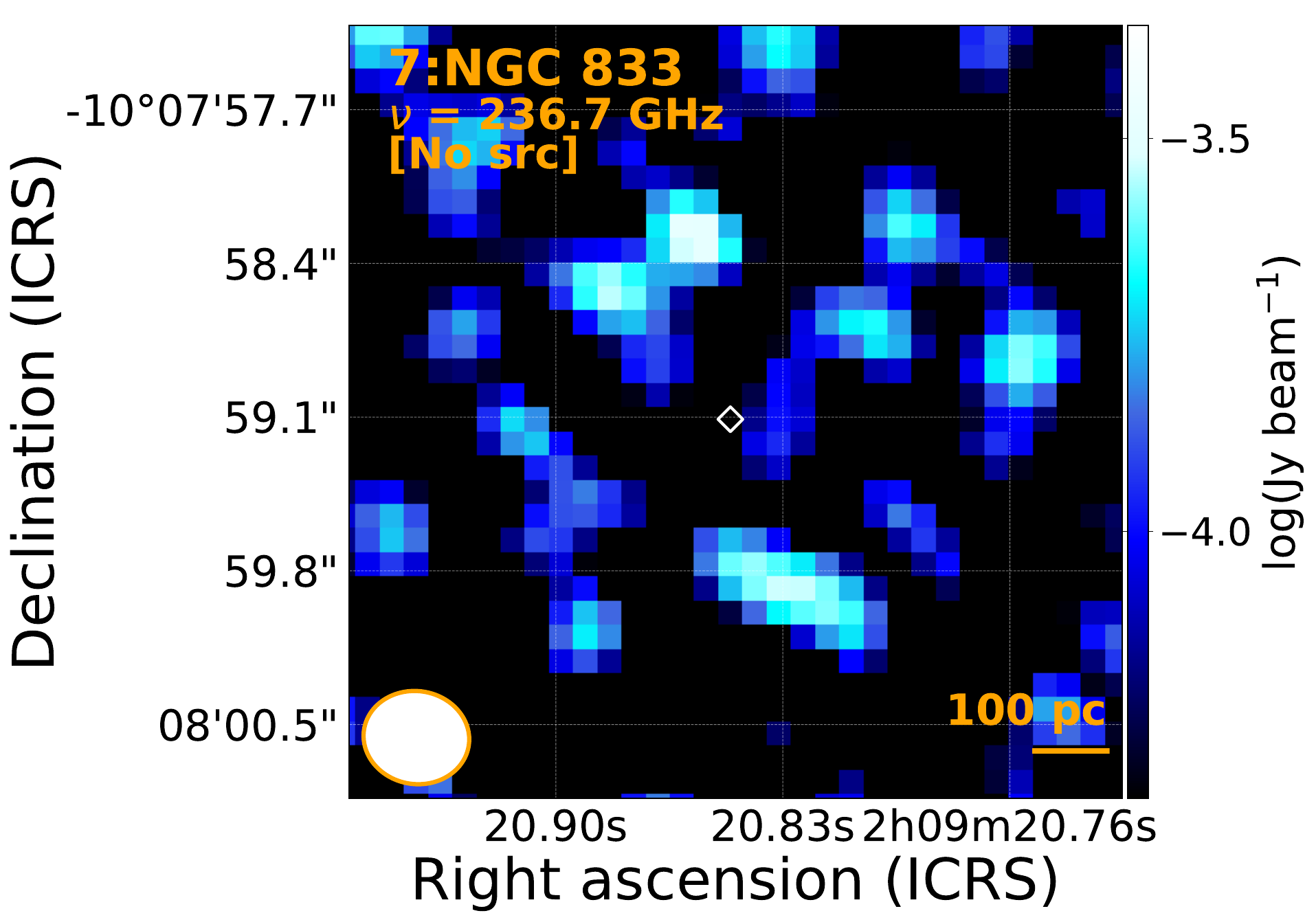}
\includegraphics[width=5.9cm]{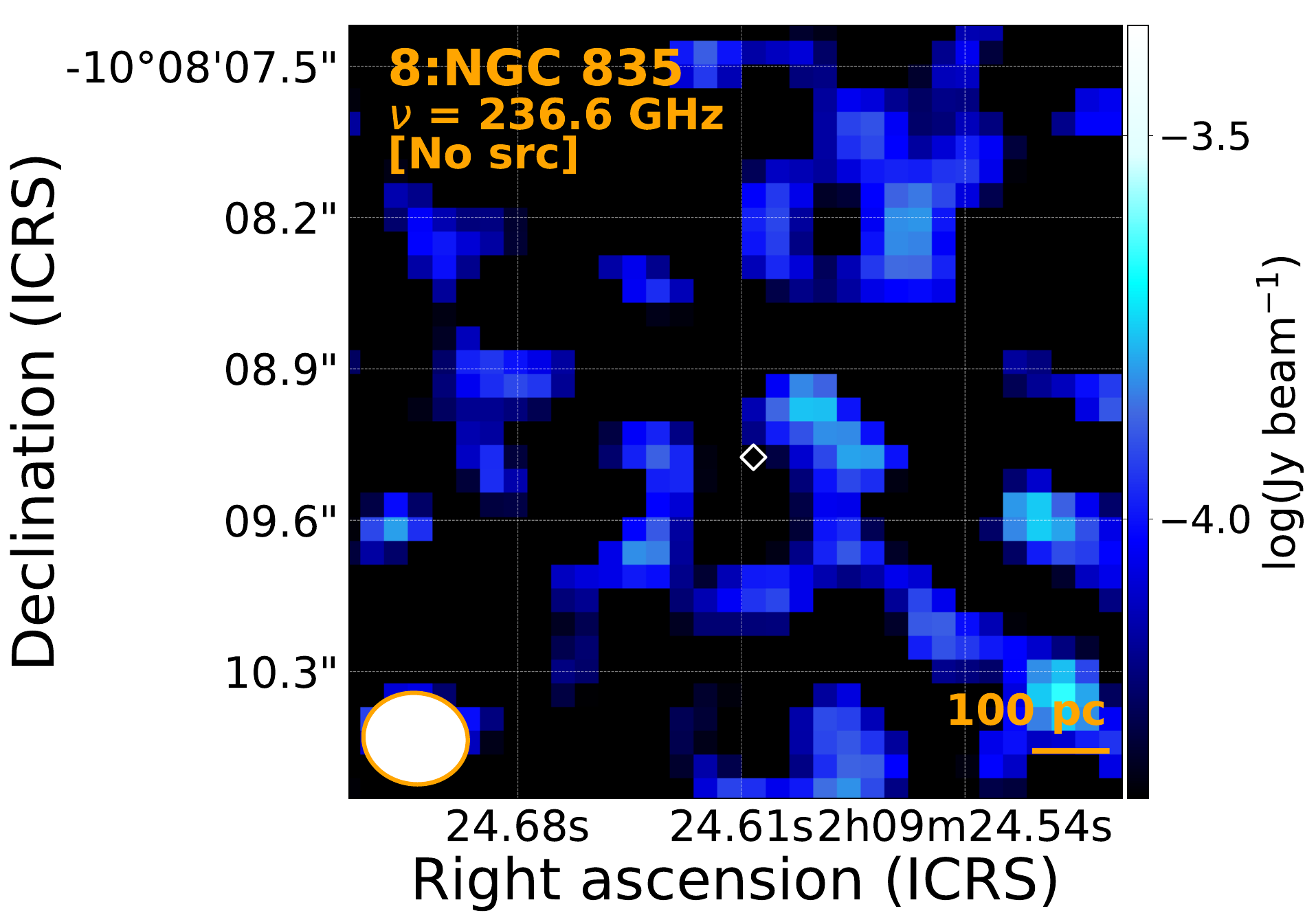}
\includegraphics[width=5.9cm]{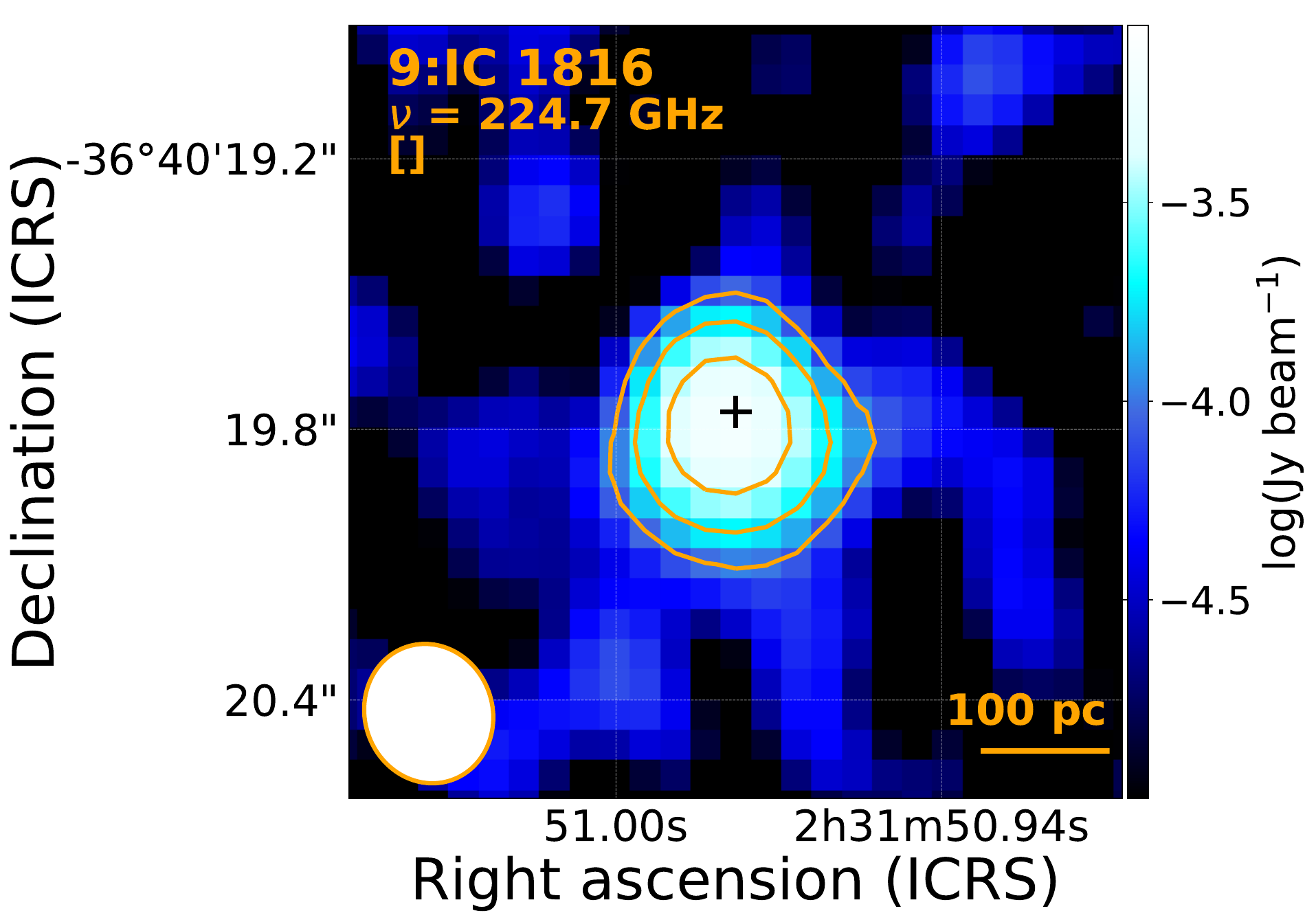}
\includegraphics[width=5.9cm]{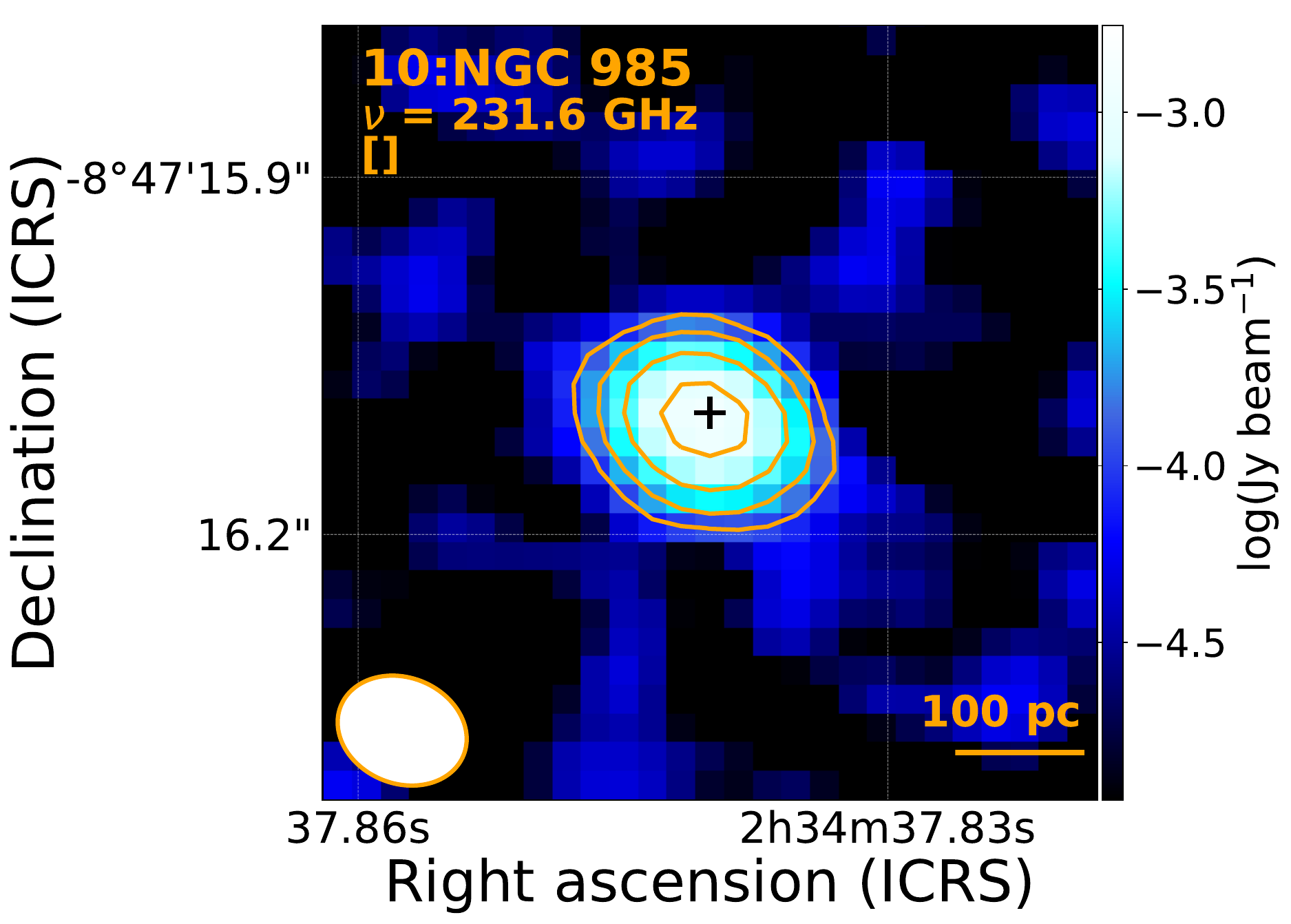}
\includegraphics[width=5.9cm]{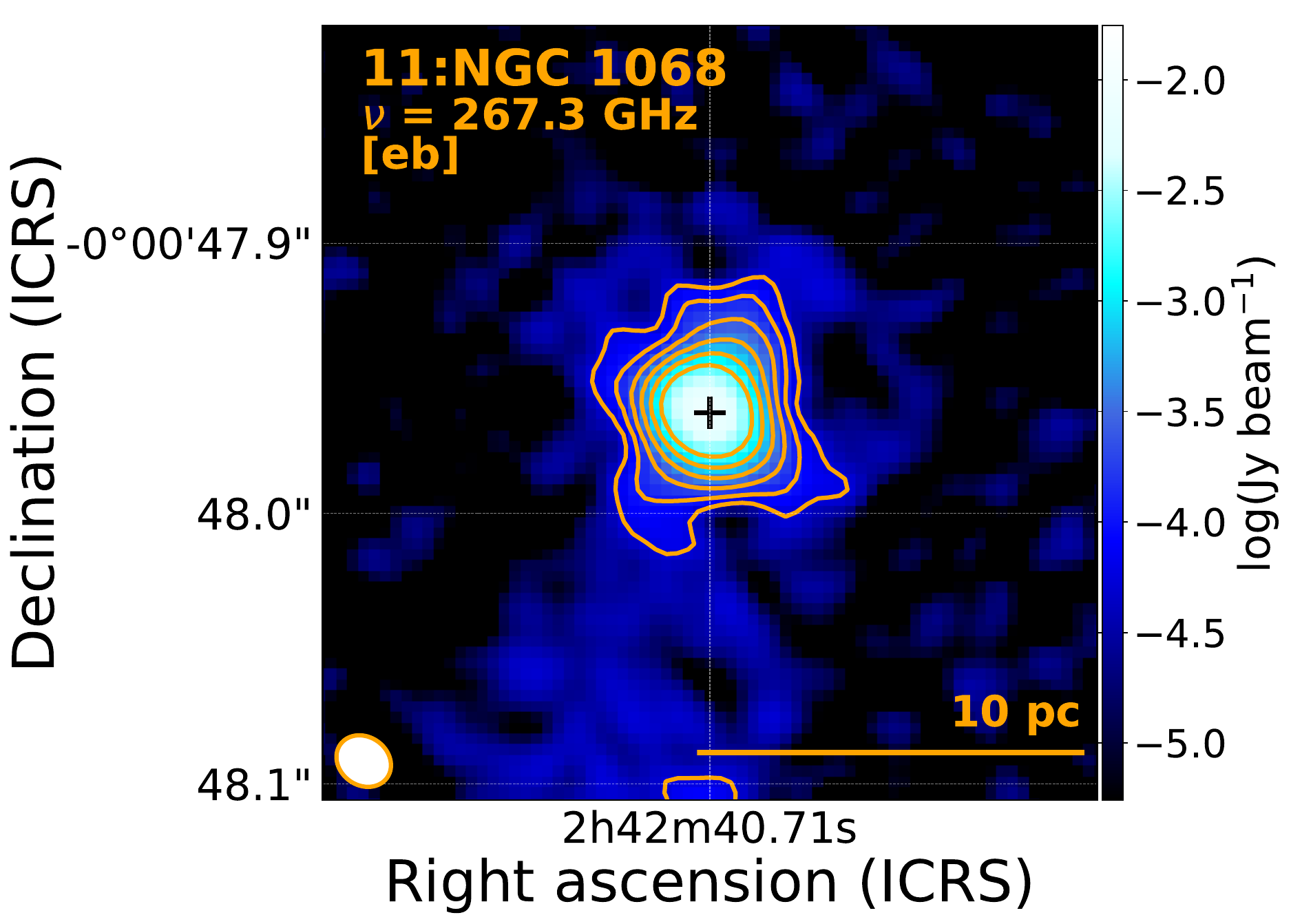}
\includegraphics[width=5.9cm]{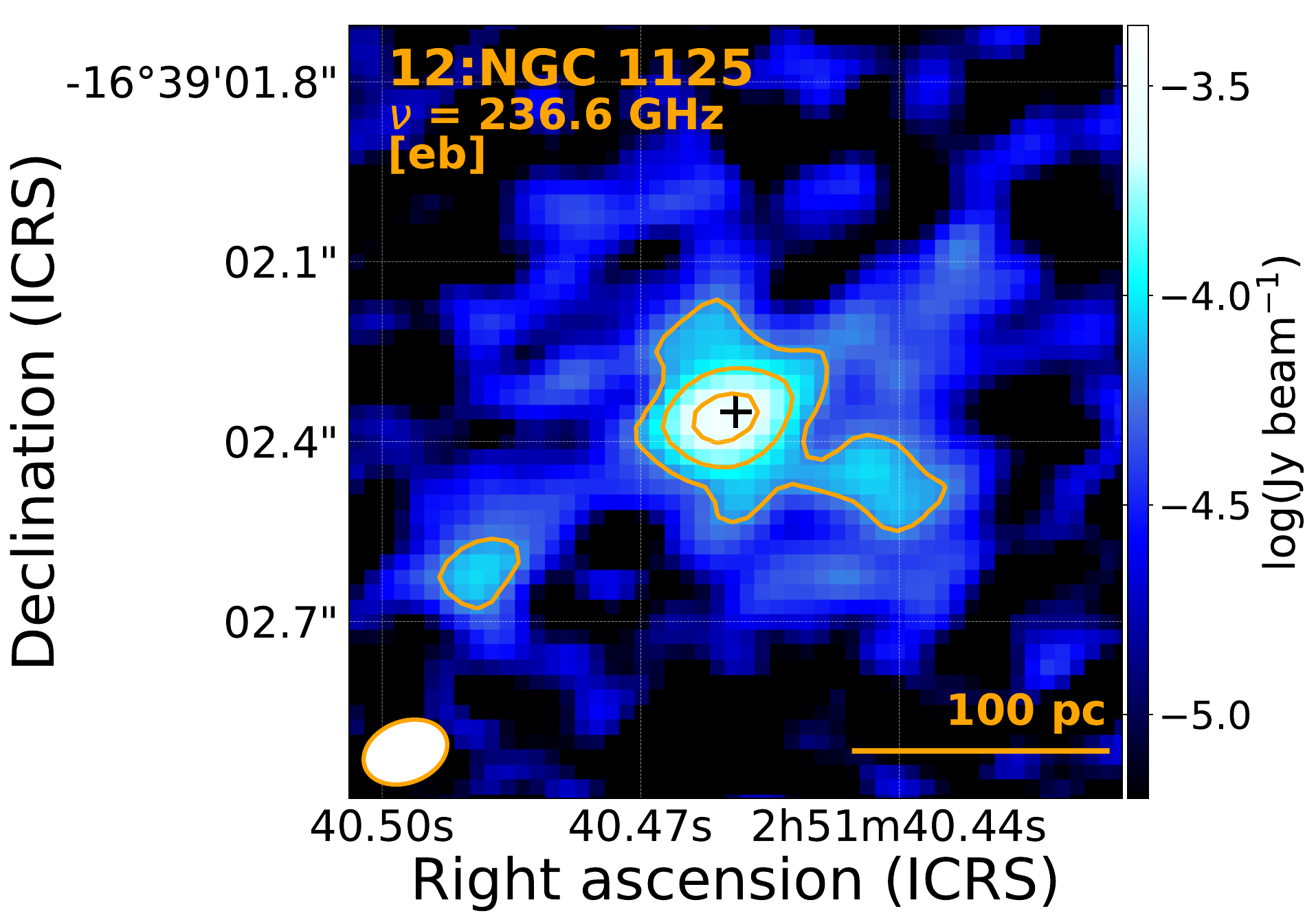}
    \caption{
    ALMA images, where north is up and east is to the left.
    In each figure, the beam size is indicated by the white ellipse at the bottom-left corner.
    The number to the left of the object name is the assigned number in this study and Paper~I. 
    The frequency is the rest-frame central frequency of the collapsed spectral window adopted, after removing
    any emission line flux. The central black cross indicates the peak of the nuclear mm-wave emission, expected to be the AGN position.
    If no mm-wave emission is detected, an unfilled diamond is shown as an AGN position inferred either from Chandra or WISE observations.
    A morphological flag, which can consist of e" and b", is also presented below the frequency.
    e" and b" indicate the presence of extended emission connected with the central un-resolved component and the presence of isolated blob-like emission, respectively.
    Colors are assigned according to flux density per beam following the color bar on the right side.
    The orange contours indicate where flux densities are $5\sigma_{\rm mm}$, $10\sigma_{\rm mm}$, $20\sigma_{\rm mm}$, $40\sigma_{\rm mm}$,
    $80\sigma_{\rm mm}$ and $160\sigma_{\rm mm}$, while dashed red ones are drawn where flux densities are $-5\sigma_{\rm mm}$, if any.
    The individual values of $\sigma_{\rm mm}$ are listed in Table~\ref{tab_app:mm_prop}.
    }\label{fig_app:images}
\end{figure*}

\addtocounter{figure}{-1}

\begin{figure*}
    \centering
    \includegraphics[width=5.9cm]{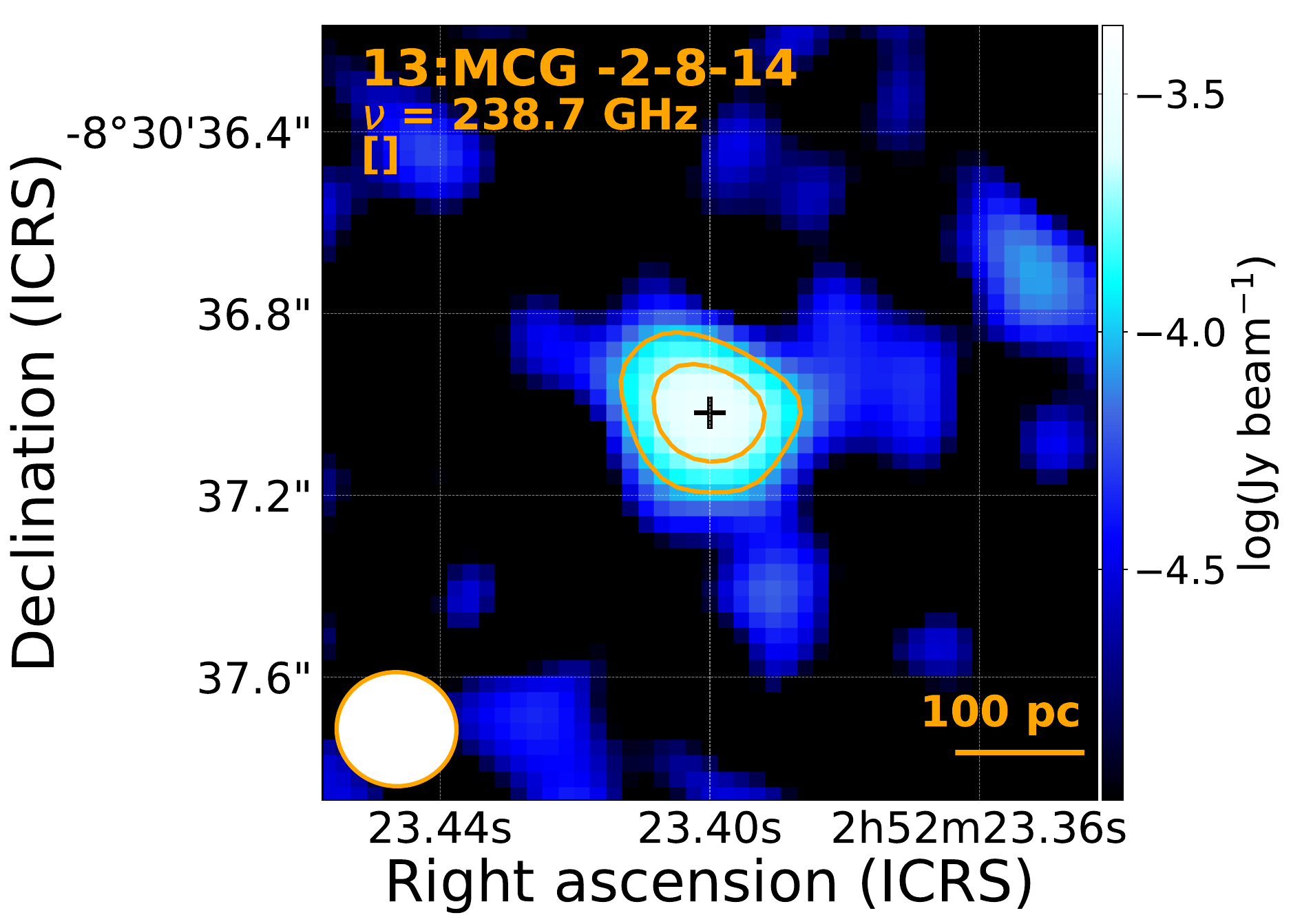}
\includegraphics[width=5.9cm]{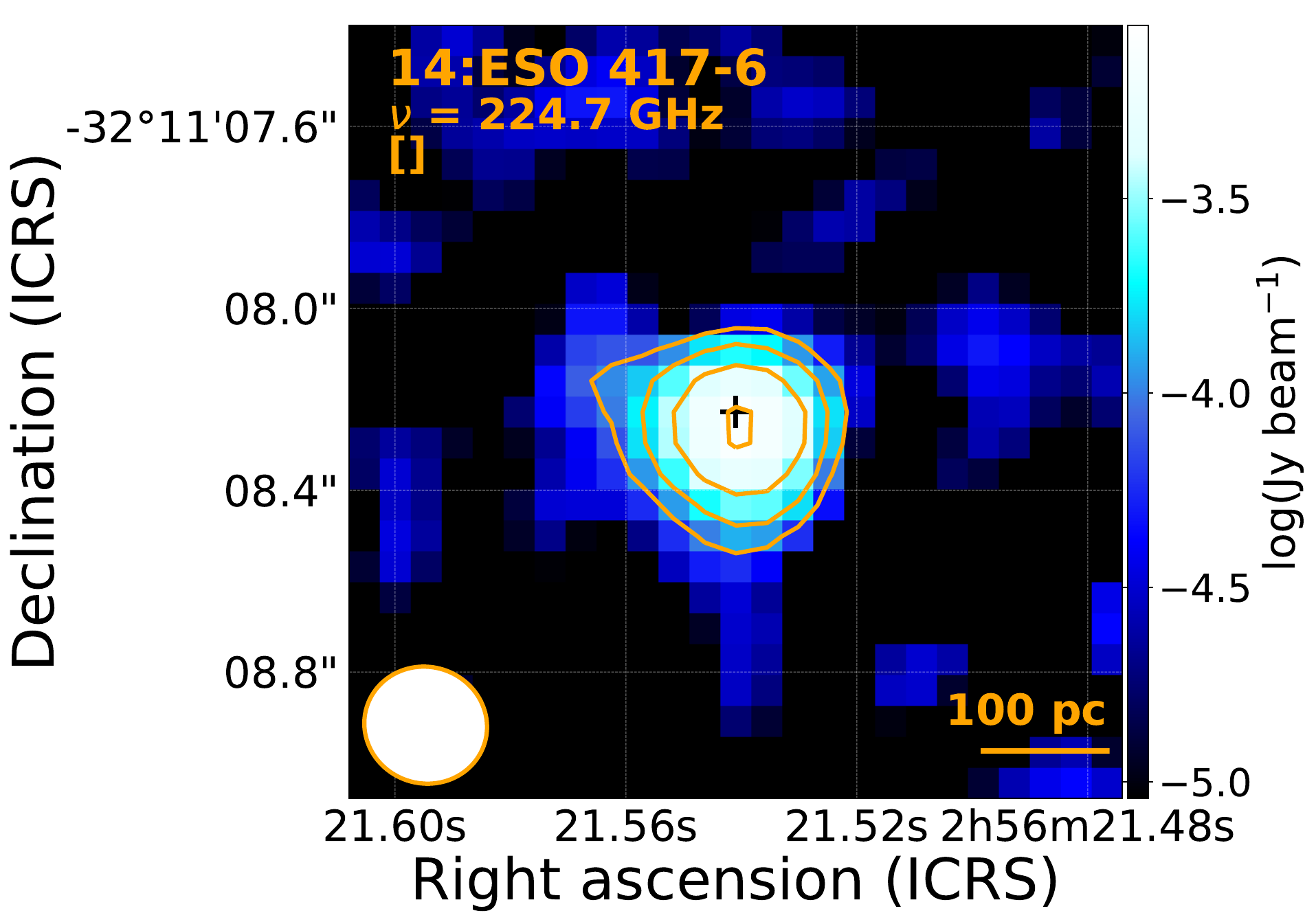}
\includegraphics[width=5.9cm]{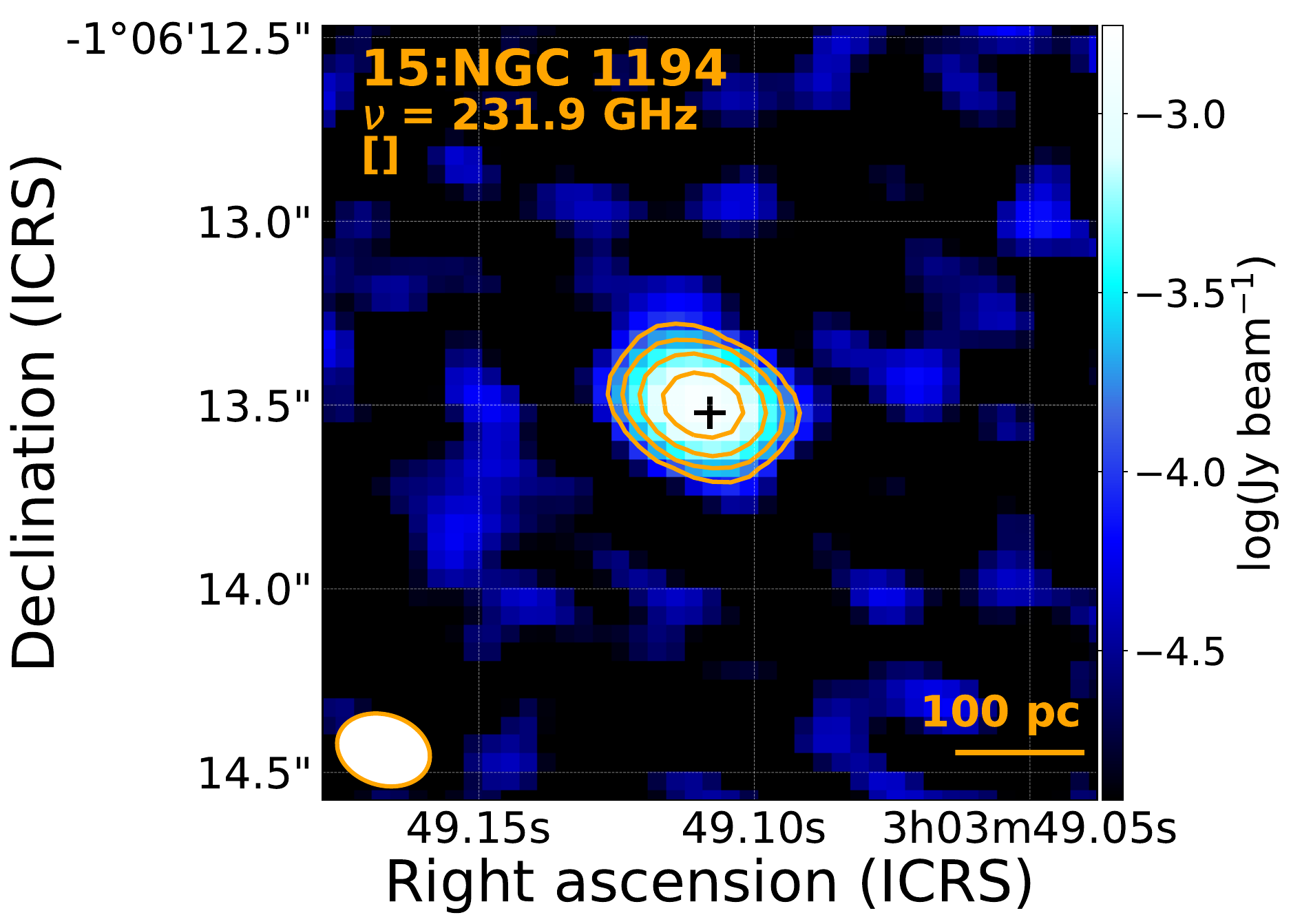}
\includegraphics[width=5.9cm]{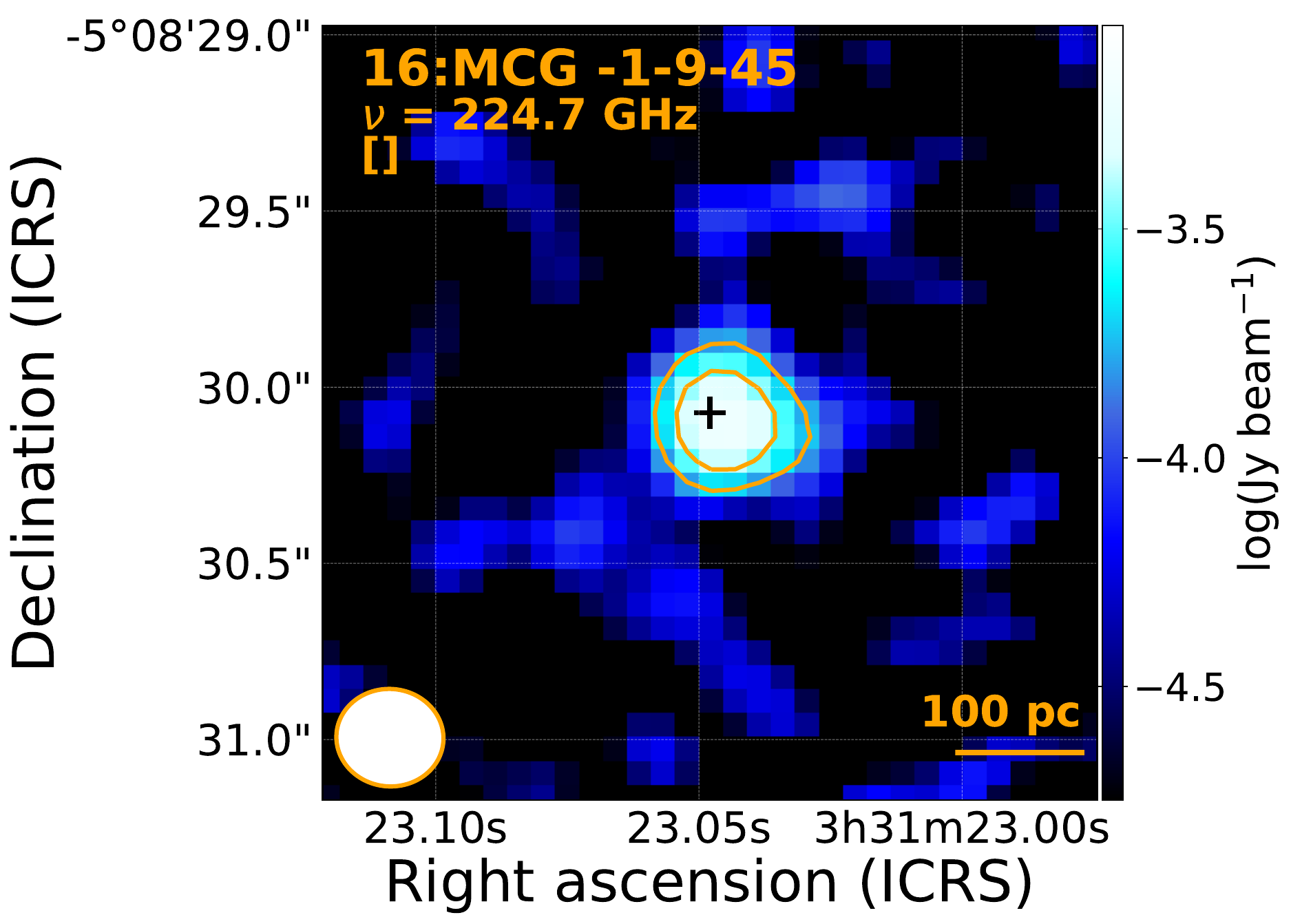}
\includegraphics[width=5.9cm]{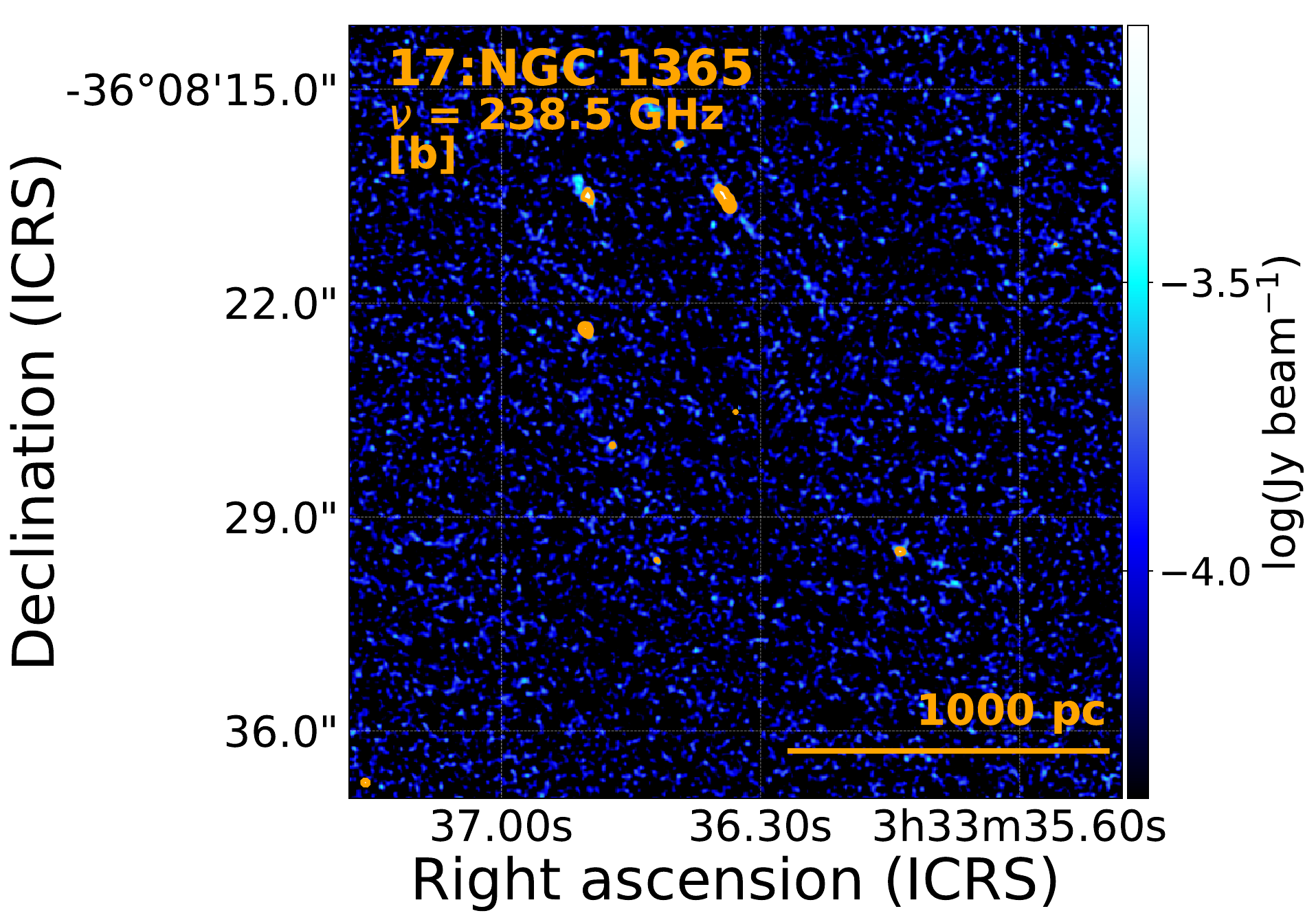}
\includegraphics[width=5.9cm]{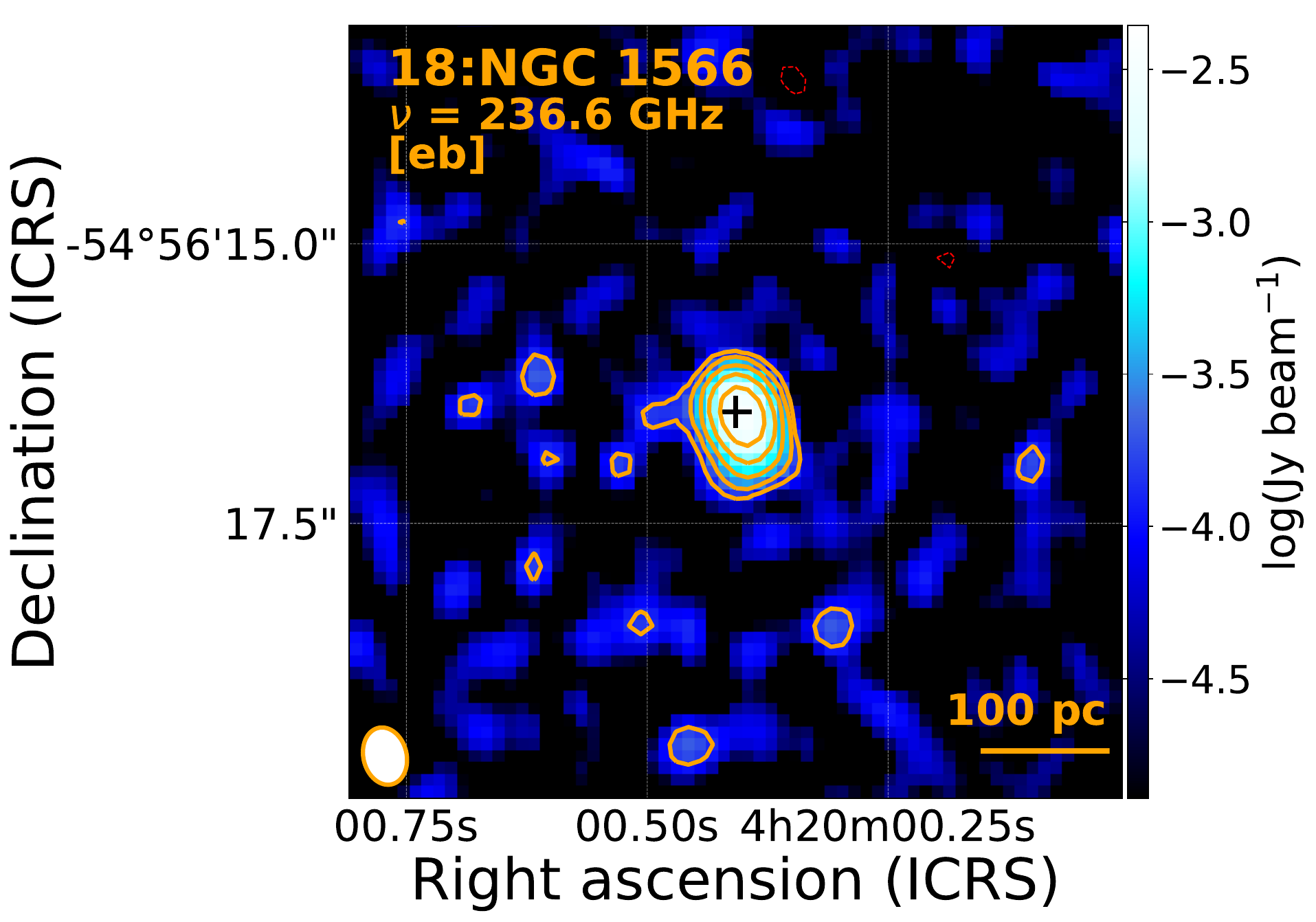}
\includegraphics[width=5.9cm]{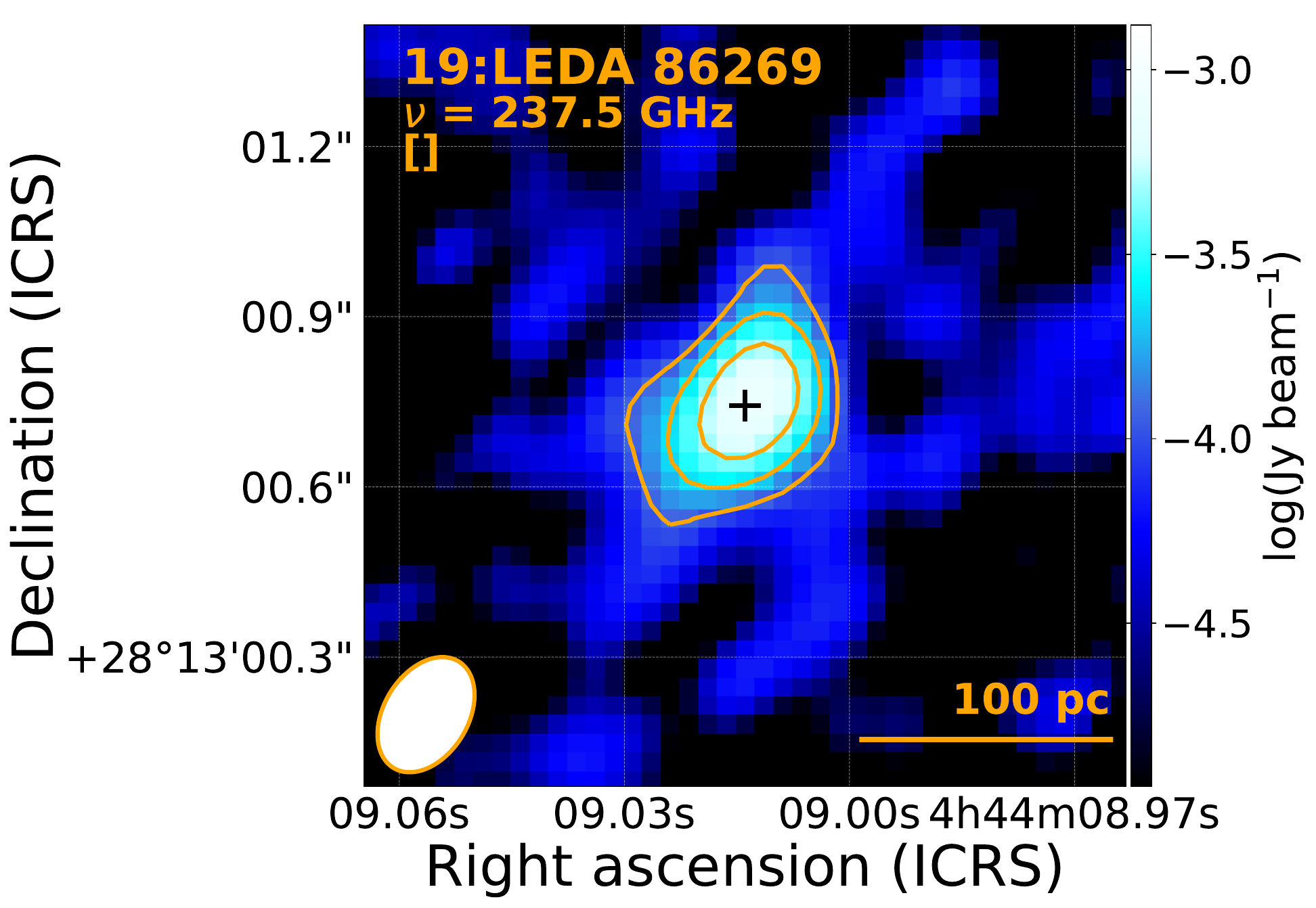}
\includegraphics[width=5.9cm]{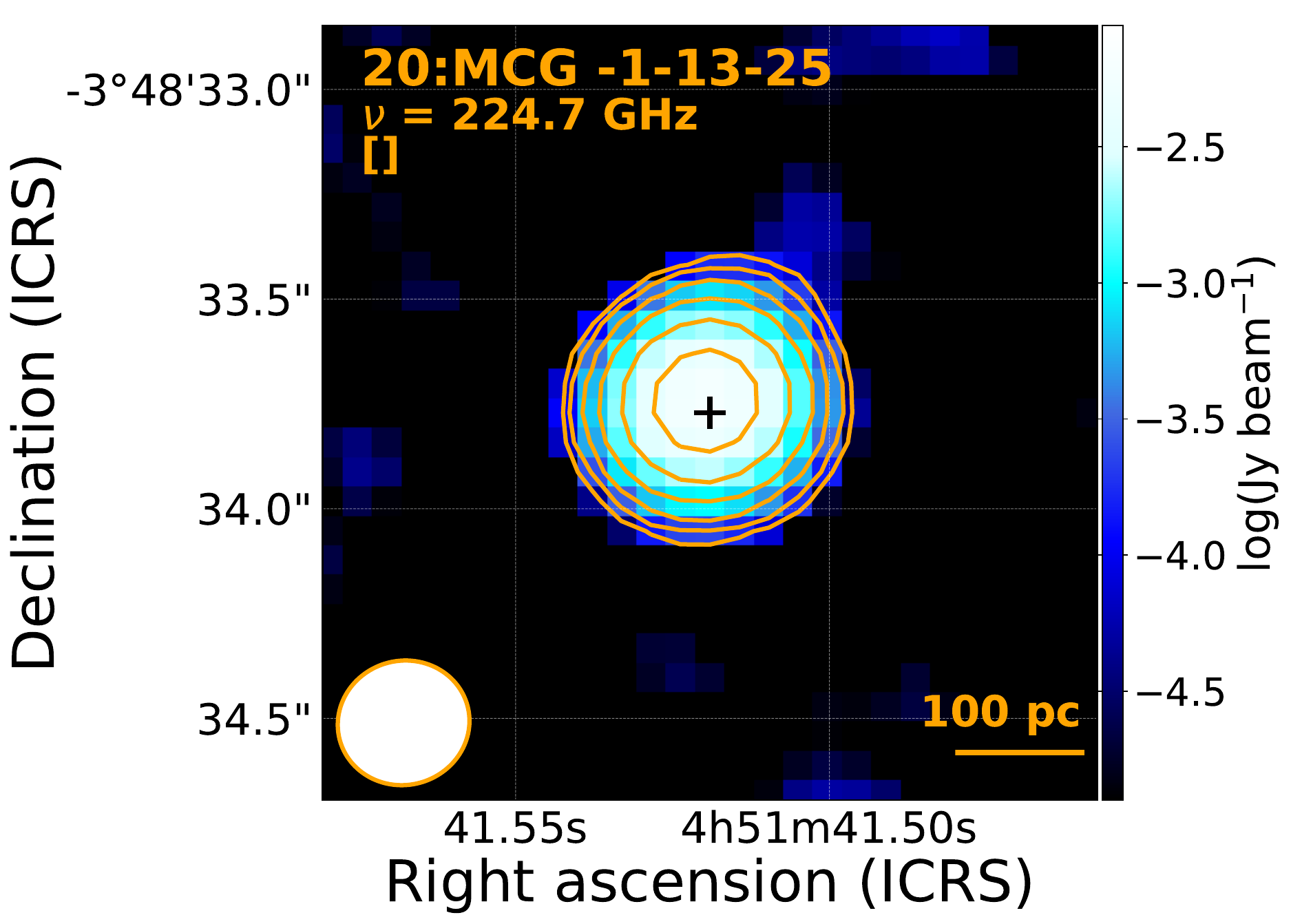}
\includegraphics[width=5.9cm]{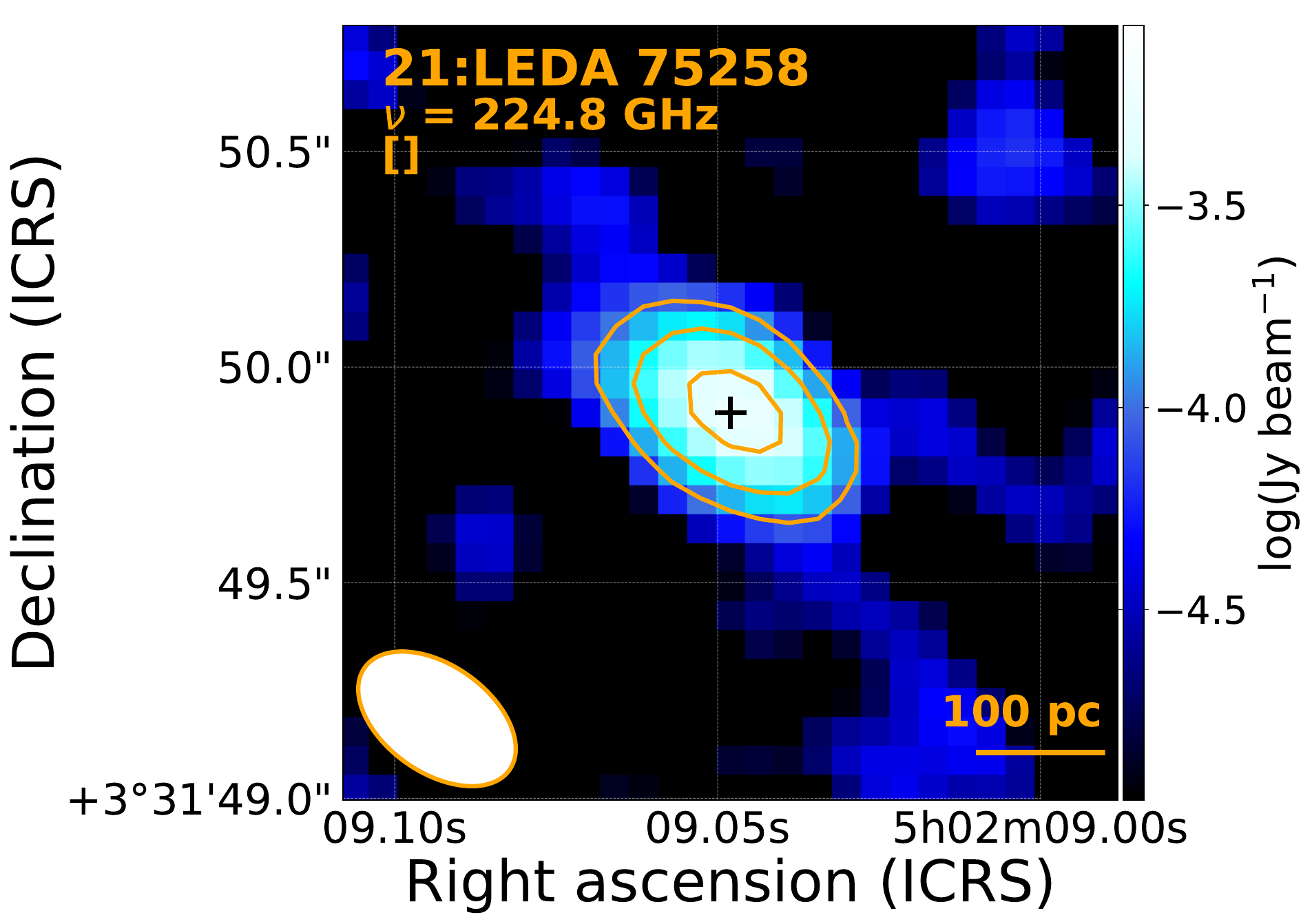}
\includegraphics[width=5.9cm]{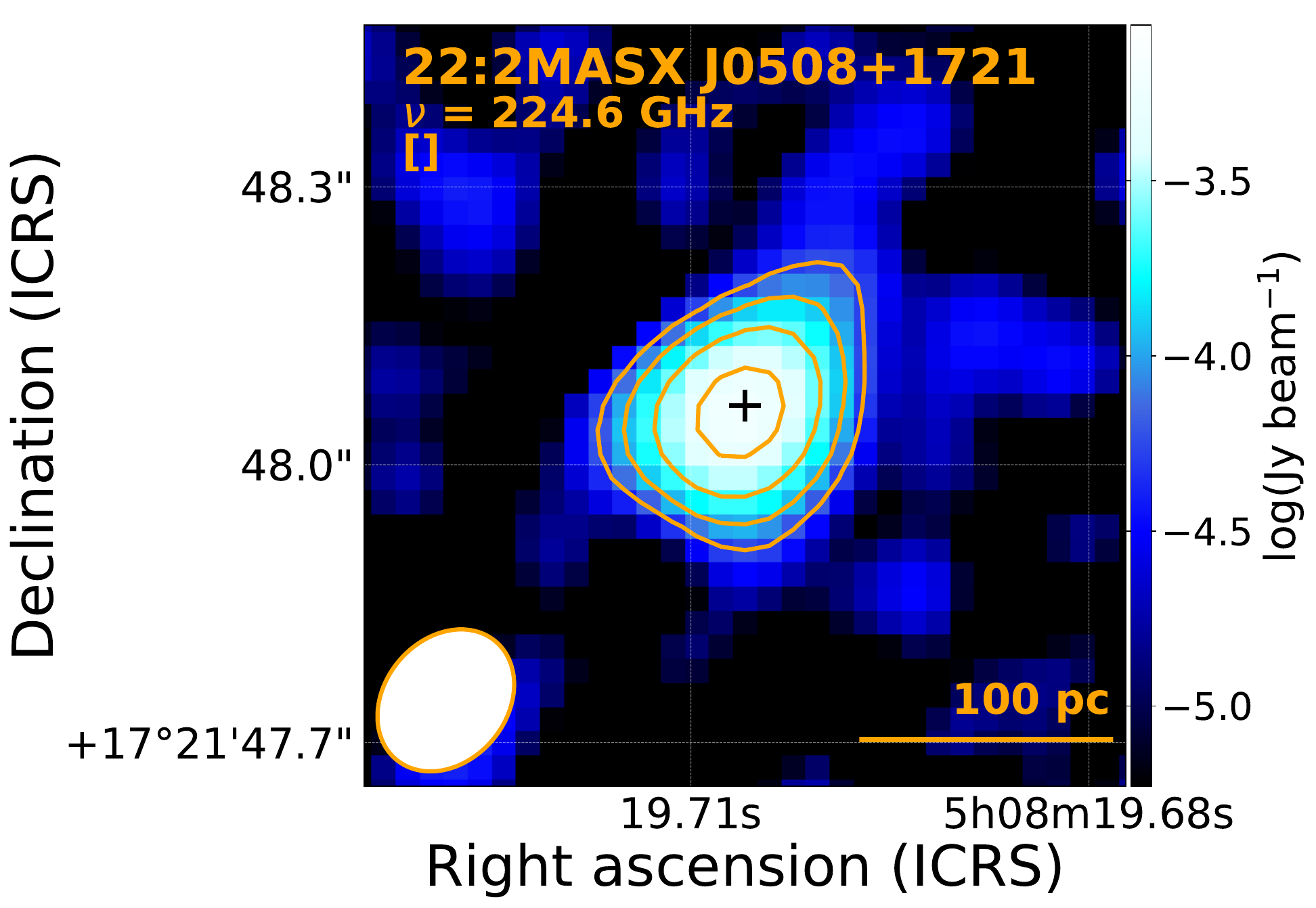}
\includegraphics[width=5.9cm]{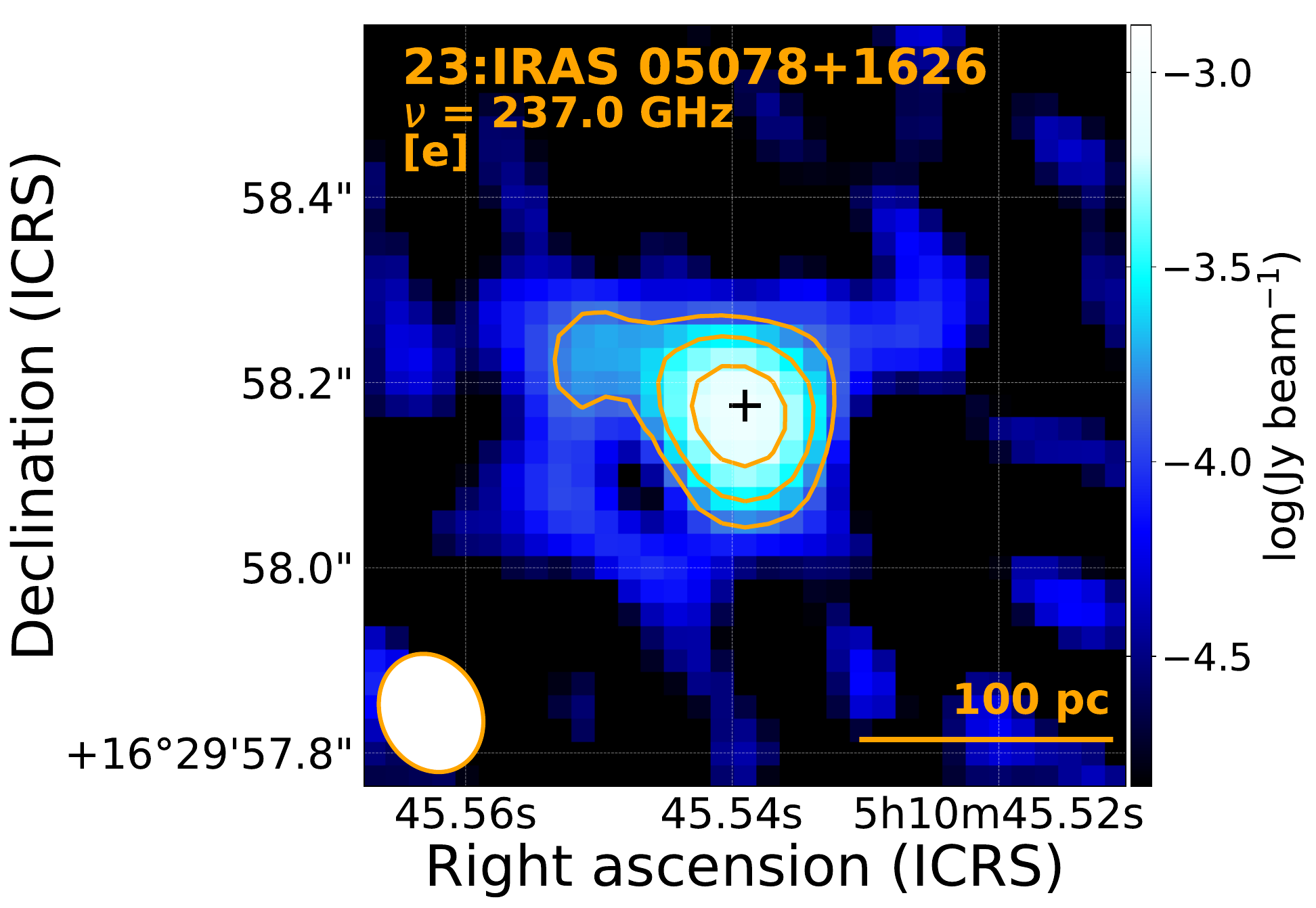}
\includegraphics[width=5.9cm]{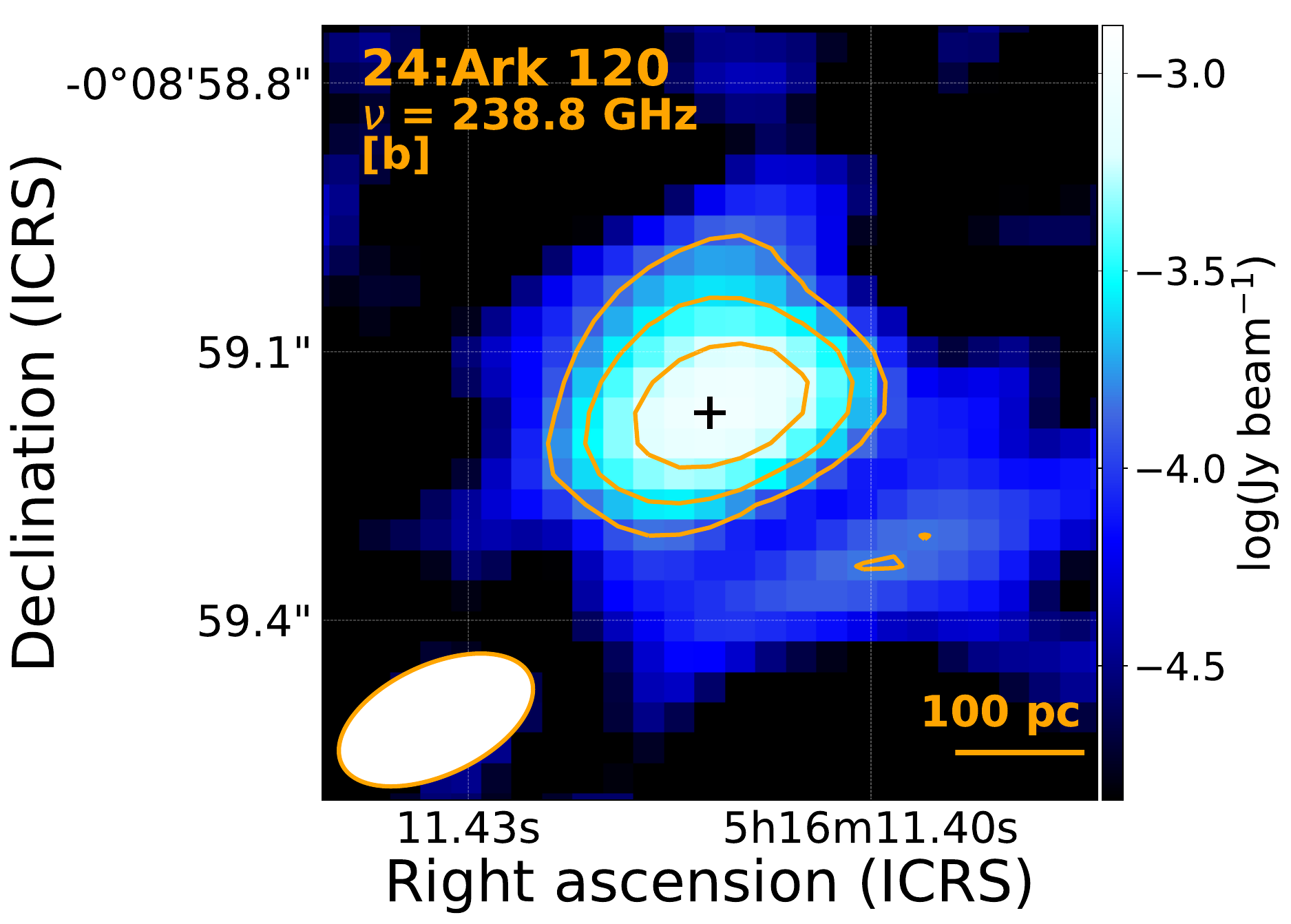}
\includegraphics[width=5.9cm]{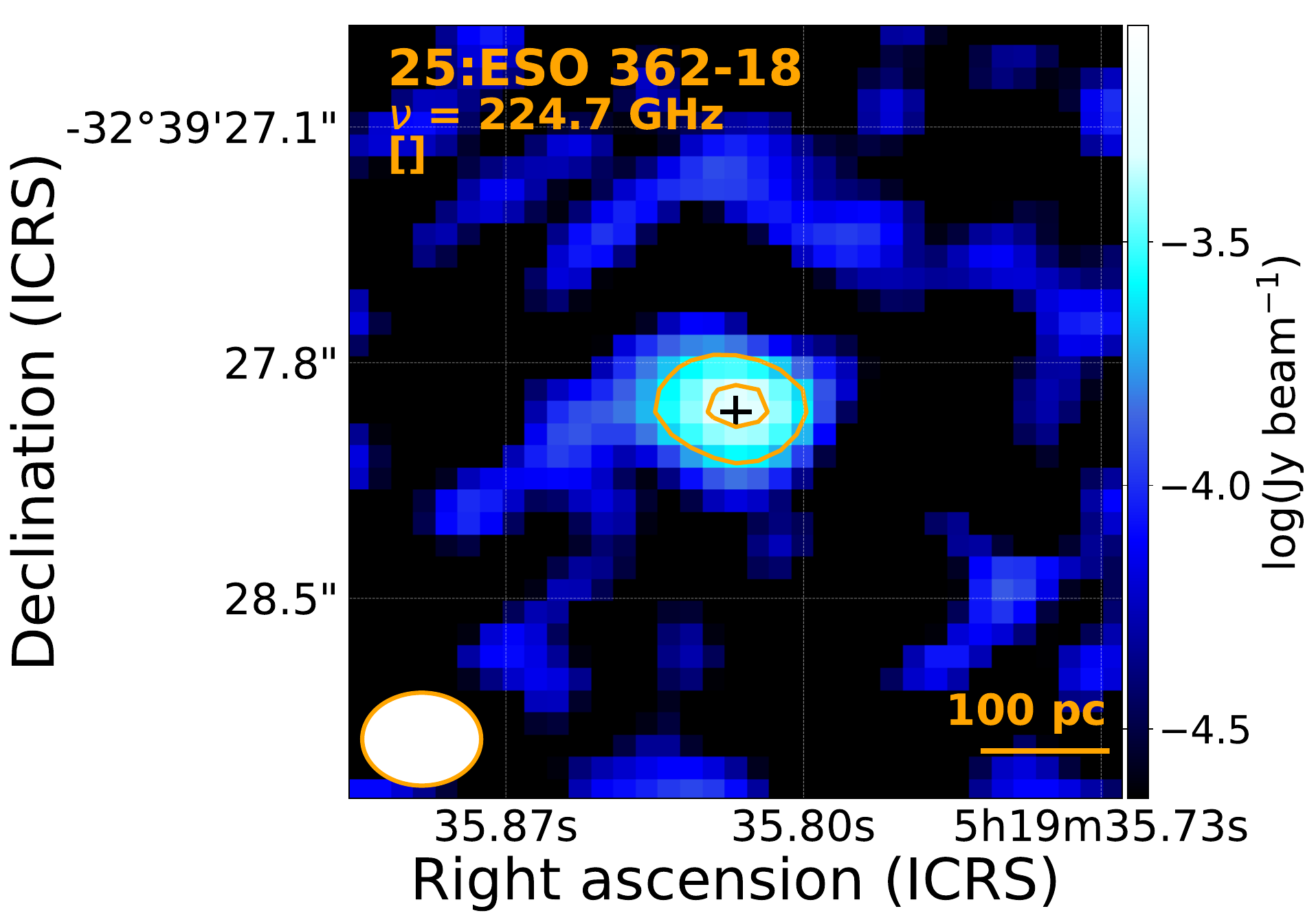}
\includegraphics[width=5.9cm]{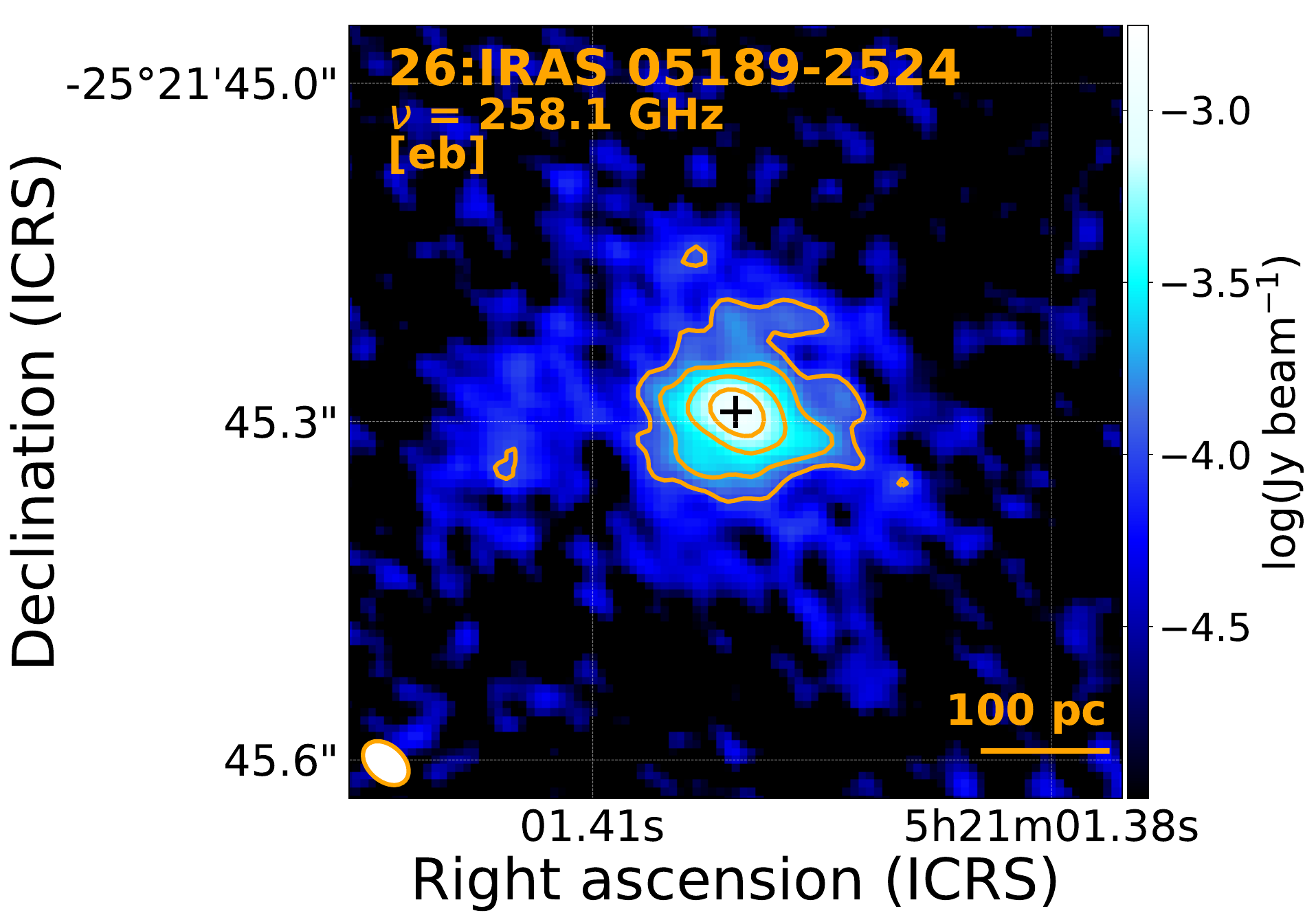}
\includegraphics[width=5.9cm]{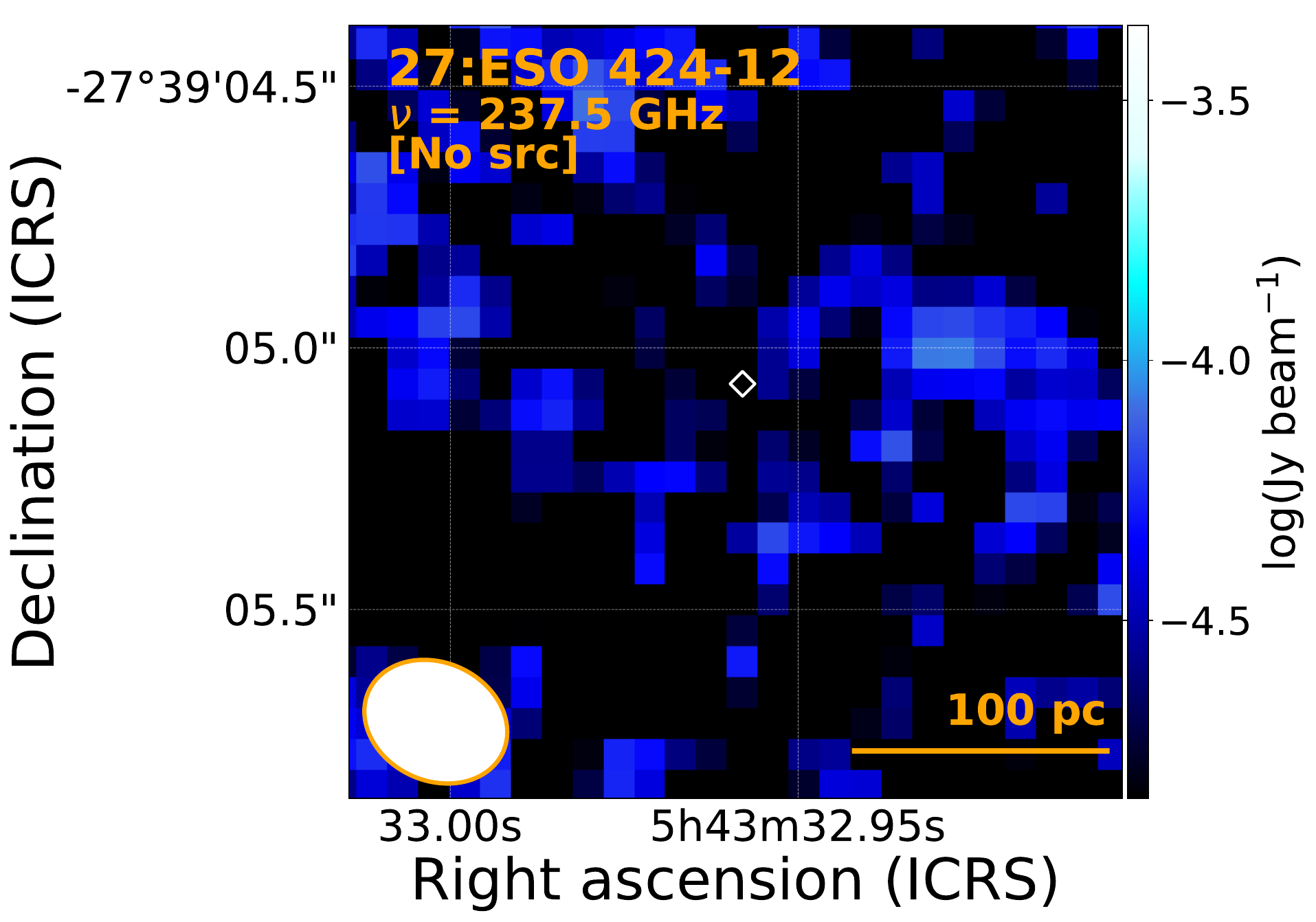}
    \caption{Continued. 
    }
\end{figure*}

\addtocounter{figure}{-1}

\begin{figure*}
    \centering
    \includegraphics[width=5.9cm]{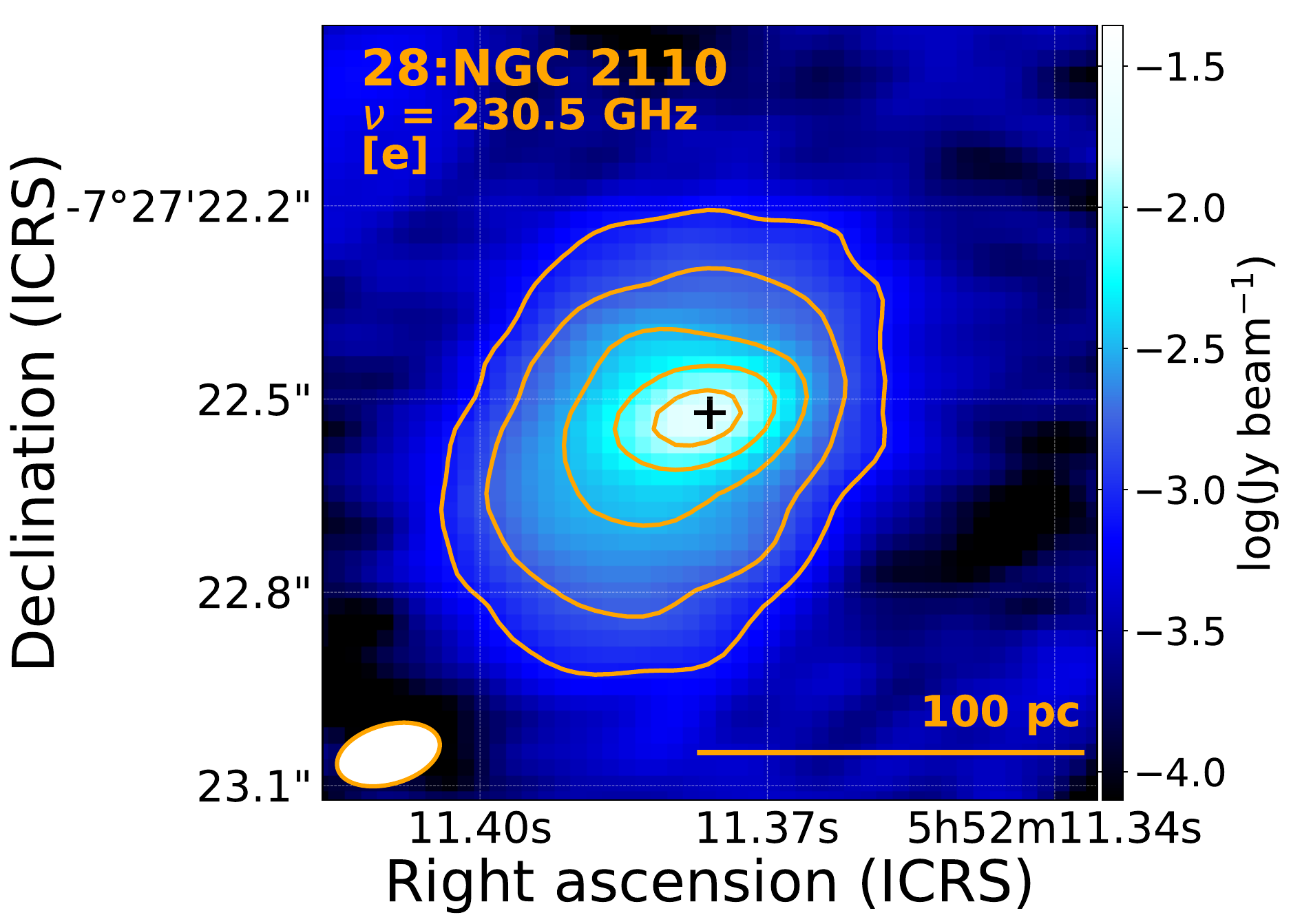}
\includegraphics[width=5.9cm]{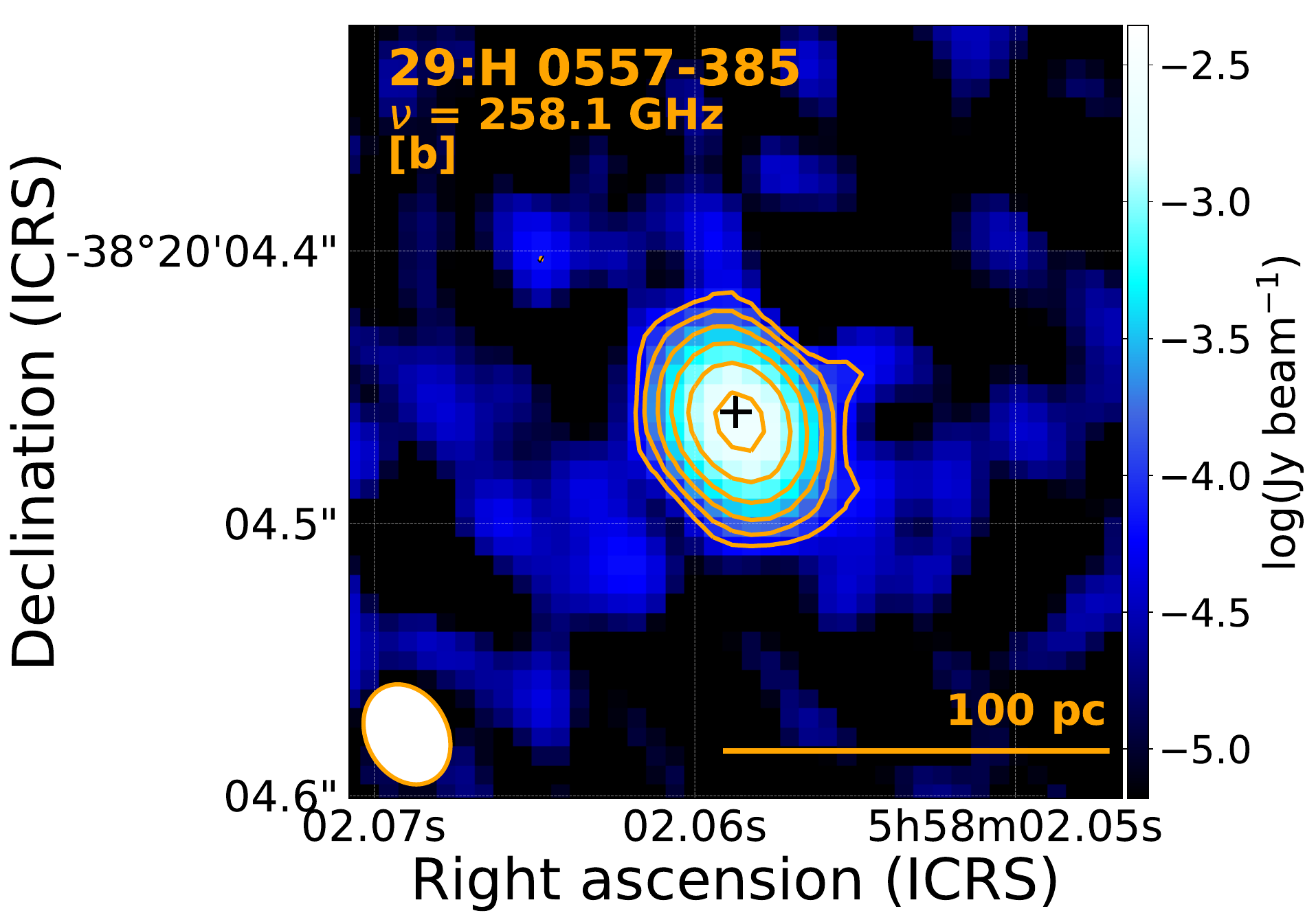}
\includegraphics[width=5.9cm]{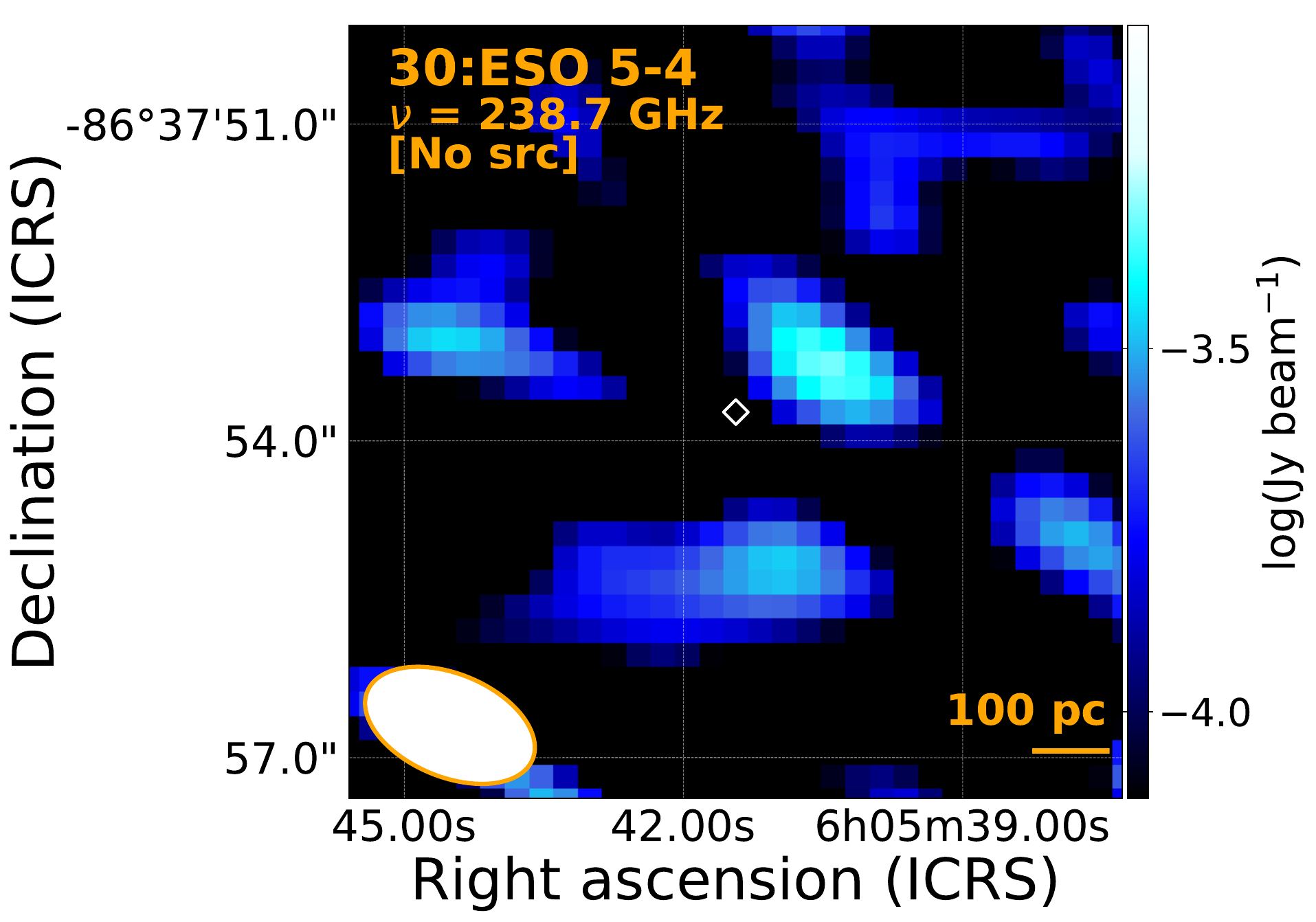}
\includegraphics[width=5.9cm]{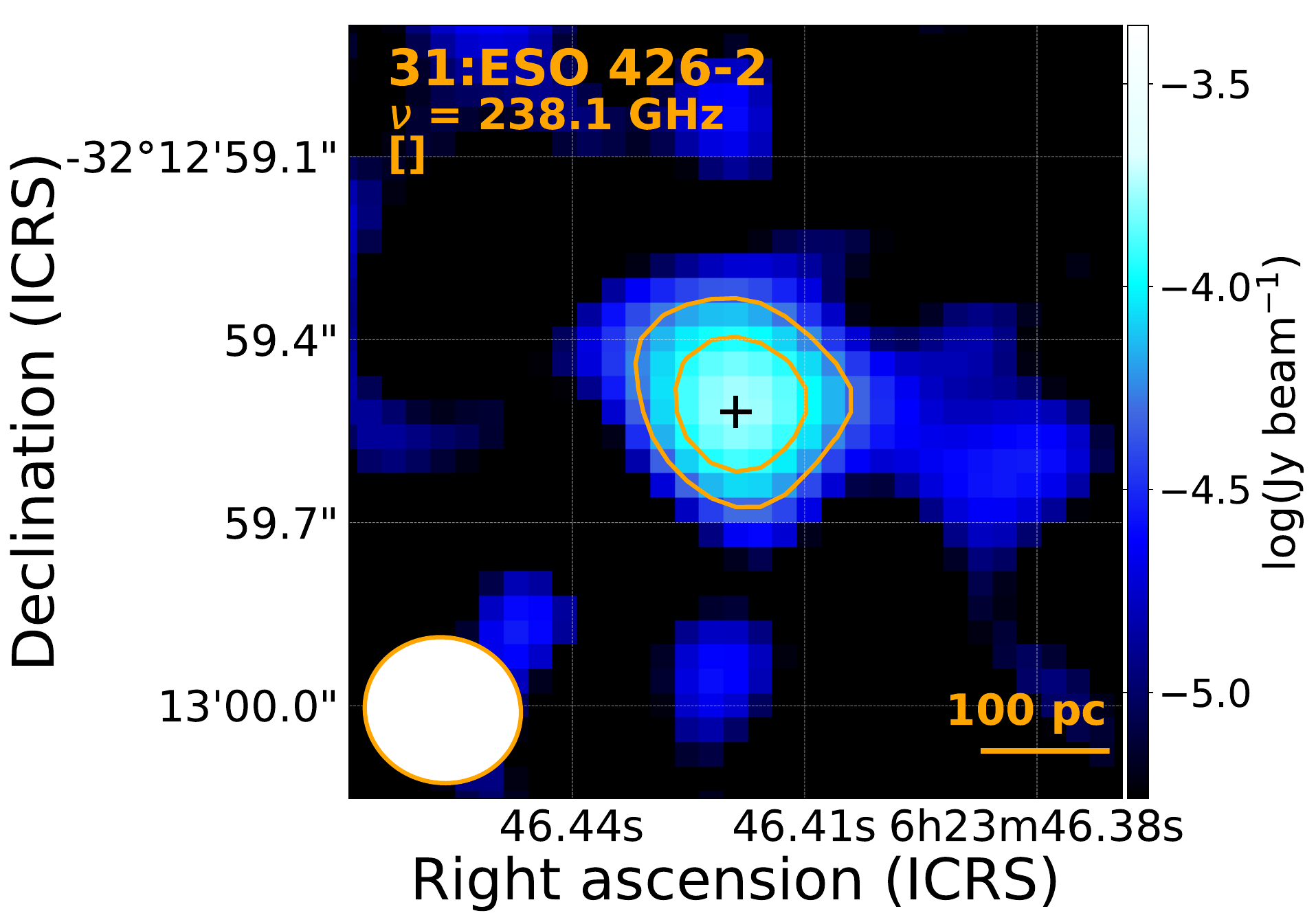}
\includegraphics[width=5.9cm]{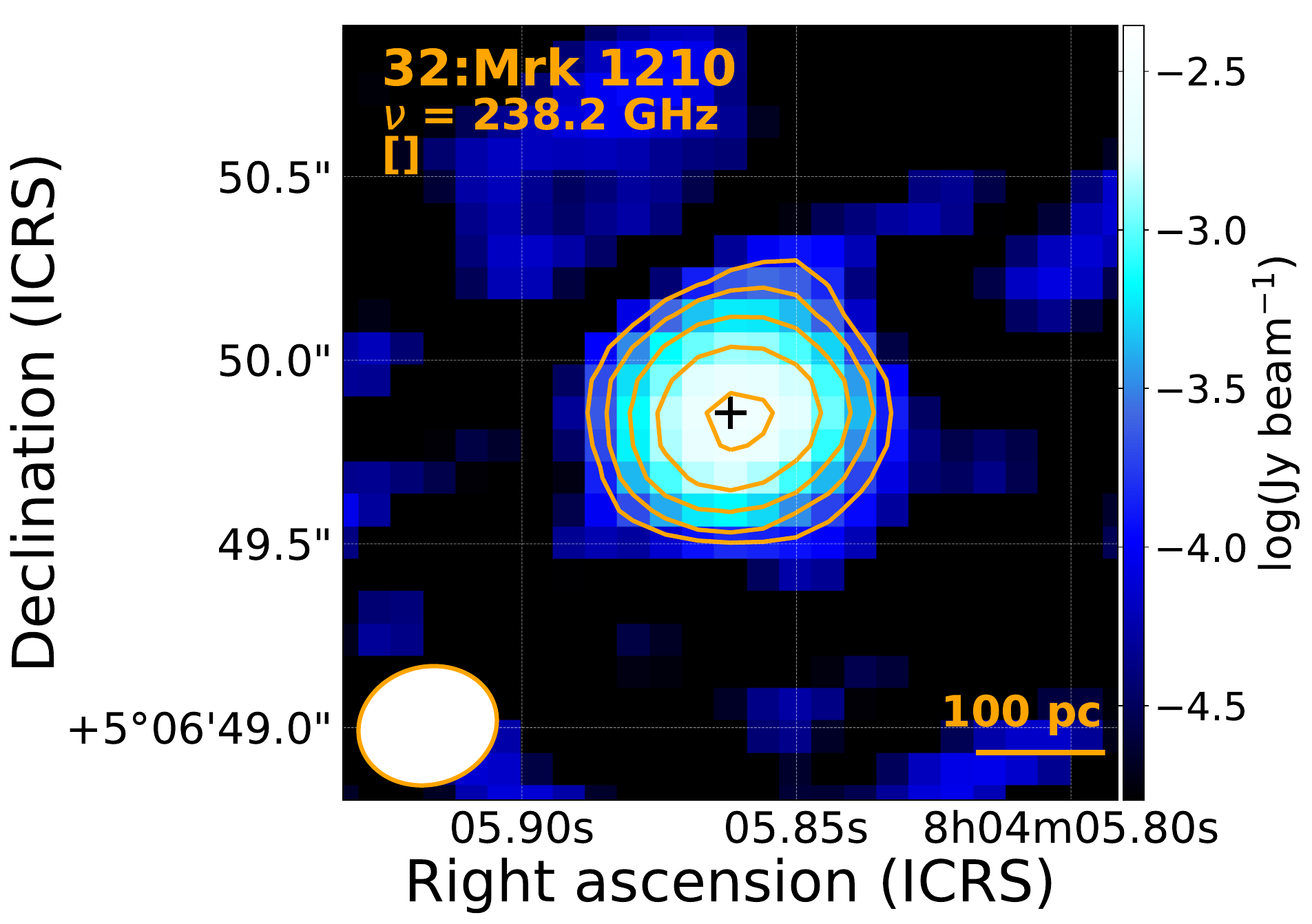}
\includegraphics[width=5.9cm]{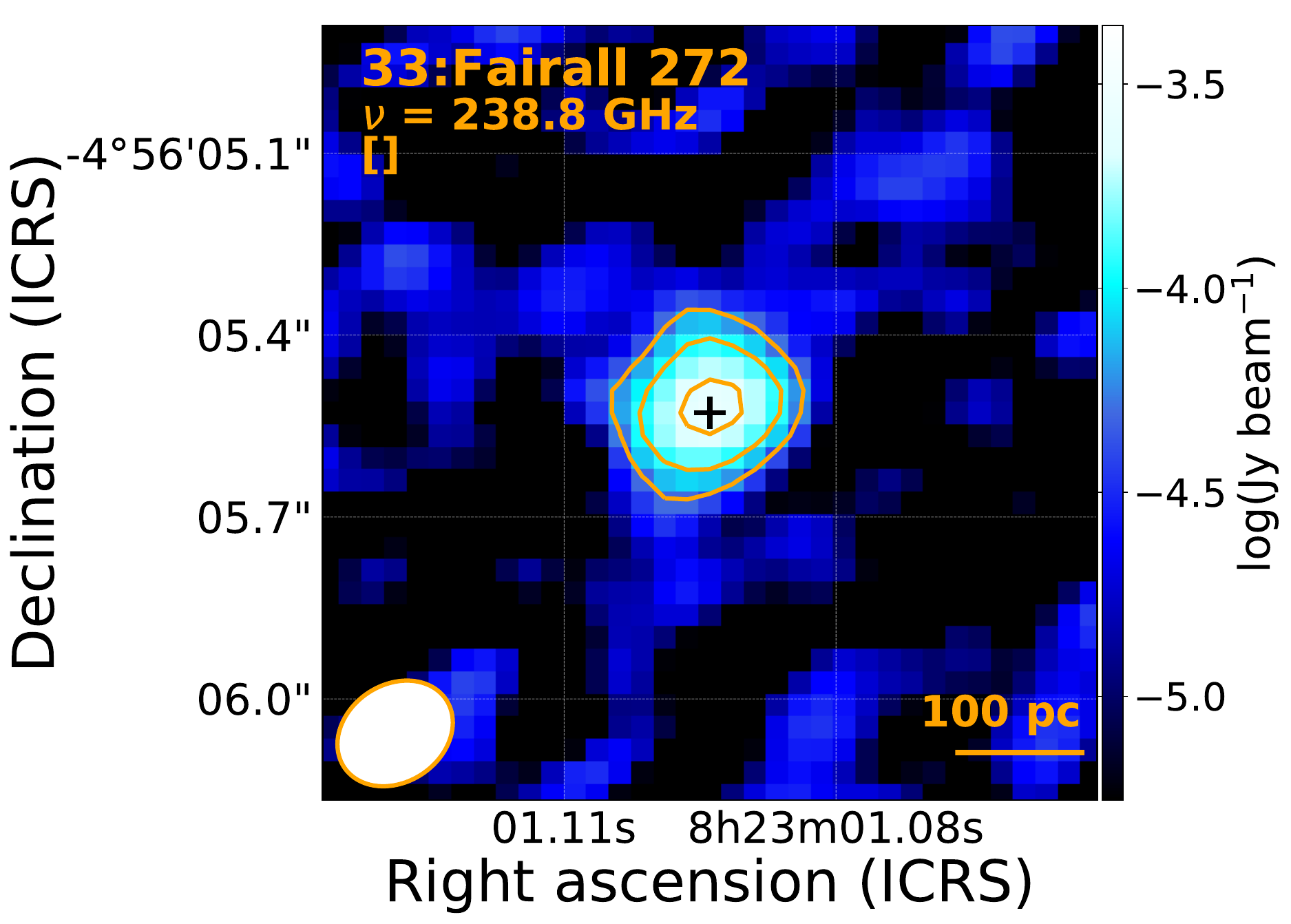}
\includegraphics[width=5.9cm]{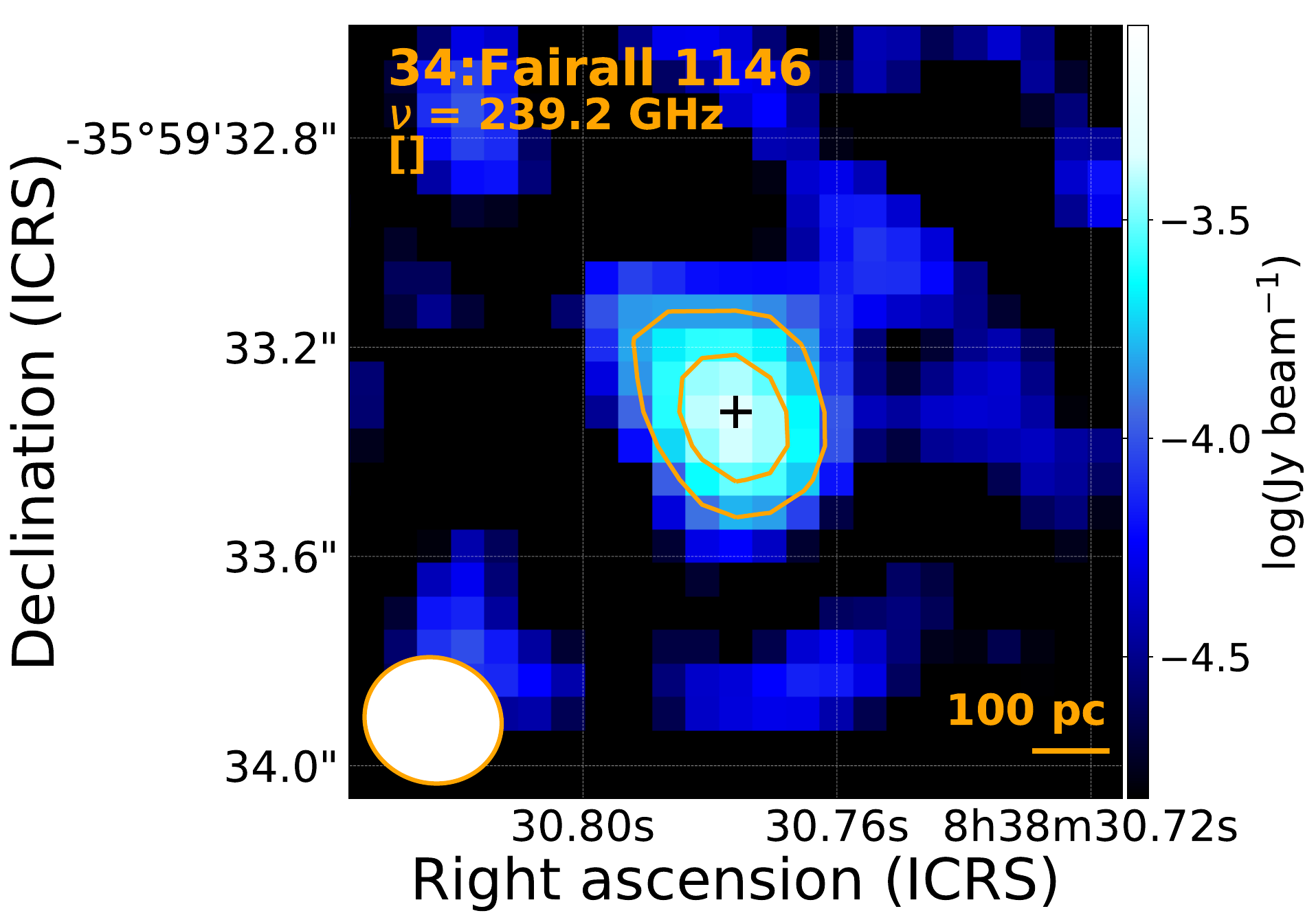}
\includegraphics[width=5.9cm]{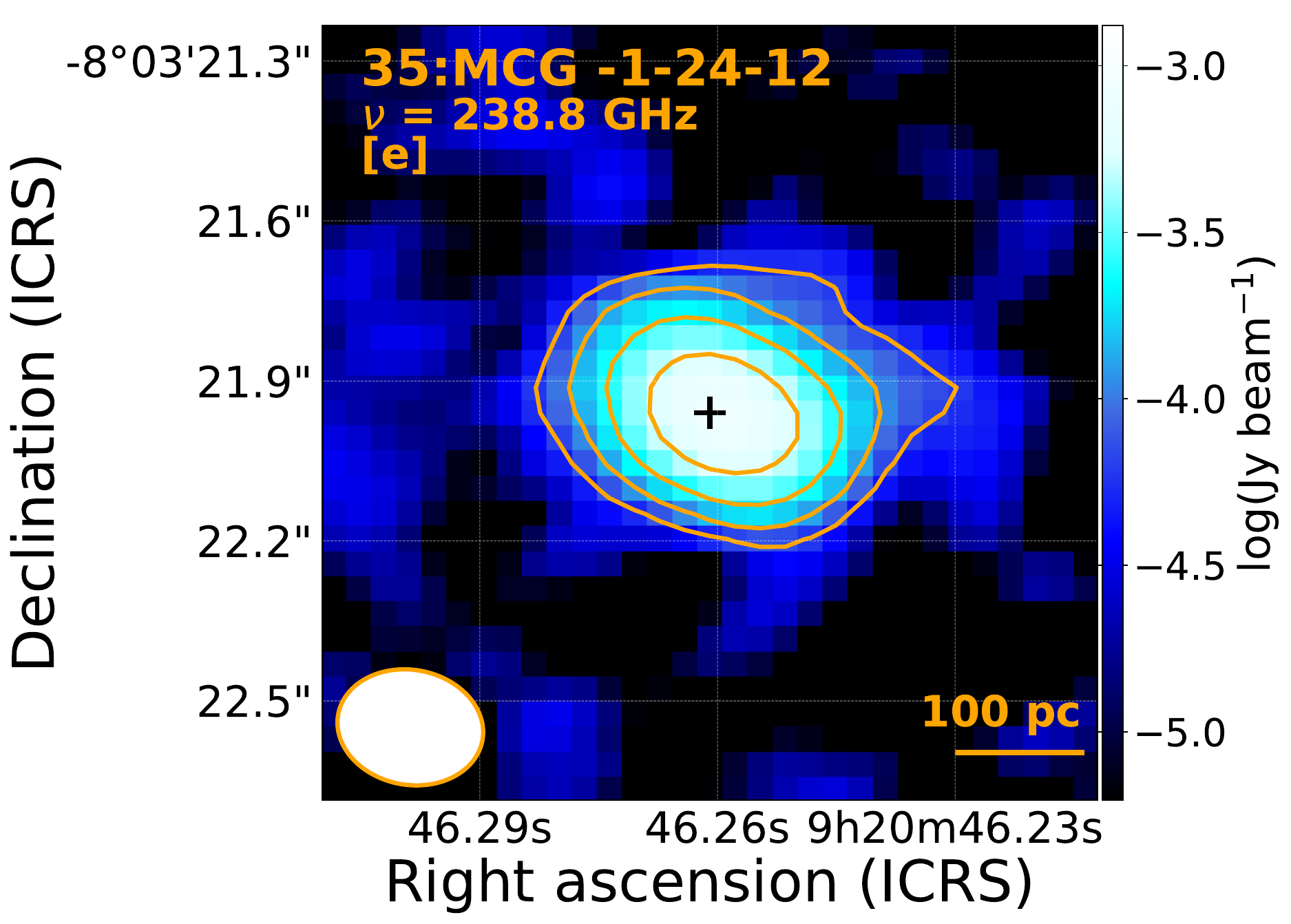}
\includegraphics[width=5.9cm]{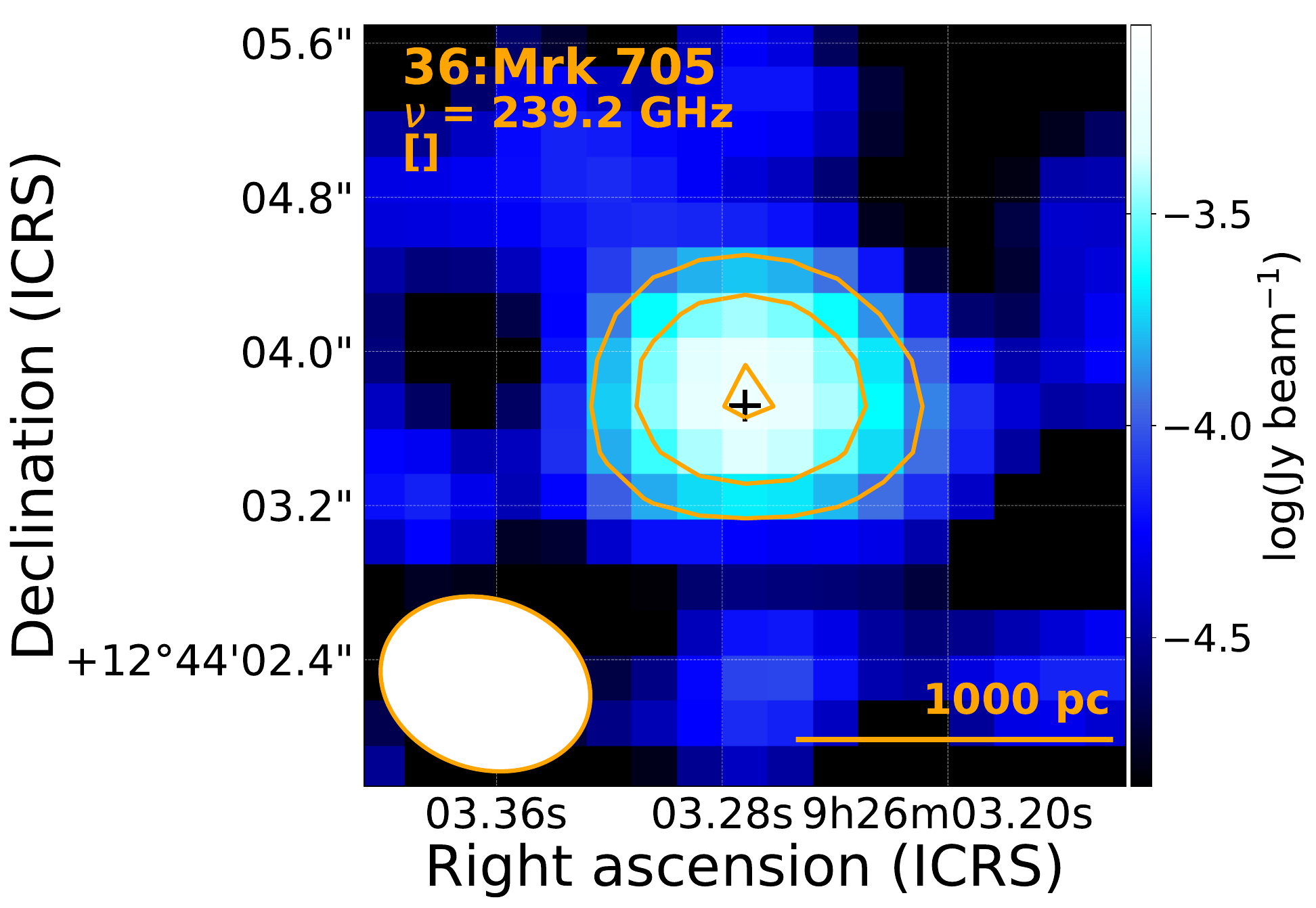}
\includegraphics[width=5.9cm]{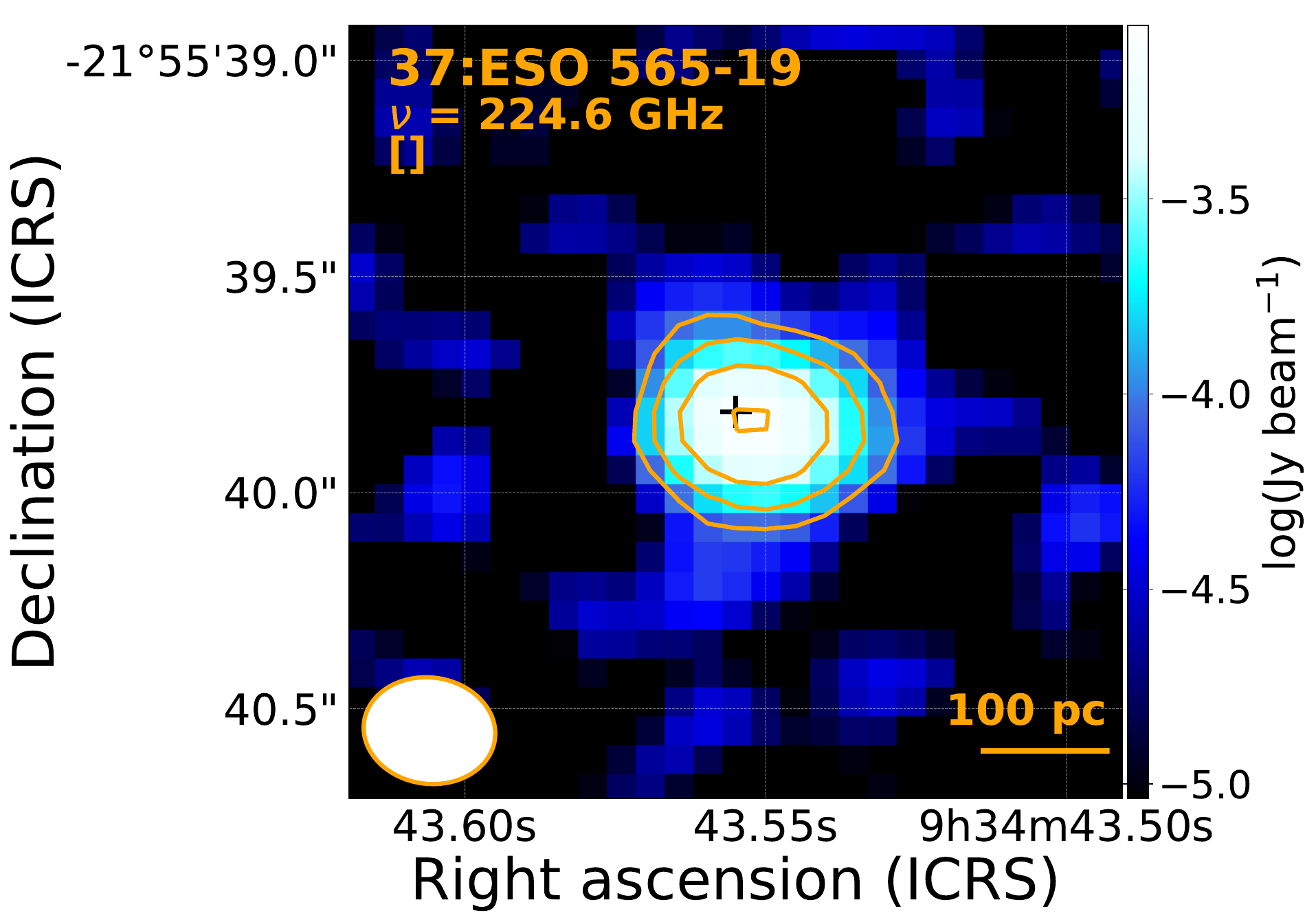}
\includegraphics[width=5.9cm]{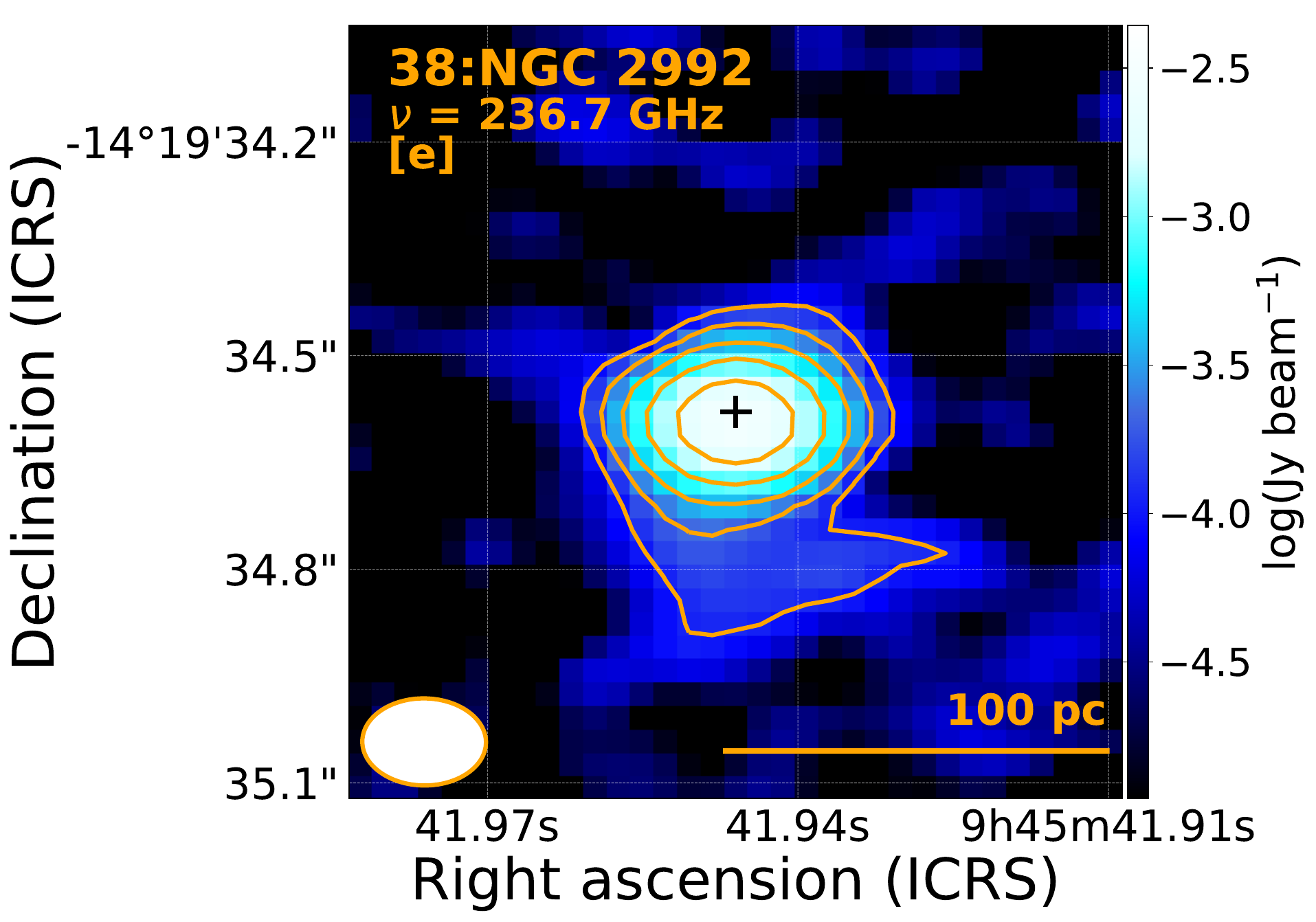}
\includegraphics[width=5.9cm]{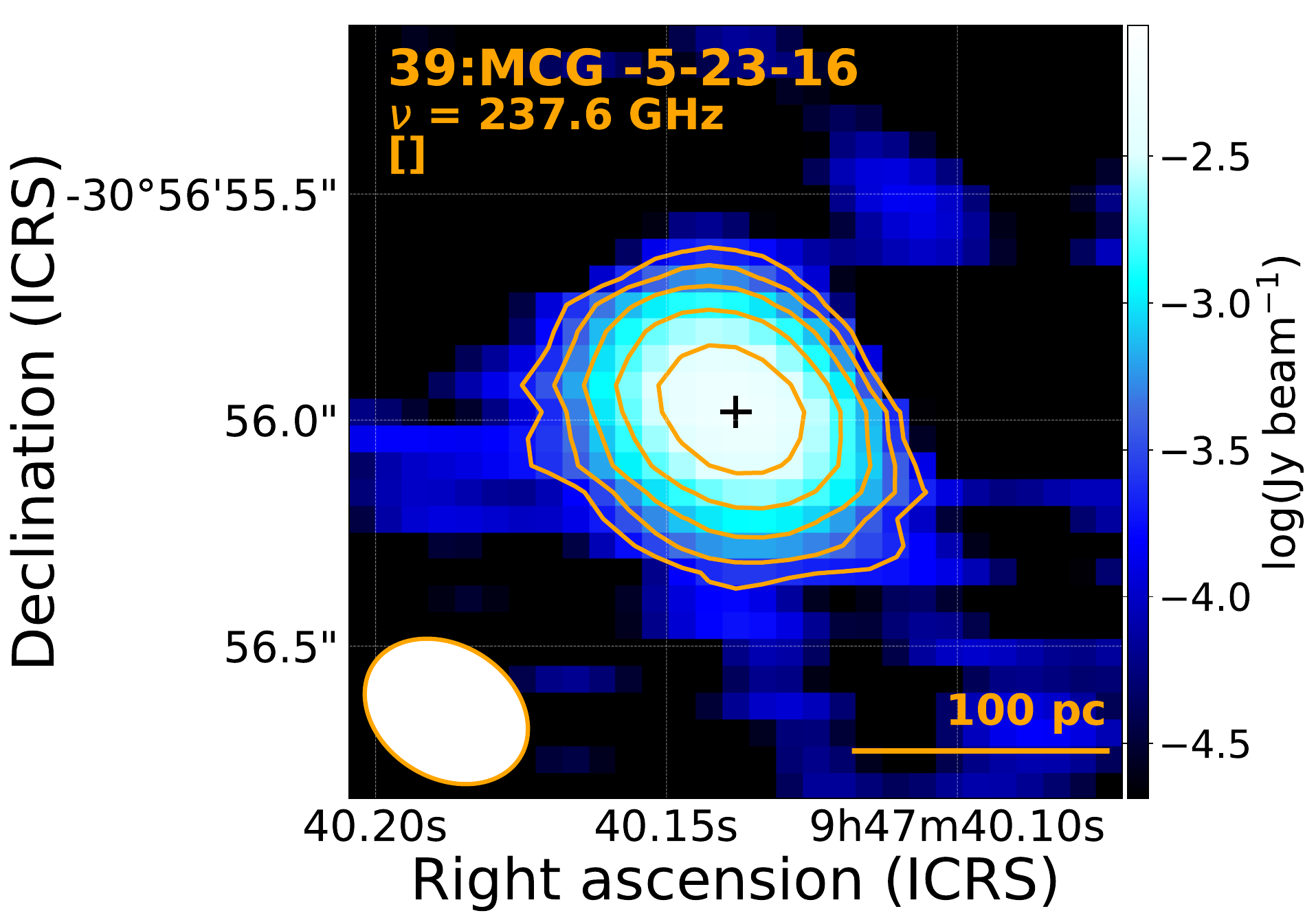}
\includegraphics[width=5.9cm]{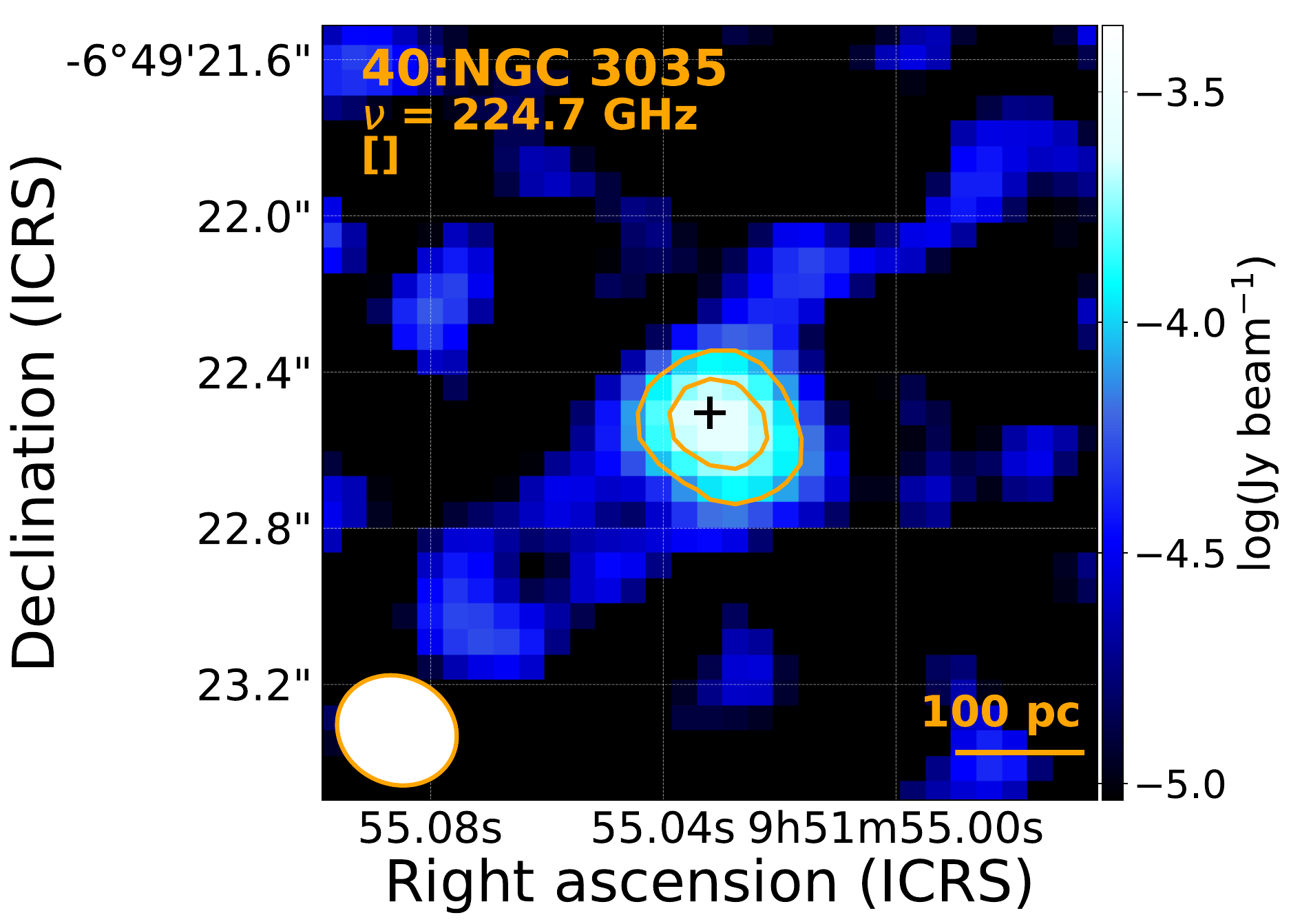}
\includegraphics[width=5.9cm]{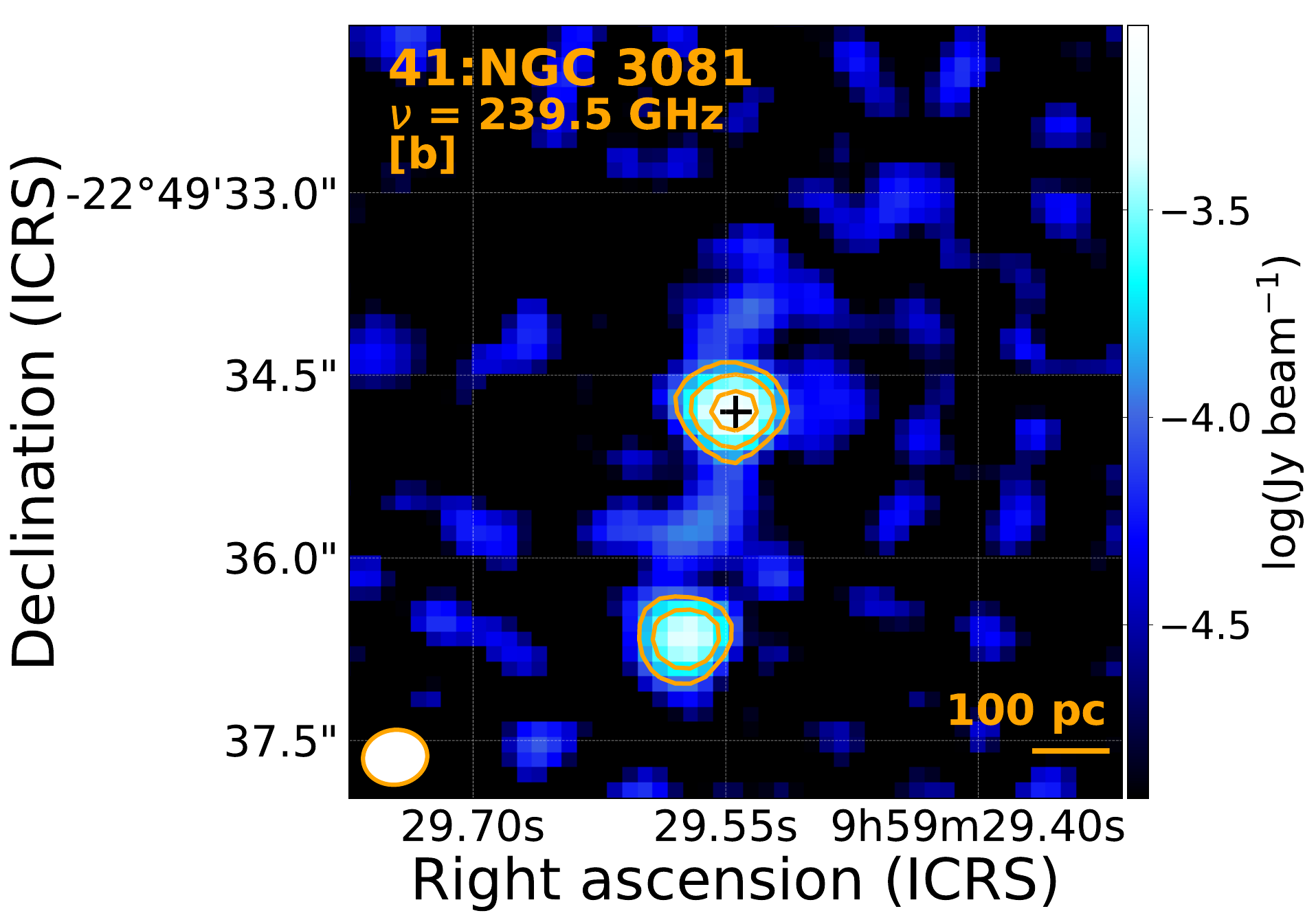}
\includegraphics[width=5.9cm]{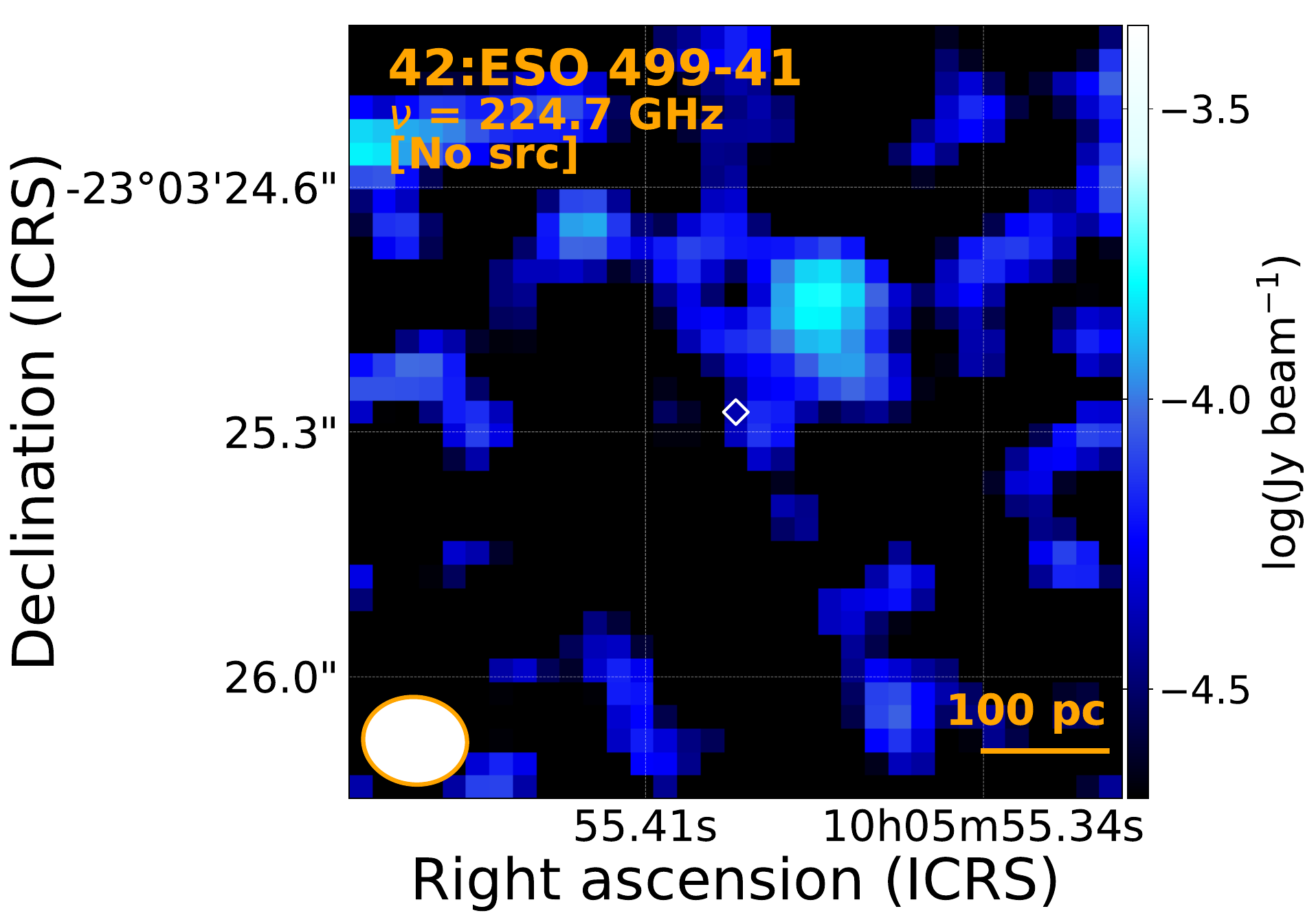}
    \caption{Continued. 
    }
\end{figure*} 

\addtocounter{figure}{-1}

\begin{figure*}
    \centering
    \includegraphics[width=5.9cm]{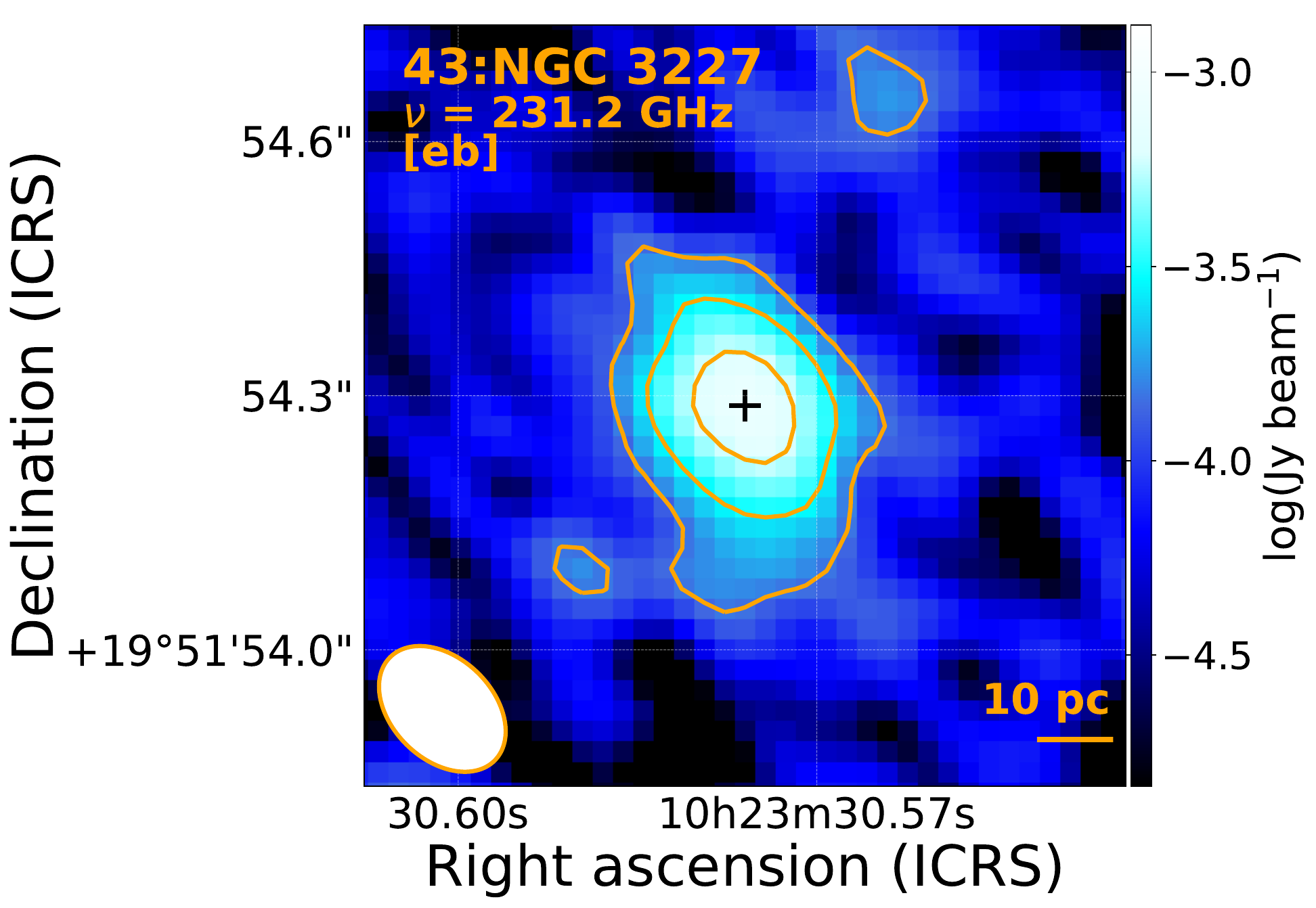}
\includegraphics[width=5.9cm]{044_NGC_3281_spwall_c_300pc.pdf}
\includegraphics[width=5.9cm]{045_NGC_3393_spwall_c_300pc.pdf}
\includegraphics[width=5.9cm]{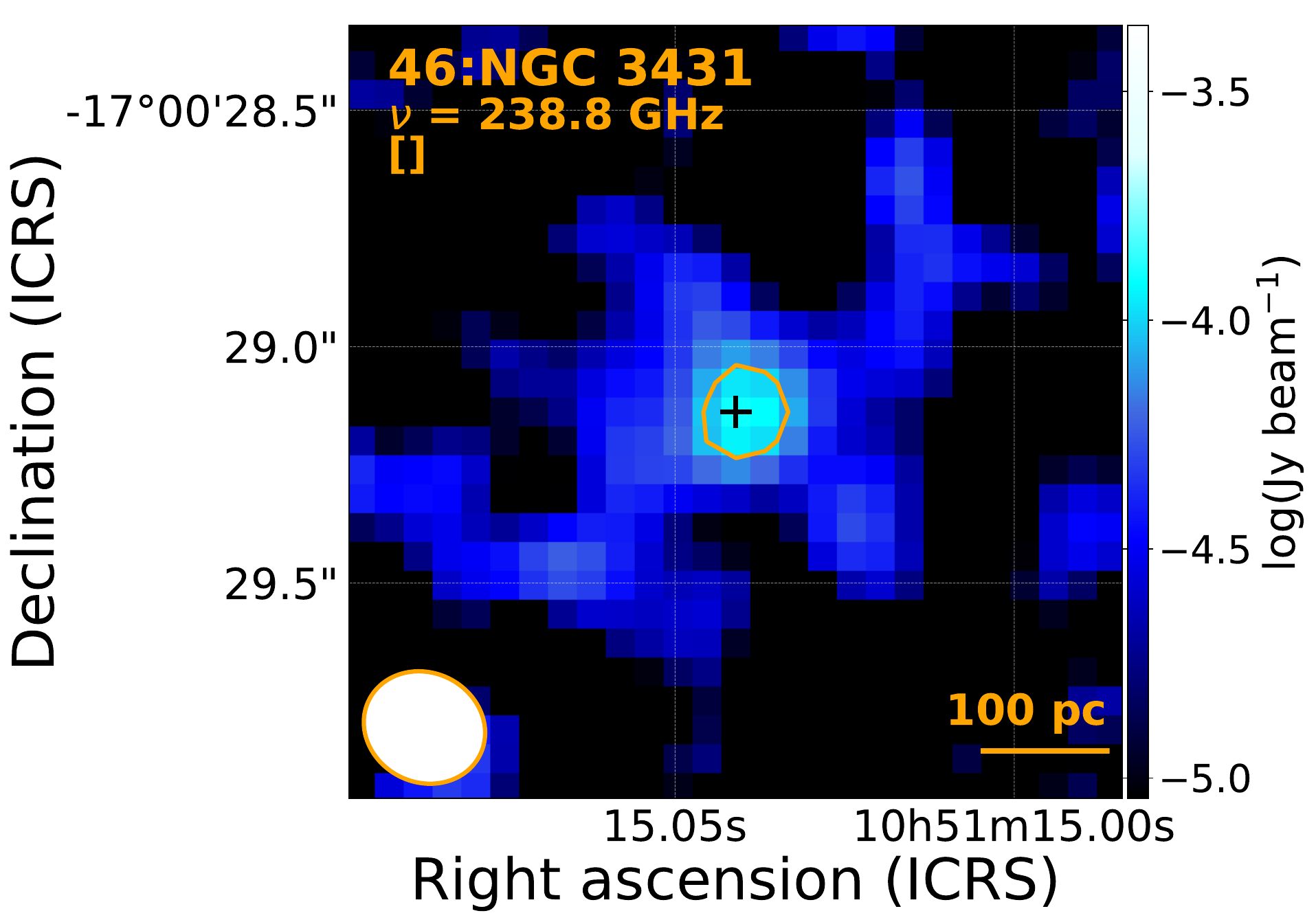}
\includegraphics[width=5.9cm]{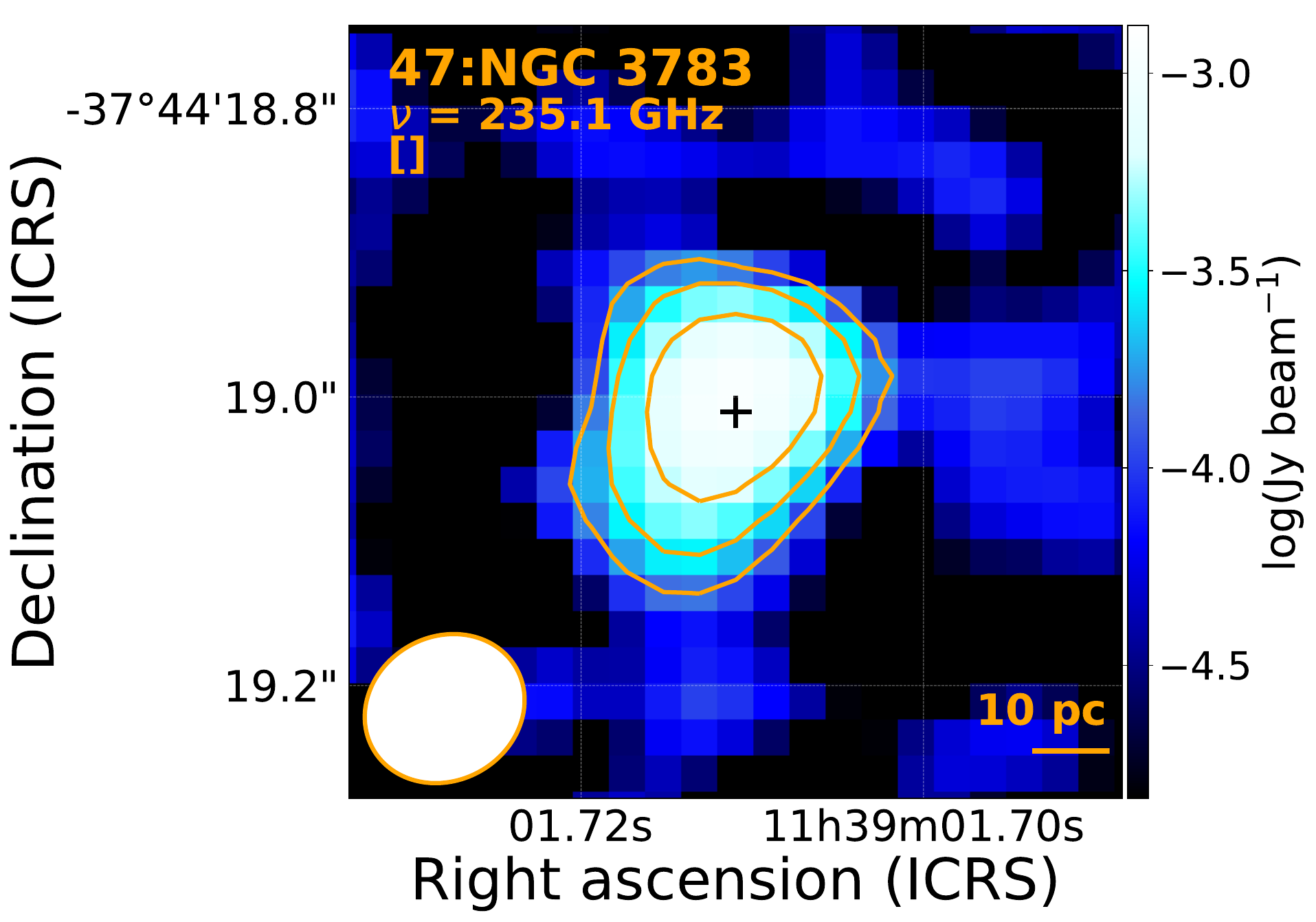}
\includegraphics[width=5.9cm]{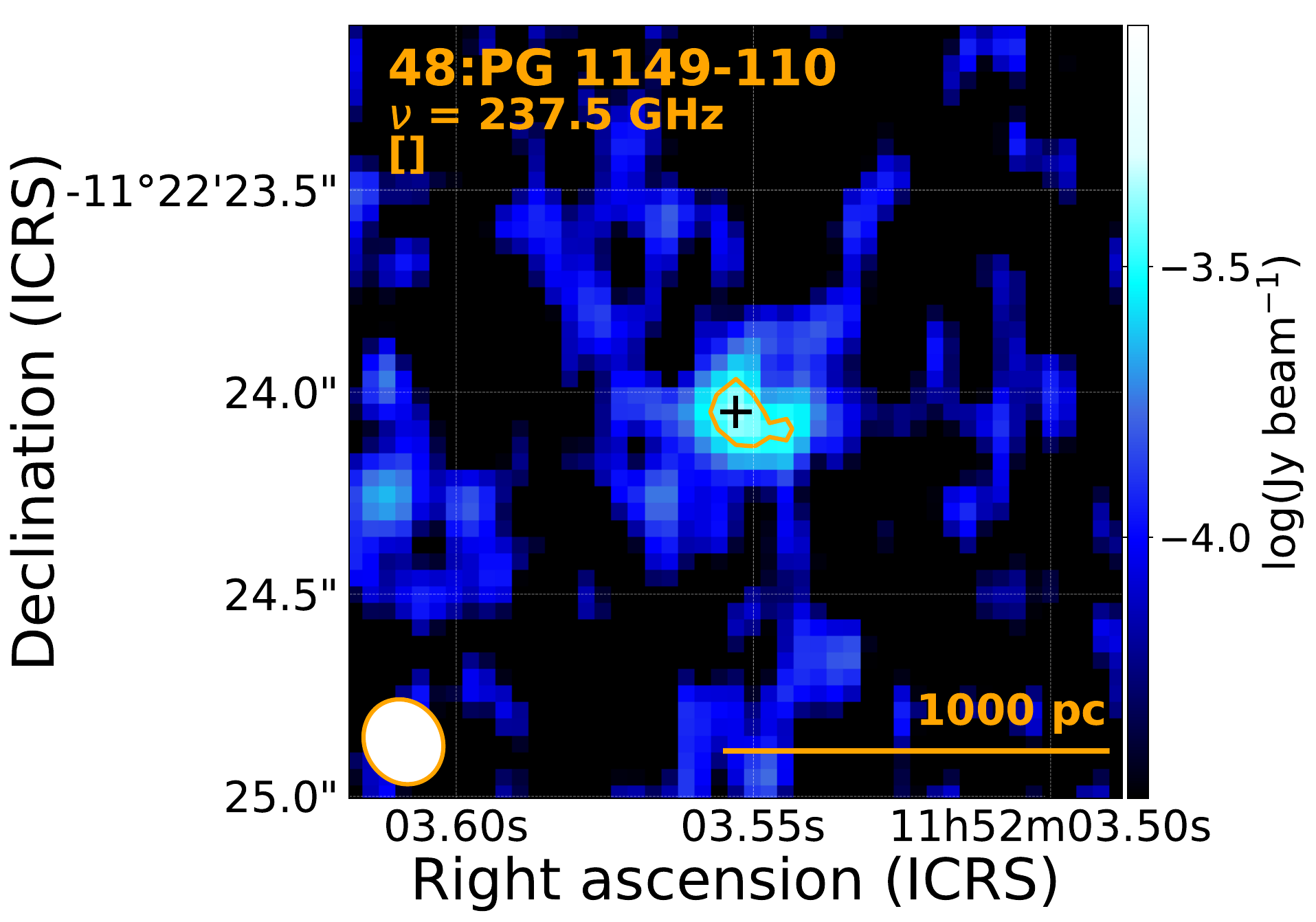}
\includegraphics[width=5.9cm]{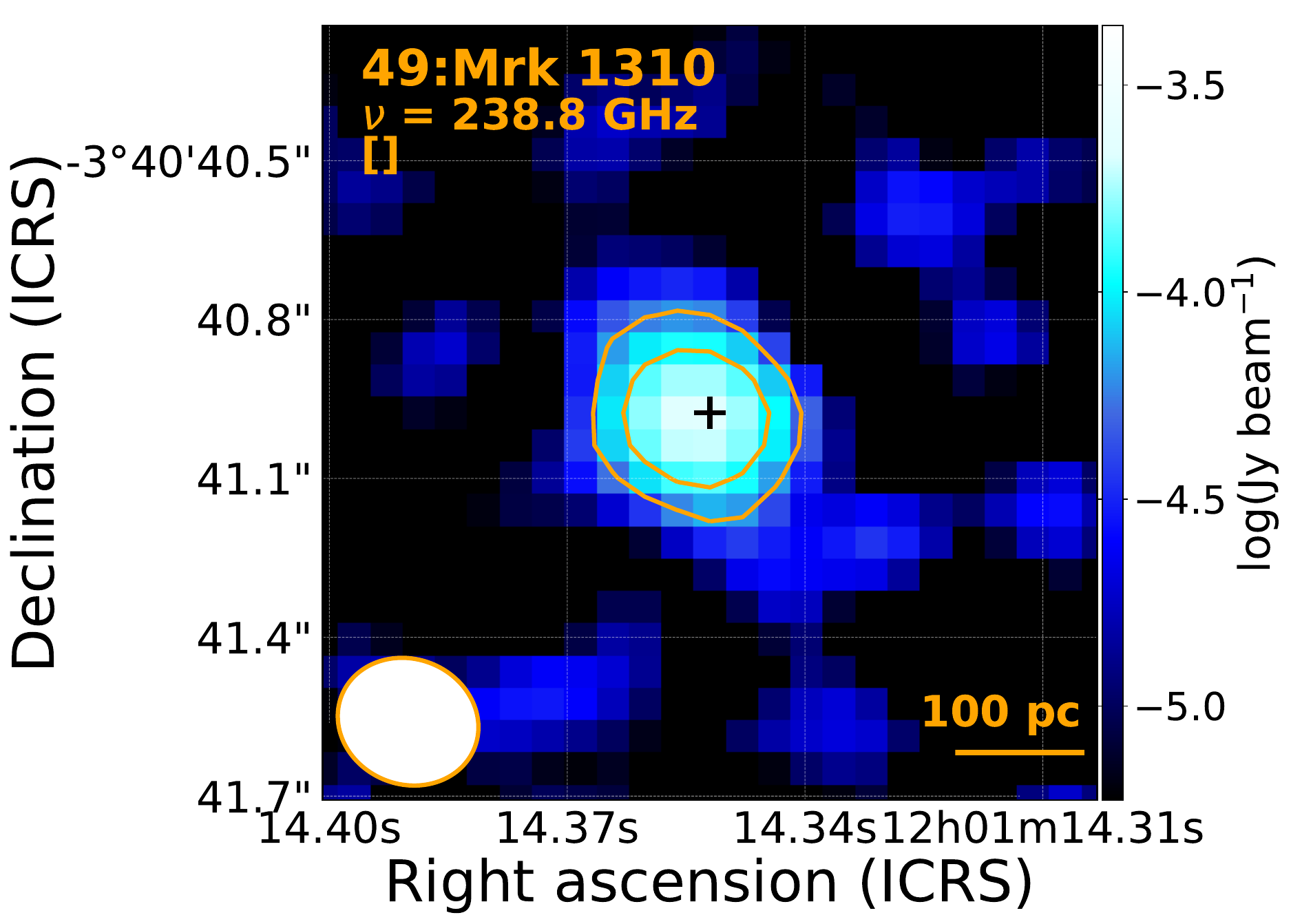}
\includegraphics[width=5.9cm]{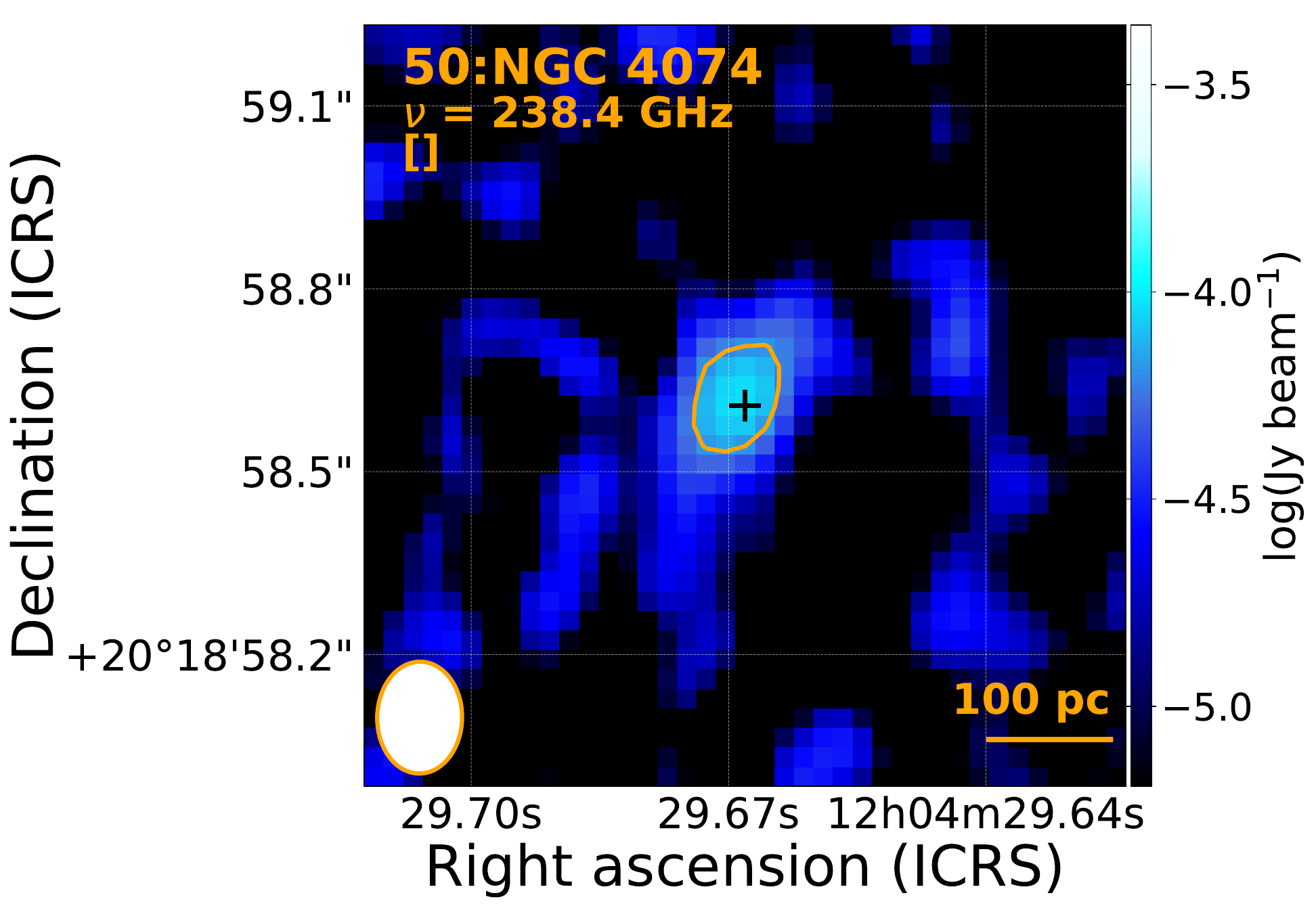}
\includegraphics[width=5.9cm]{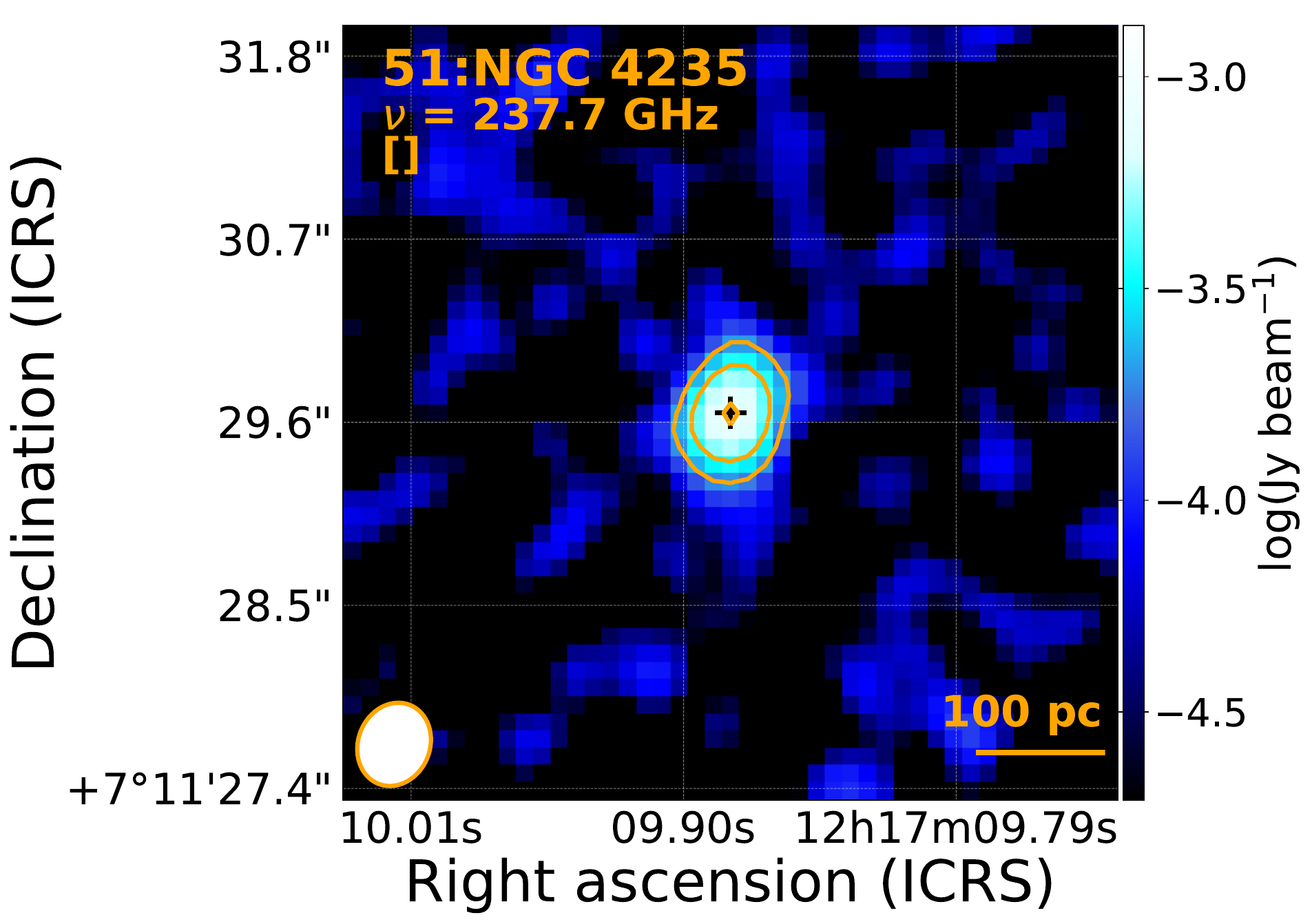}
\includegraphics[width=5.9cm]{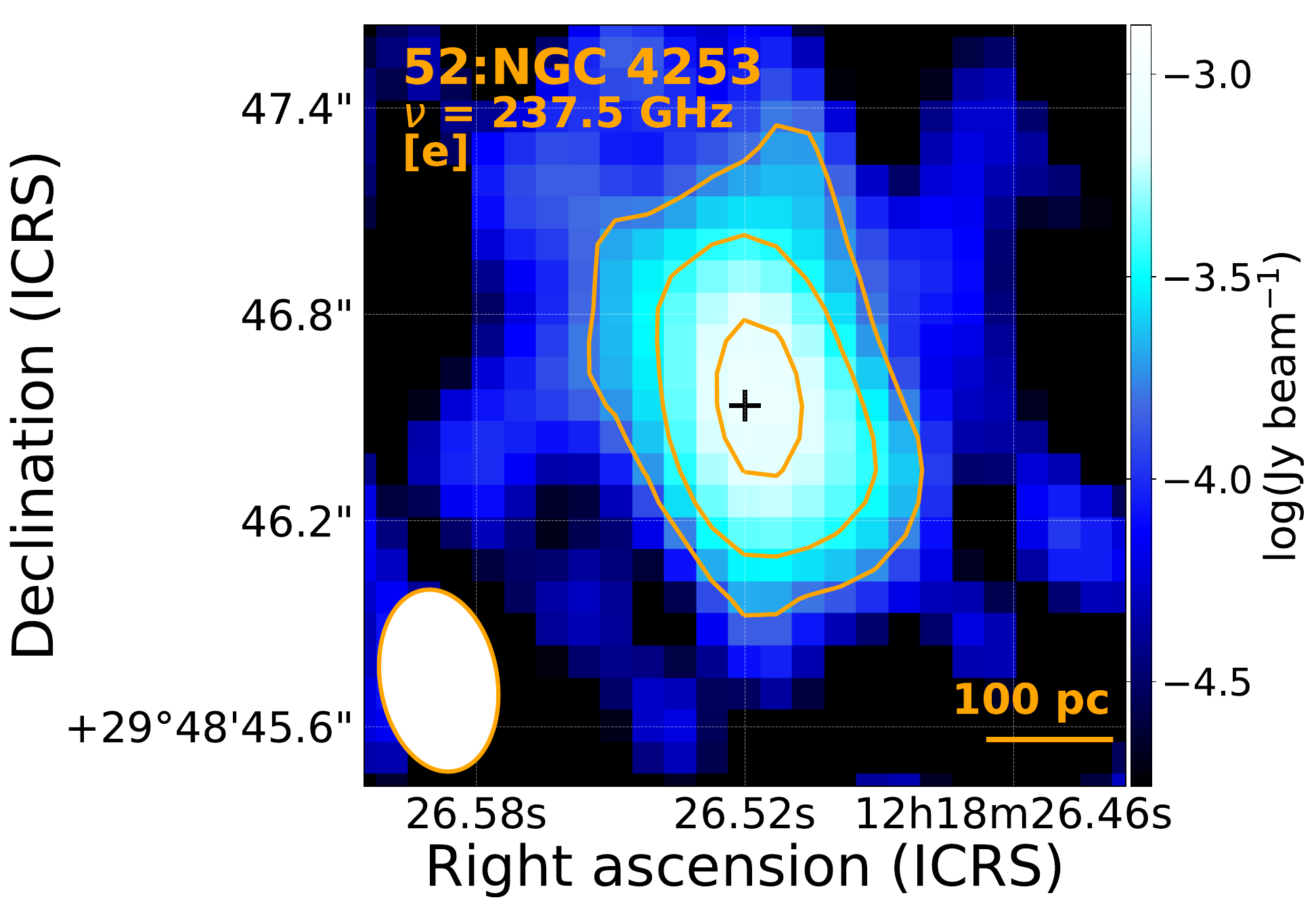}
\includegraphics[width=5.9cm]{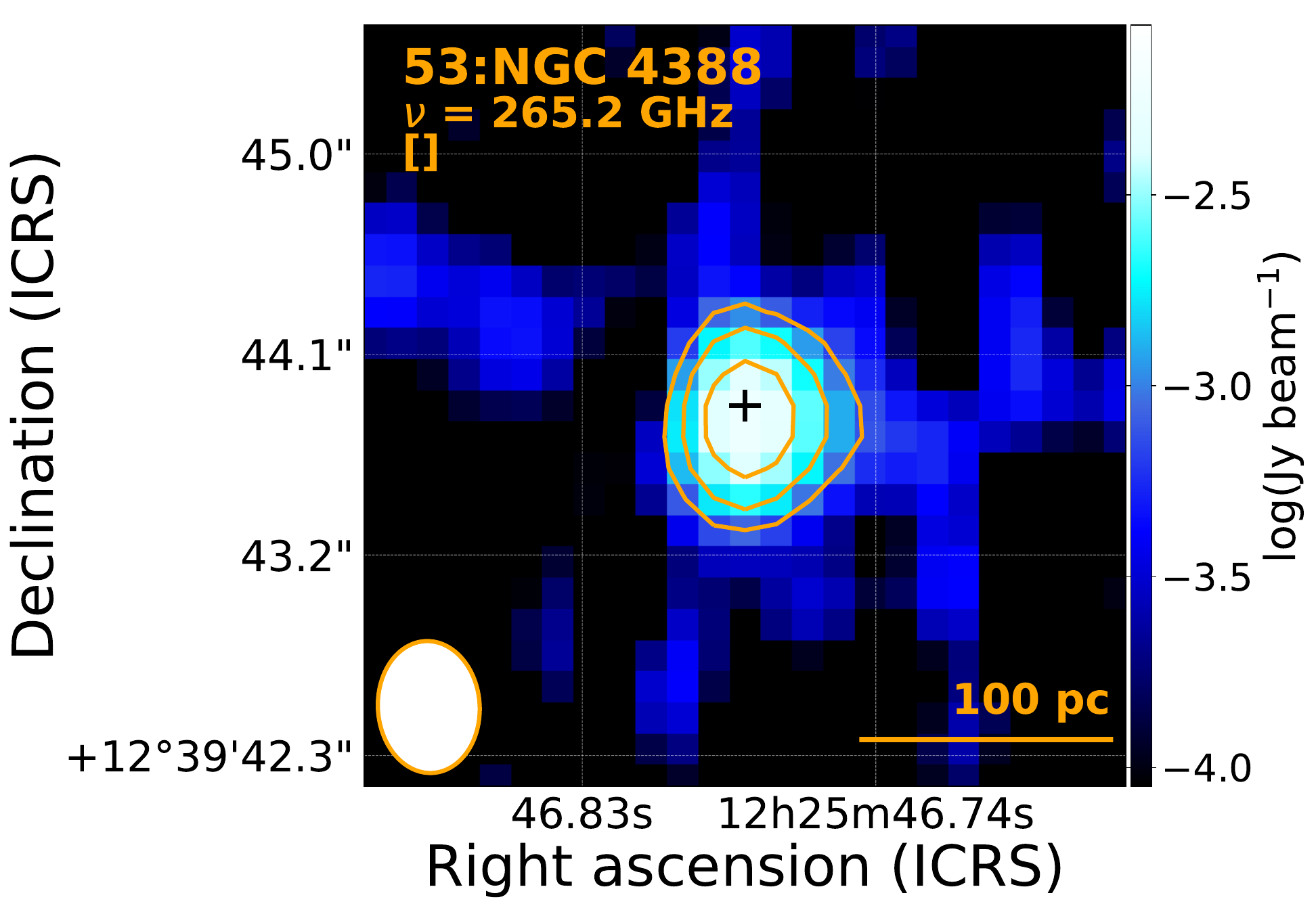}
\includegraphics[width=5.9cm]{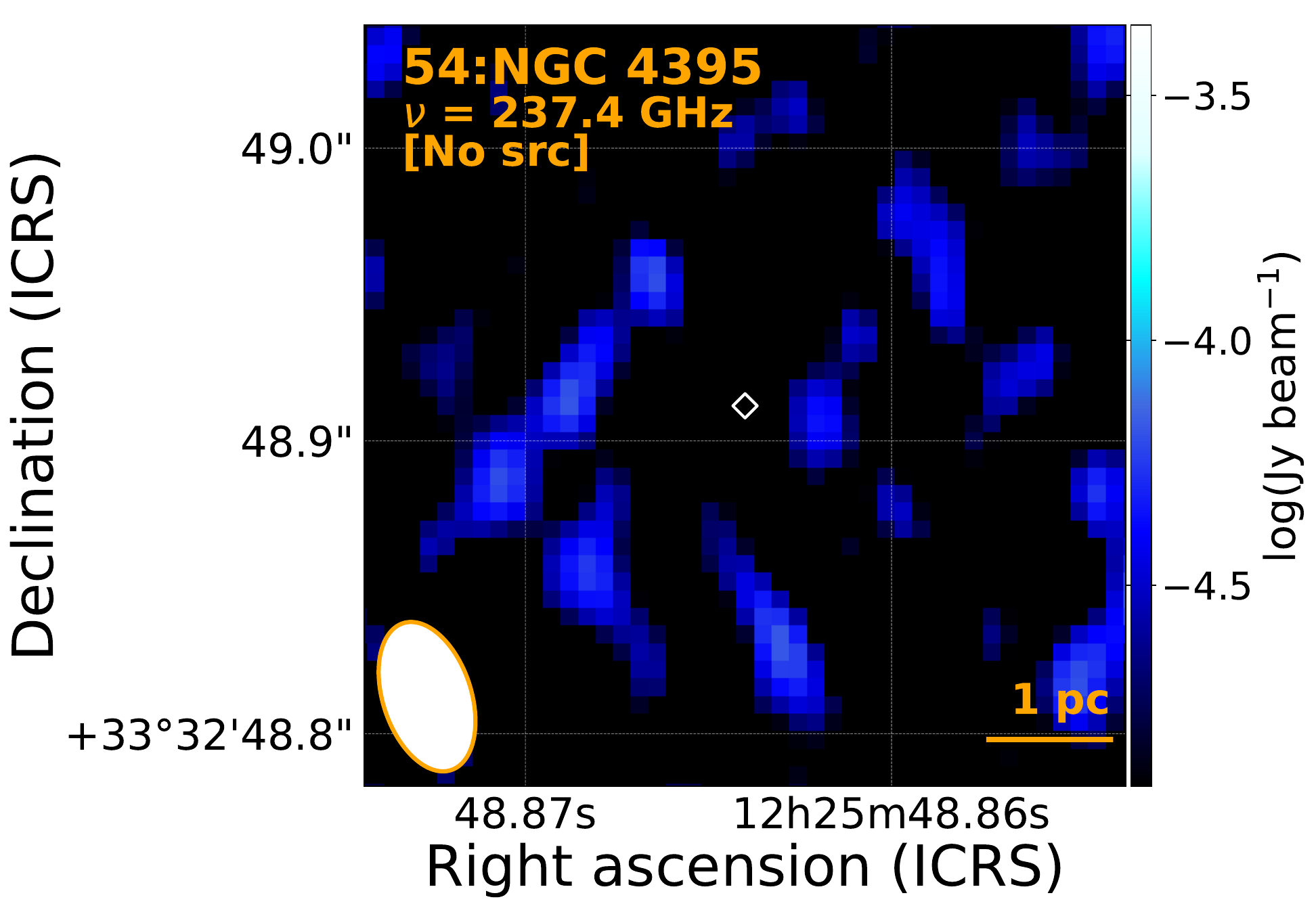}
\includegraphics[width=5.9cm]{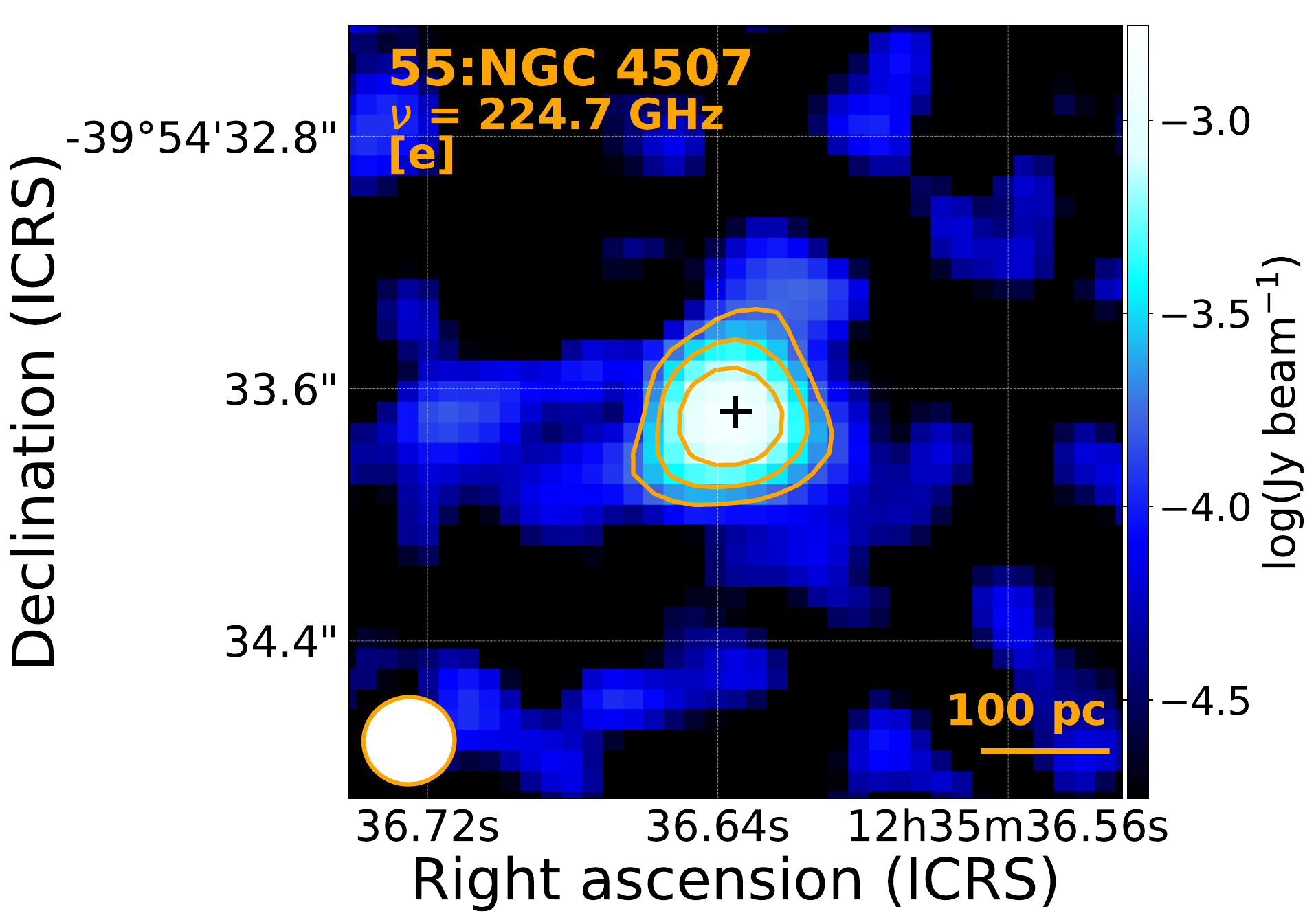}
\includegraphics[width=5.9cm]{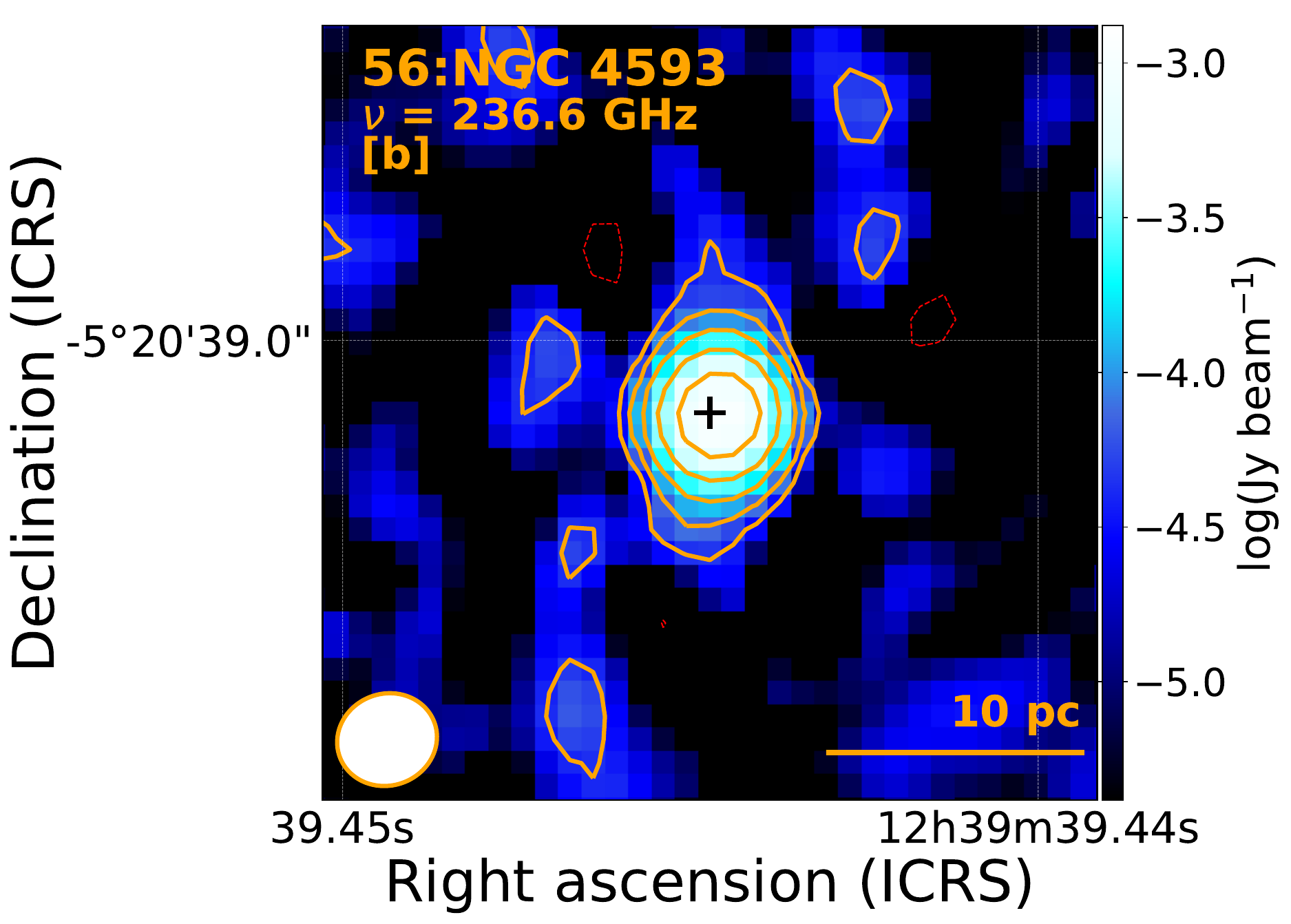}
\includegraphics[width=5.9cm]{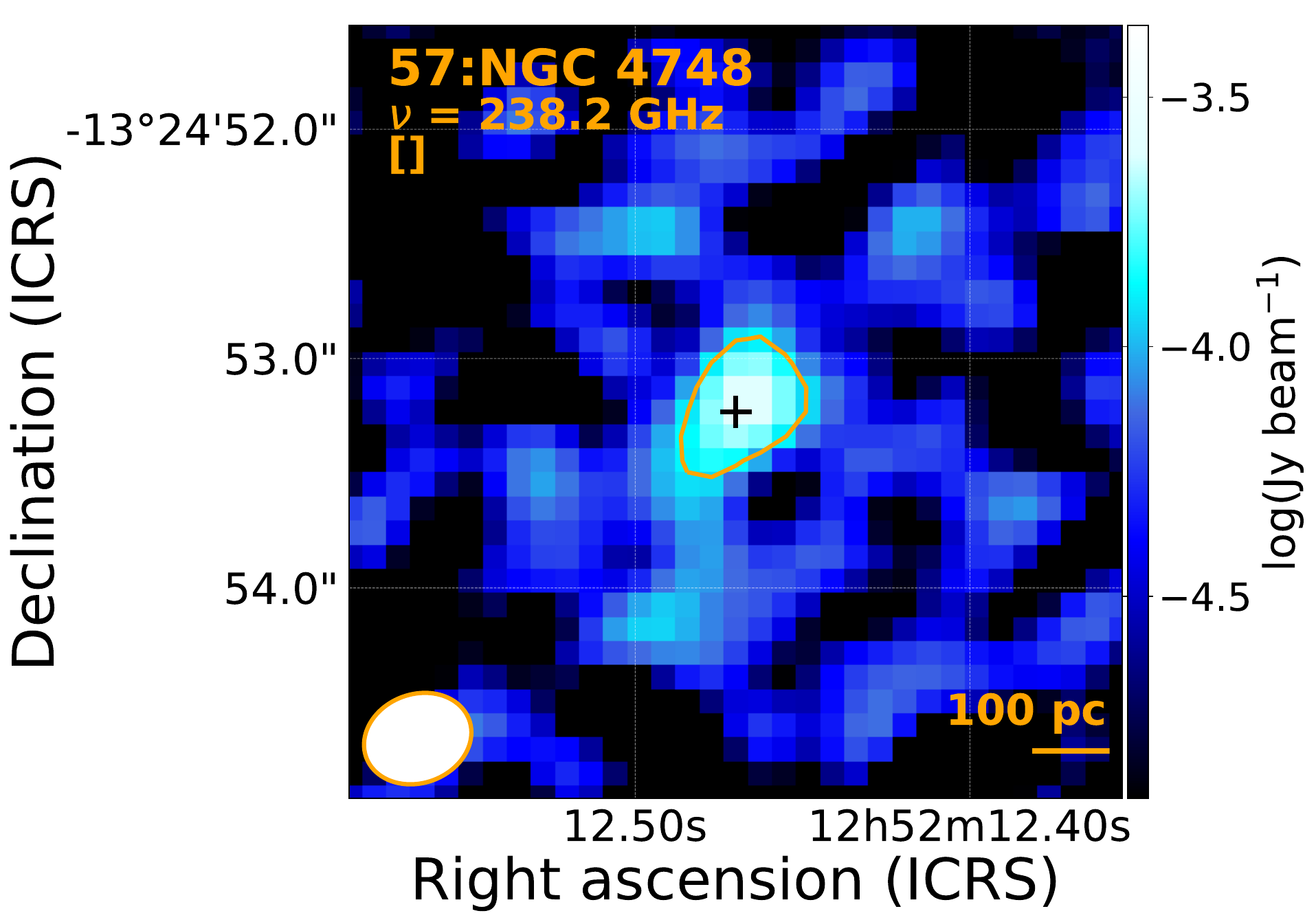}
    \caption{Continued. 
    }
\end{figure*}

\addtocounter{figure}{-1}

\begin{figure*}
    \centering
    \includegraphics[width=5.9cm]{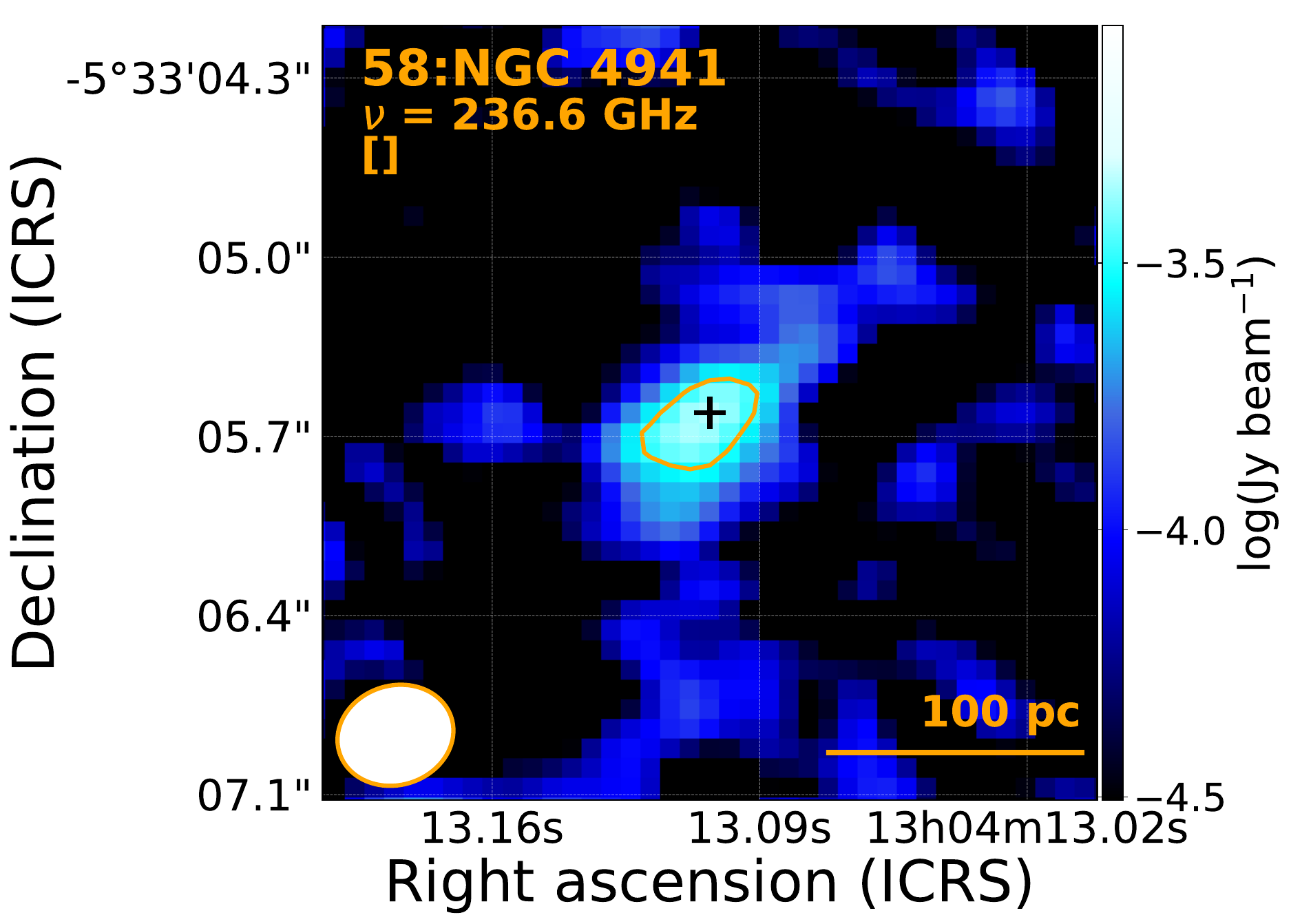}
\includegraphics[width=5.9cm]{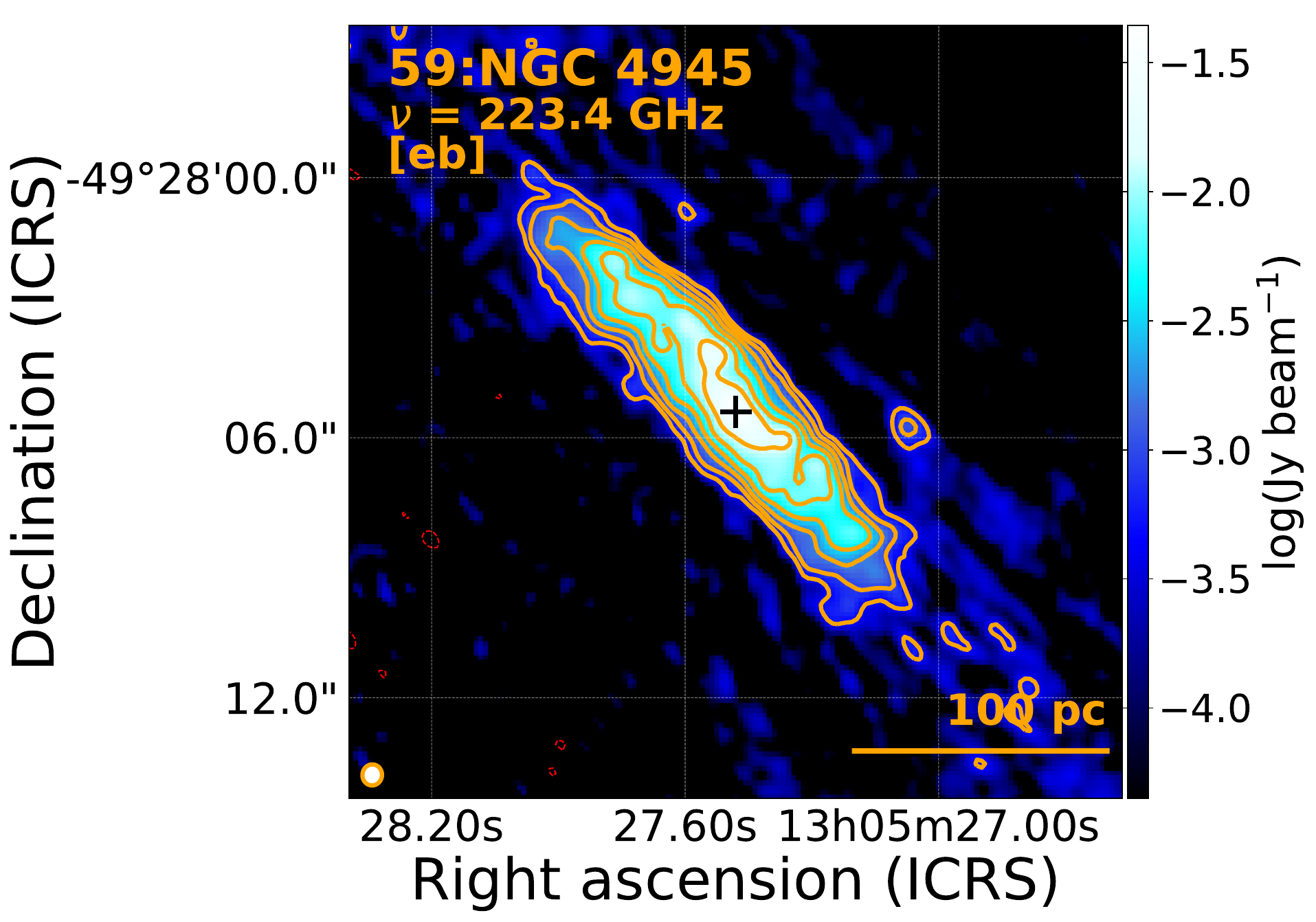}
\includegraphics[width=5.9cm]{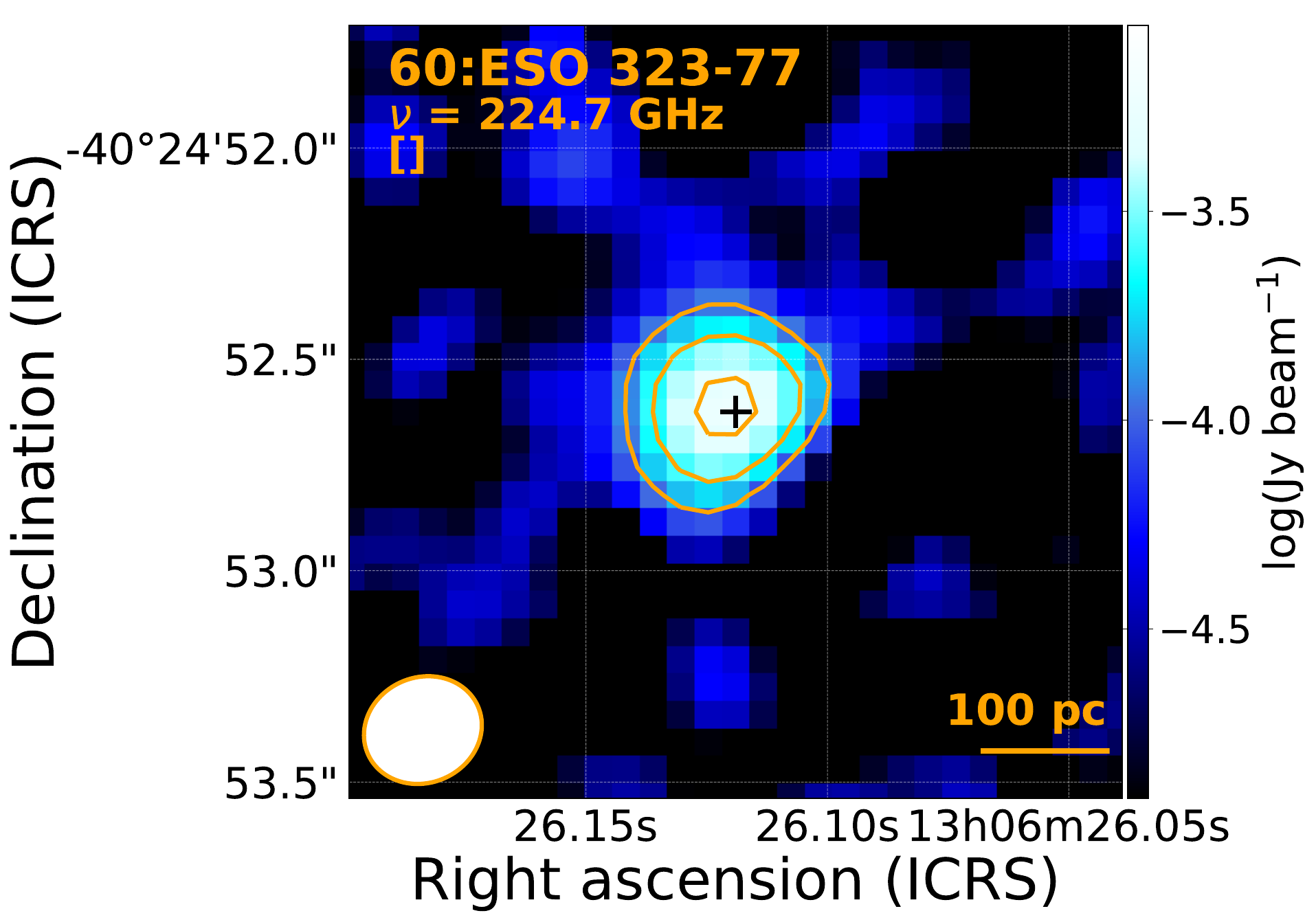}
\includegraphics[width=5.9cm]{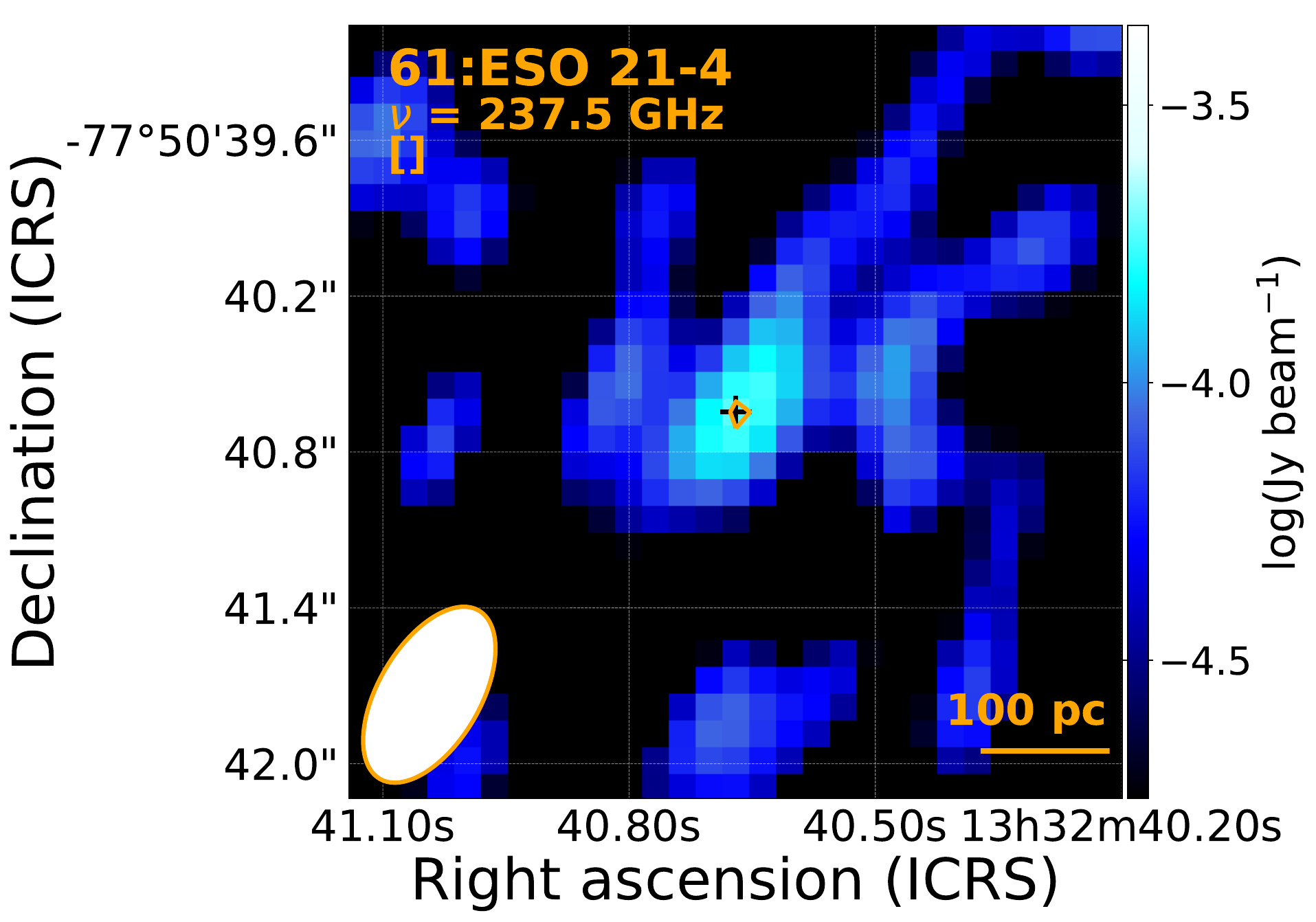}
\includegraphics[width=5.9cm]{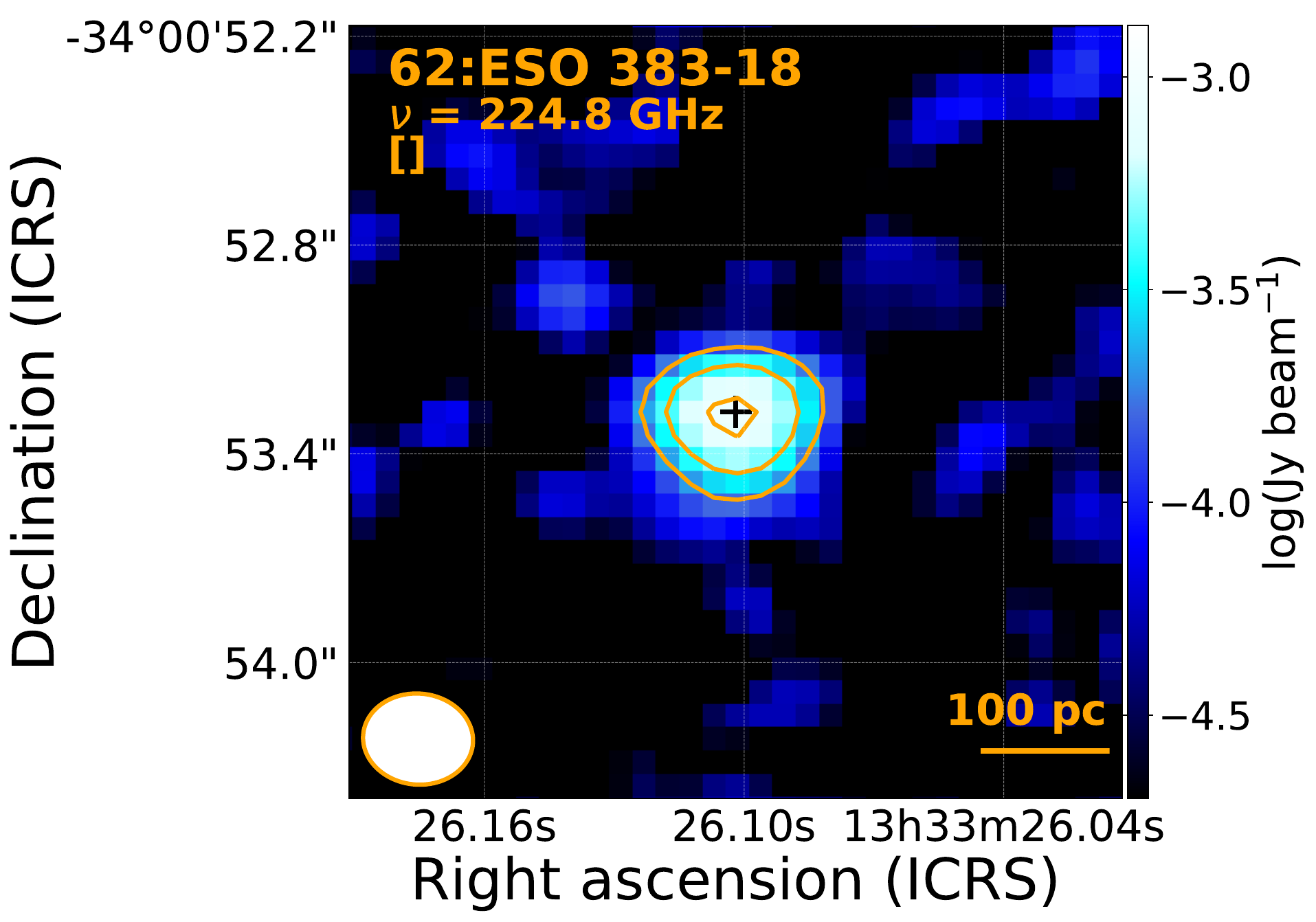}
\includegraphics[width=5.9cm]{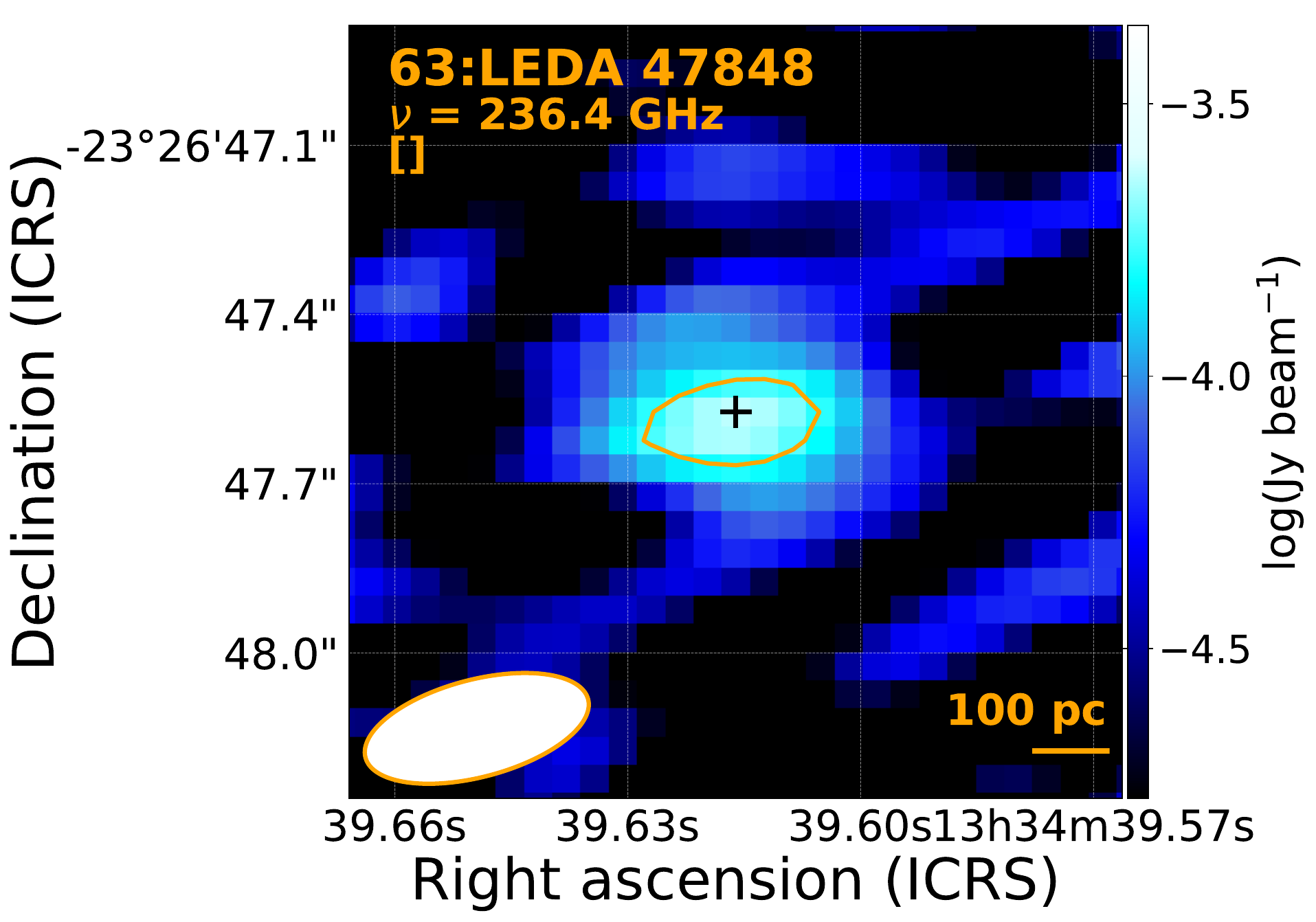}
\includegraphics[width=5.9cm]{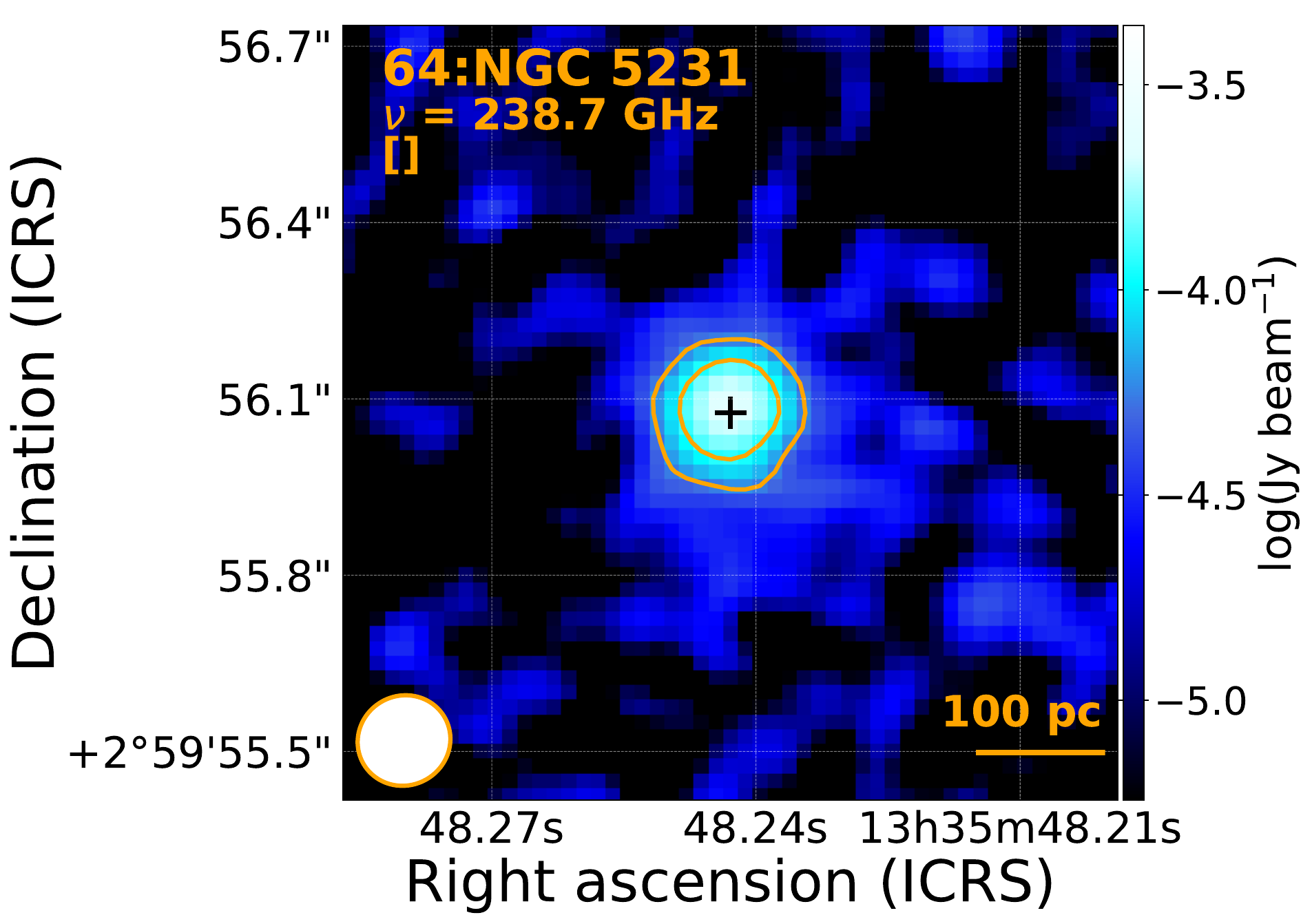}
\includegraphics[width=5.9cm]{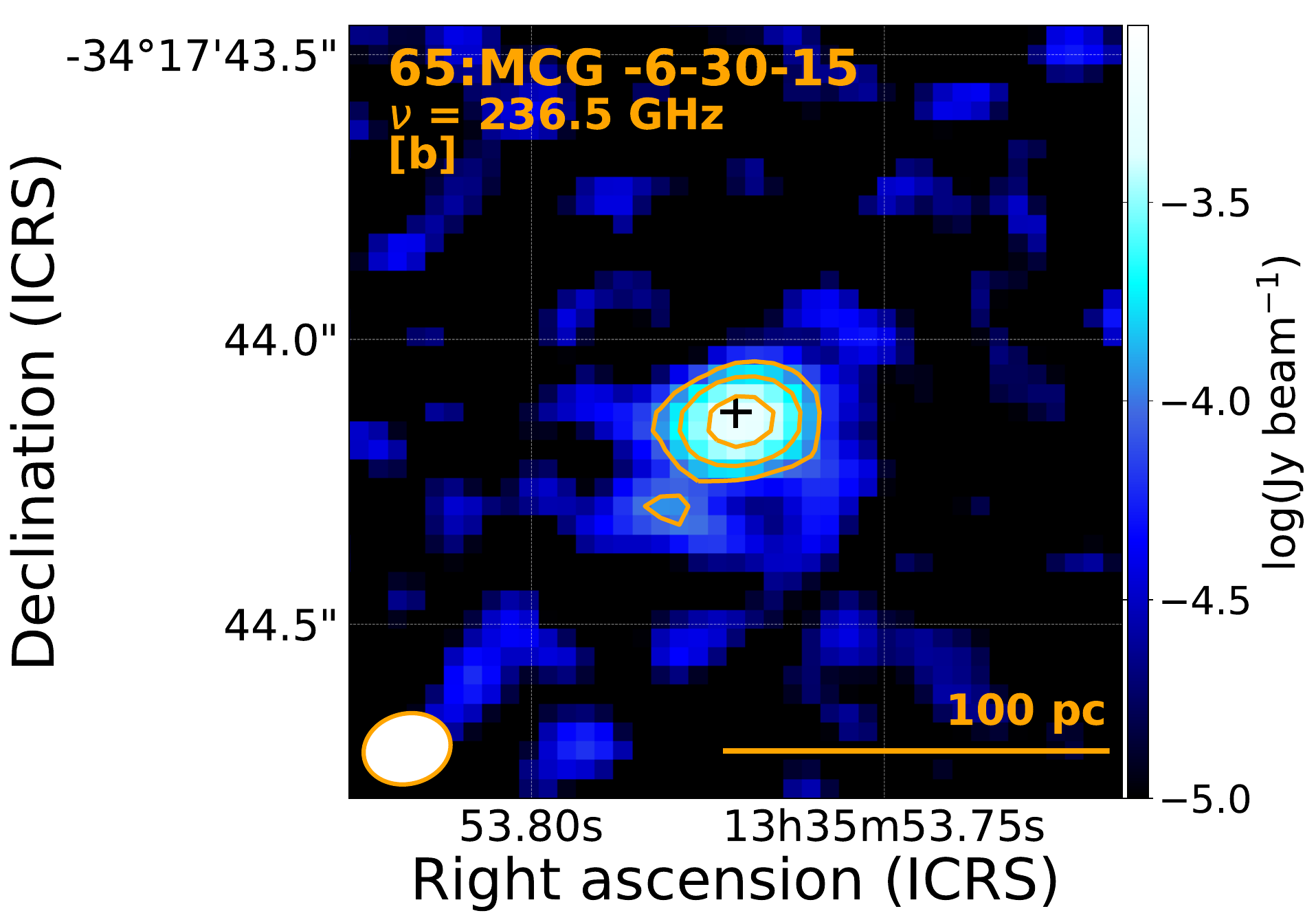}
\includegraphics[width=5.9cm]{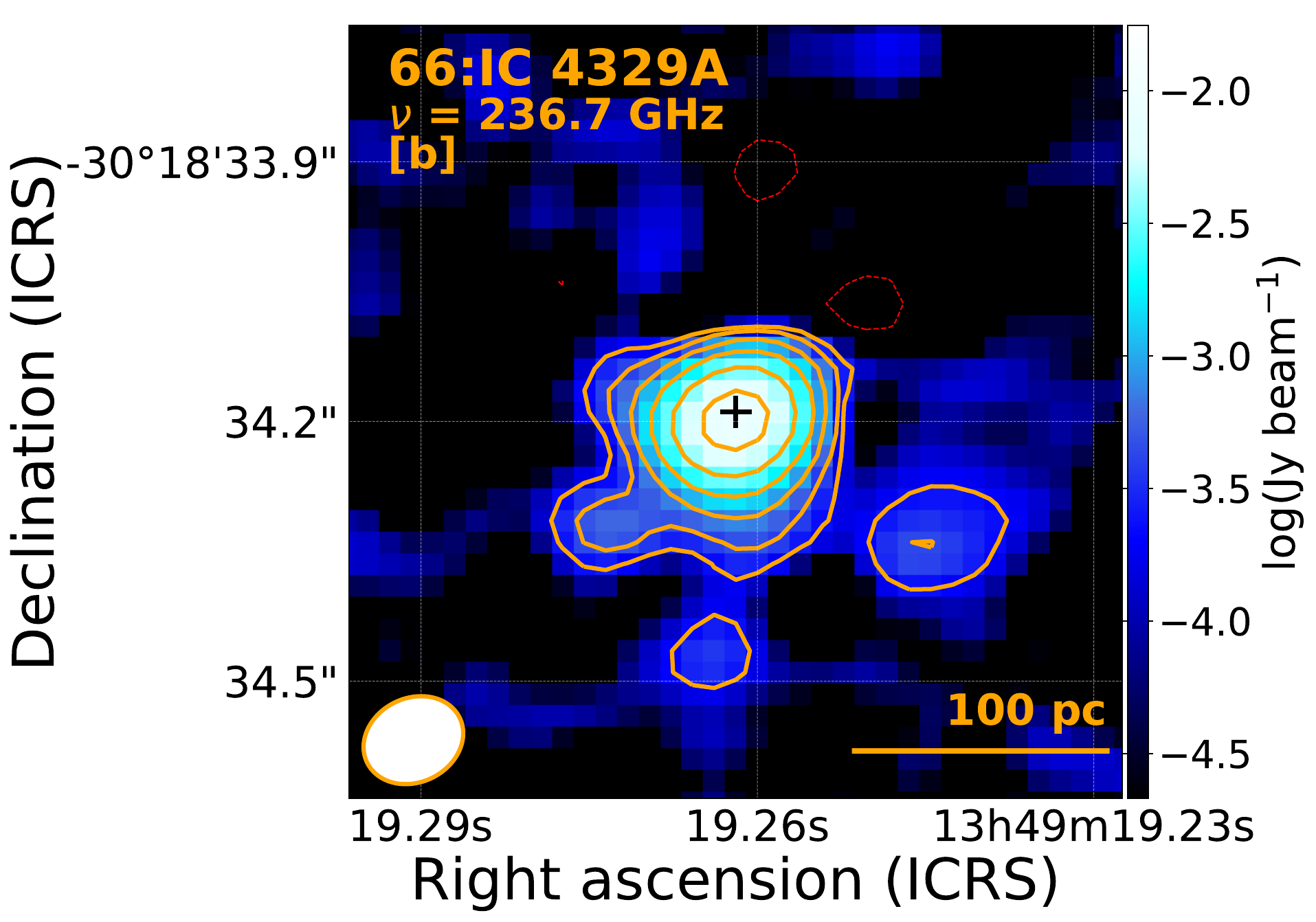}
\includegraphics[width=5.9cm]{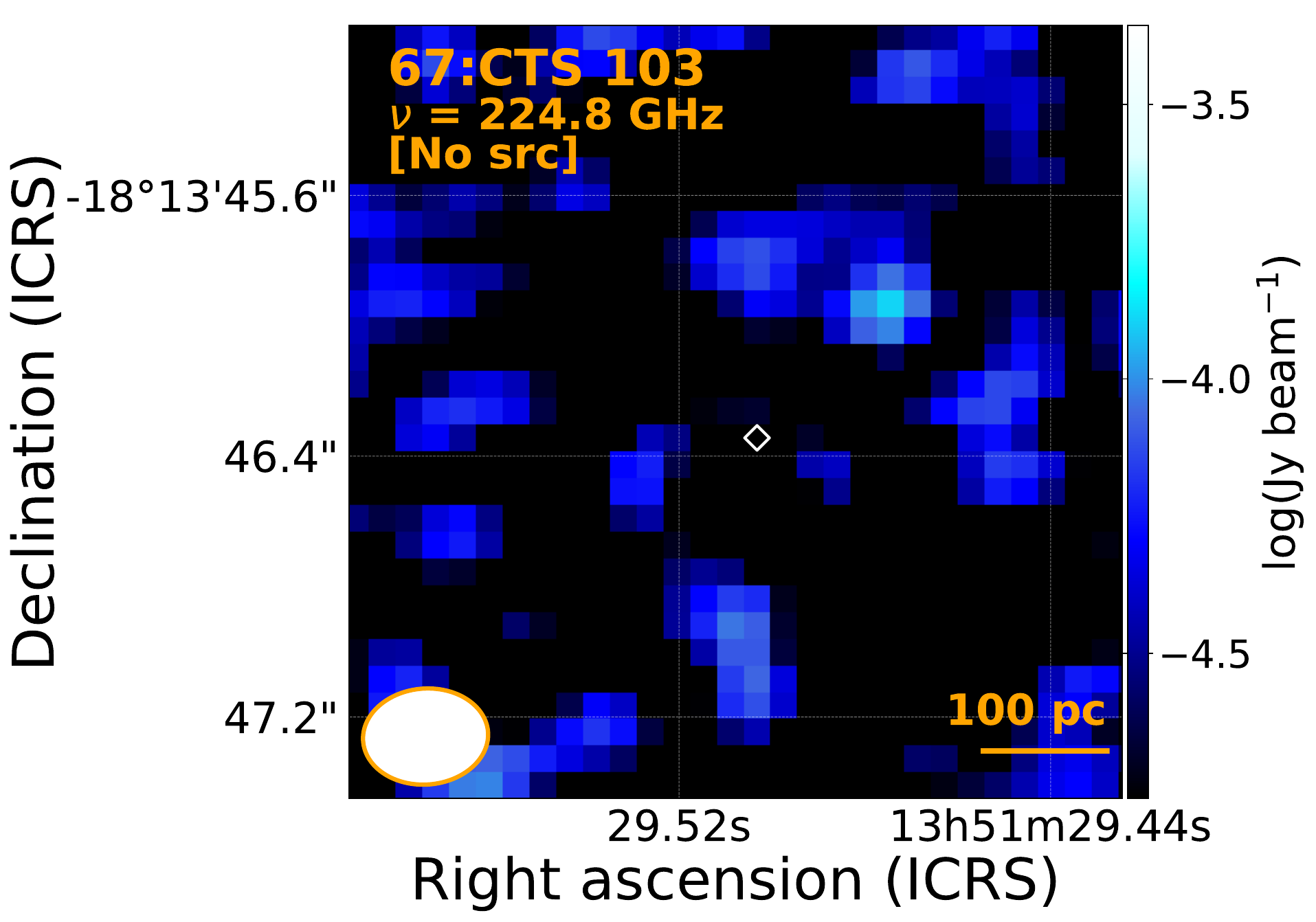}
\includegraphics[width=5.9cm]{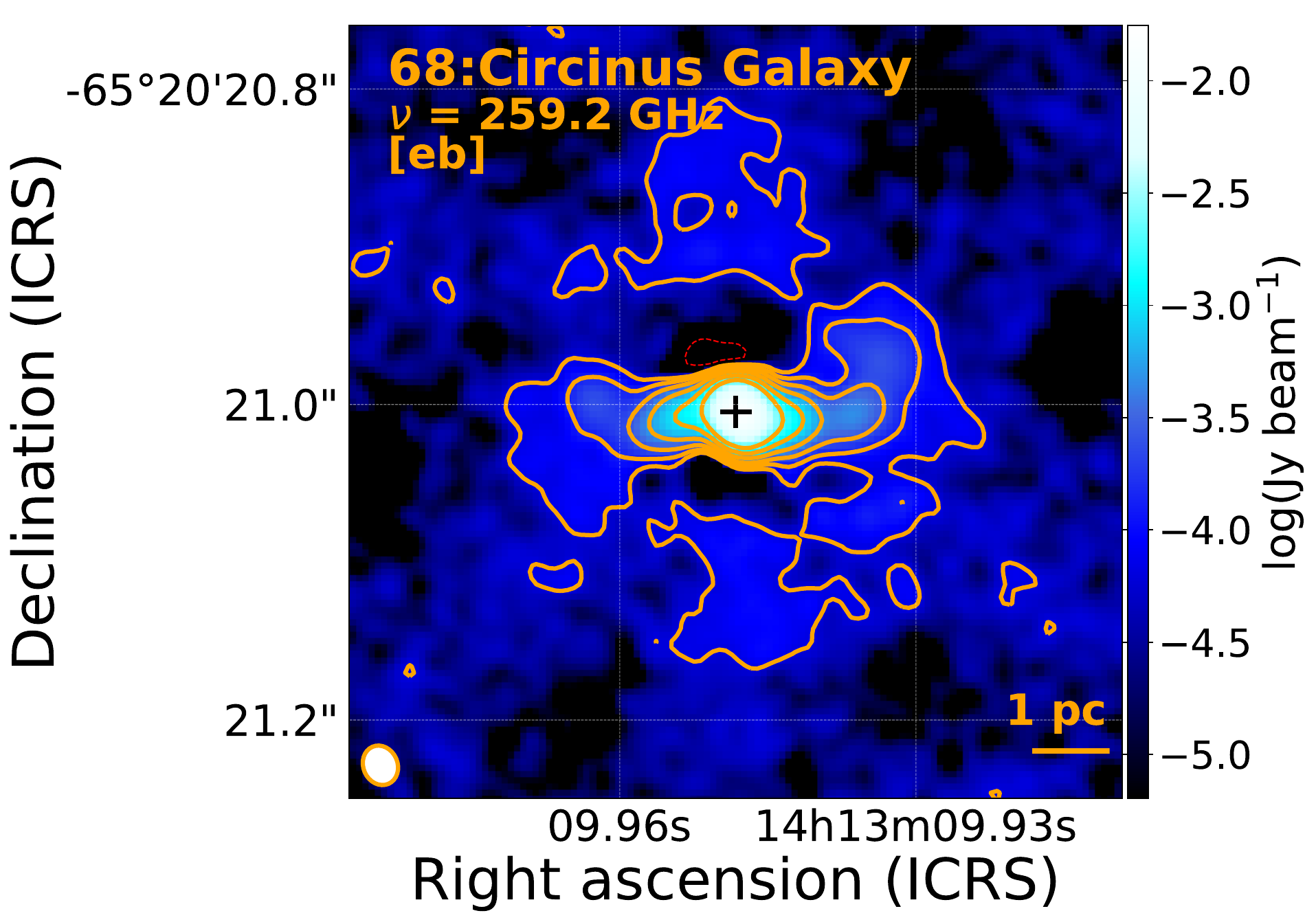}
\includegraphics[width=5.9cm]{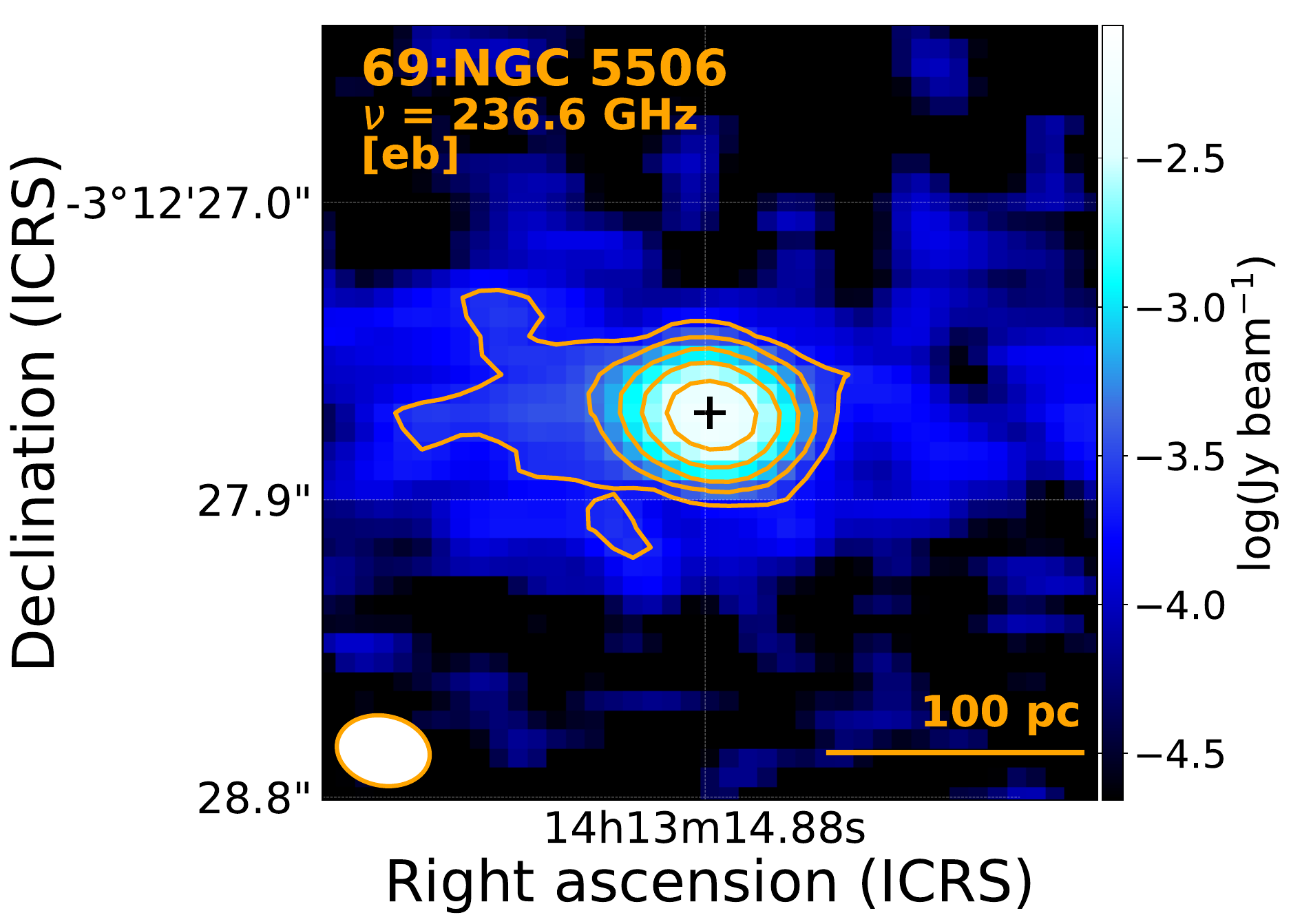}
\includegraphics[width=5.9cm]{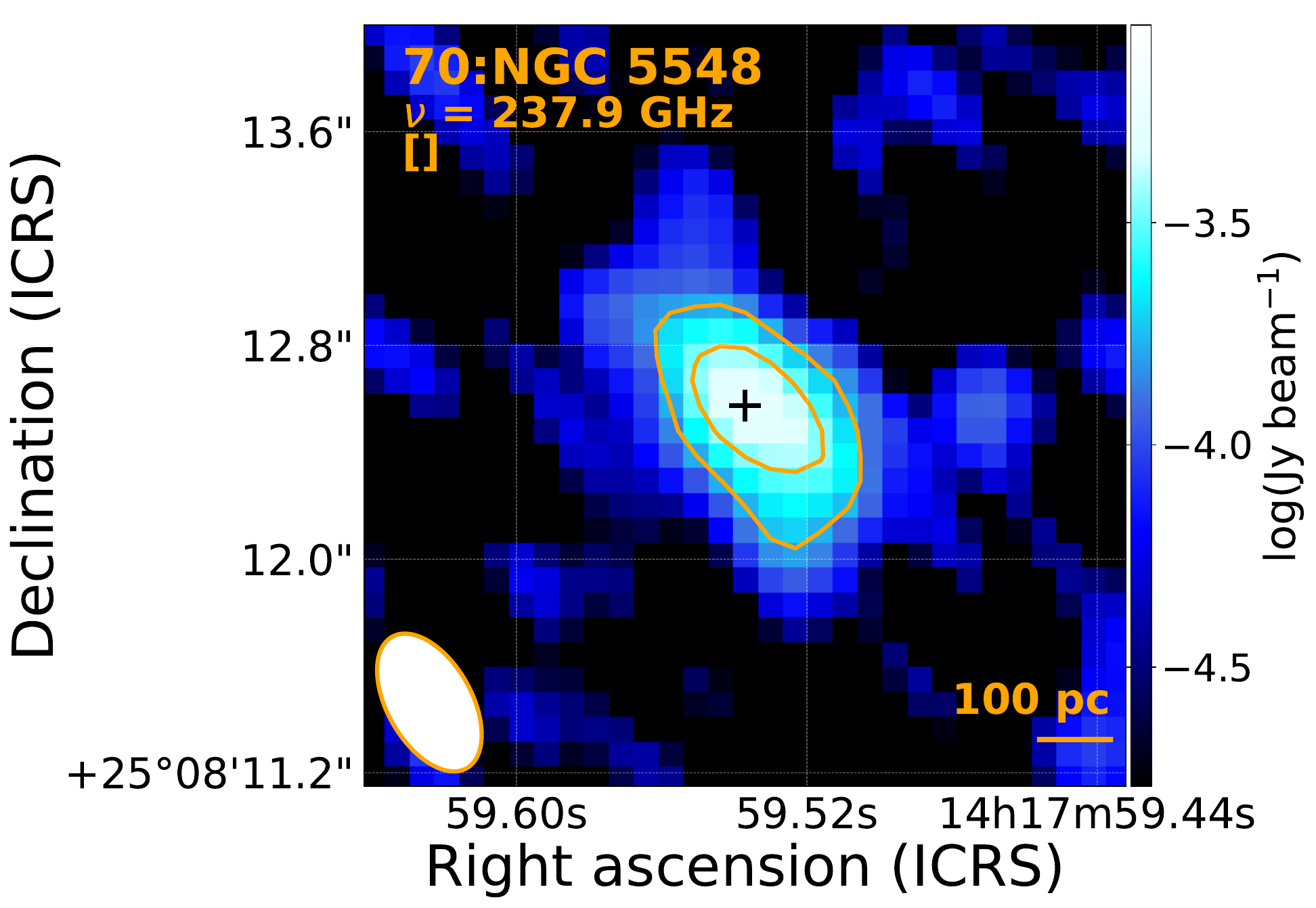}
\includegraphics[width=5.9cm]{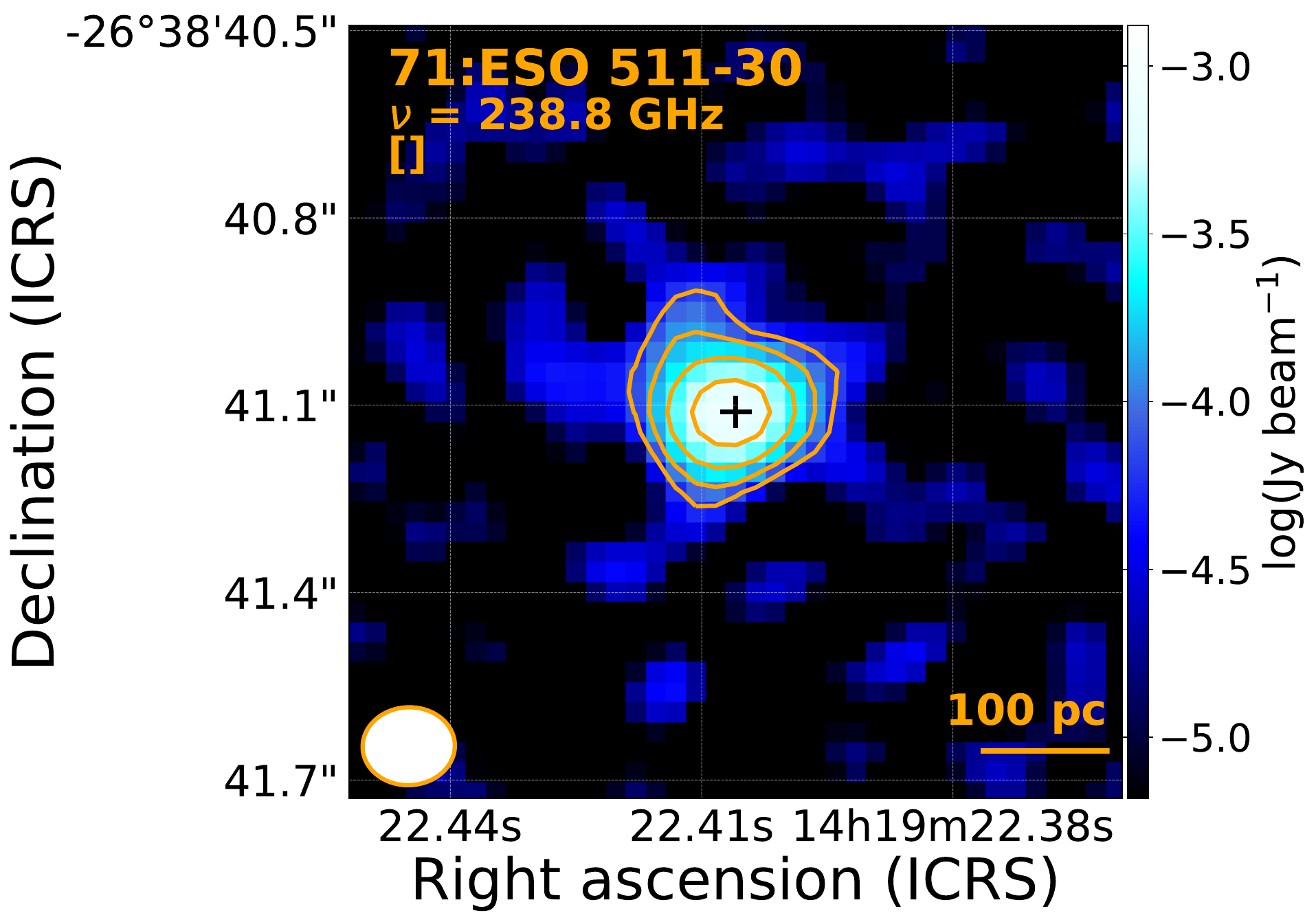}
\includegraphics[width=5.9cm]{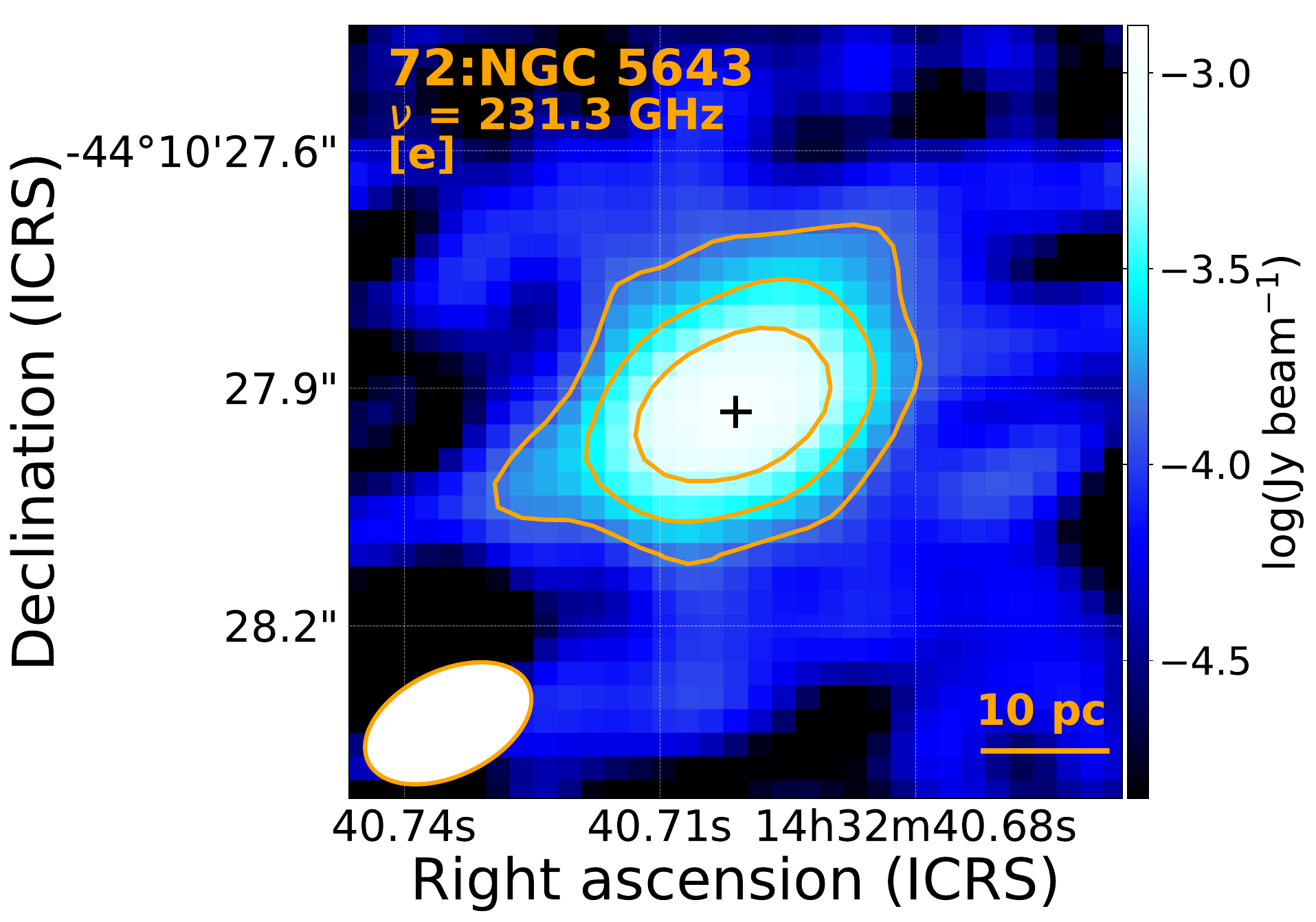}
    \caption{Continued. 
    }
\end{figure*}    
    
\addtocounter{figure}{-1}

\begin{figure*}
    \centering    
    \includegraphics[width=5.9cm]{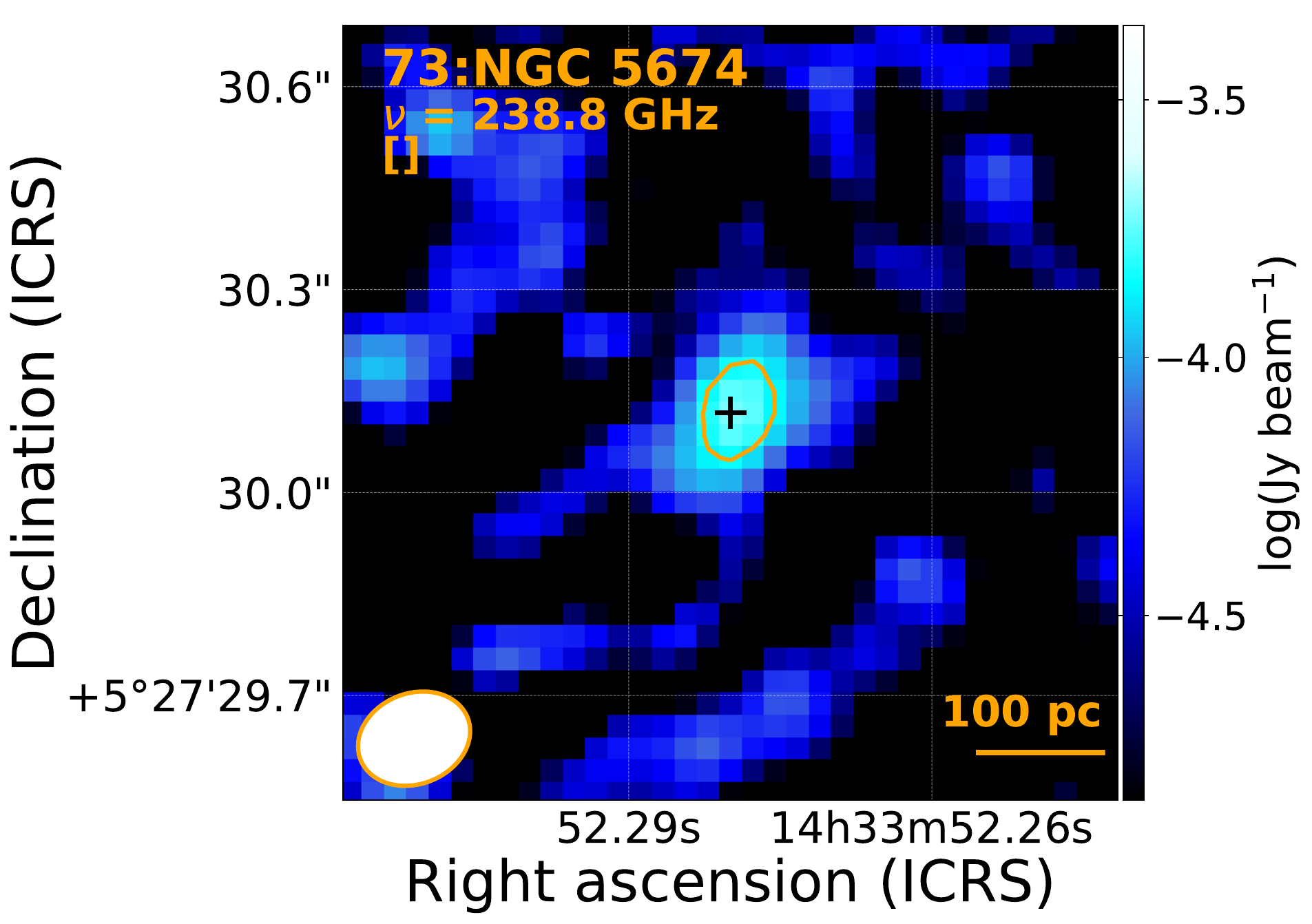}
\includegraphics[width=5.9cm]{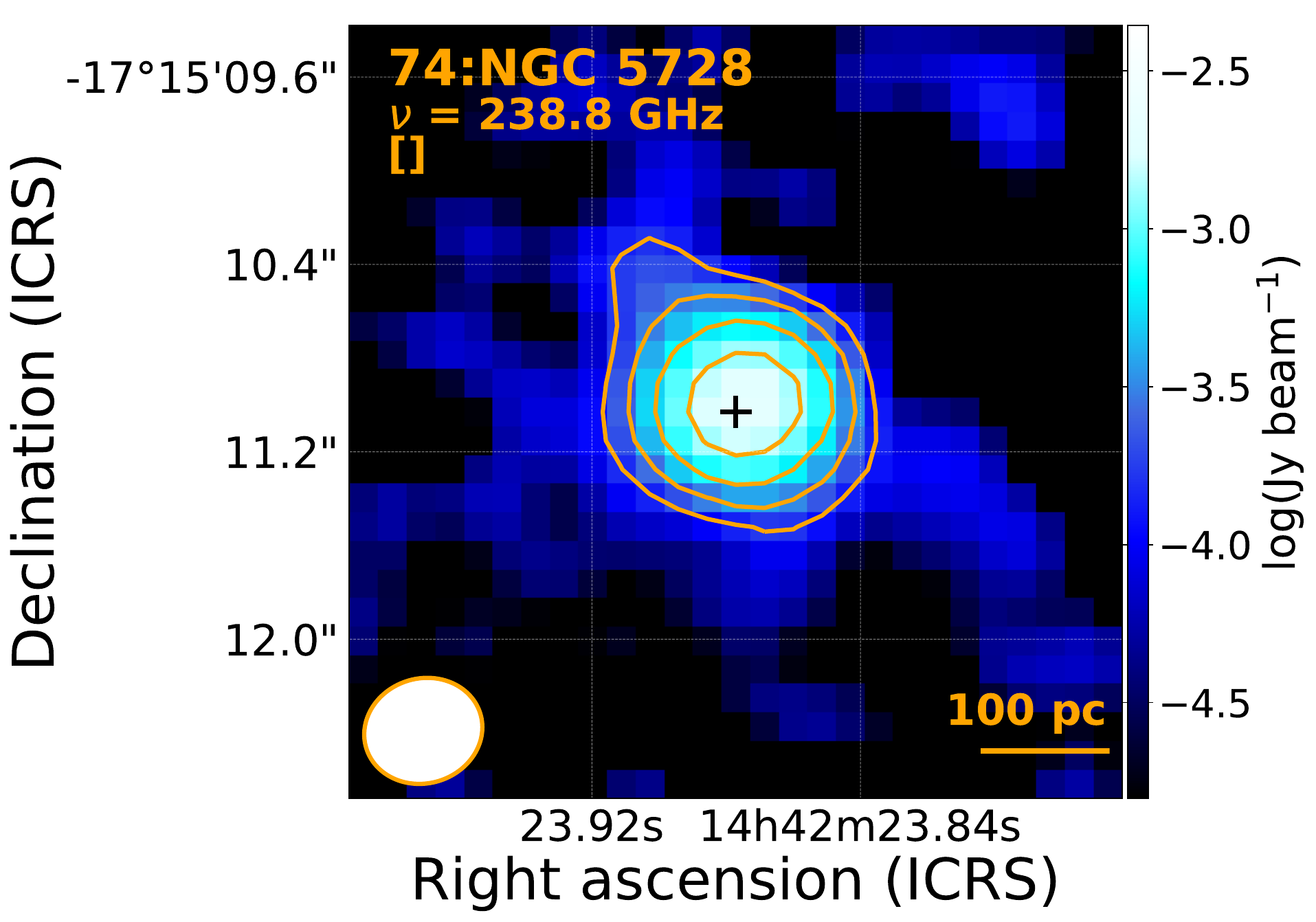}
\includegraphics[width=5.9cm]{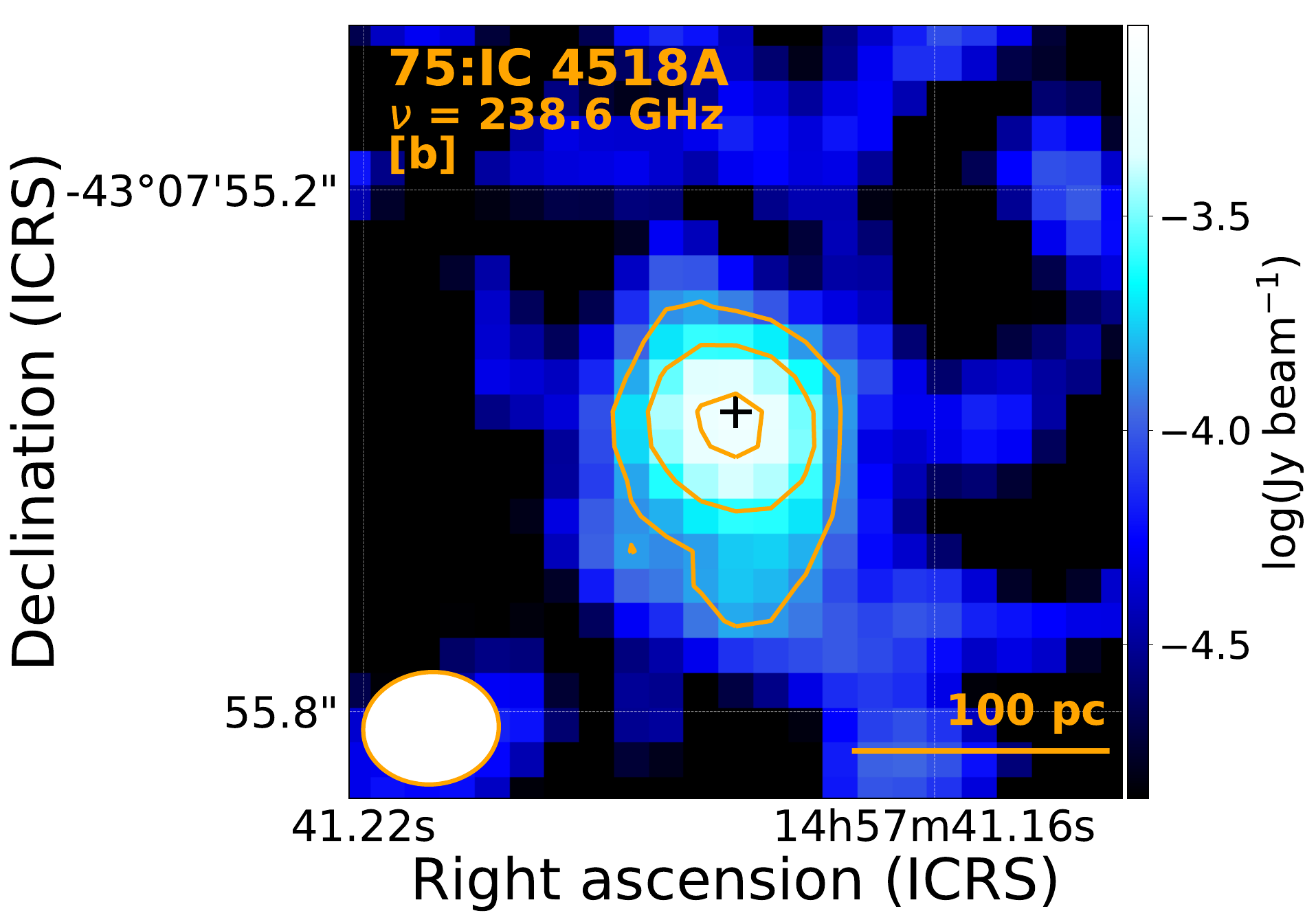}
\includegraphics[width=5.9cm]{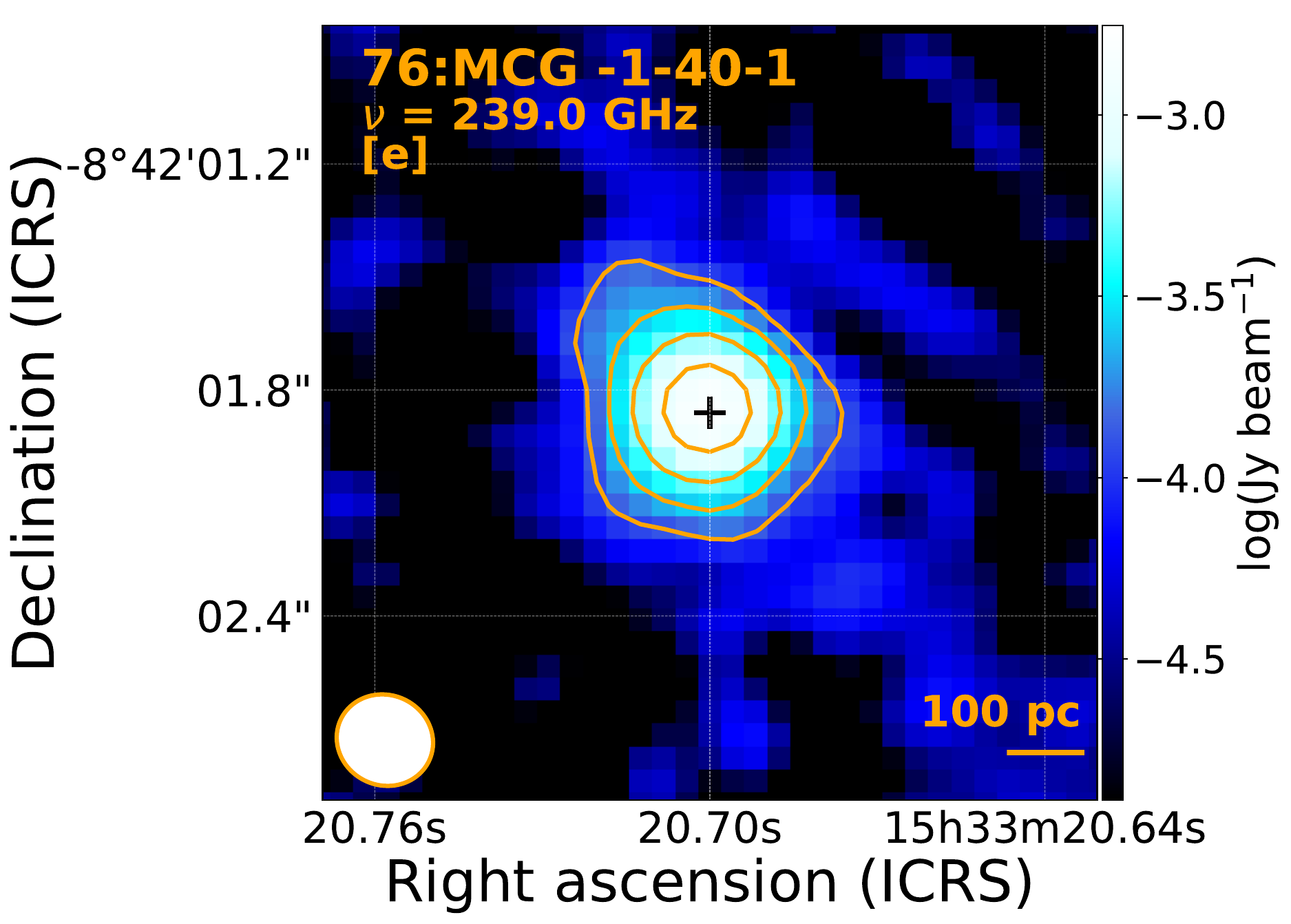}
\includegraphics[width=5.9cm]{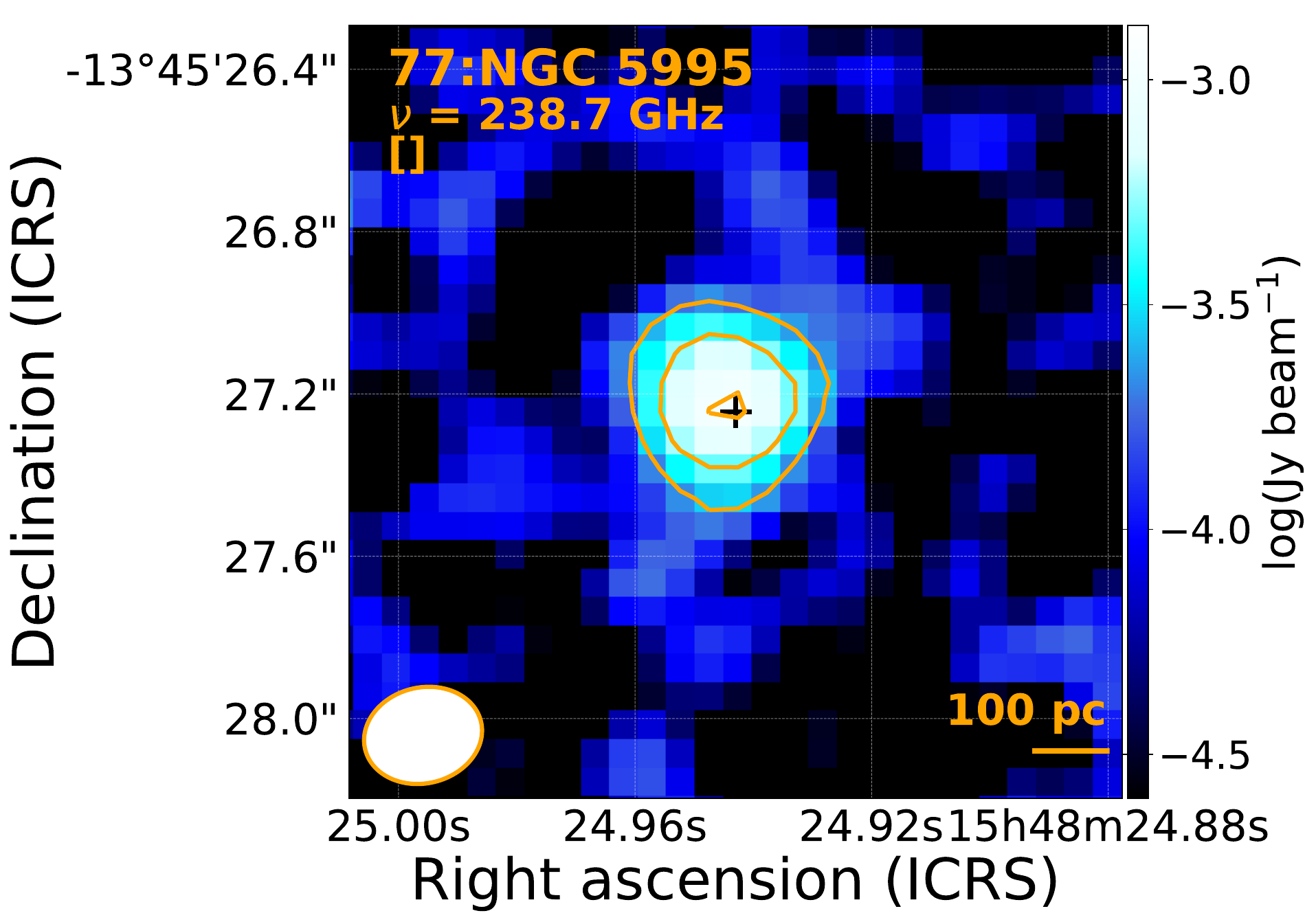}
\includegraphics[width=5.9cm]{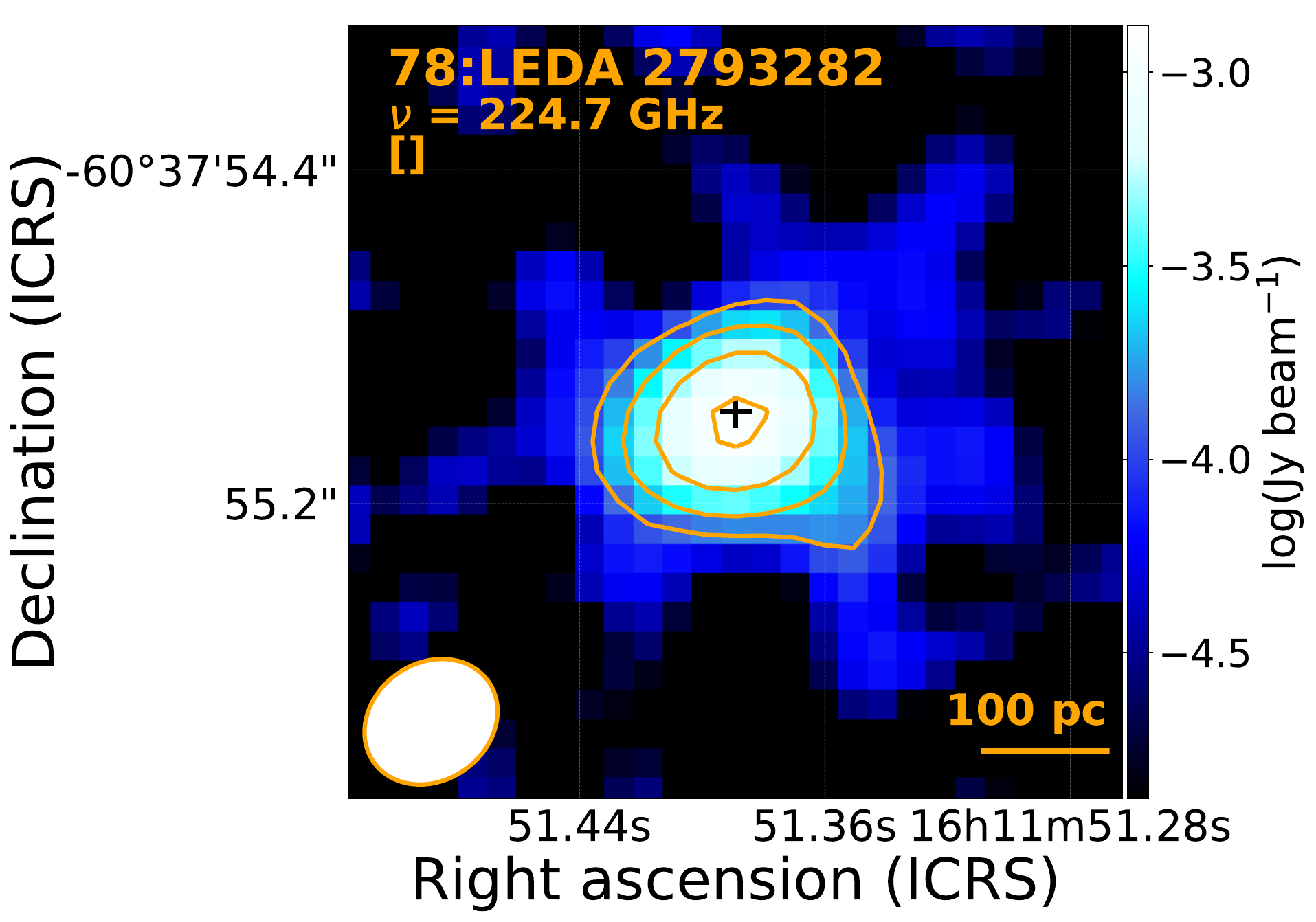}
\includegraphics[width=5.9cm]{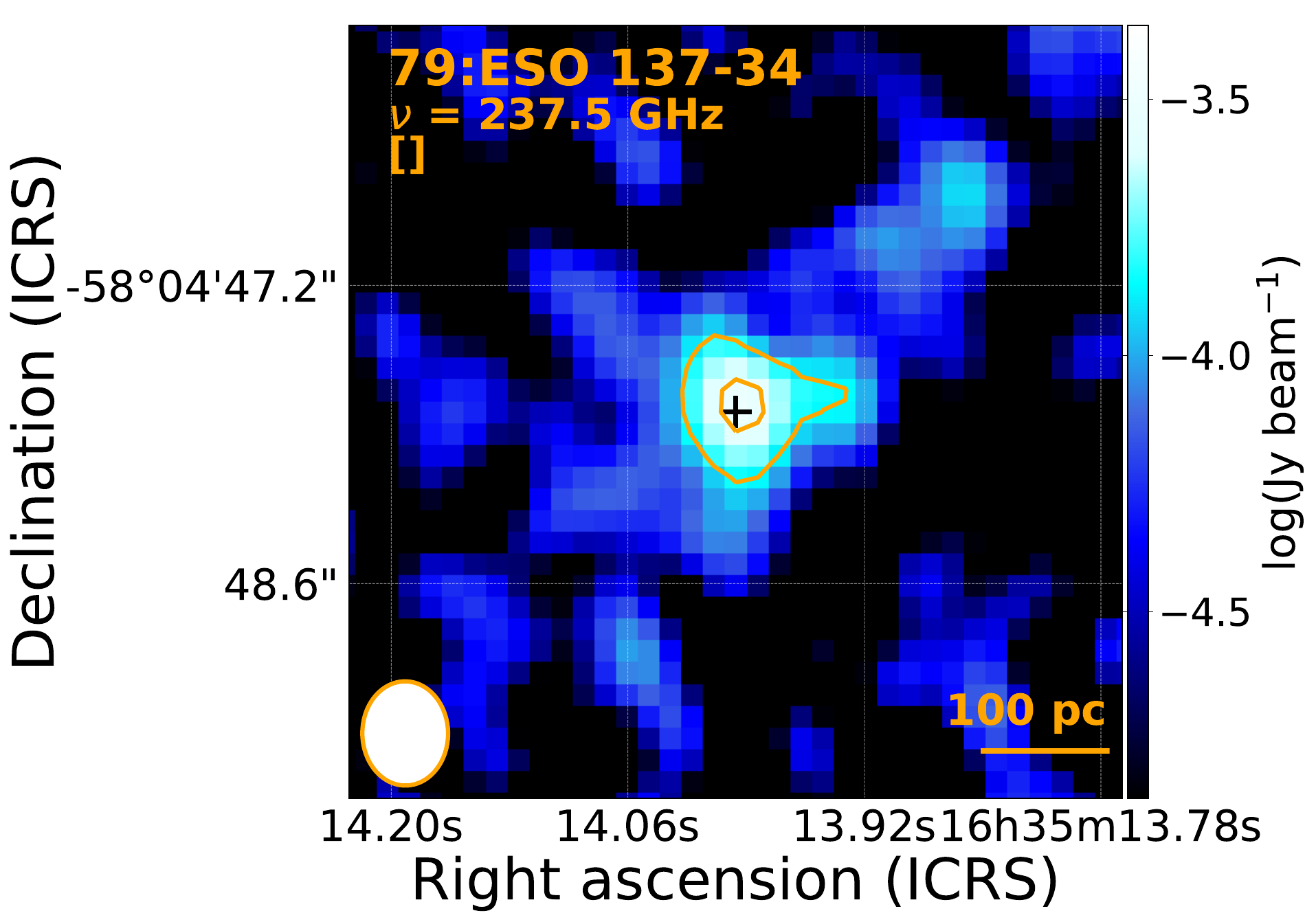}
\includegraphics[width=5.9cm]{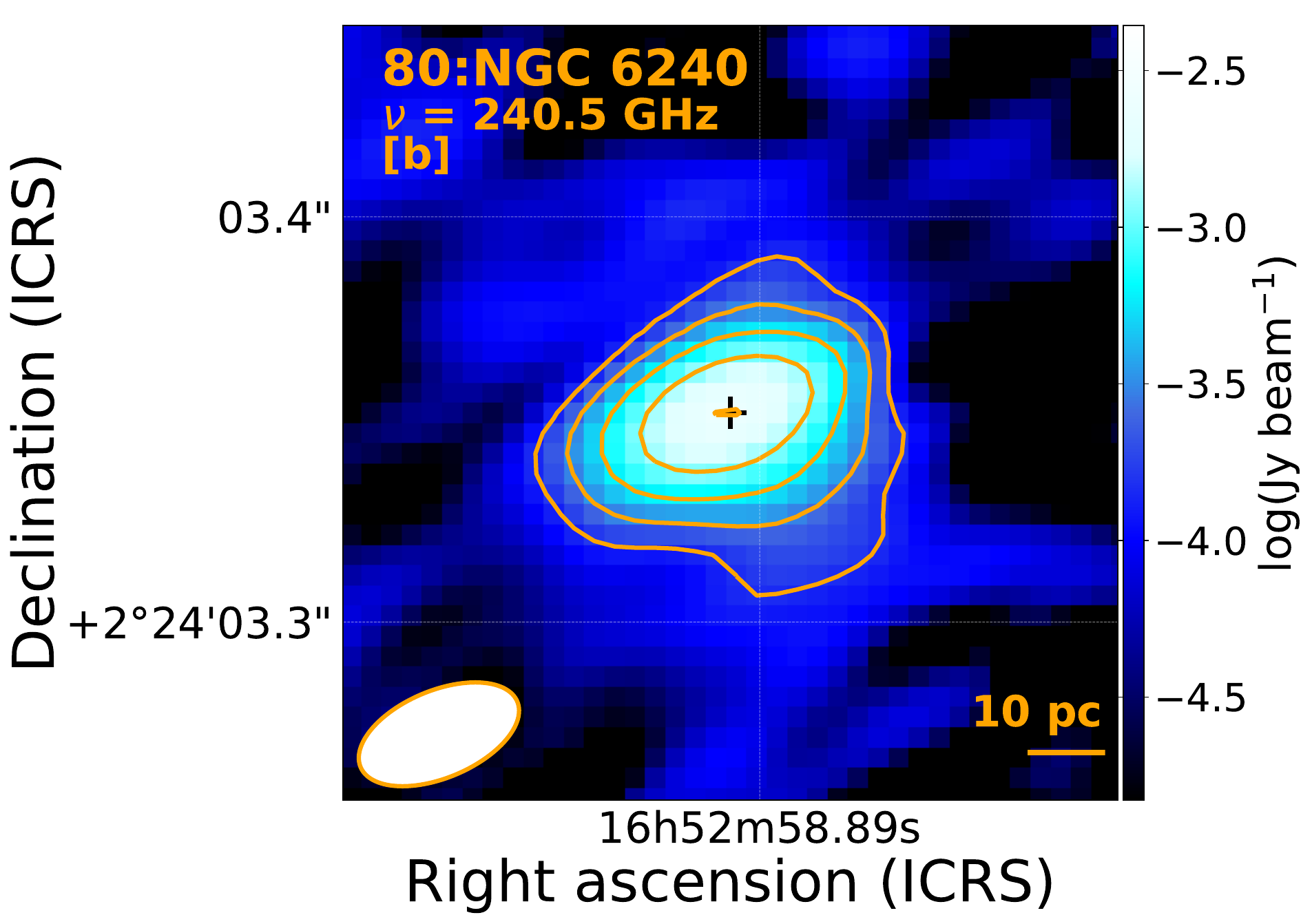}
\includegraphics[width=5.9cm]{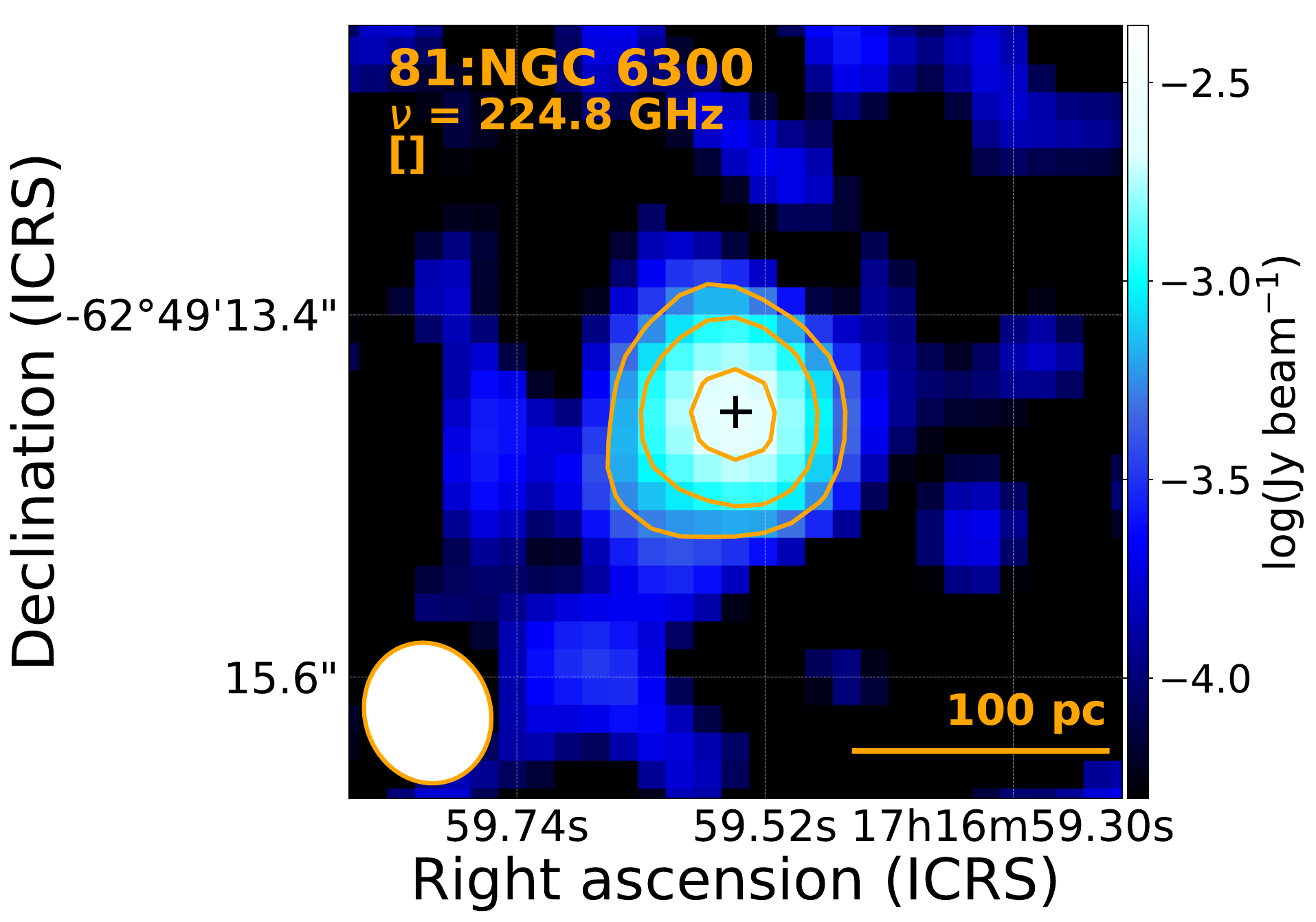}
\includegraphics[width=5.9cm]{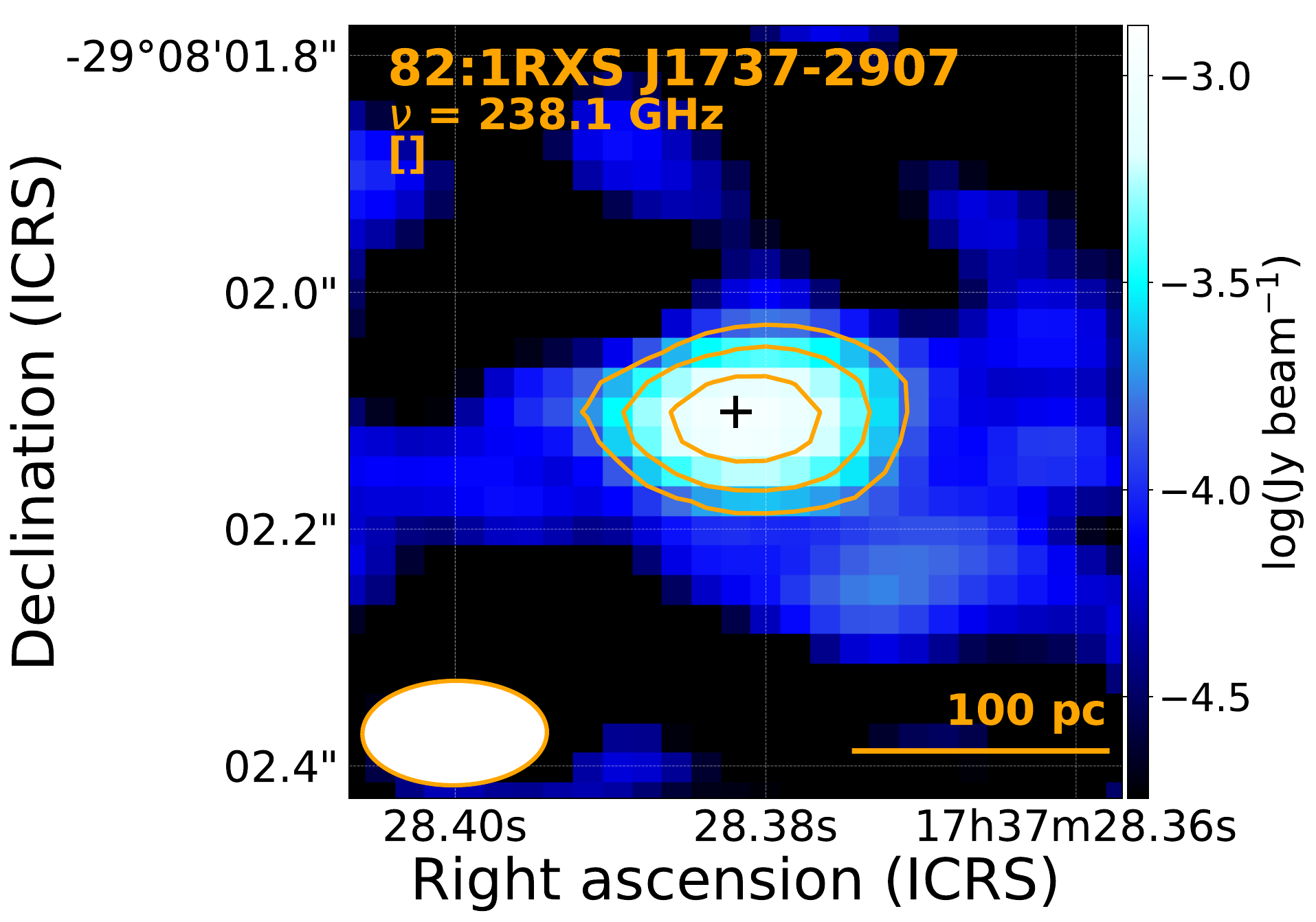}
\includegraphics[width=5.9cm]{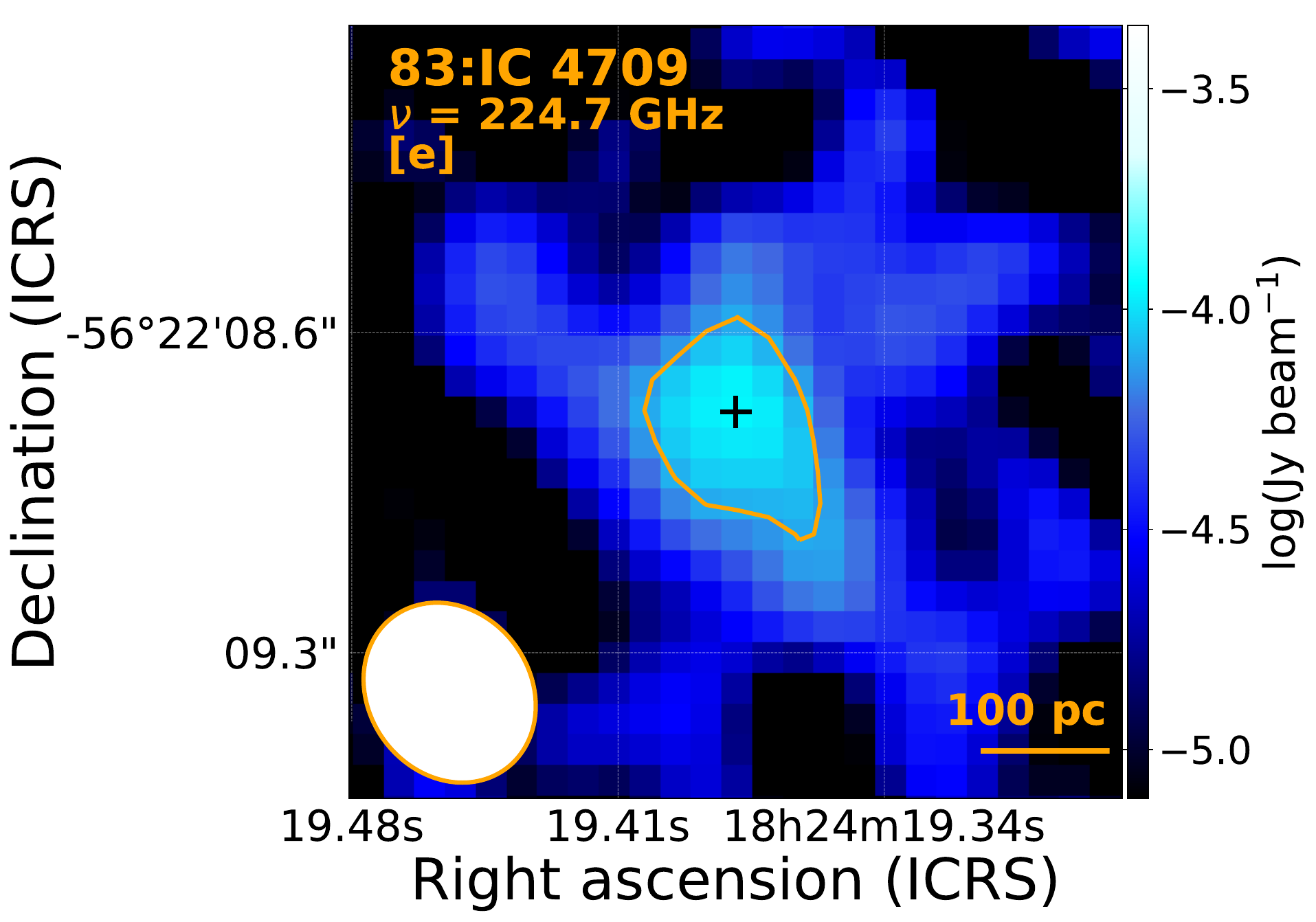}
\includegraphics[width=5.9cm]{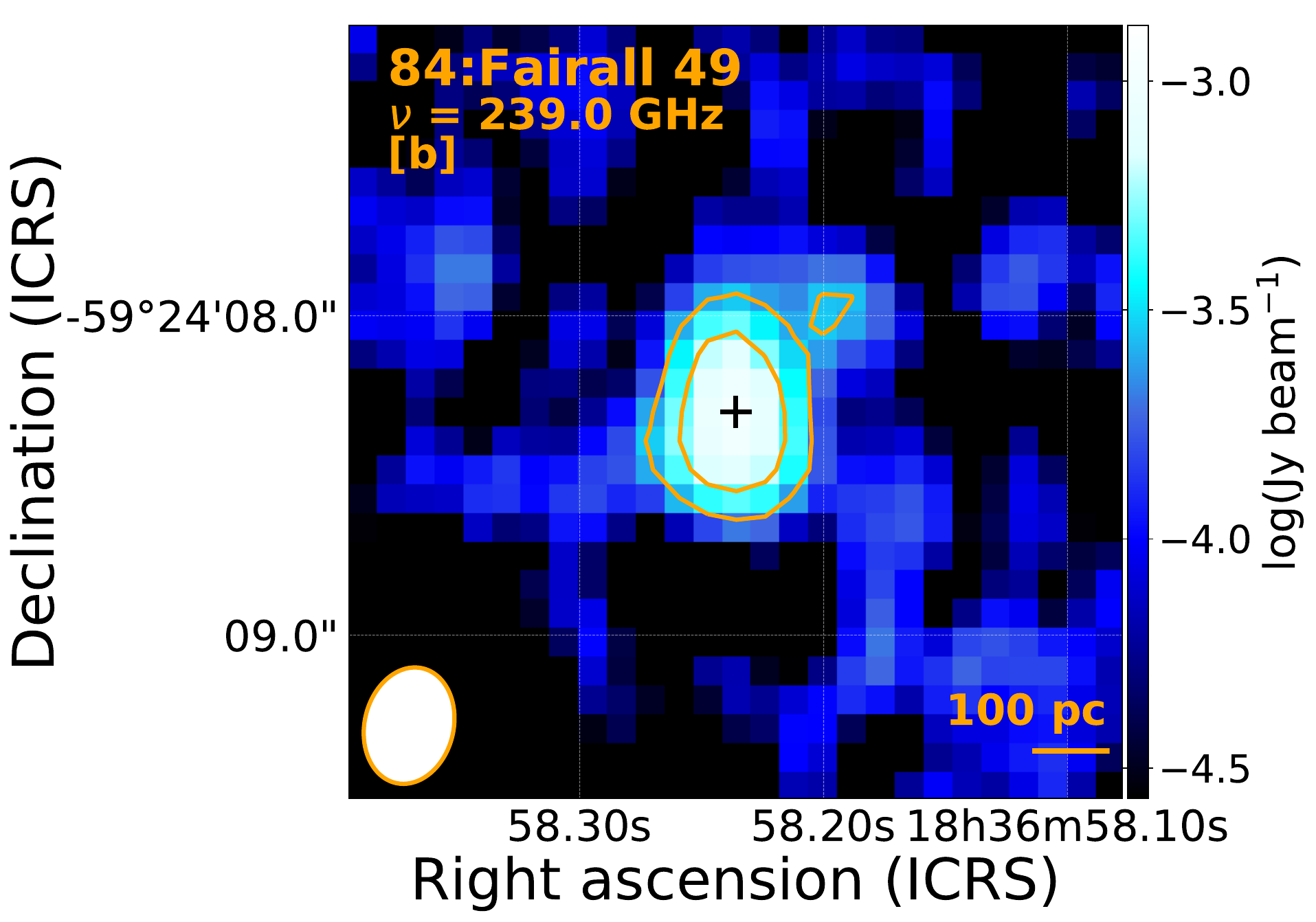}
\includegraphics[width=5.9cm]{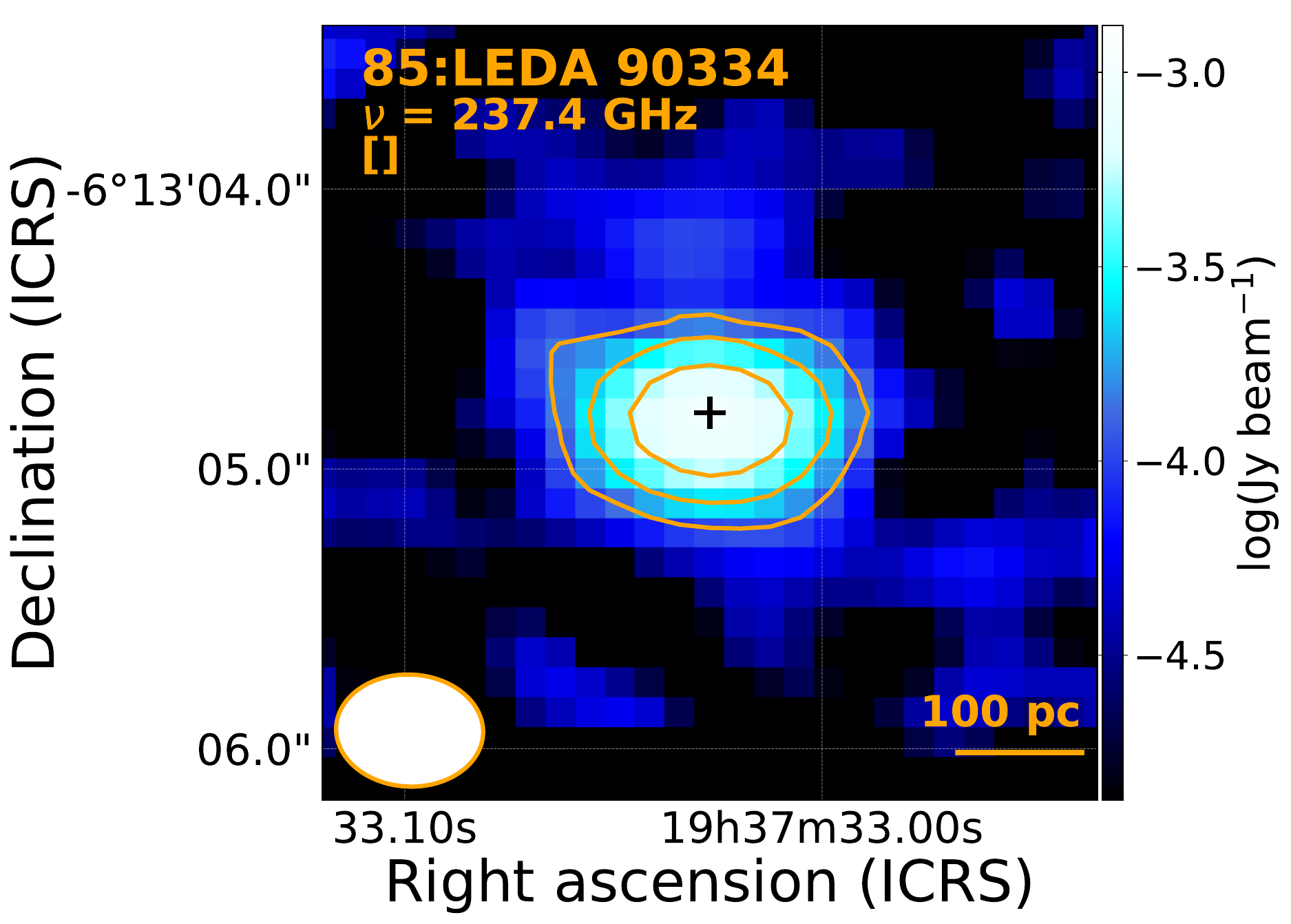}
\includegraphics[width=5.9cm]{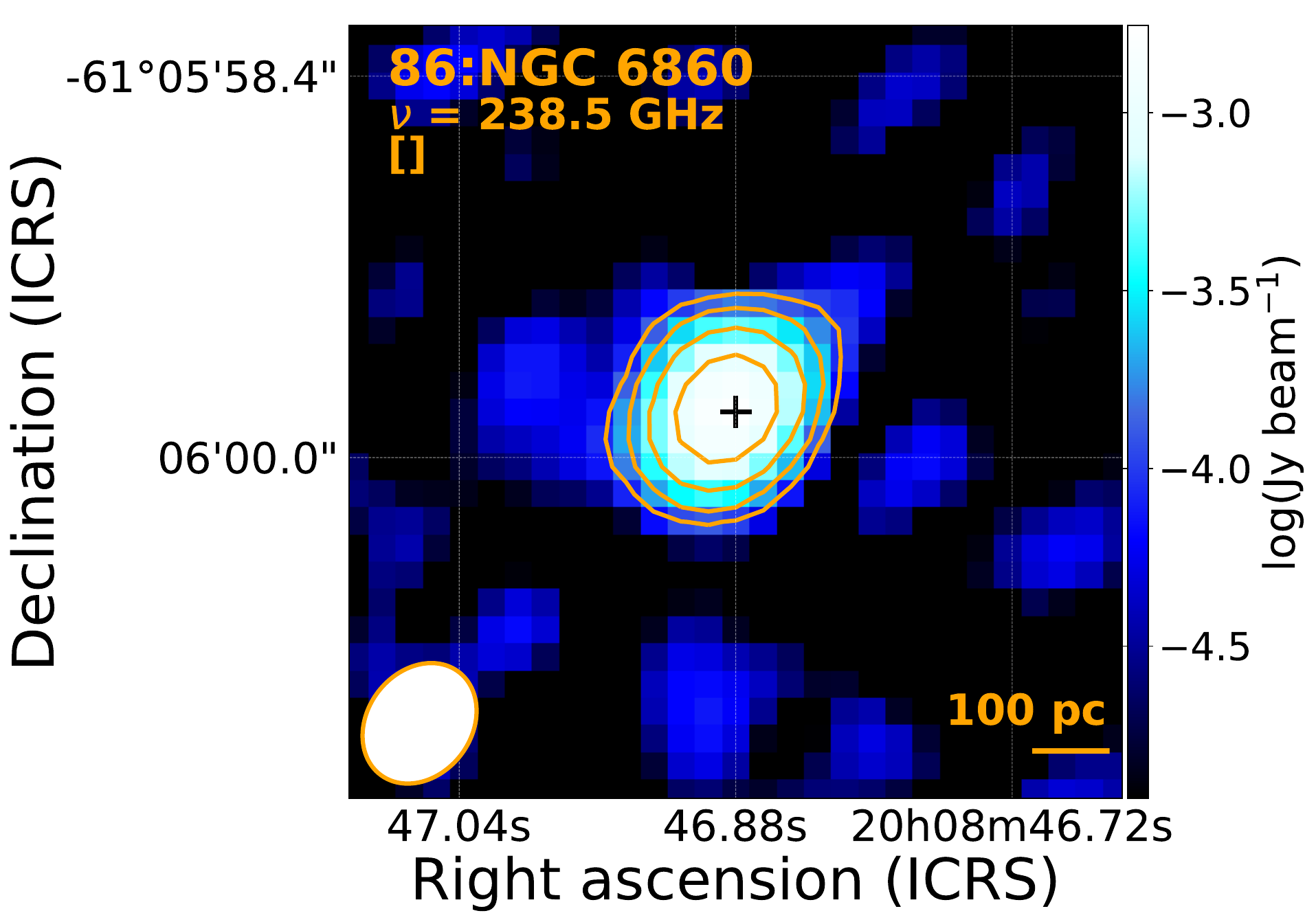}
\includegraphics[width=5.9cm]{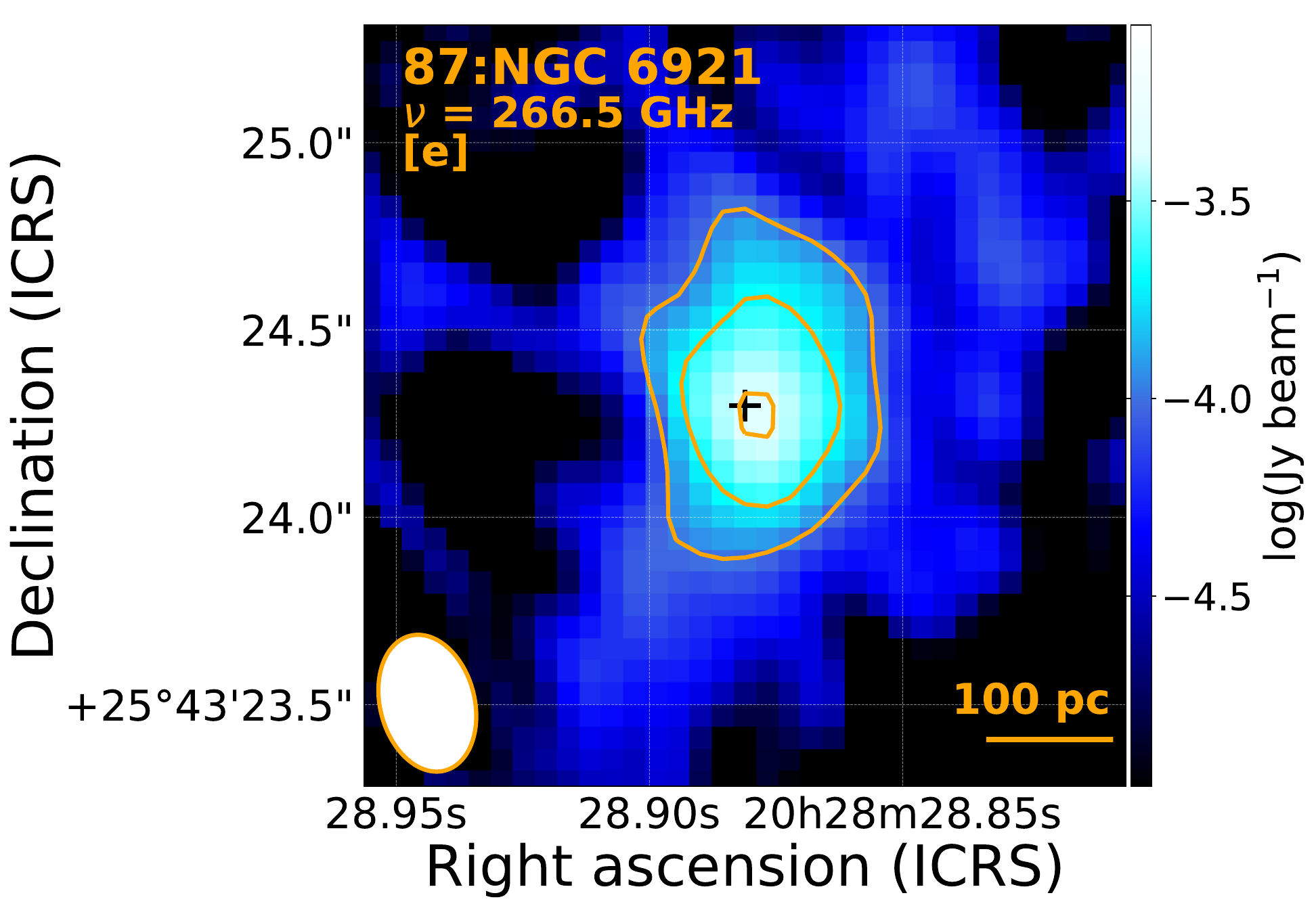}
\caption{Continued. 
    }
\end{figure*}

\addtocounter{figure}{-1}

\begin{figure*}
    \centering
    \includegraphics[width=5.9cm]{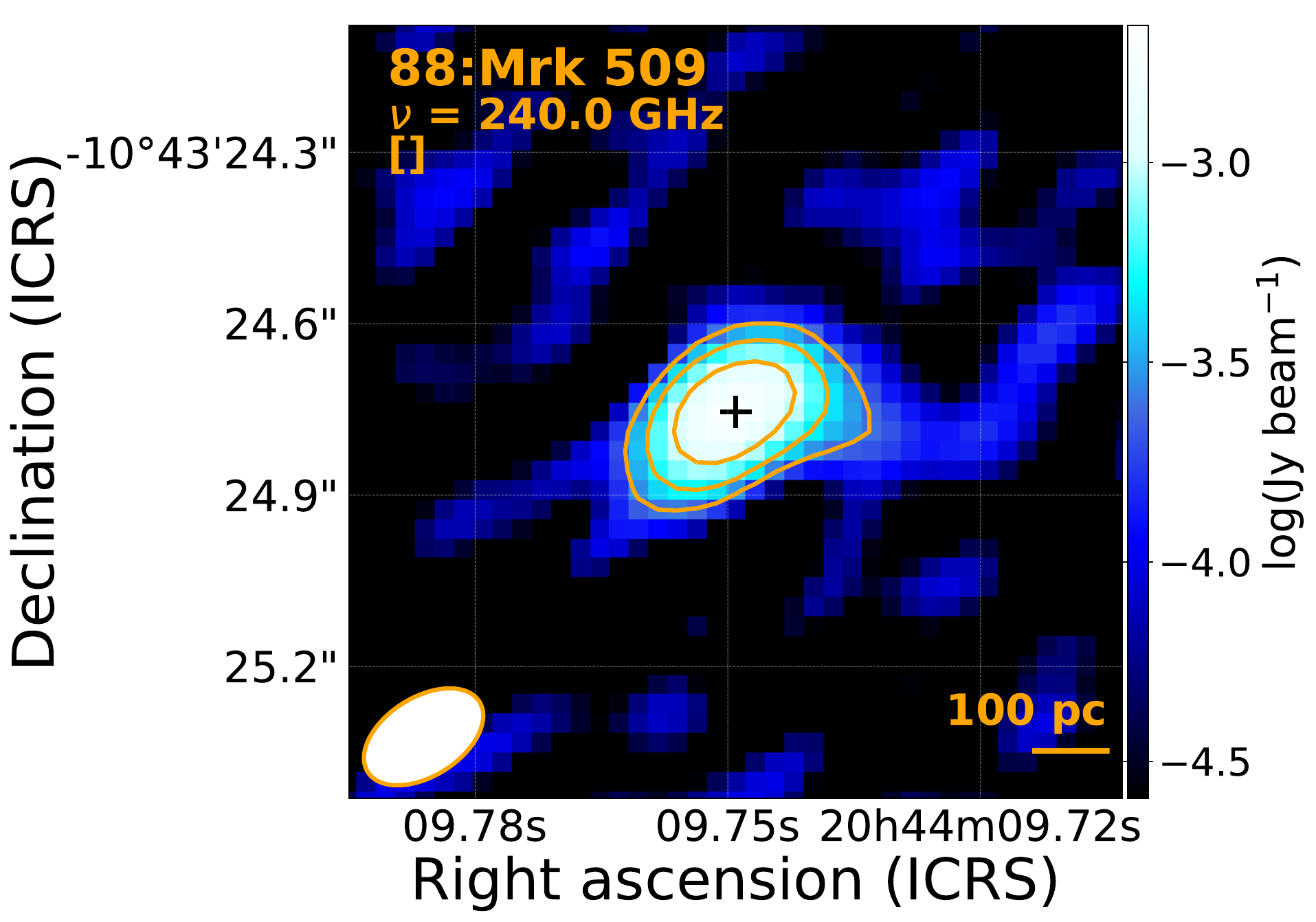}
\includegraphics[width=5.9cm]{089_IC_5063_spwall_c_500pc.pdf}
\includegraphics[width=5.9cm]{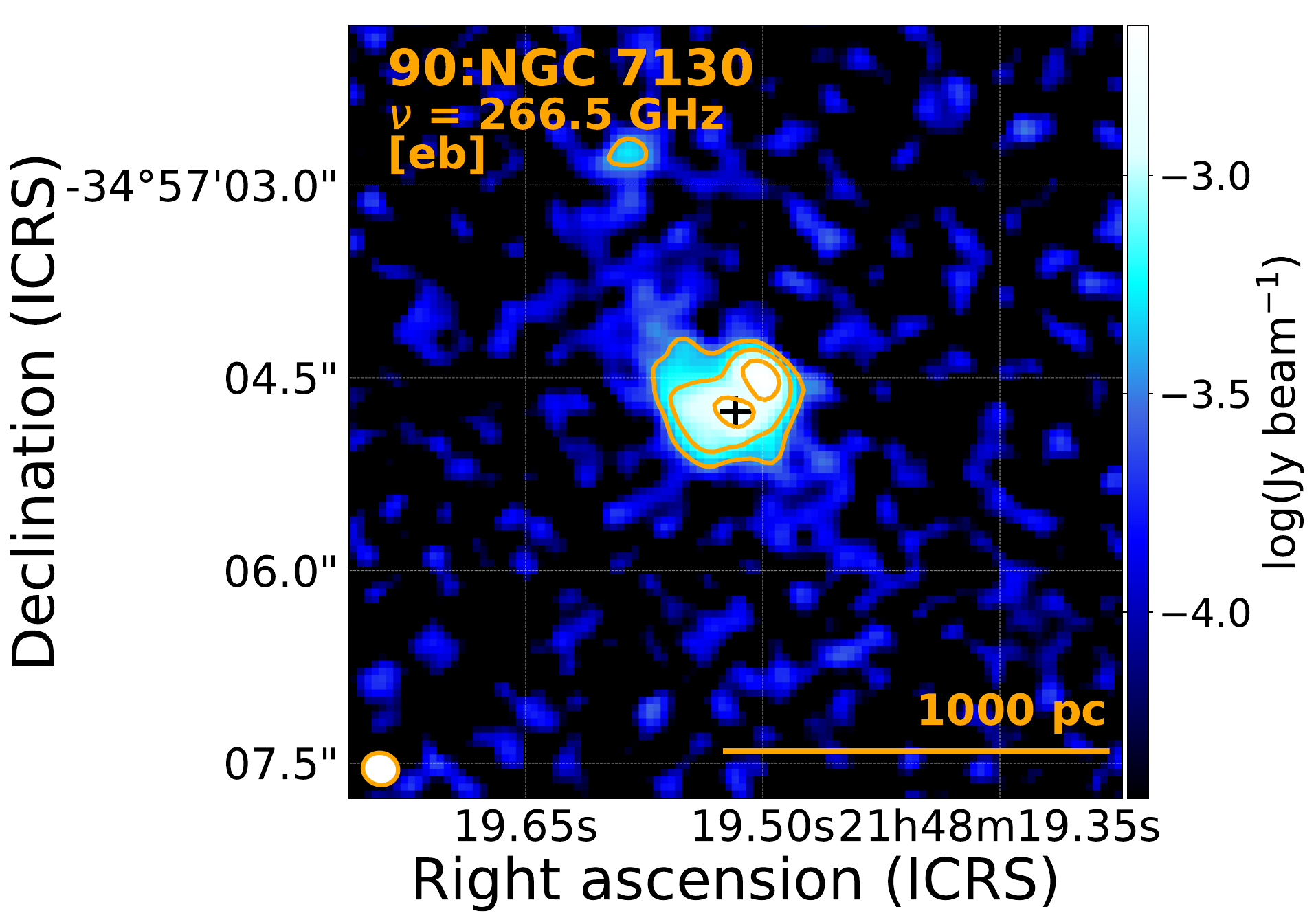}
\includegraphics[width=5.9cm]{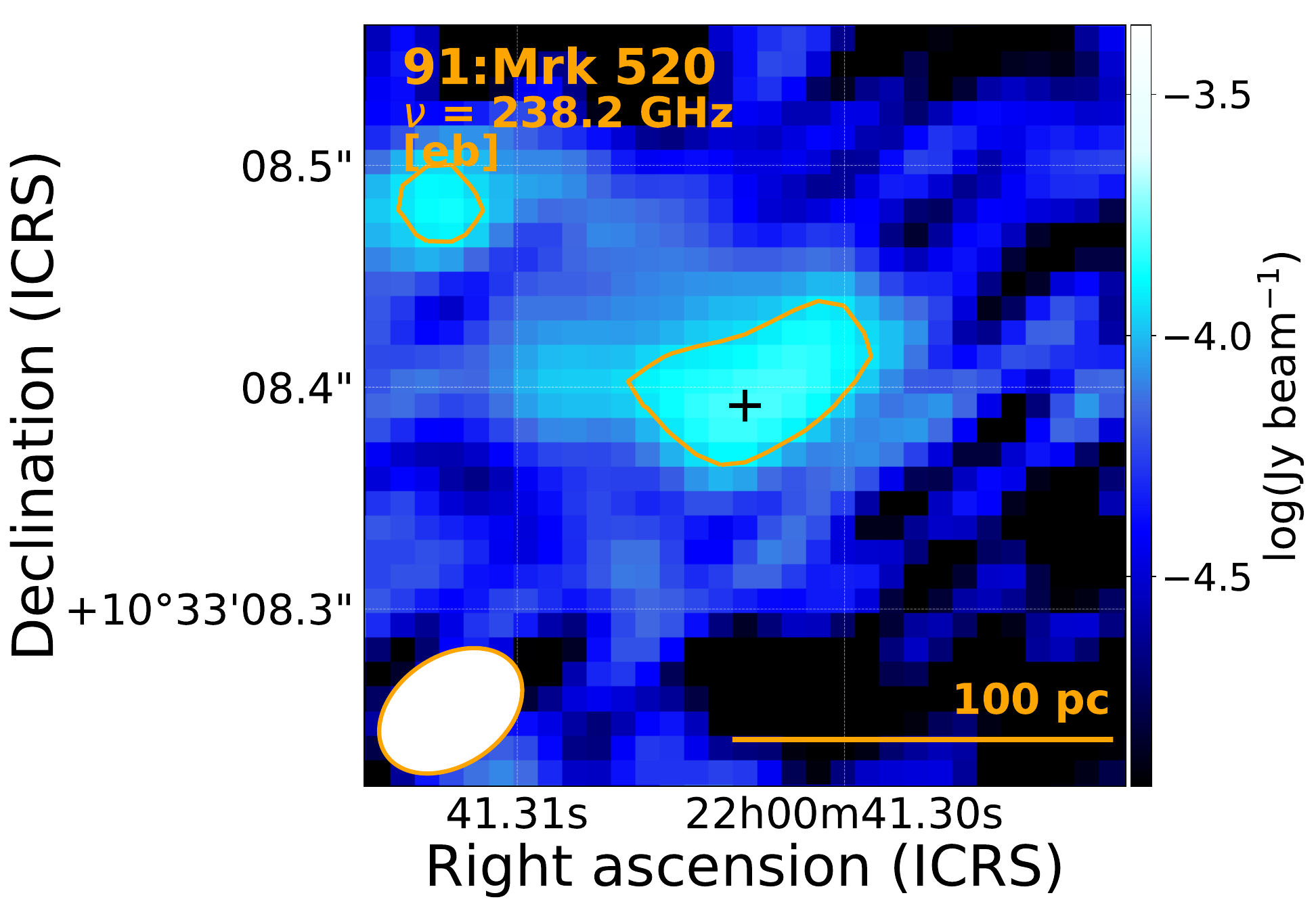}
\includegraphics[width=5.9cm]{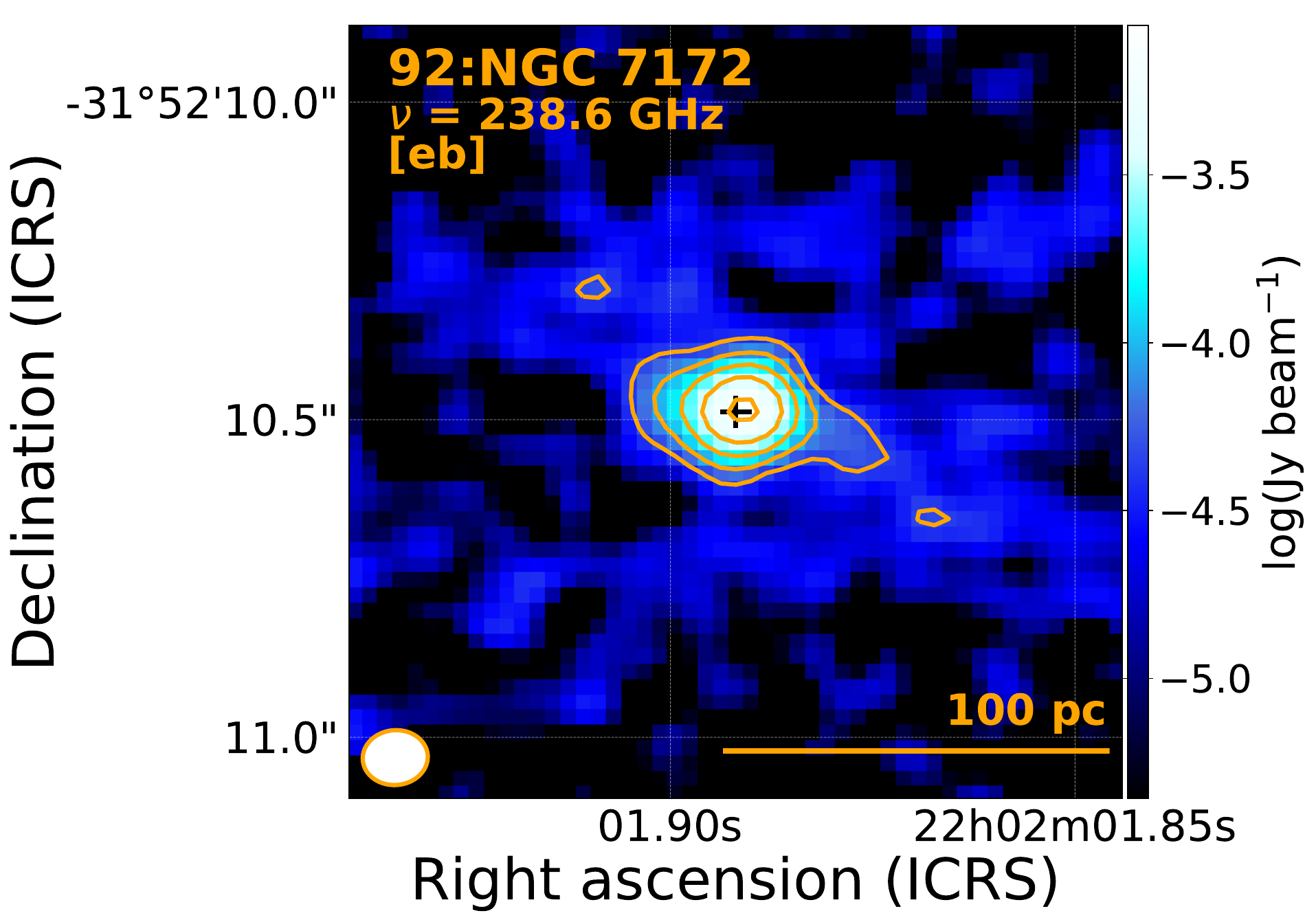}
\includegraphics[width=5.9cm]{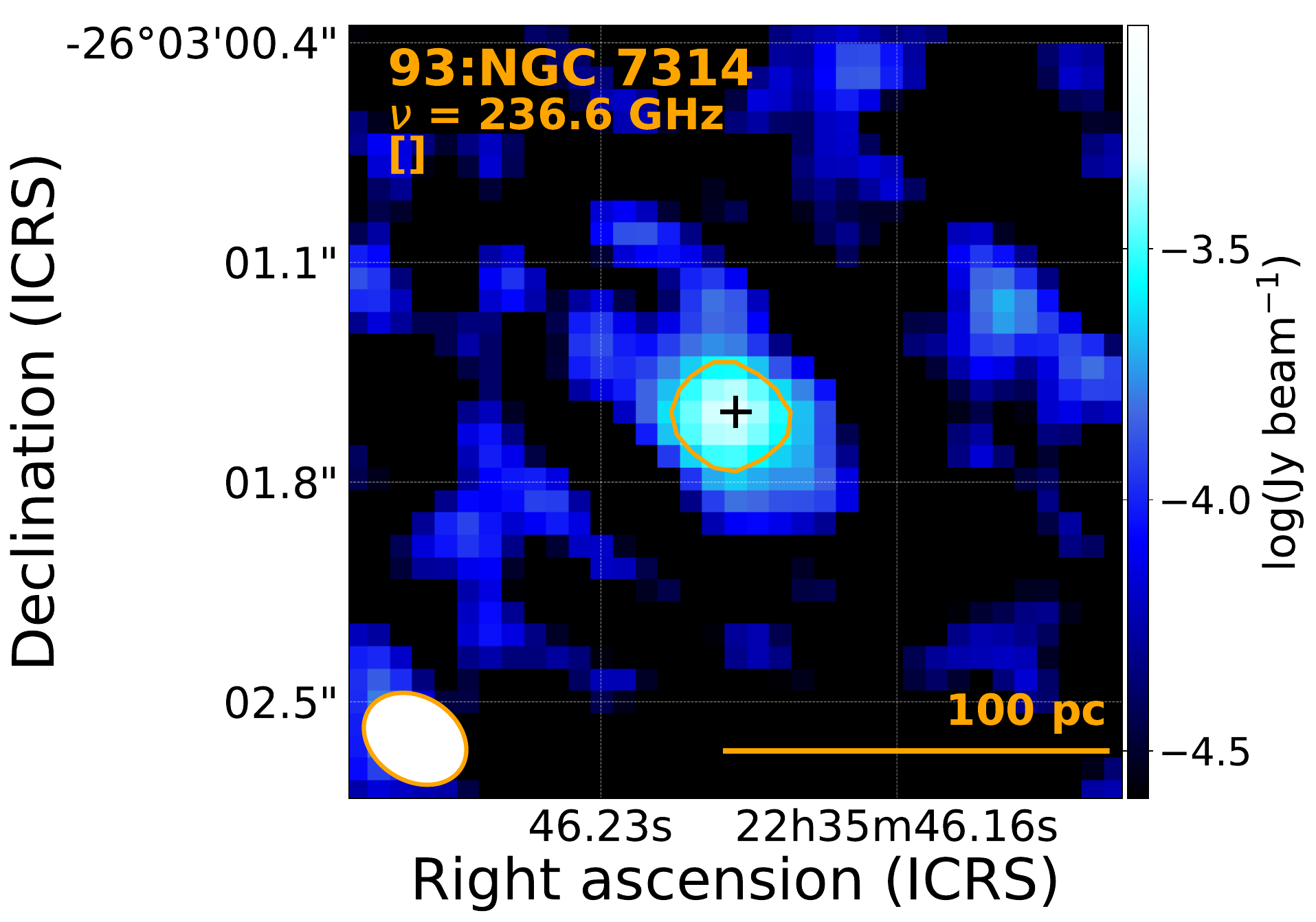}
\includegraphics[width=5.9cm]{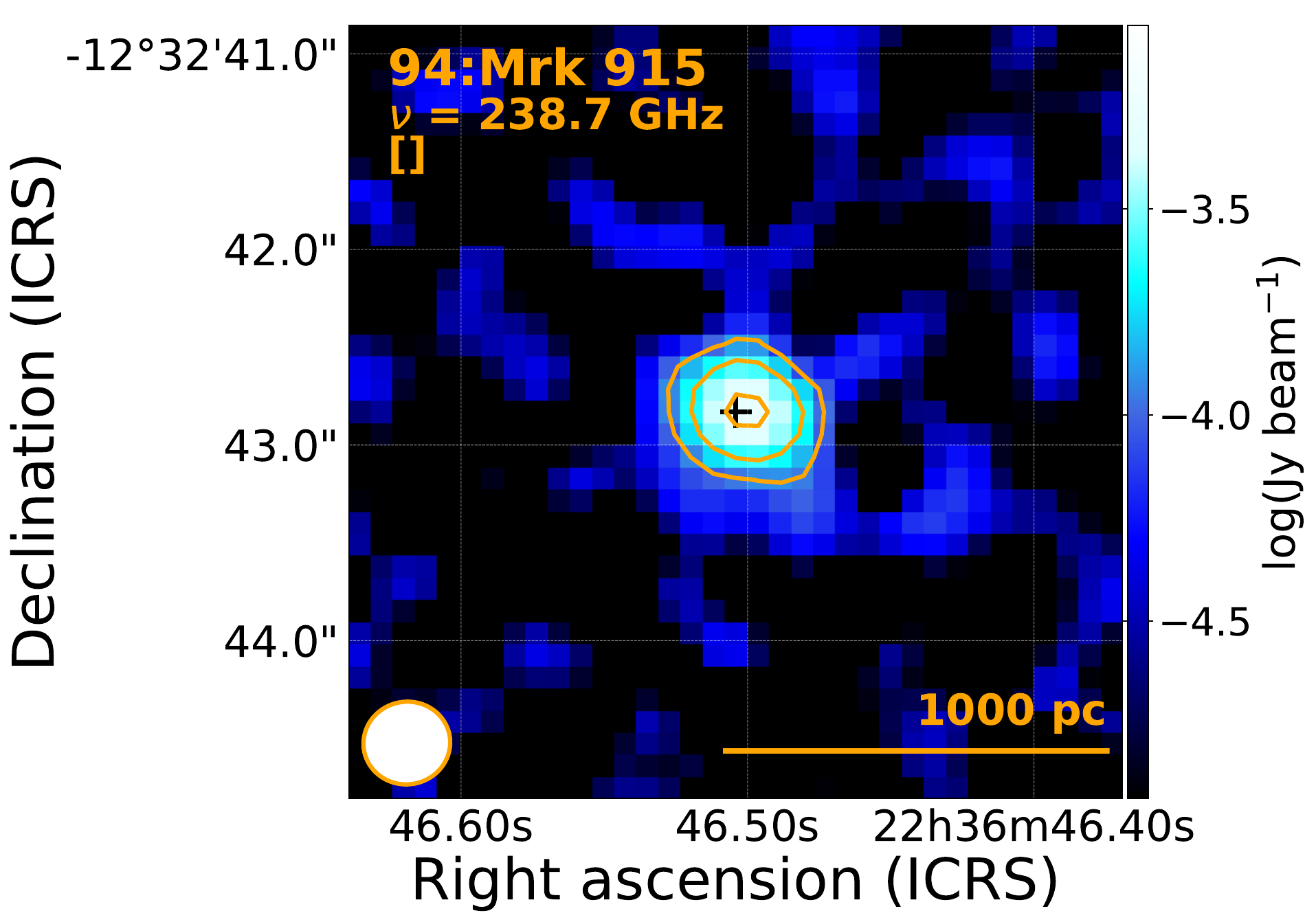}
\includegraphics[width=5.9cm]{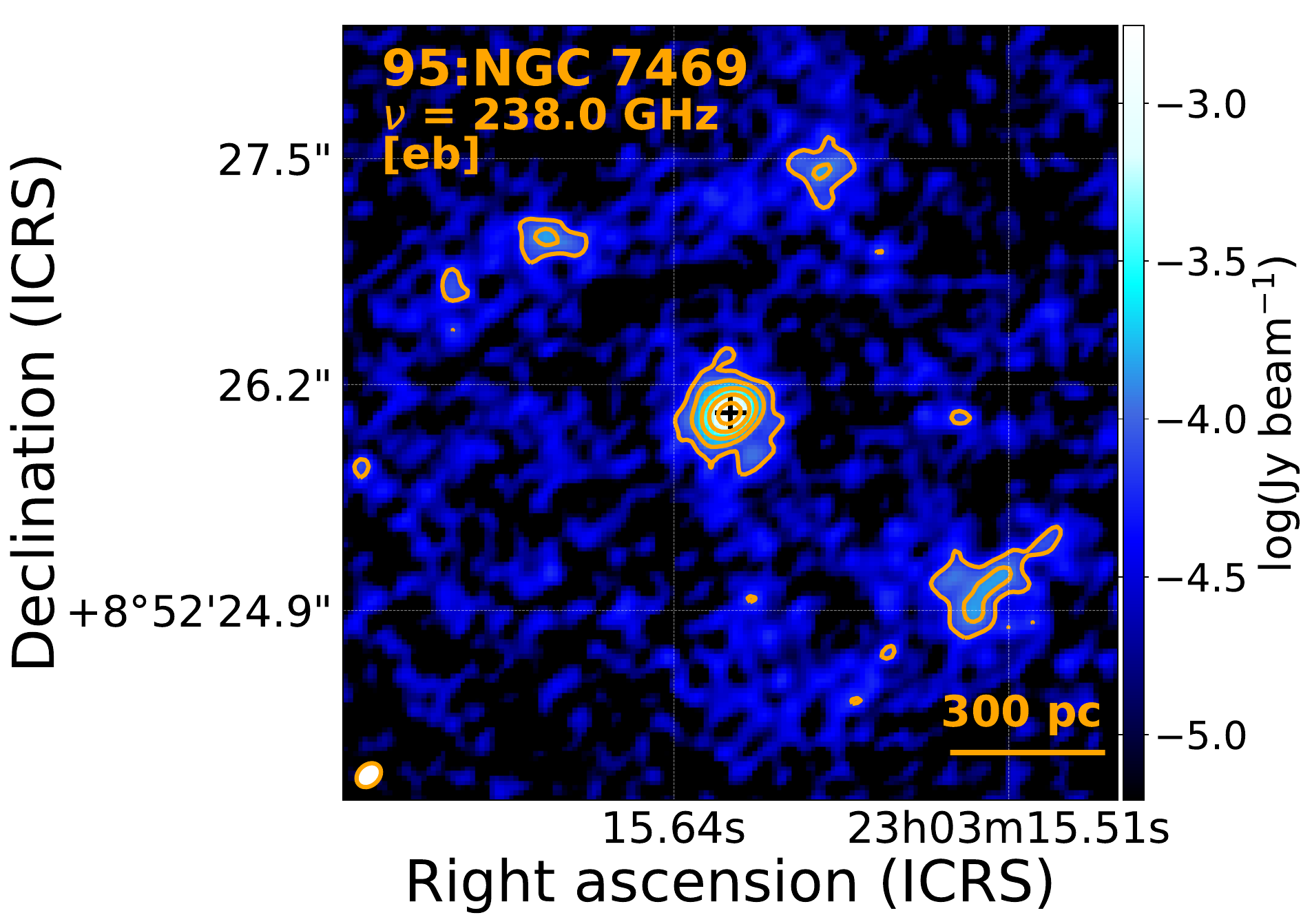}
\includegraphics[width=5.9cm]{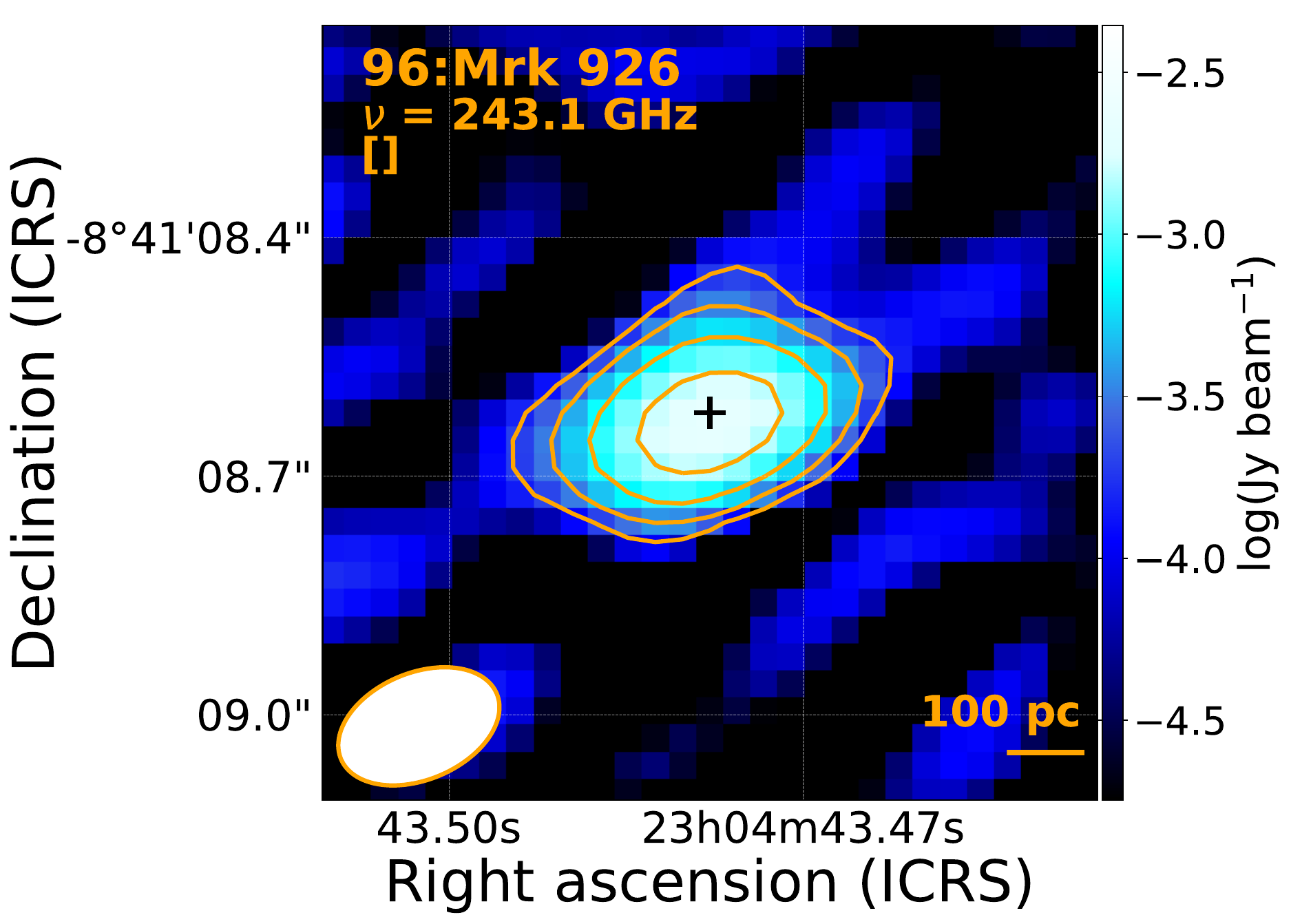}
\includegraphics[width=5.9cm]{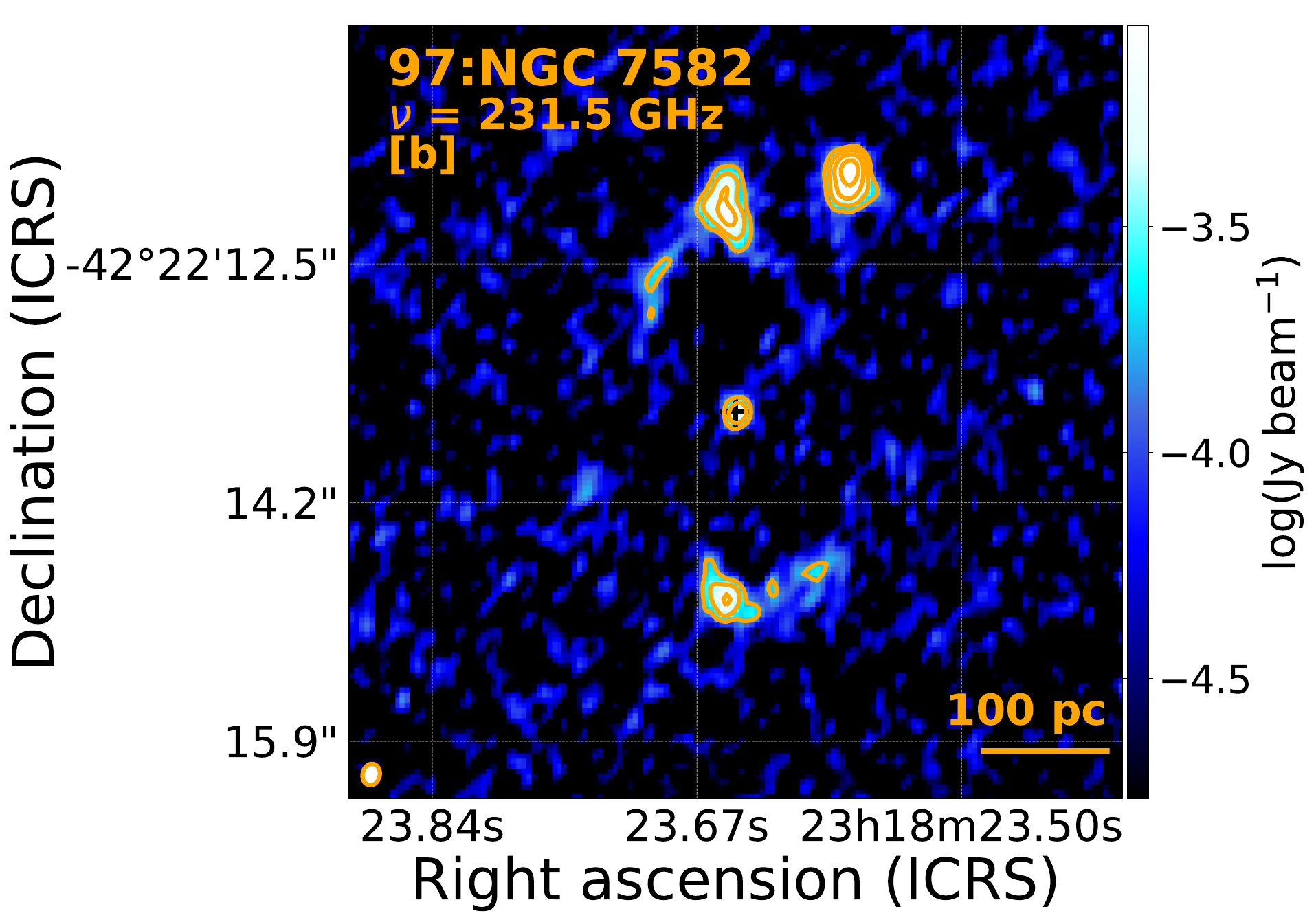}
\includegraphics[width=5.9cm]{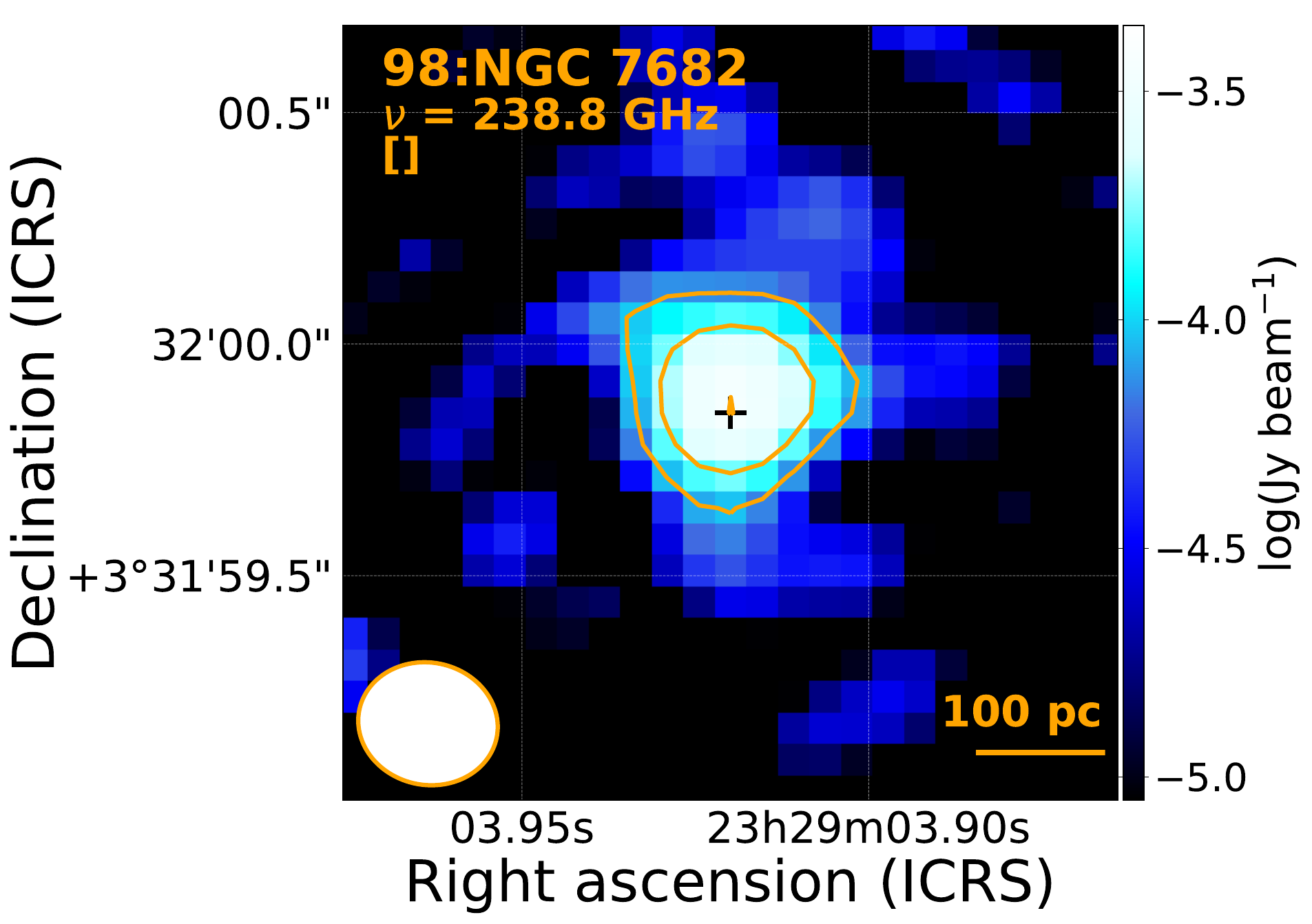}
    \caption{Continued. 
    }
\end{figure*}

\begin{figure*}
    \centering
\includegraphics[width=5.9cm]{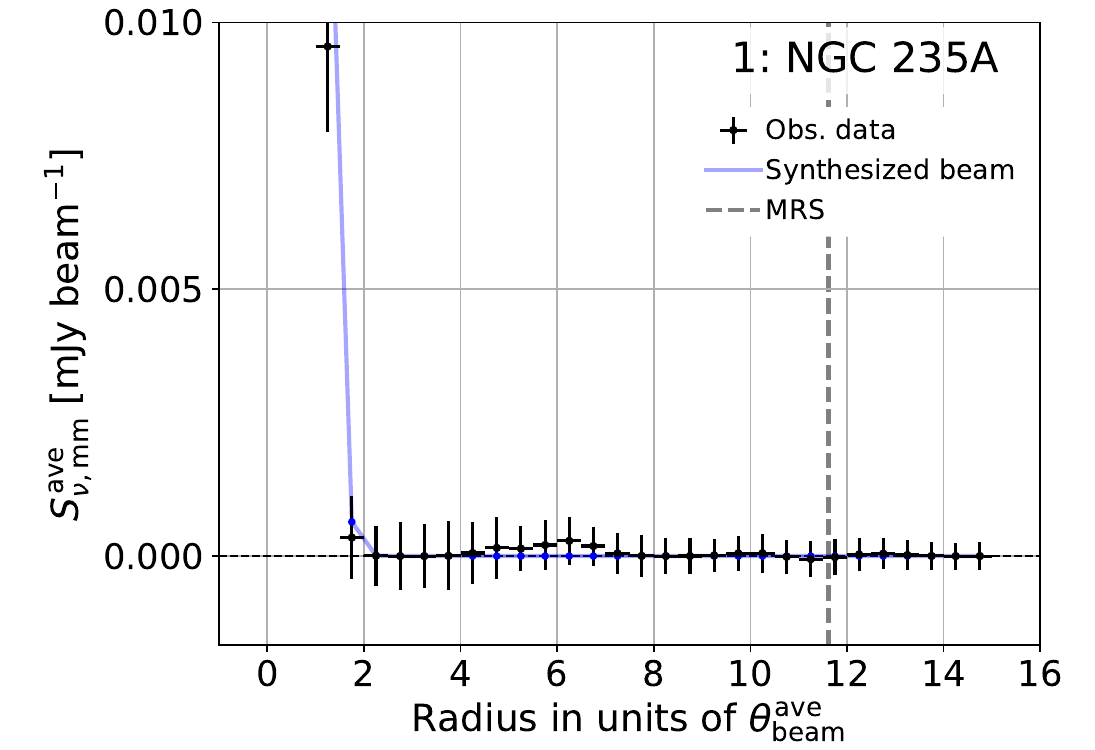}
\includegraphics[width=5.9cm]{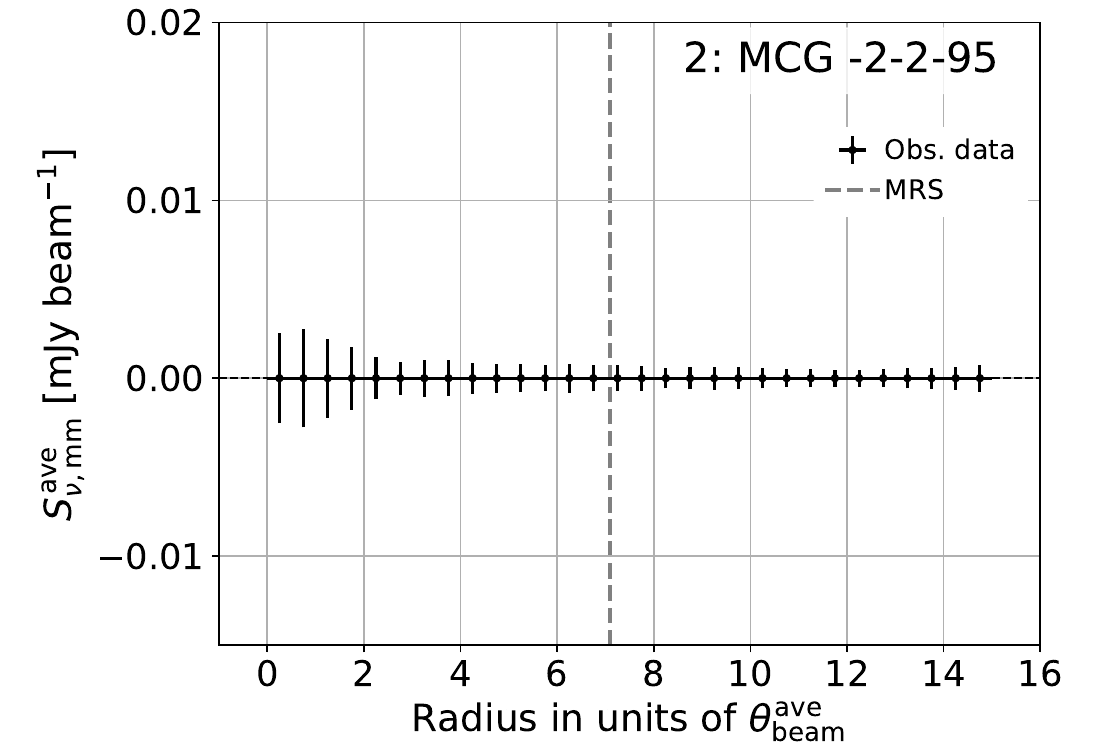}
\includegraphics[width=5.9cm]{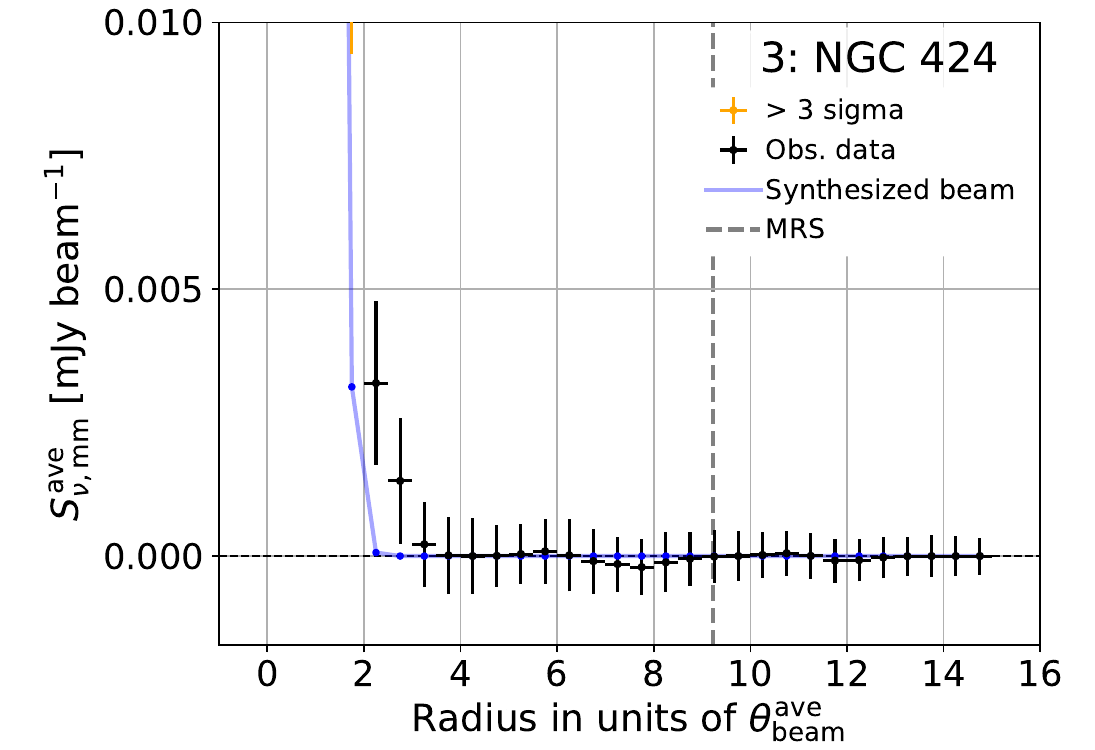}
\includegraphics[width=5.9cm]{004_NGC_526A_radpro.pdf}
\includegraphics[width=5.9cm]{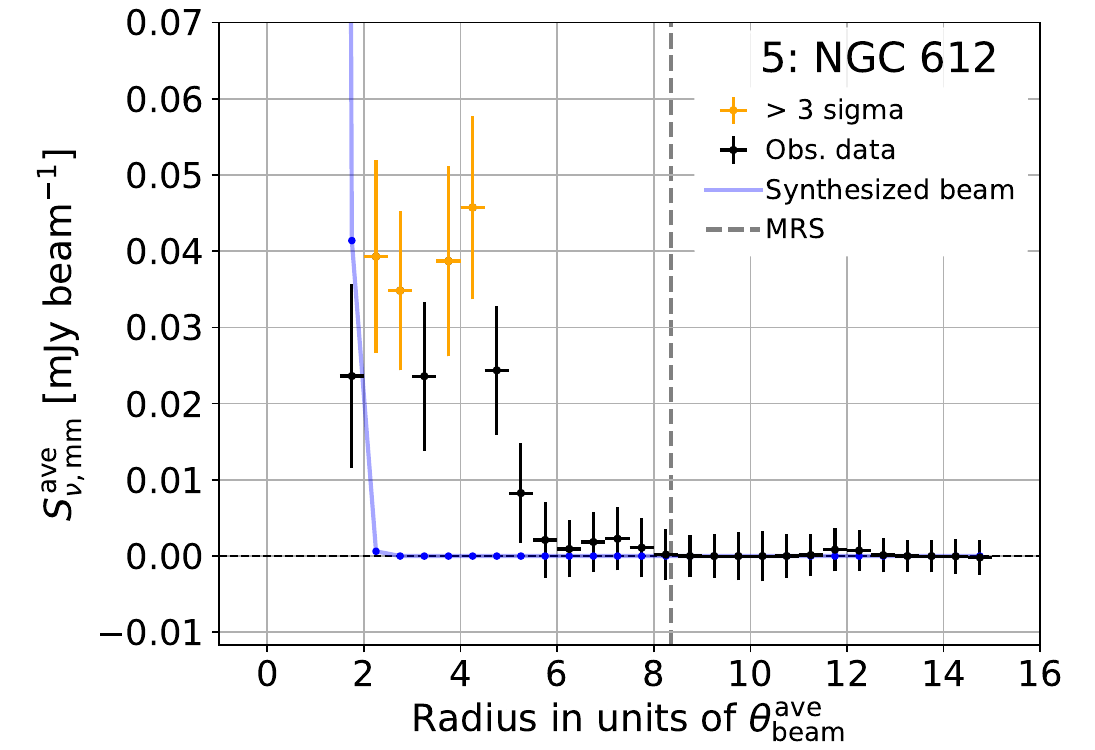}
\includegraphics[width=5.9cm]{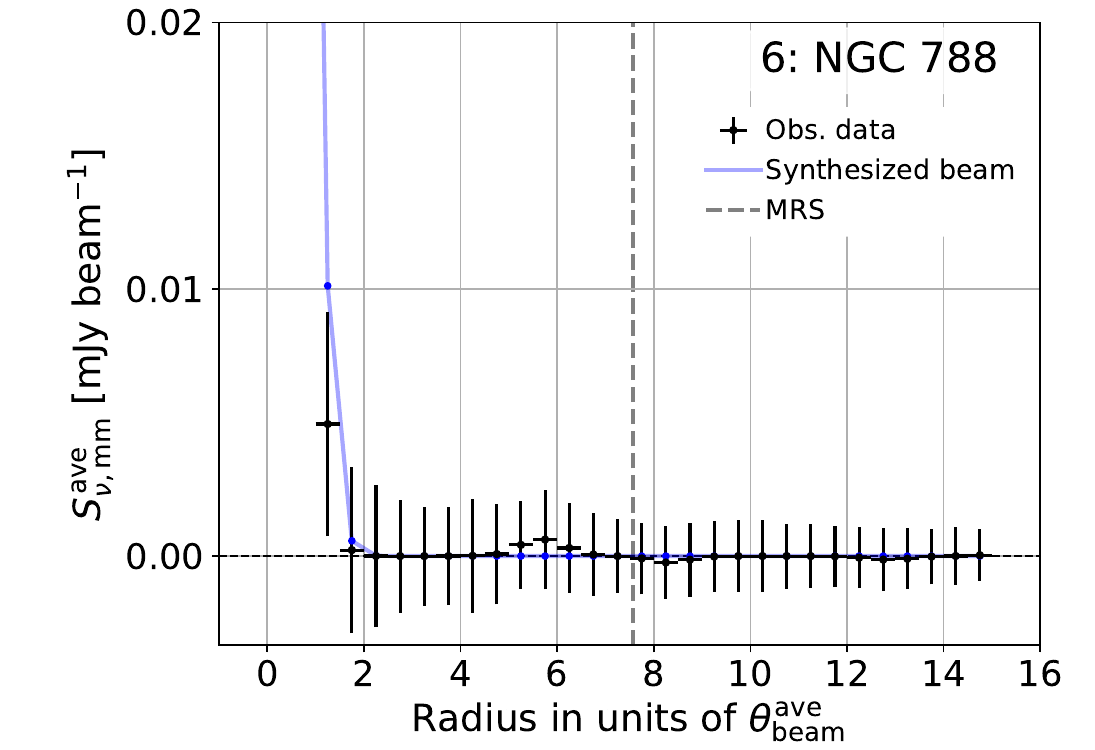}
\includegraphics[width=5.9cm]{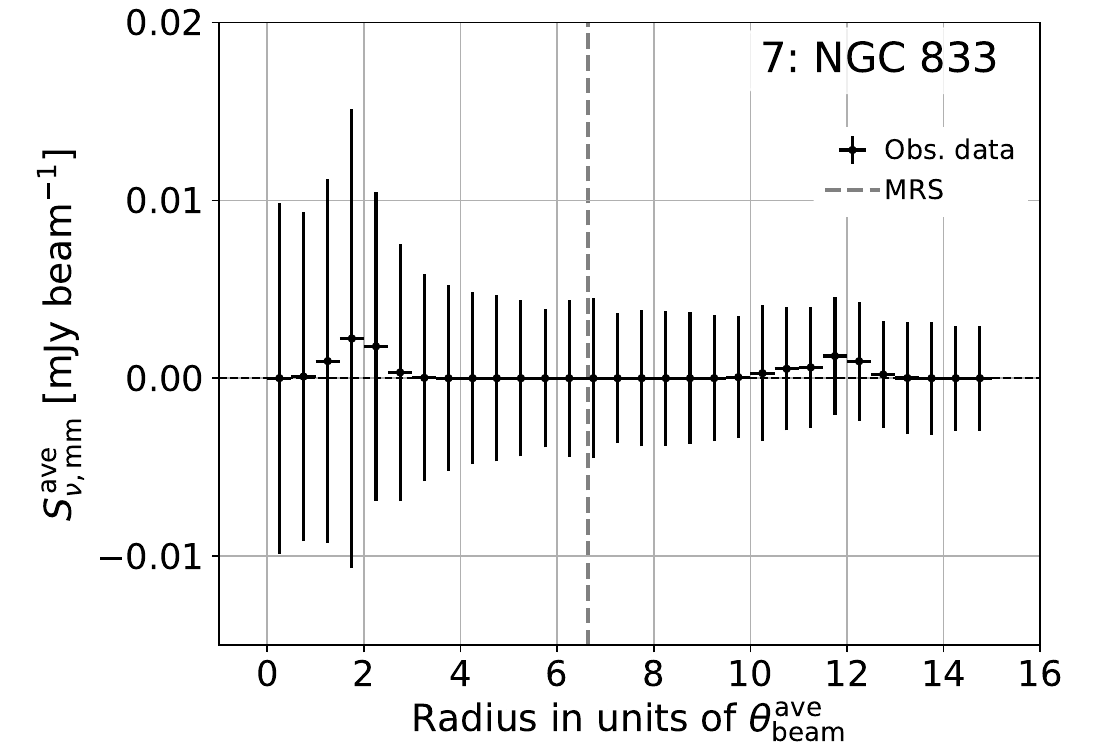}
\includegraphics[width=5.9cm]{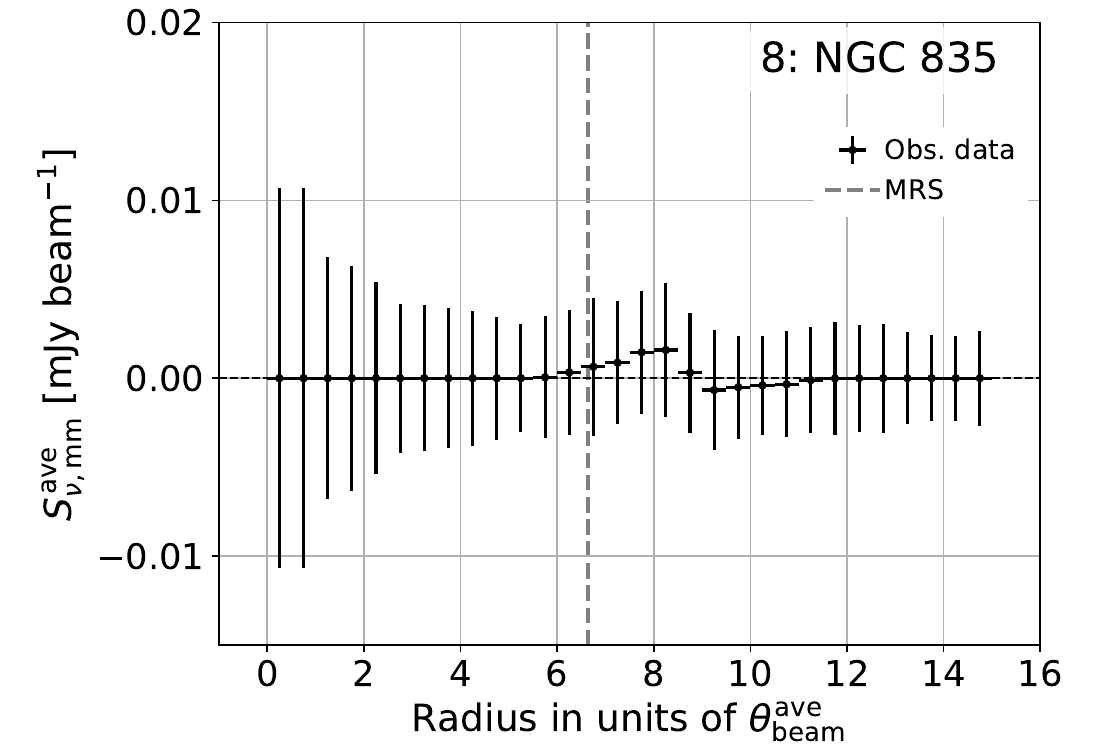}
\includegraphics[width=5.9cm]{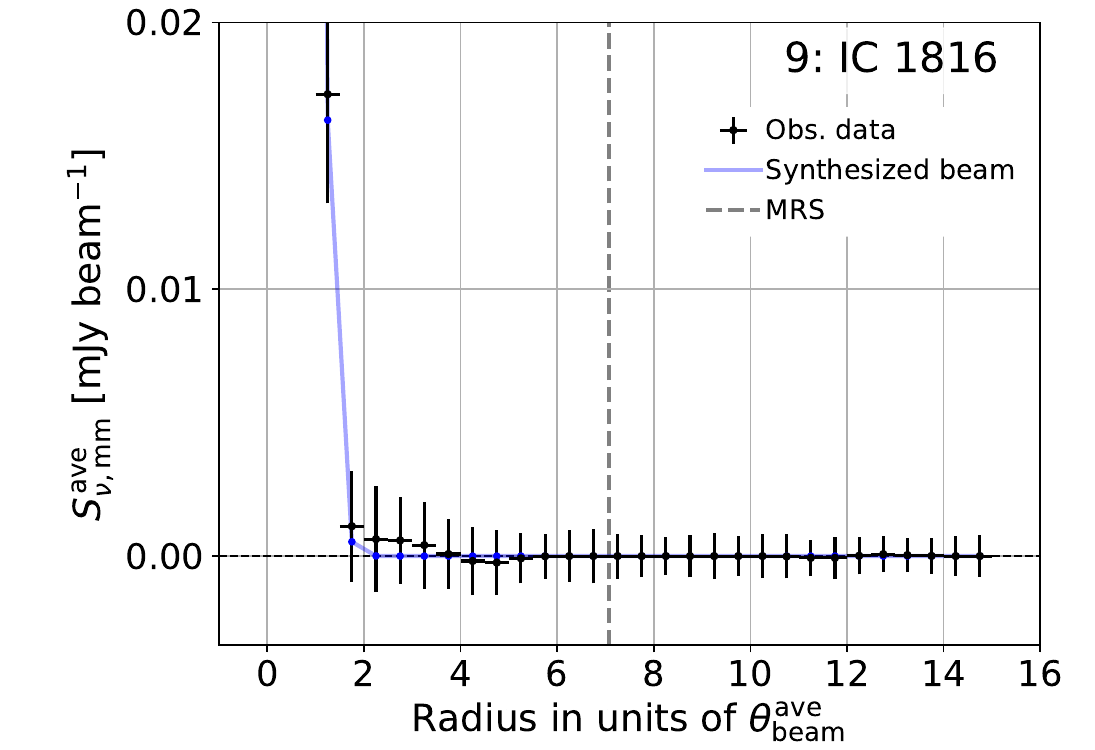}
\includegraphics[width=5.9cm]{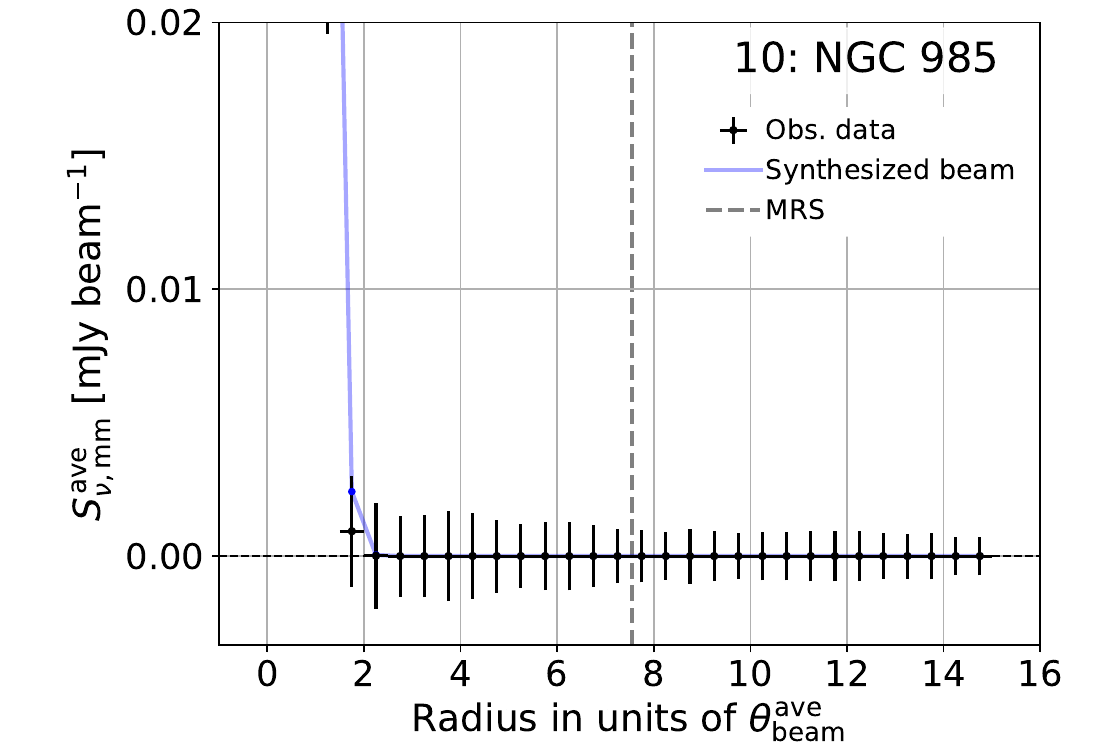}
\includegraphics[width=5.9cm]{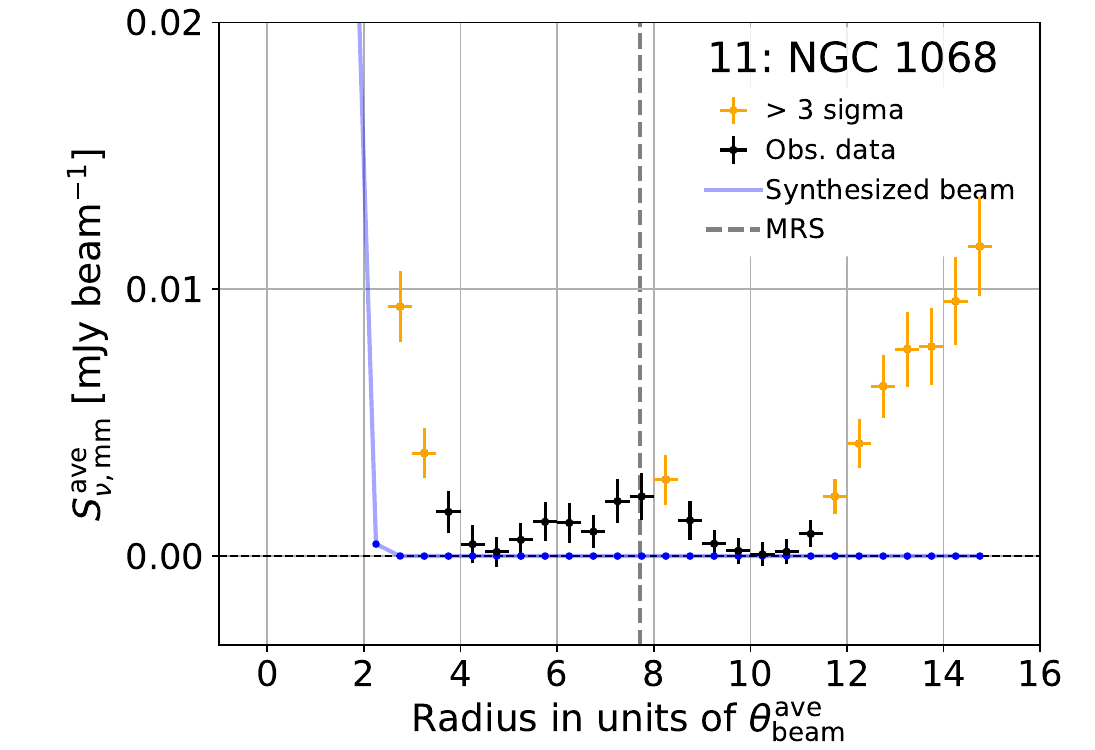}
\includegraphics[width=5.9cm]{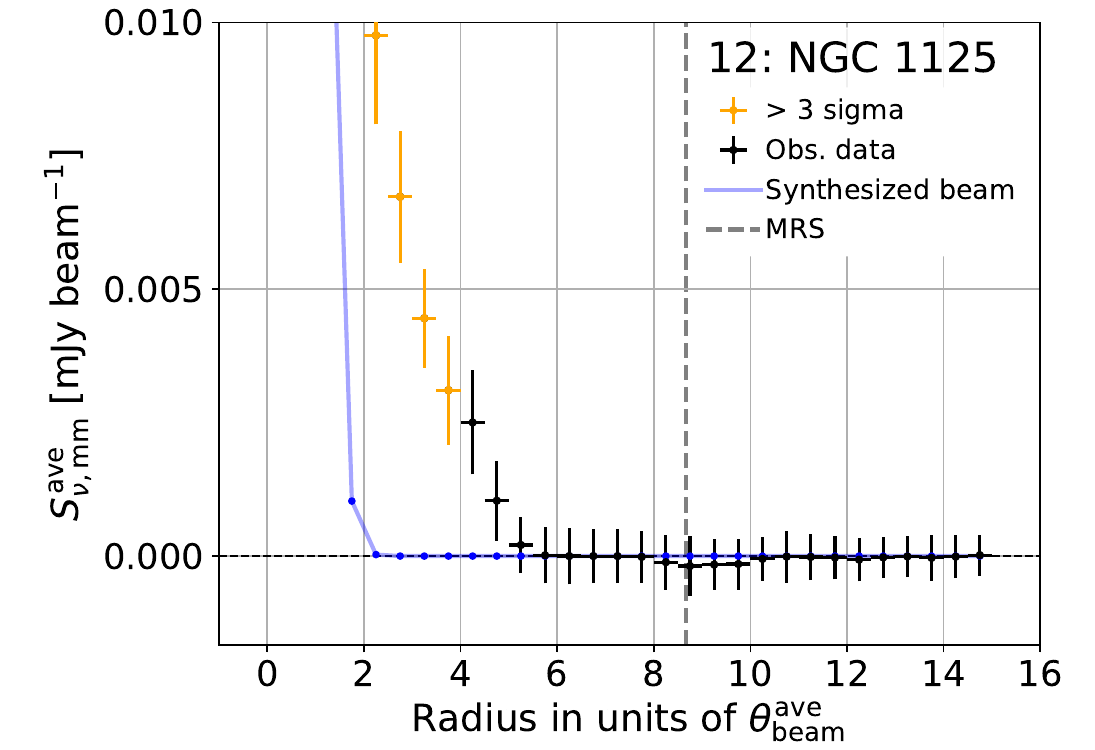}
\includegraphics[width=5.9cm]{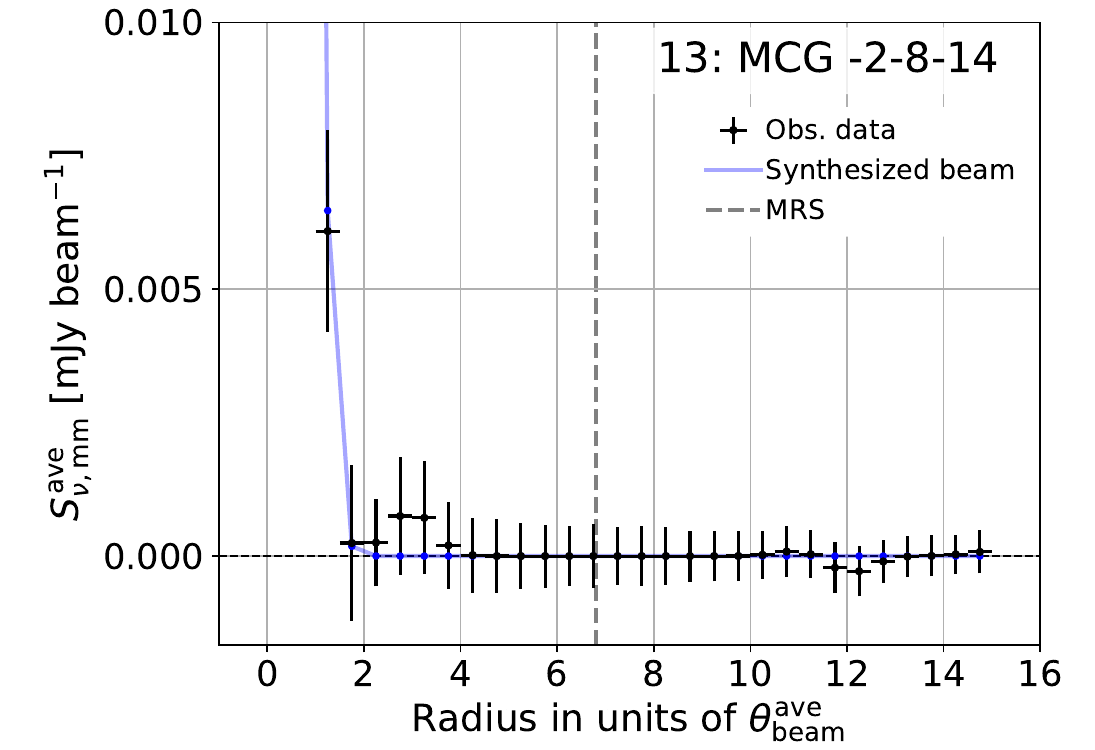}
\includegraphics[width=5.9cm]{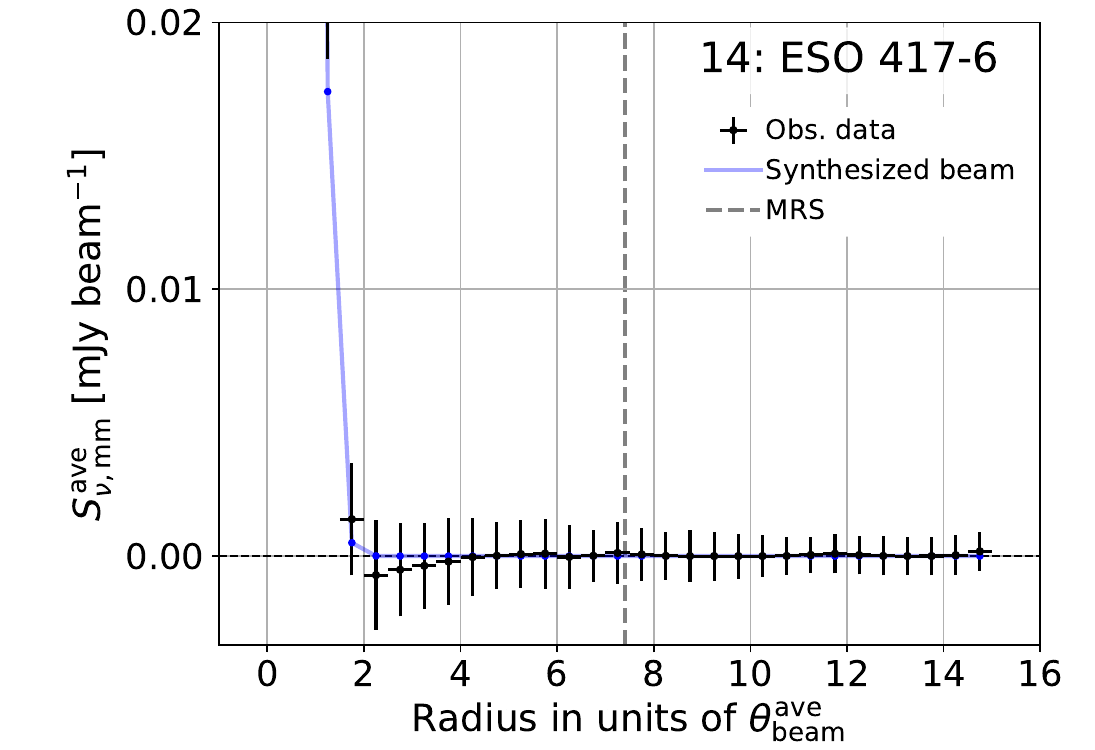}
\includegraphics[width=5.9cm]{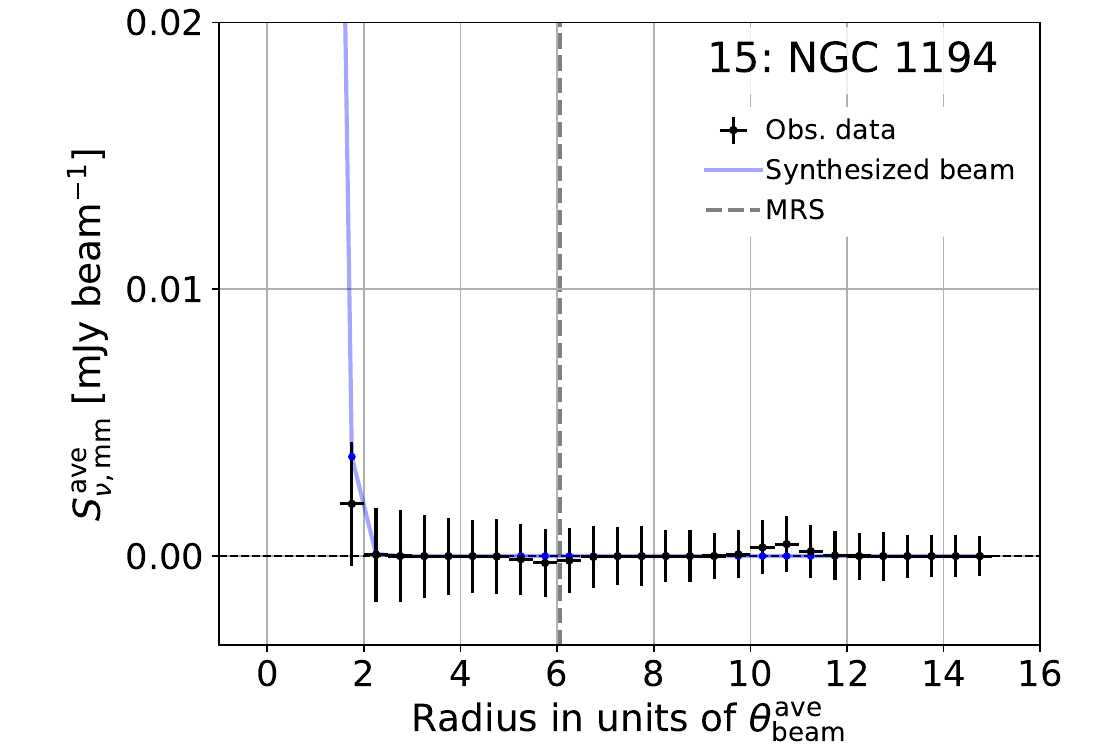}
    \caption{
    Radial surface brightness profiles. The radius is in units of the averaged beam size ($\theta^{\rm ave}_{\rm beam}$). In each figure, 
observed values are indicated with black error bars, but particularly those detected at significances more than three times the SEs with respect to an un-resolved component are shown in orange. The blue line represents the profile of the synthesized beam whose innermost value is adjusted to the observed one. The dashed gray line indicates the MRS.
    }\label{fig_app:rads}
\end{figure*}

\addtocounter{figure}{-1}

\begin{figure*}
    \centering    
\includegraphics[width=5.9cm]{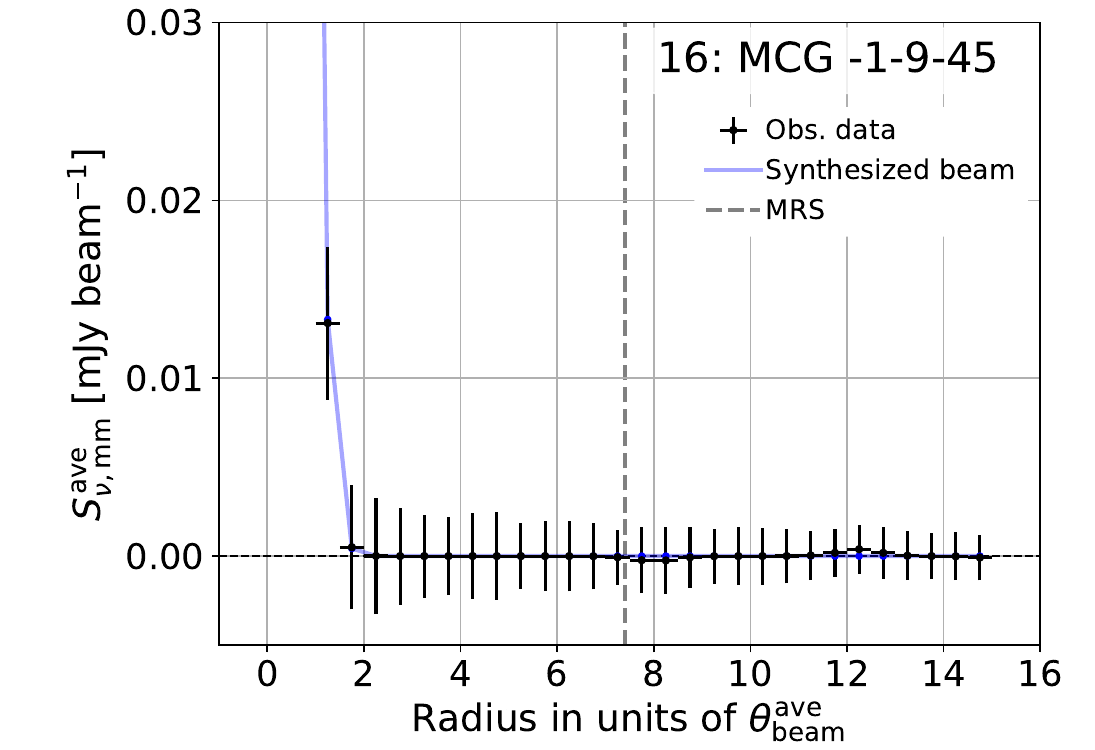}
\includegraphics[width=5.9cm]{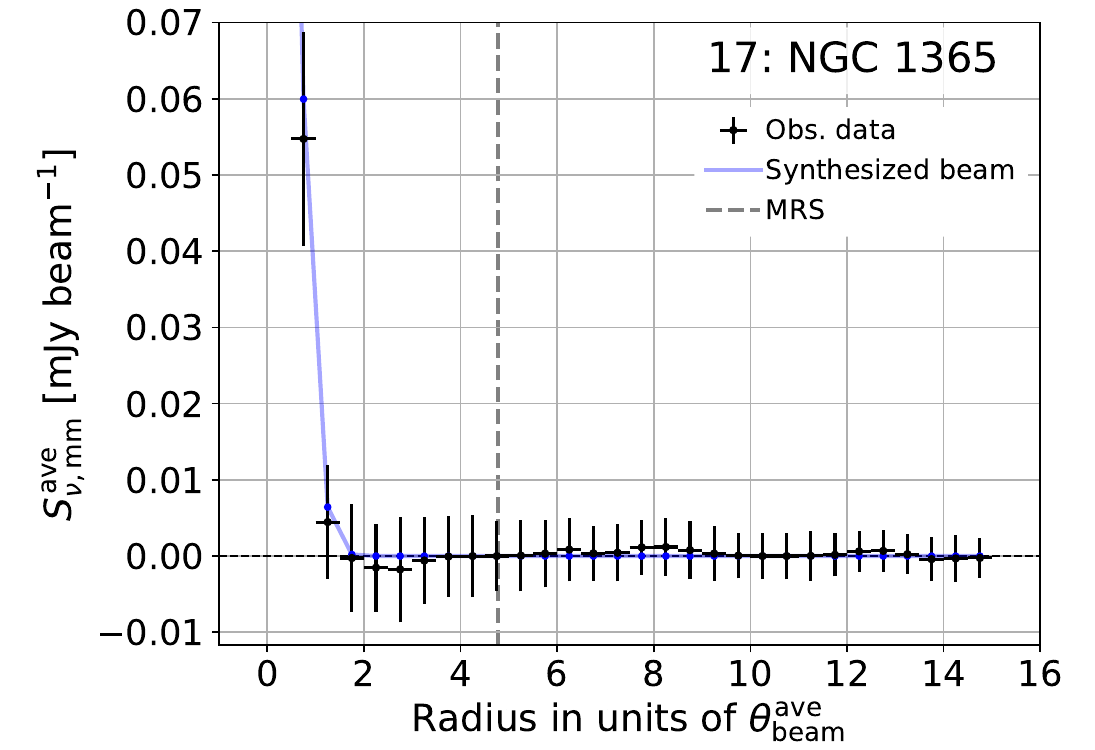}
\includegraphics[width=5.9cm]{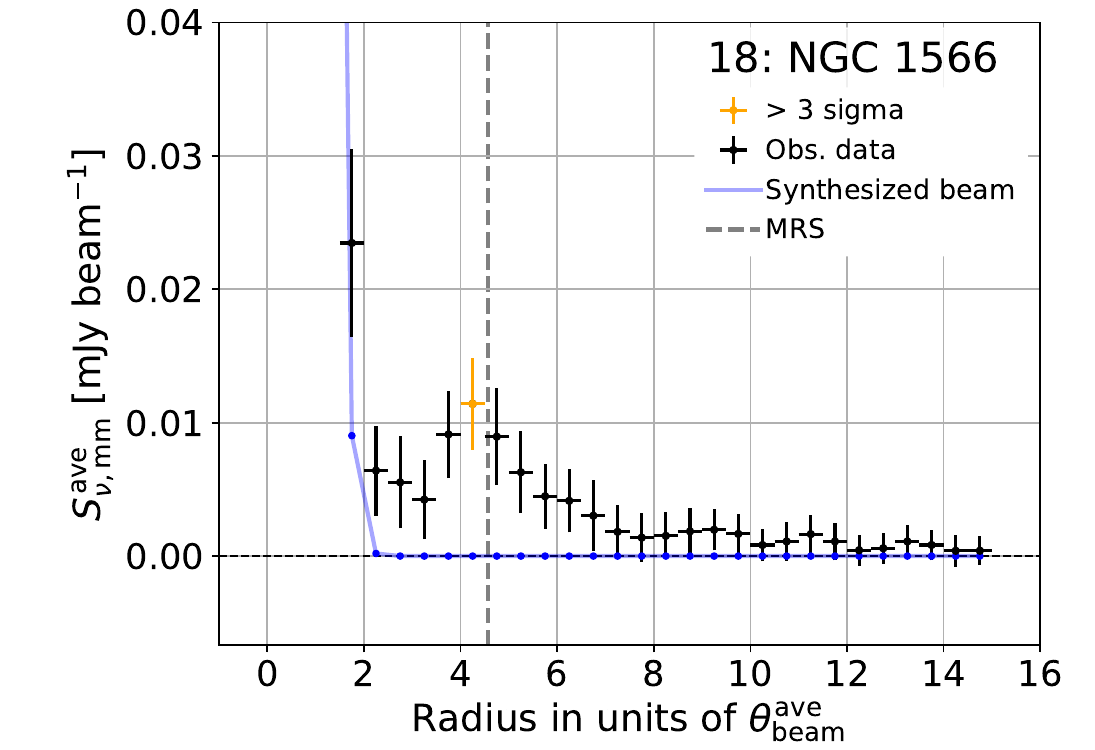}
\includegraphics[width=5.9cm]{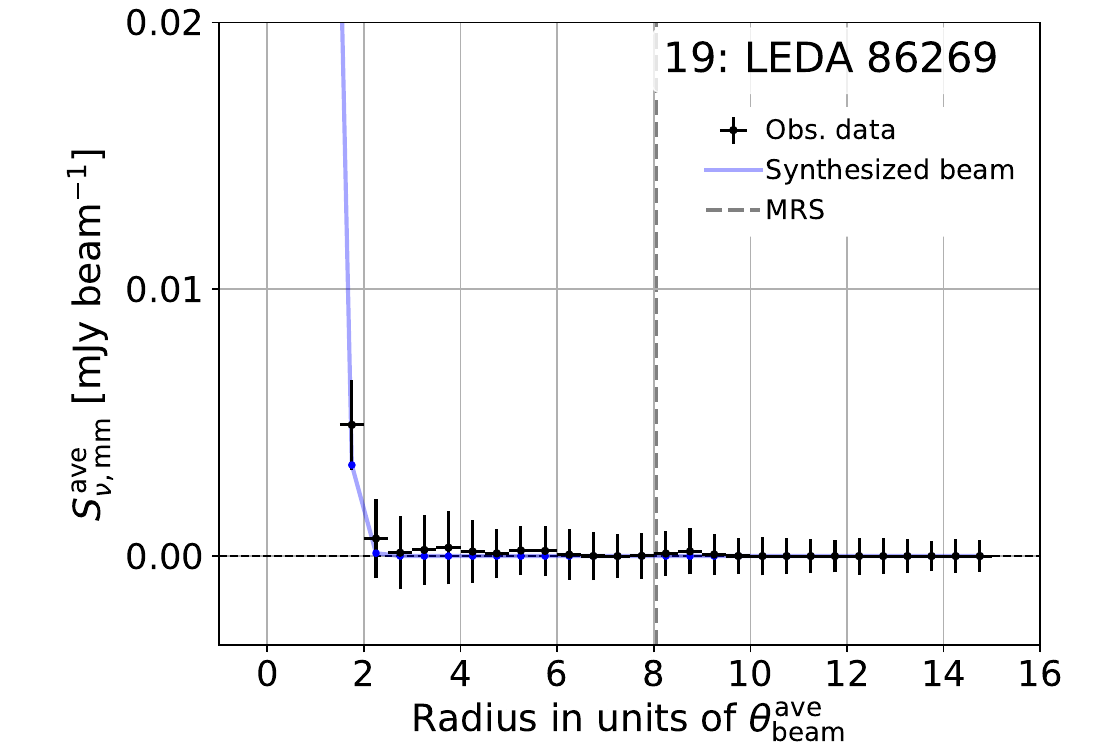}
\includegraphics[width=5.9cm]{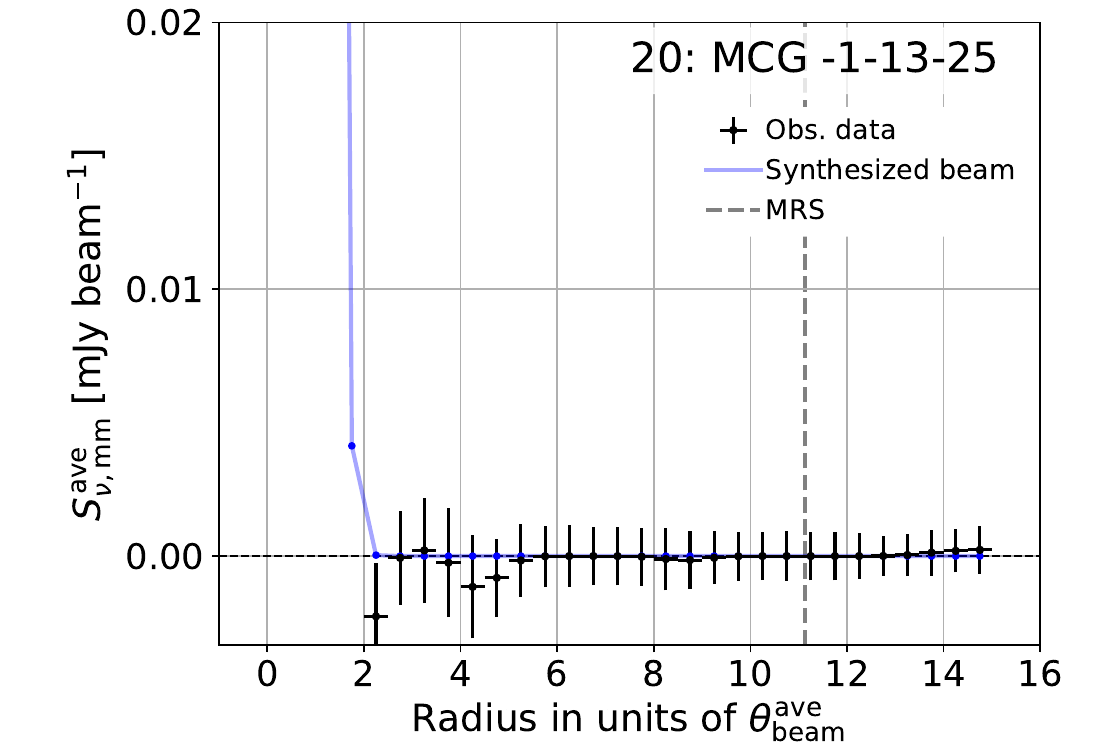}
\includegraphics[width=5.9cm]{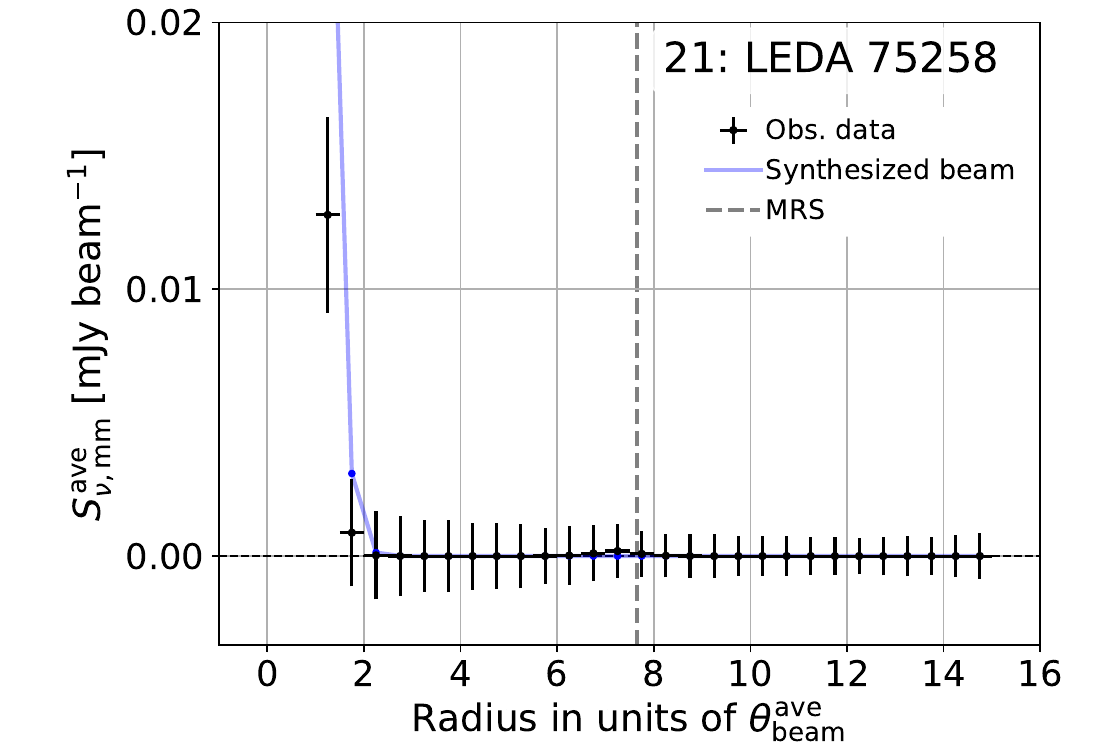}
\includegraphics[width=5.9cm]{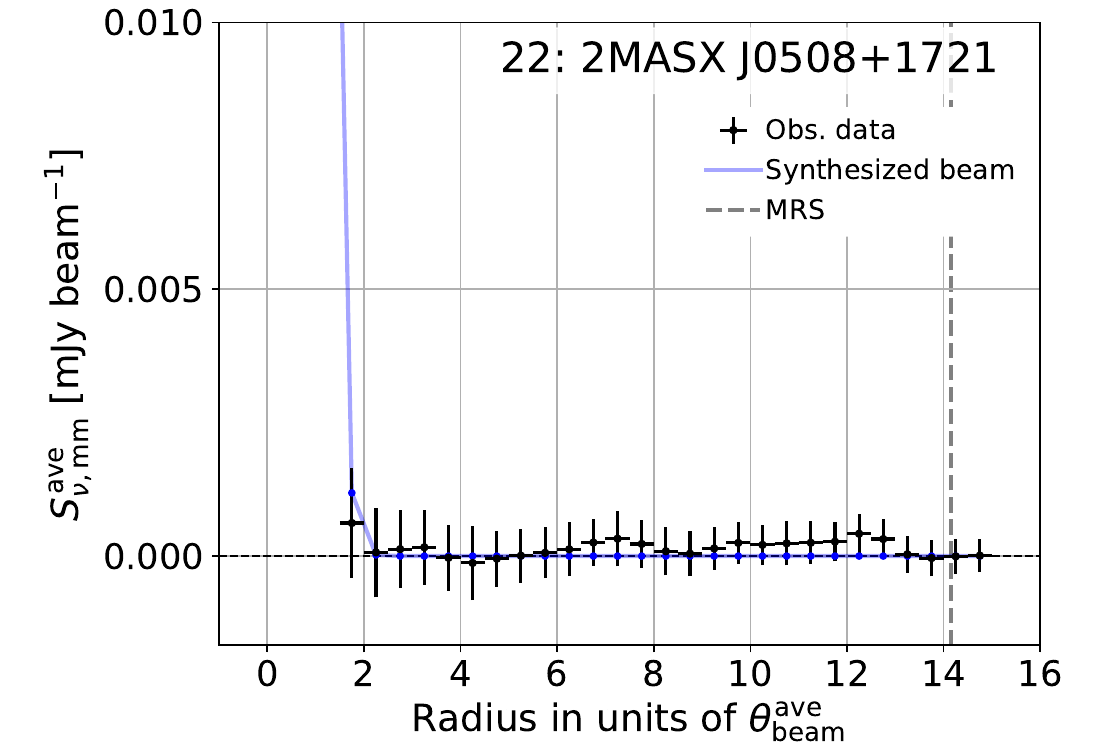}
\includegraphics[width=5.9cm]{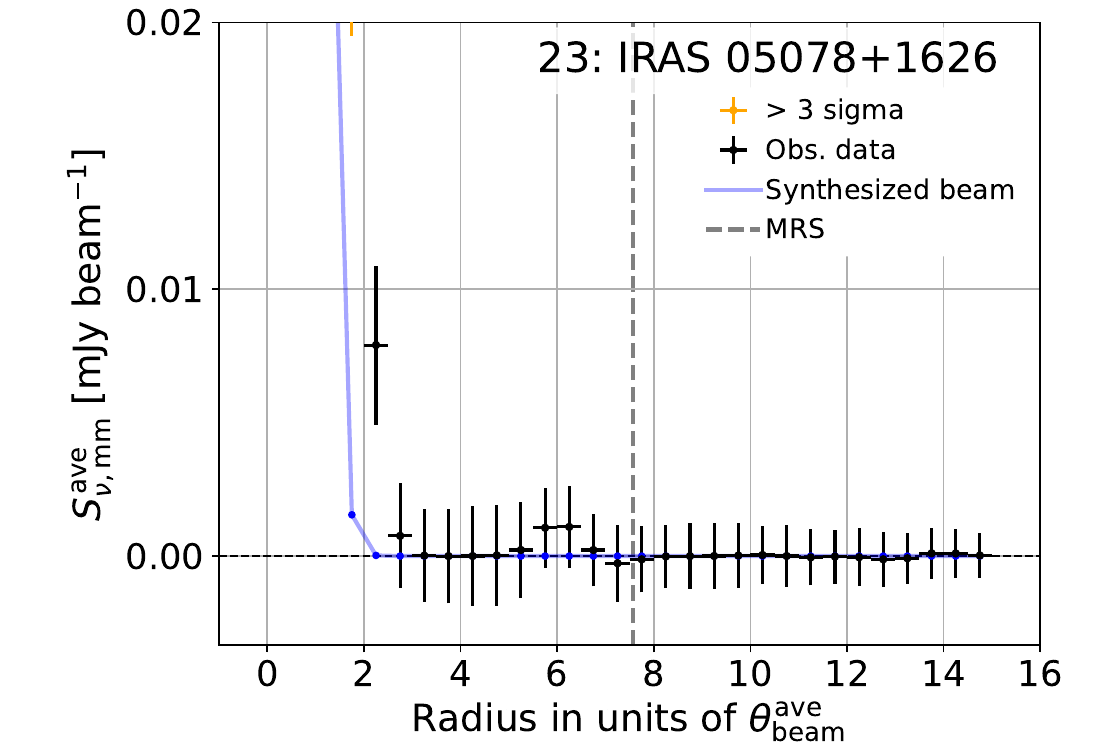}
\includegraphics[width=5.9cm]{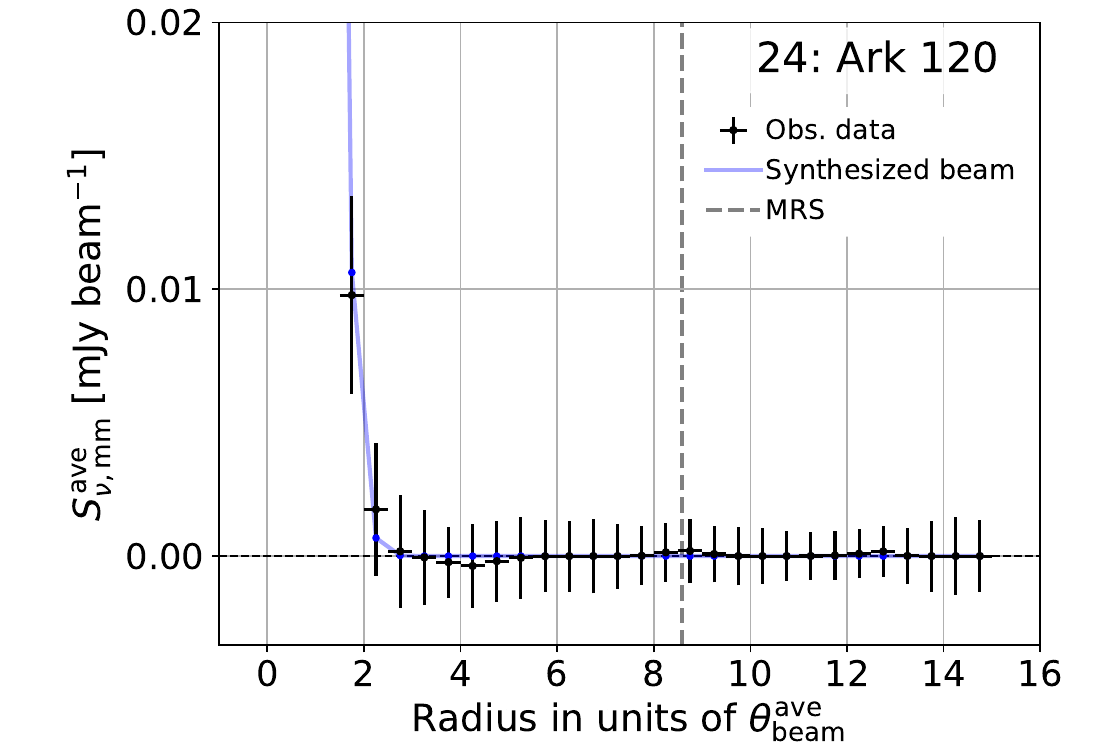}
\includegraphics[width=5.9cm]{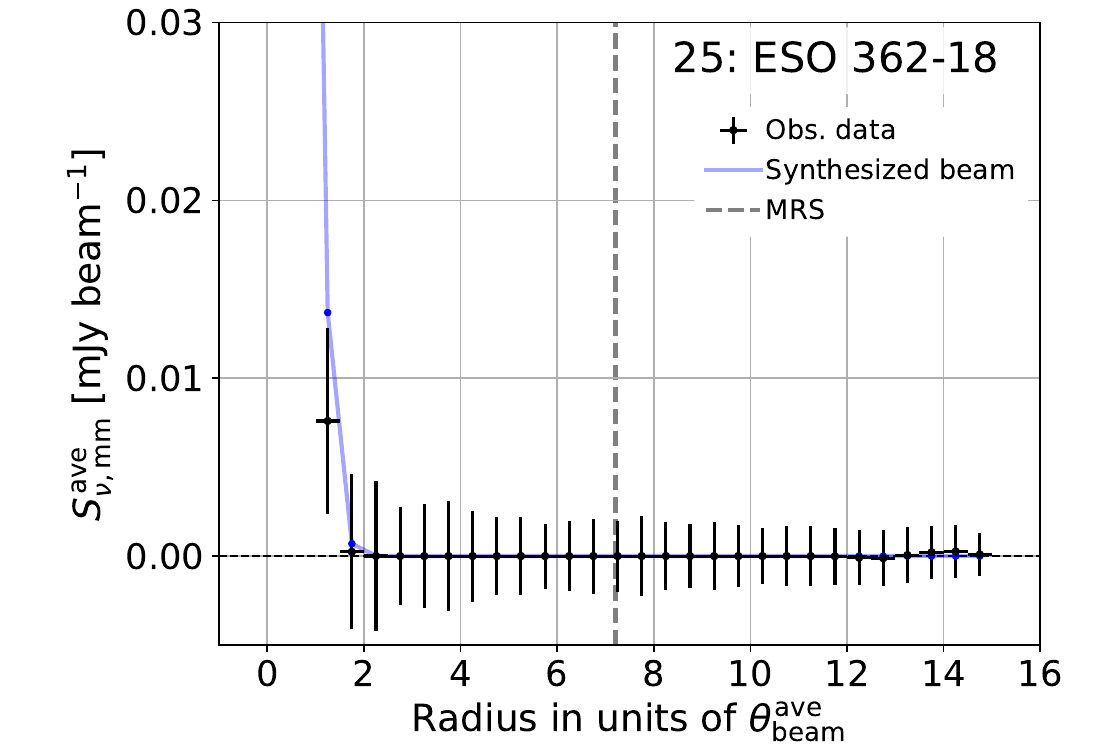}
\includegraphics[width=5.9cm]{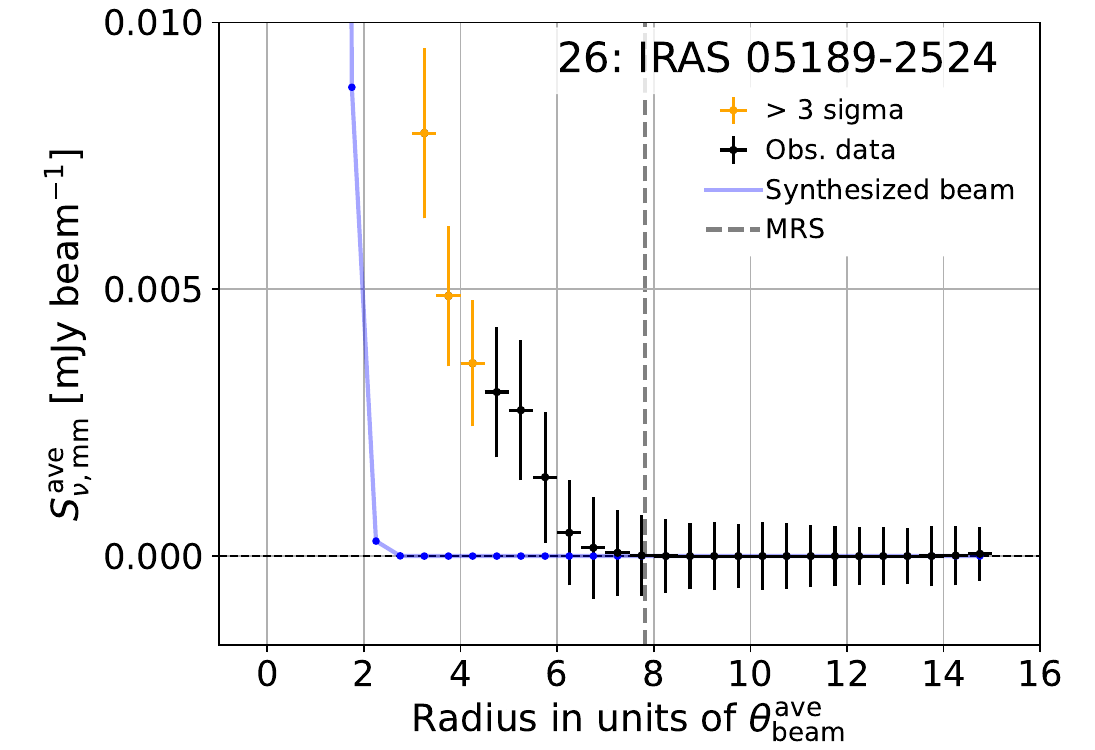}
\includegraphics[width=5.9cm]{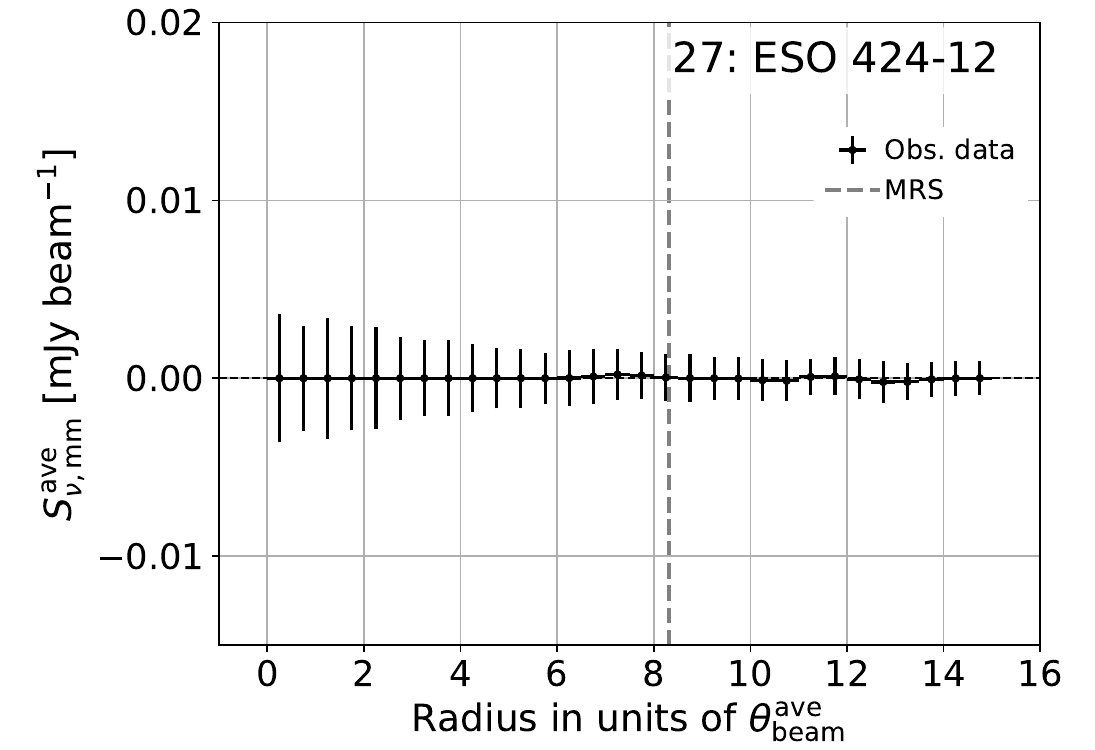}
\includegraphics[width=5.9cm]{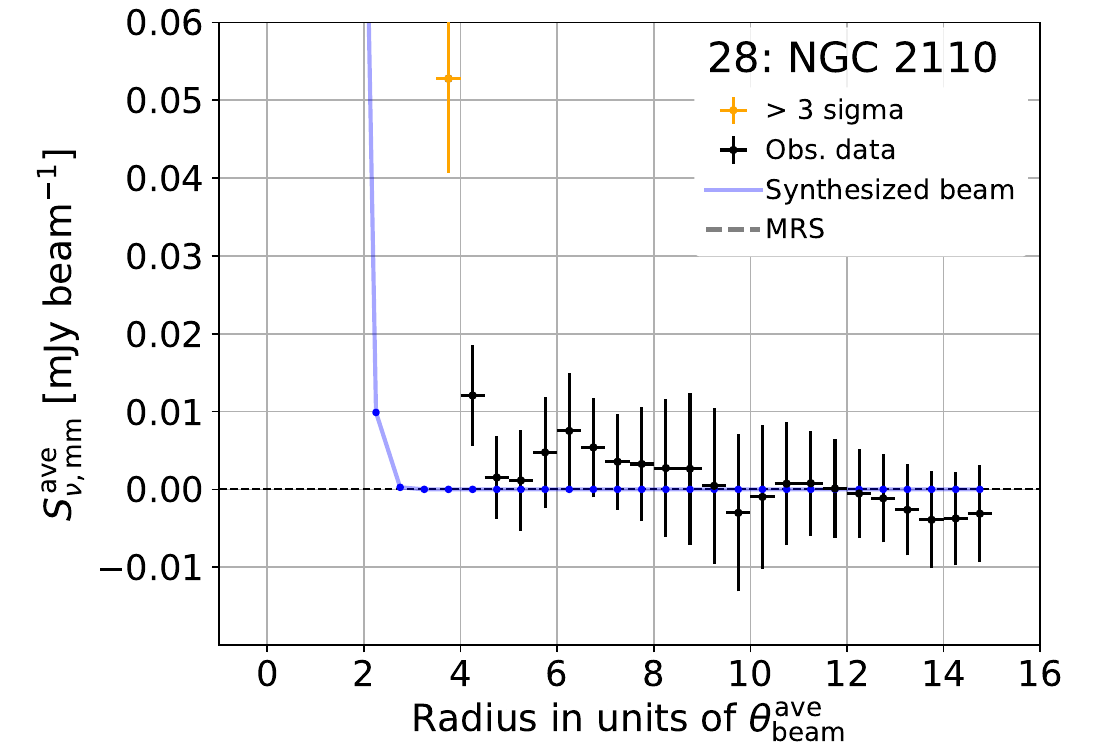}
\includegraphics[width=5.9cm]{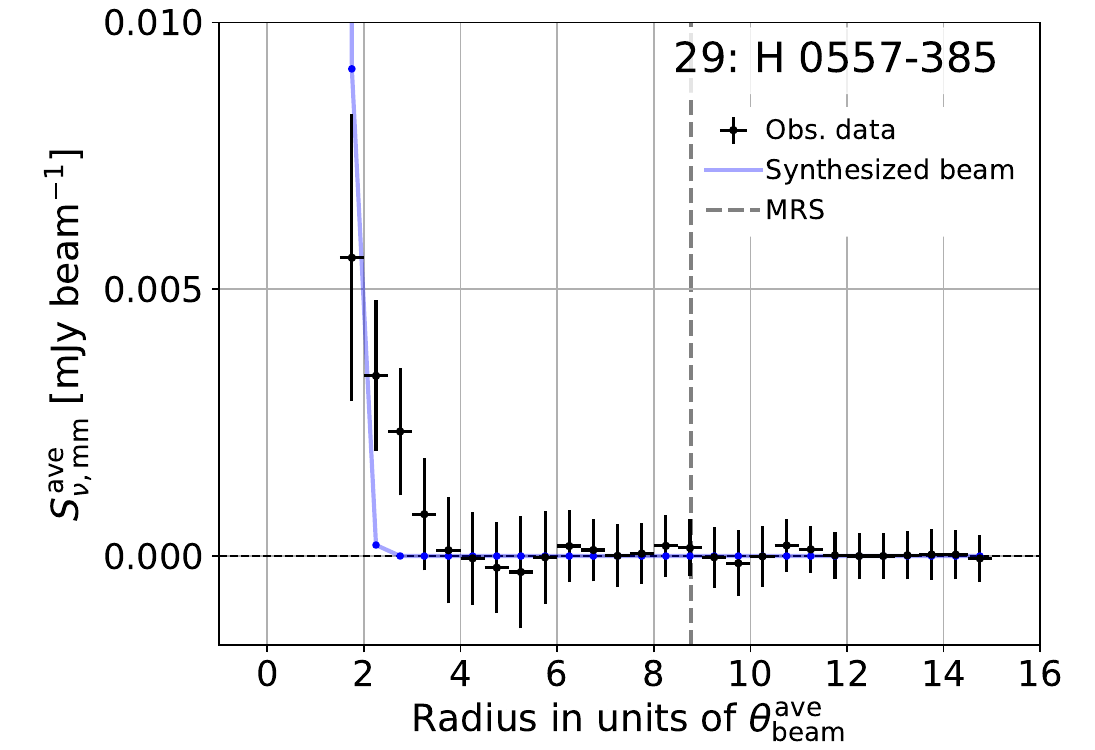}
\includegraphics[width=5.9cm]{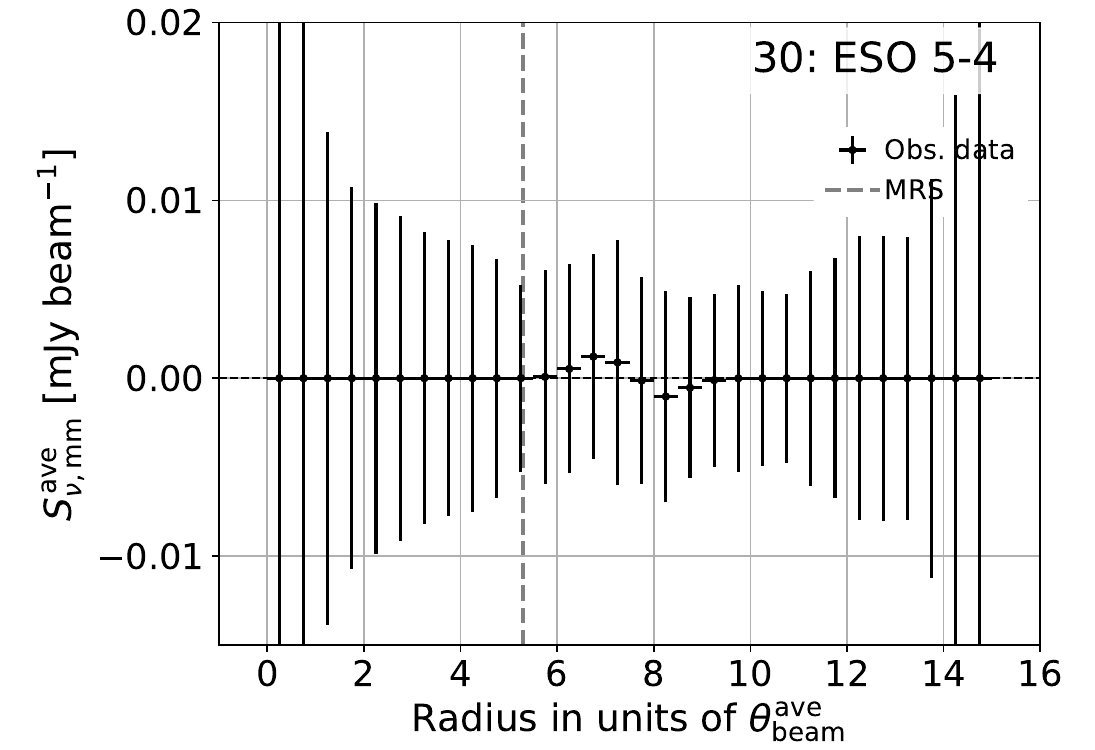}
\caption{Continued. 
    }
\end{figure*}

\addtocounter{figure}{-1}

\begin{figure*}
    \centering    
\includegraphics[width=5.9cm]{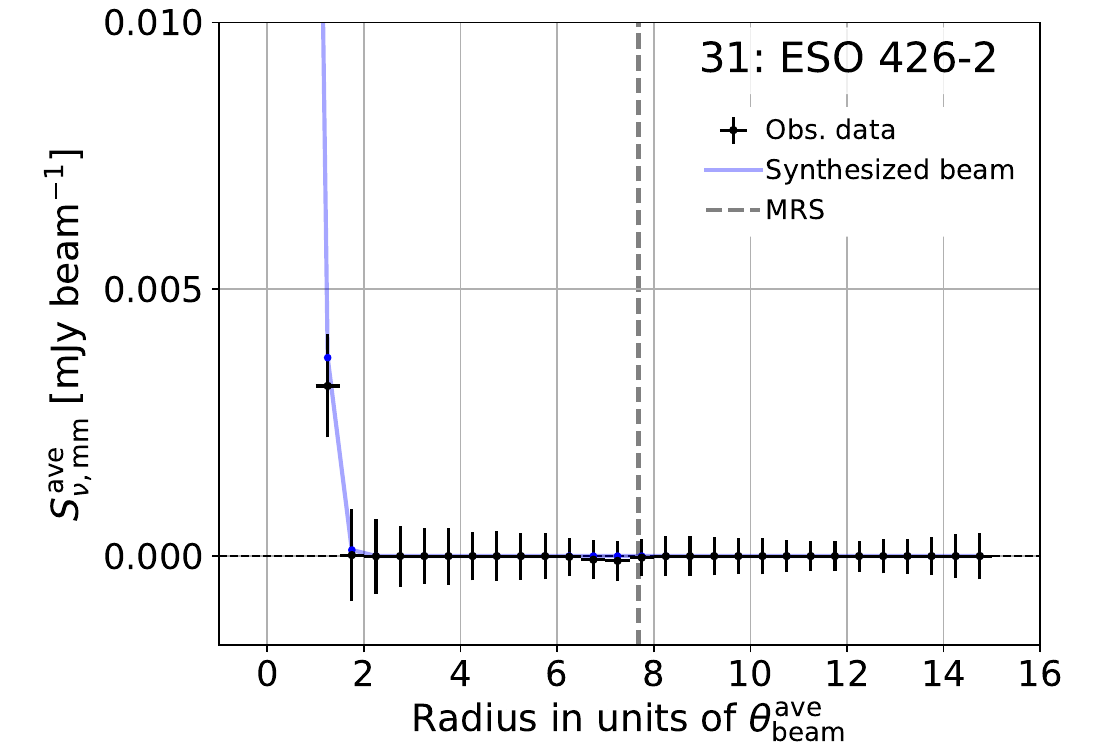}
\includegraphics[width=5.9cm]{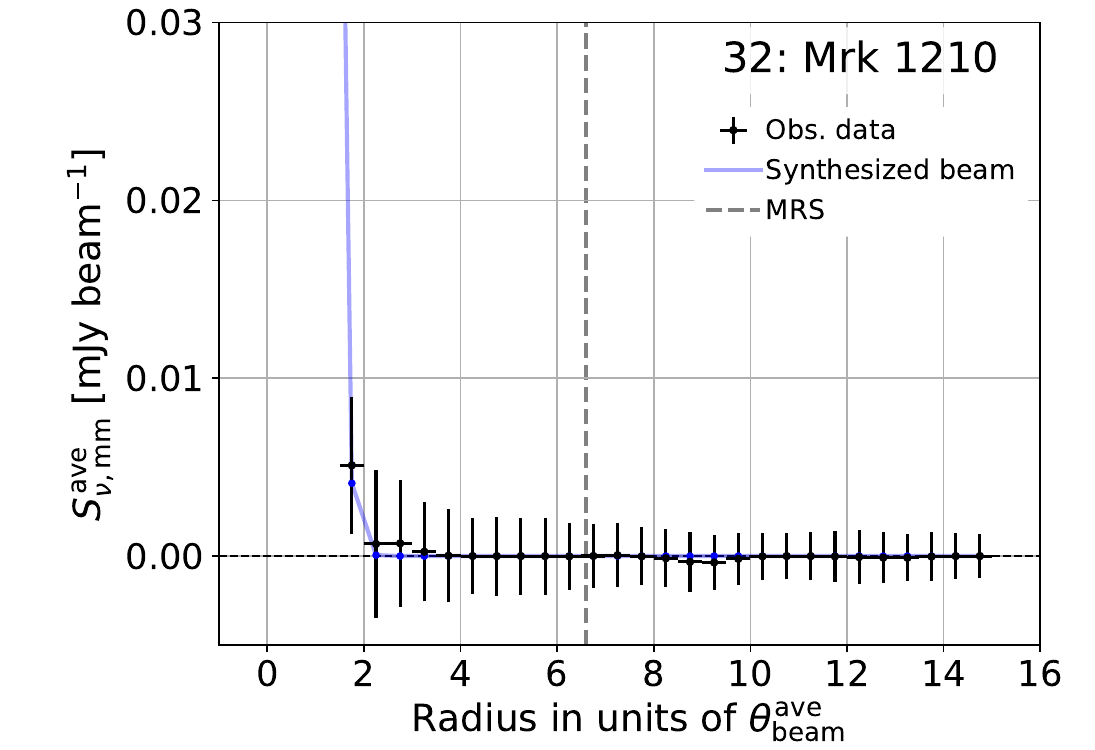}
\includegraphics[width=5.9cm]{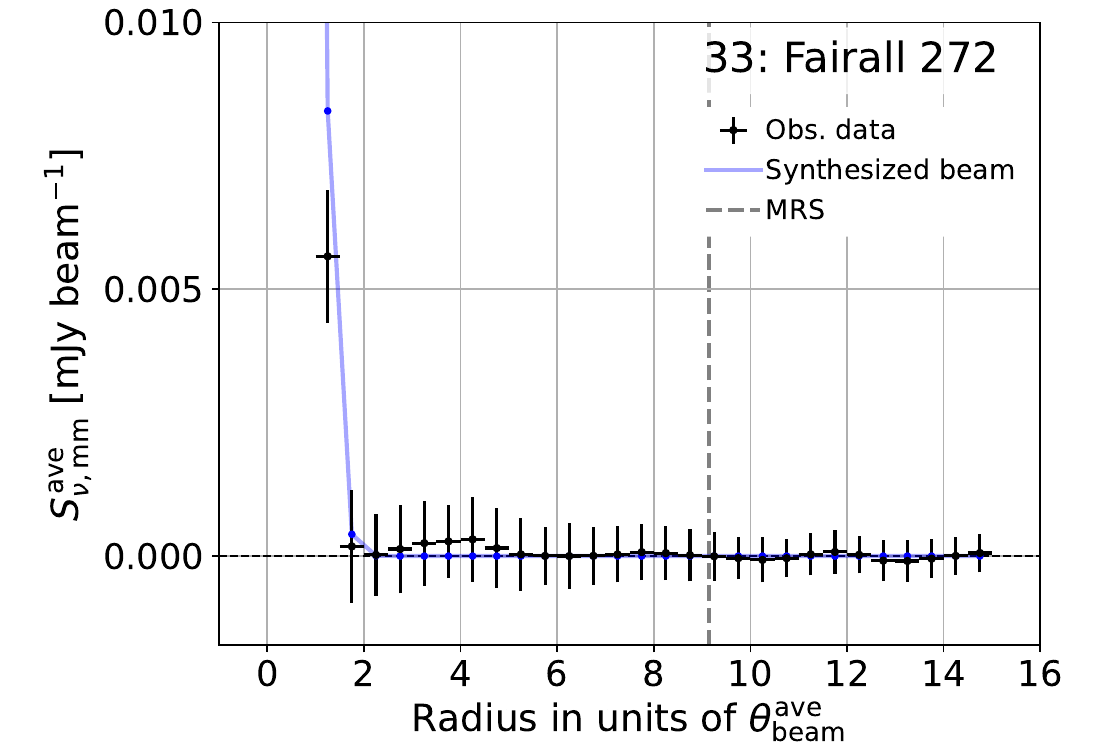}
\includegraphics[width=5.9cm]{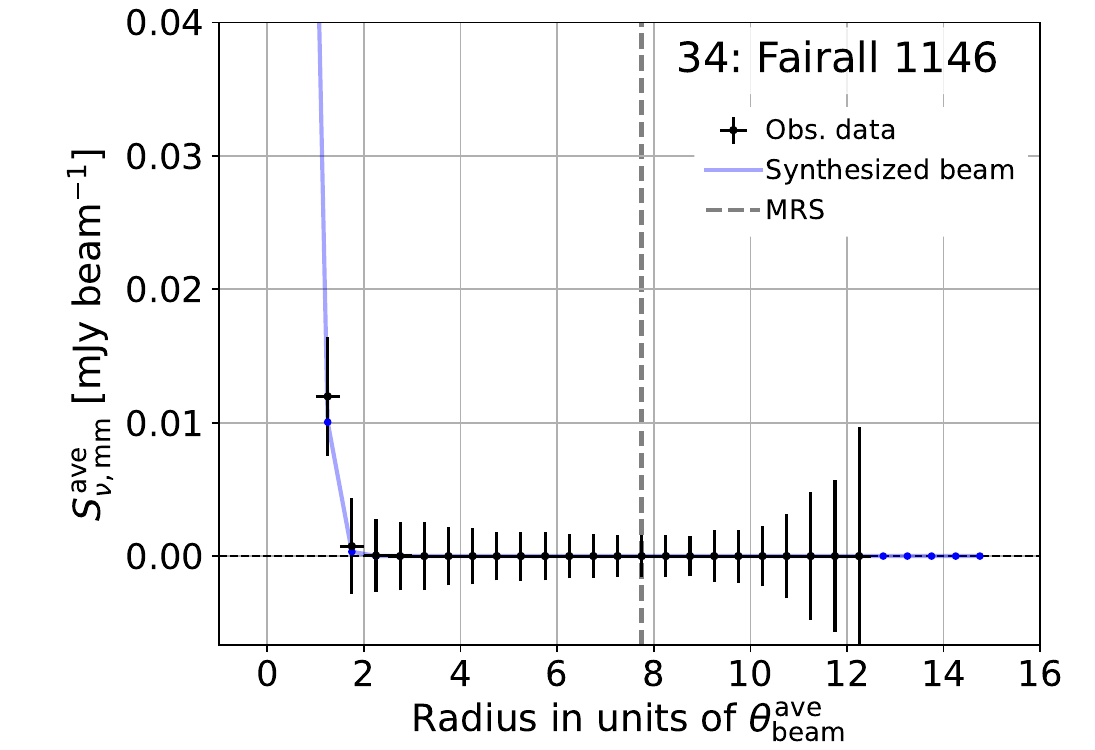}
\includegraphics[width=5.9cm]{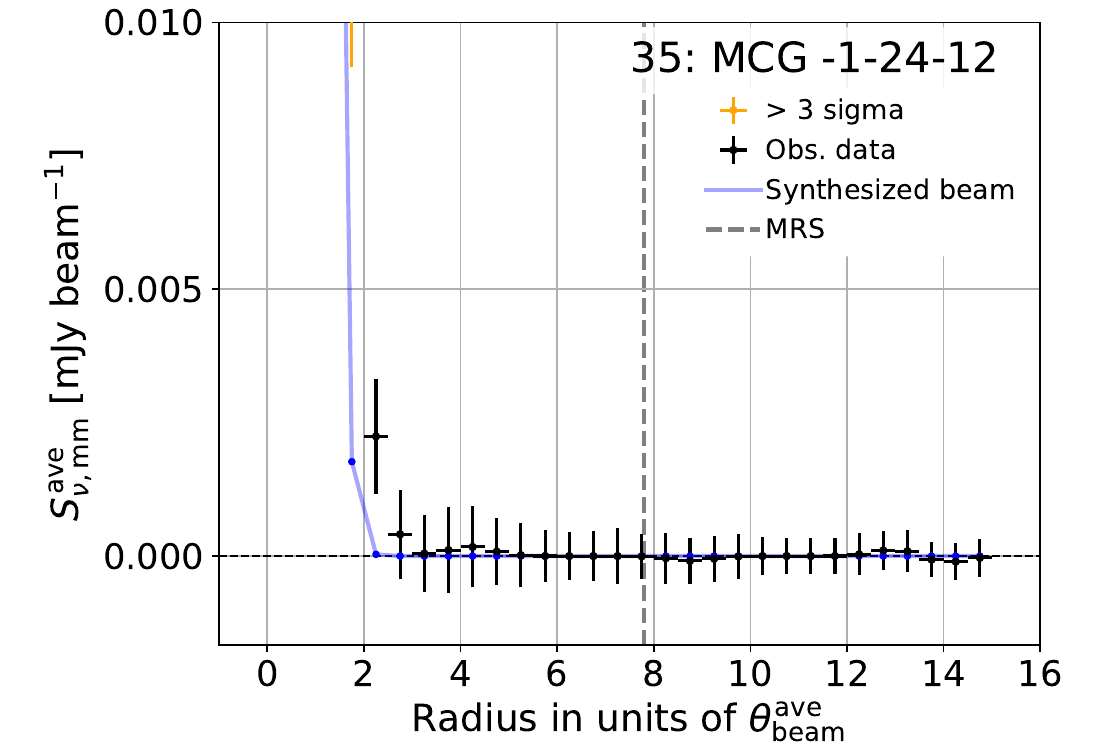}
\includegraphics[width=5.9cm]{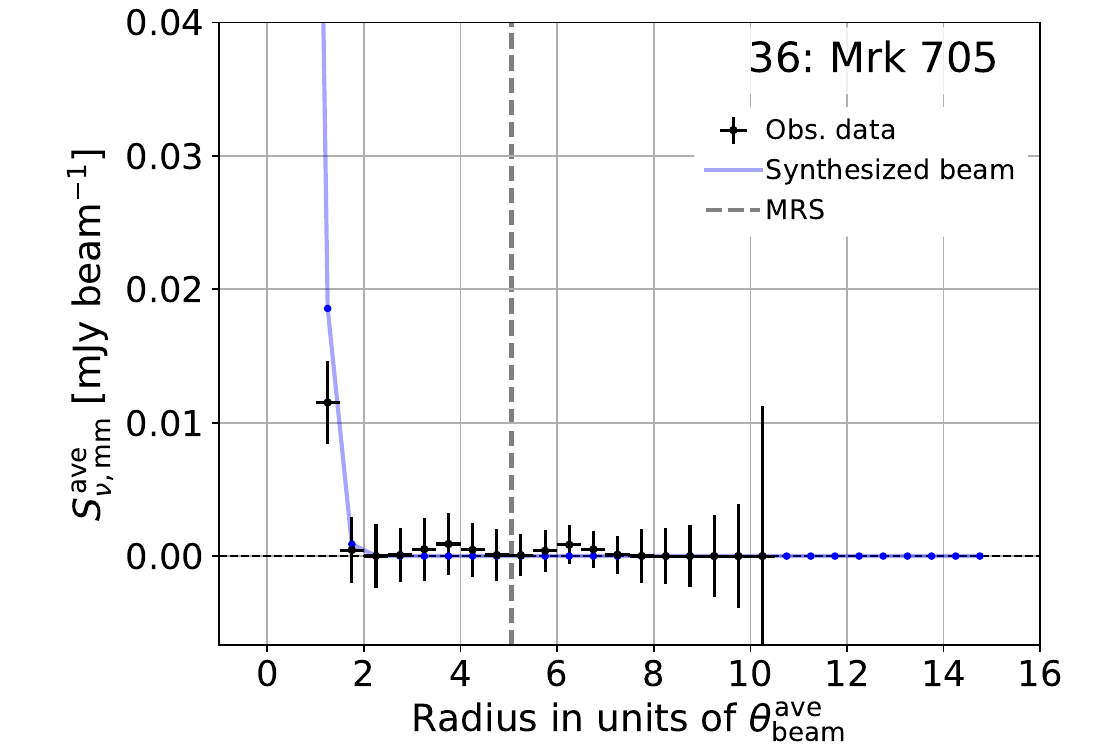}
\includegraphics[width=5.9cm]{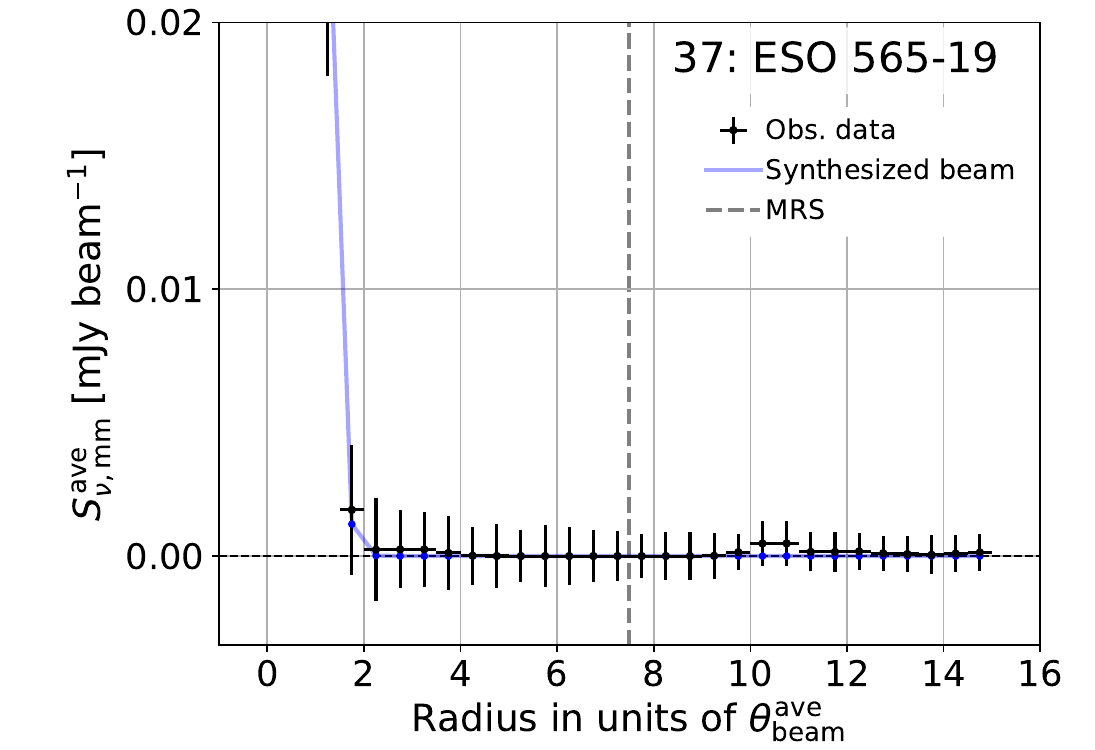}
\includegraphics[width=5.9cm]{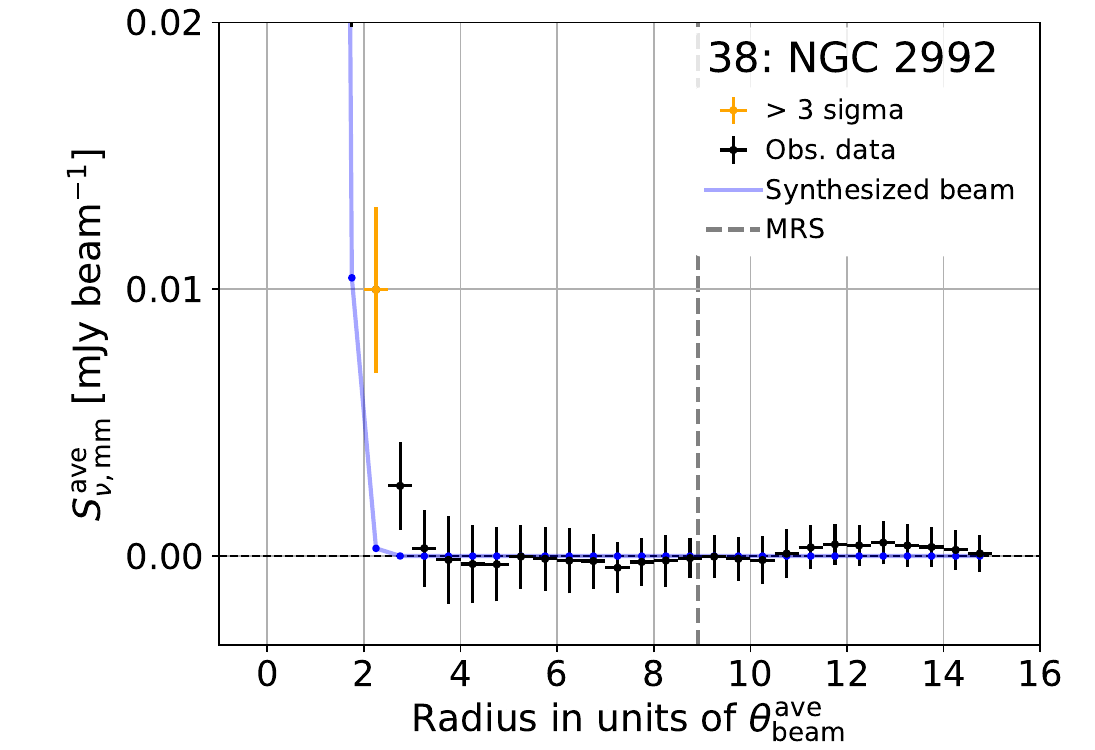}
\includegraphics[width=5.9cm]{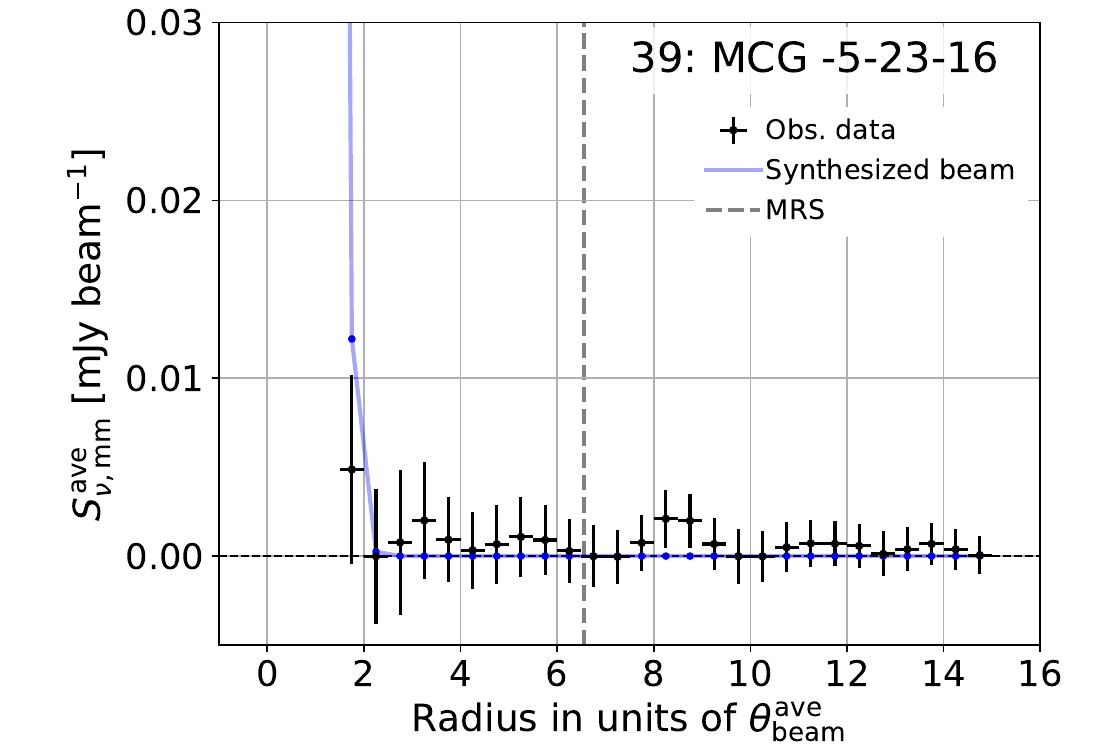}
\includegraphics[width=5.9cm]{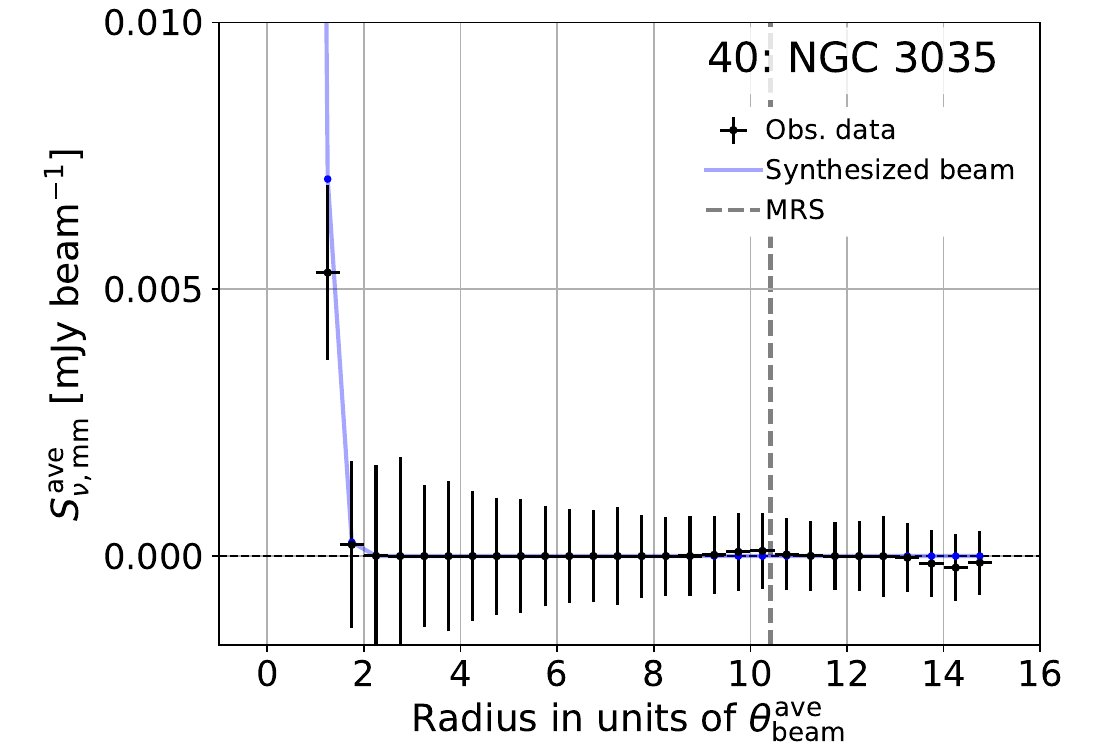}
\includegraphics[width=5.9cm]{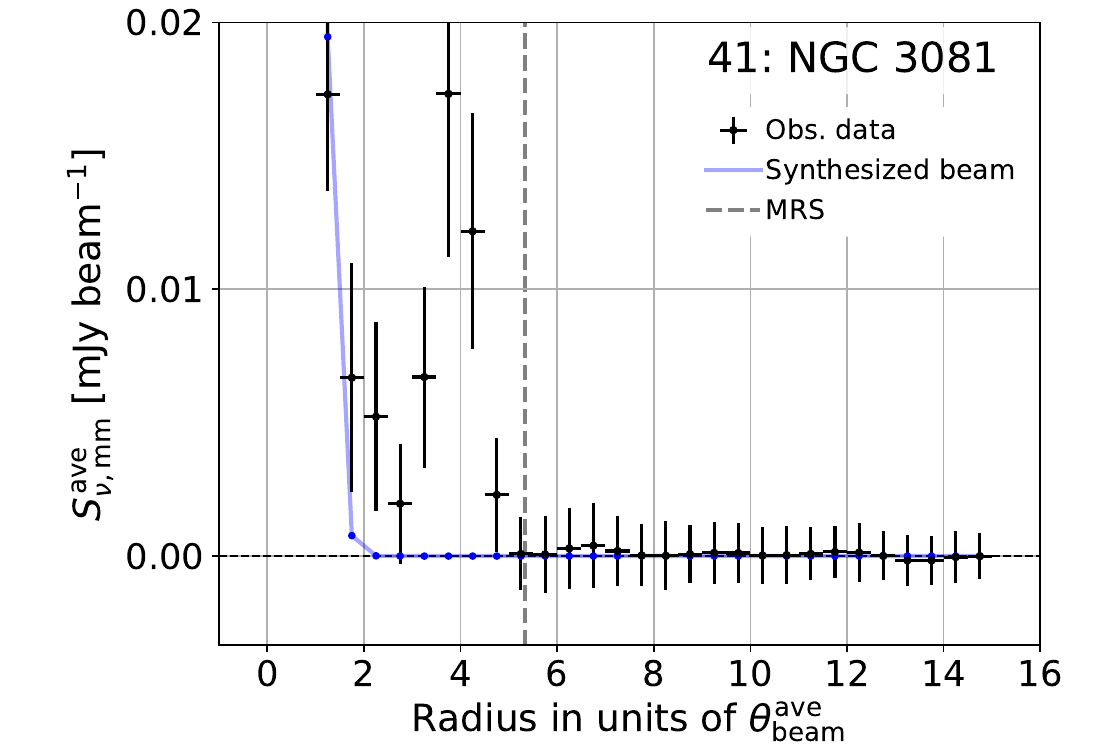}
\includegraphics[width=5.9cm]{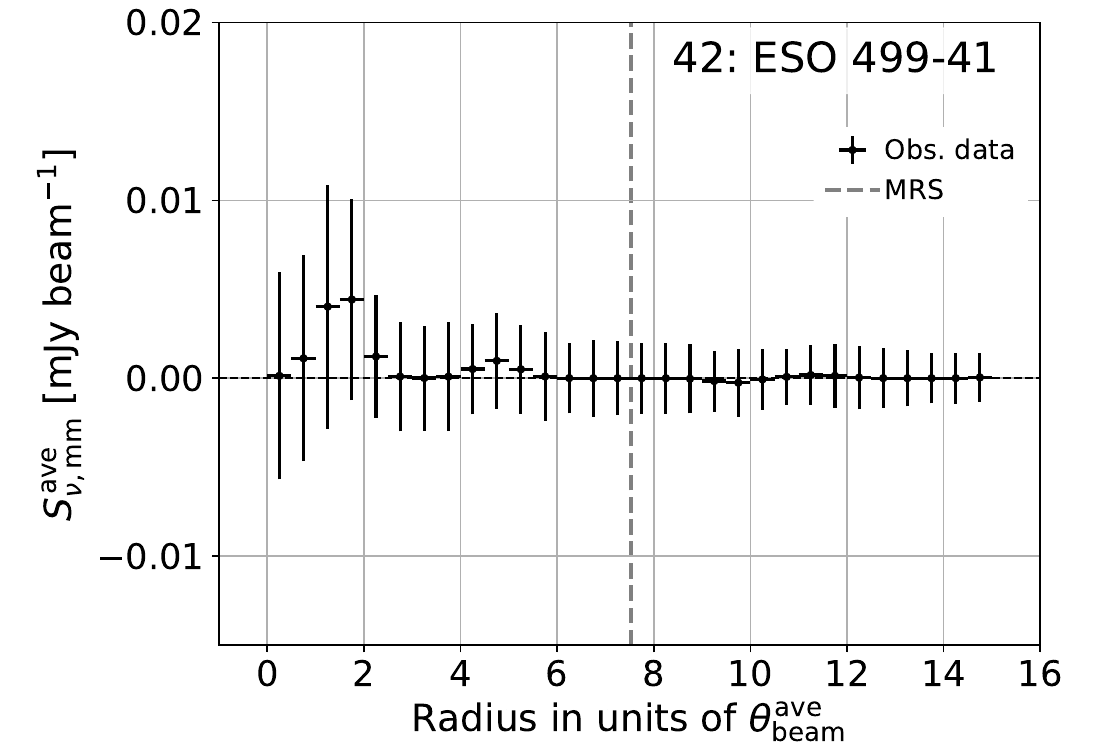}
\includegraphics[width=5.9cm]{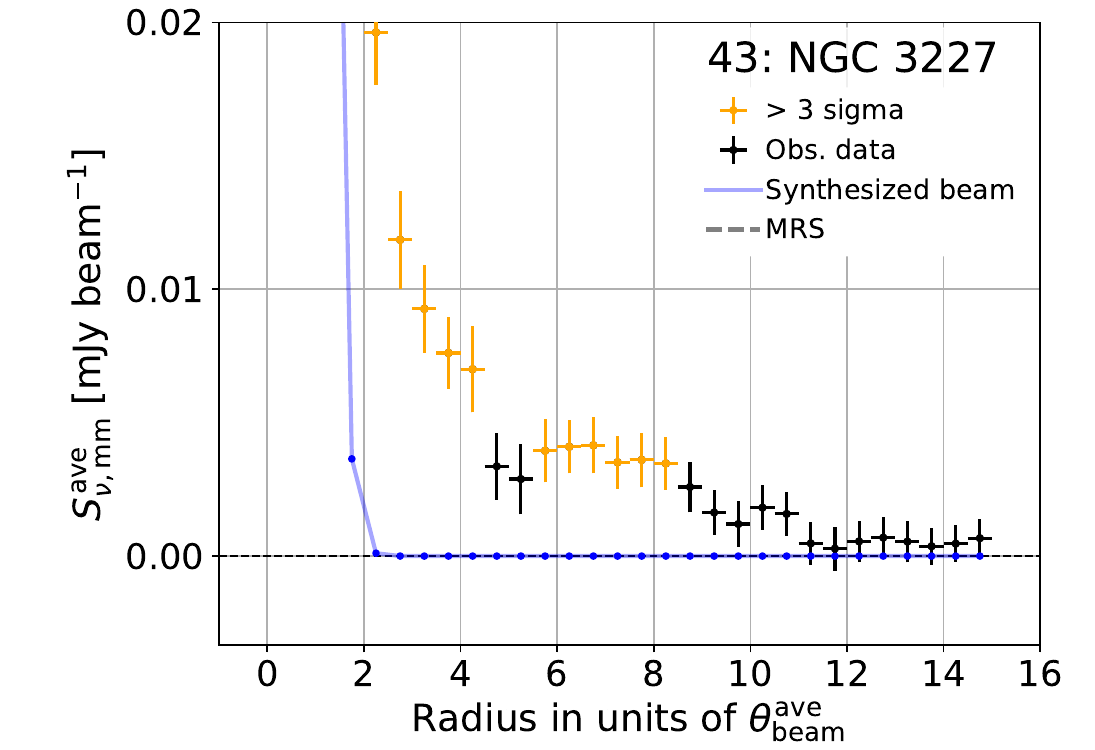}
\includegraphics[width=5.9cm]{044_NGC_3281_radpro.pdf}
\includegraphics[width=5.9cm]{045_NGC_3393_radpro.pdf}
\caption{Continued. 
    }
\end{figure*}

\addtocounter{figure}{-1}

\begin{figure*}
    \centering    
\includegraphics[width=5.9cm]{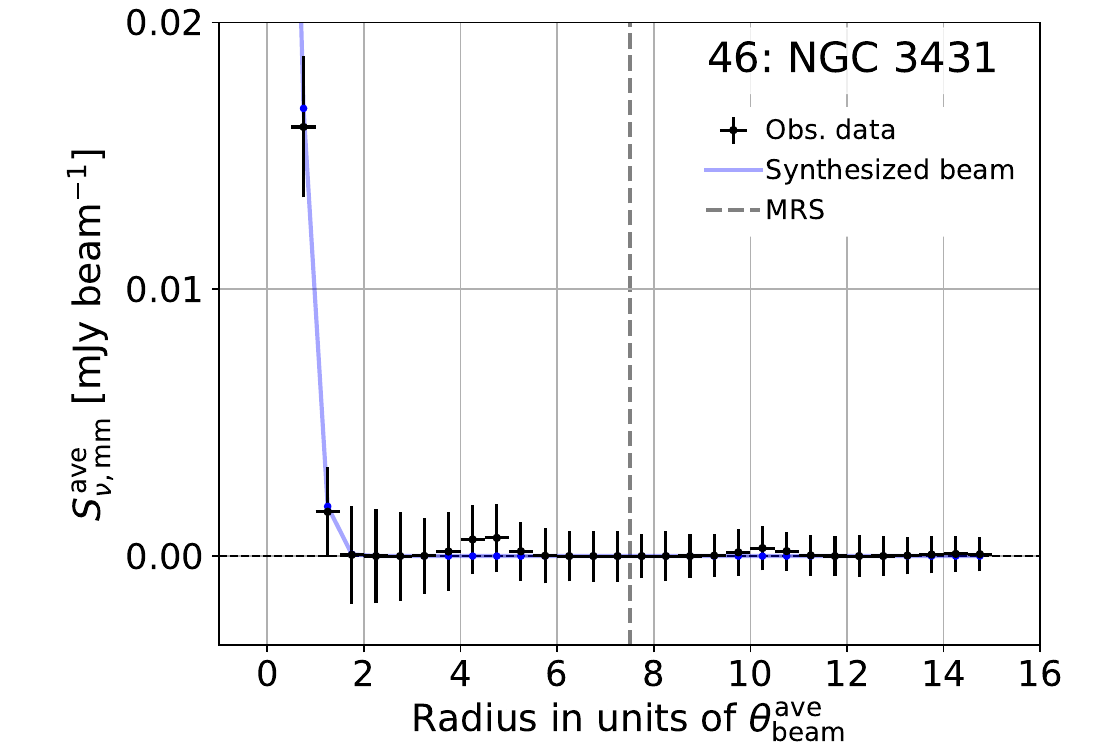}
\includegraphics[width=5.9cm]{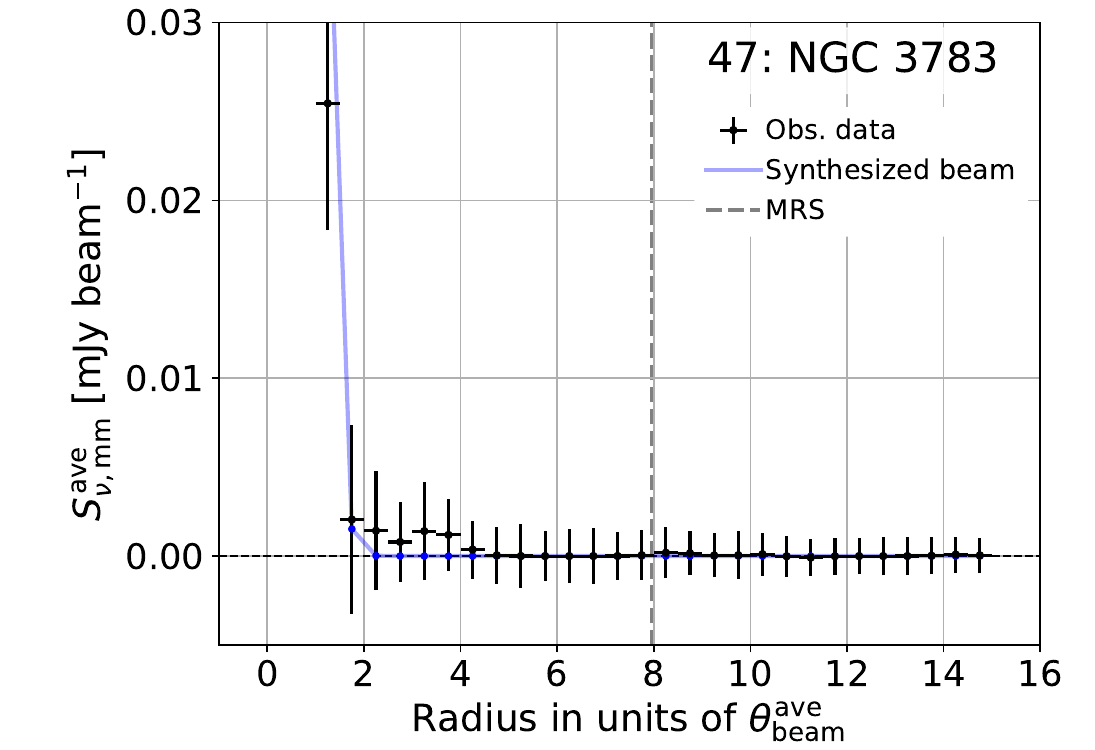}
\includegraphics[width=5.9cm]{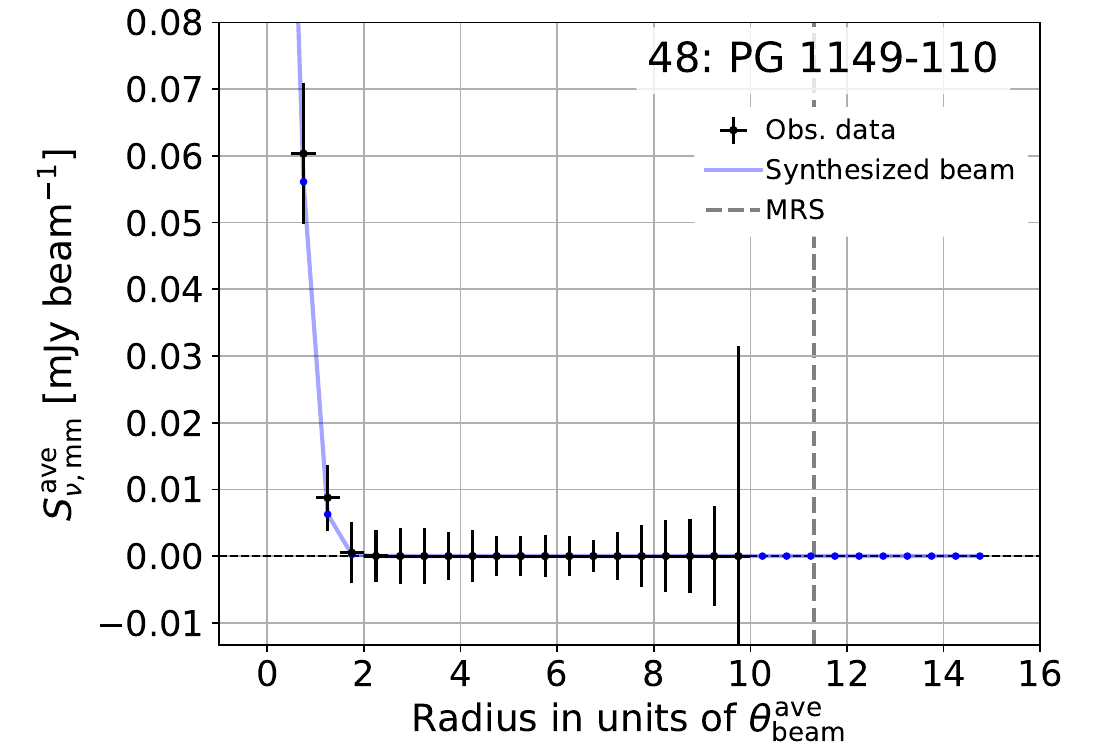}
\includegraphics[width=5.9cm]{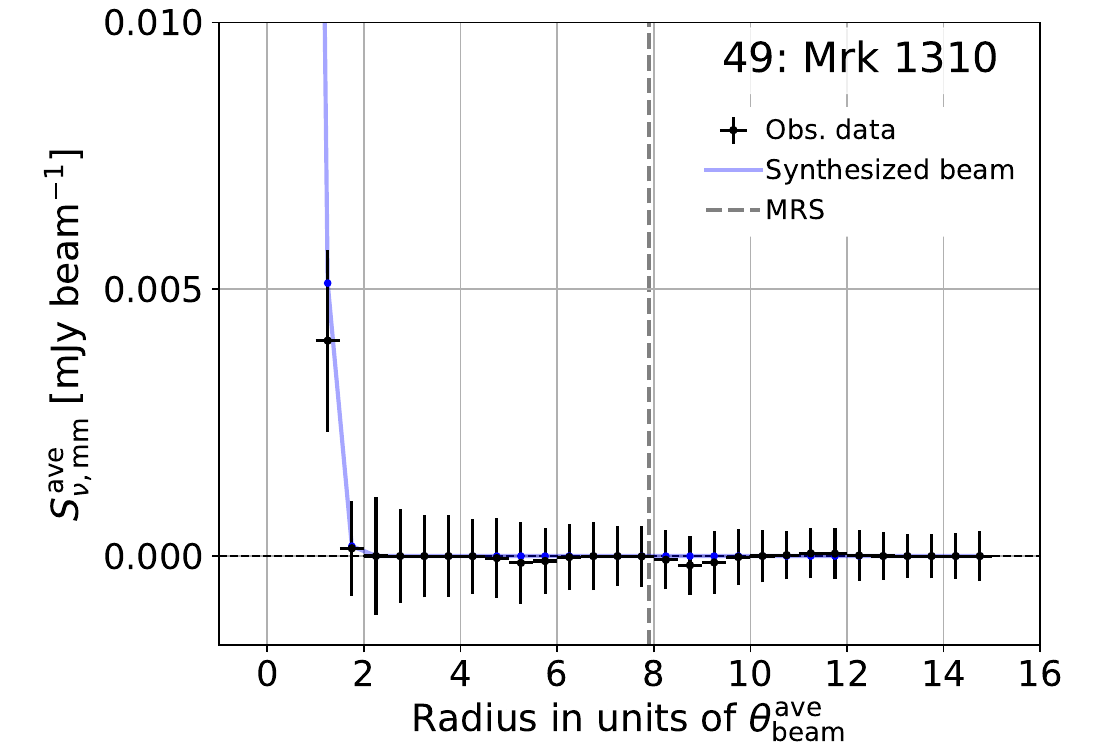}
\includegraphics[width=5.9cm]{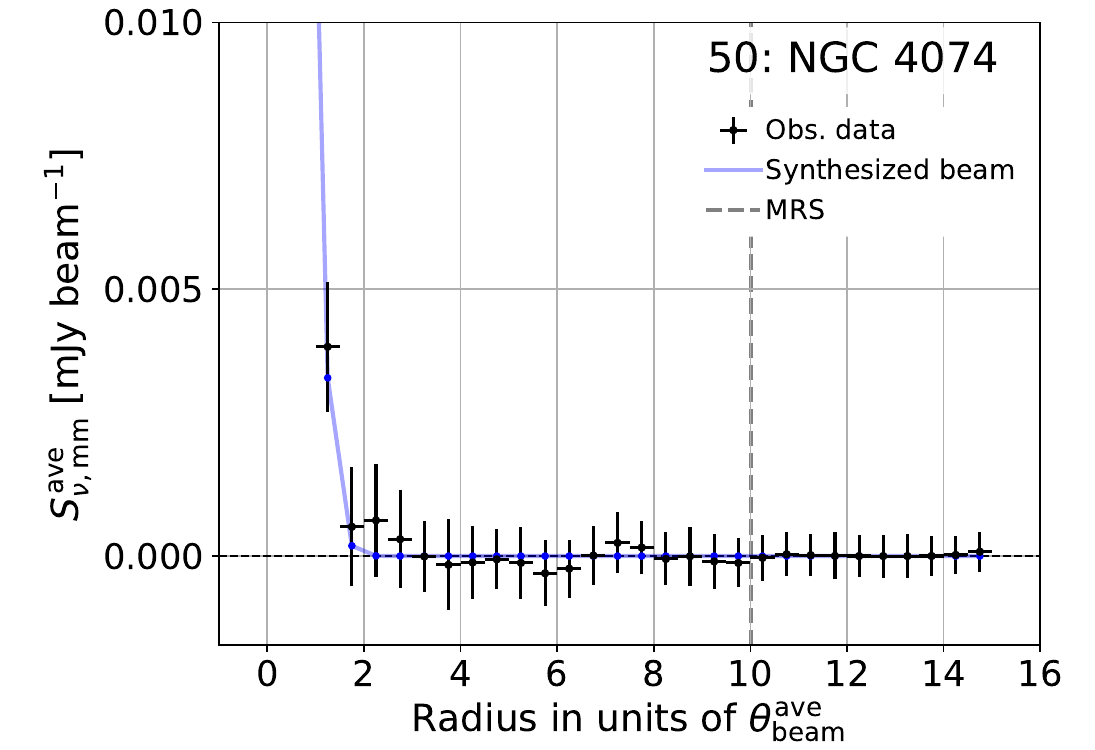}
\includegraphics[width=5.9cm]{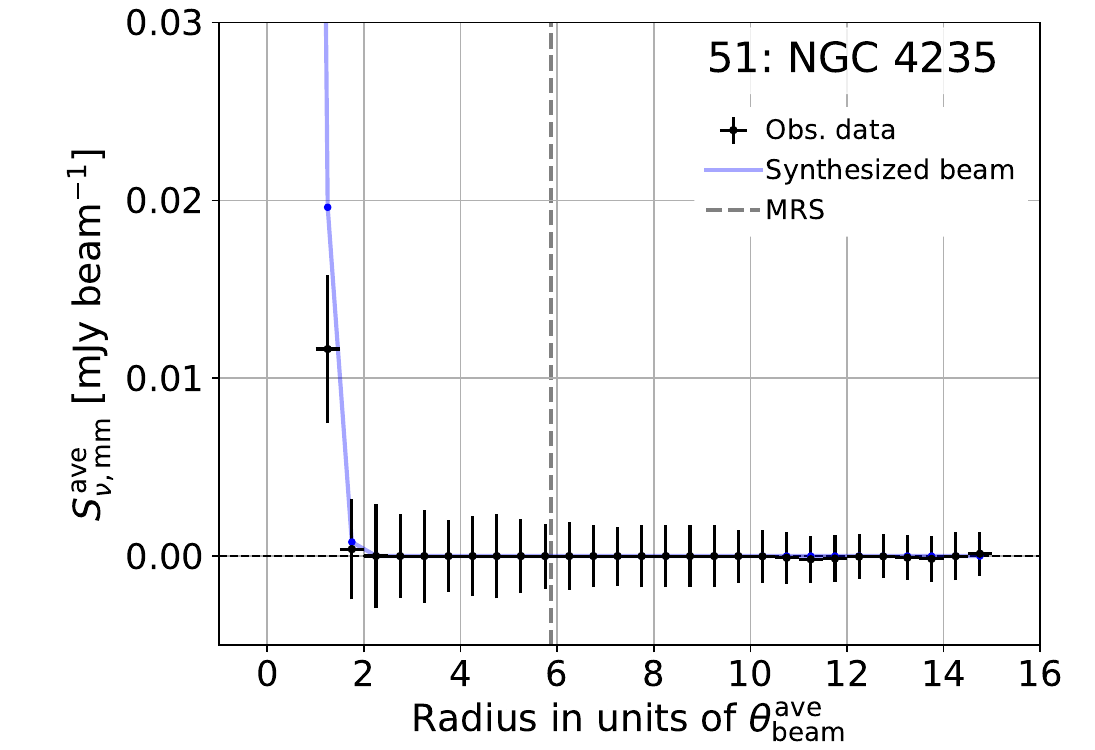}
\includegraphics[width=5.9cm]{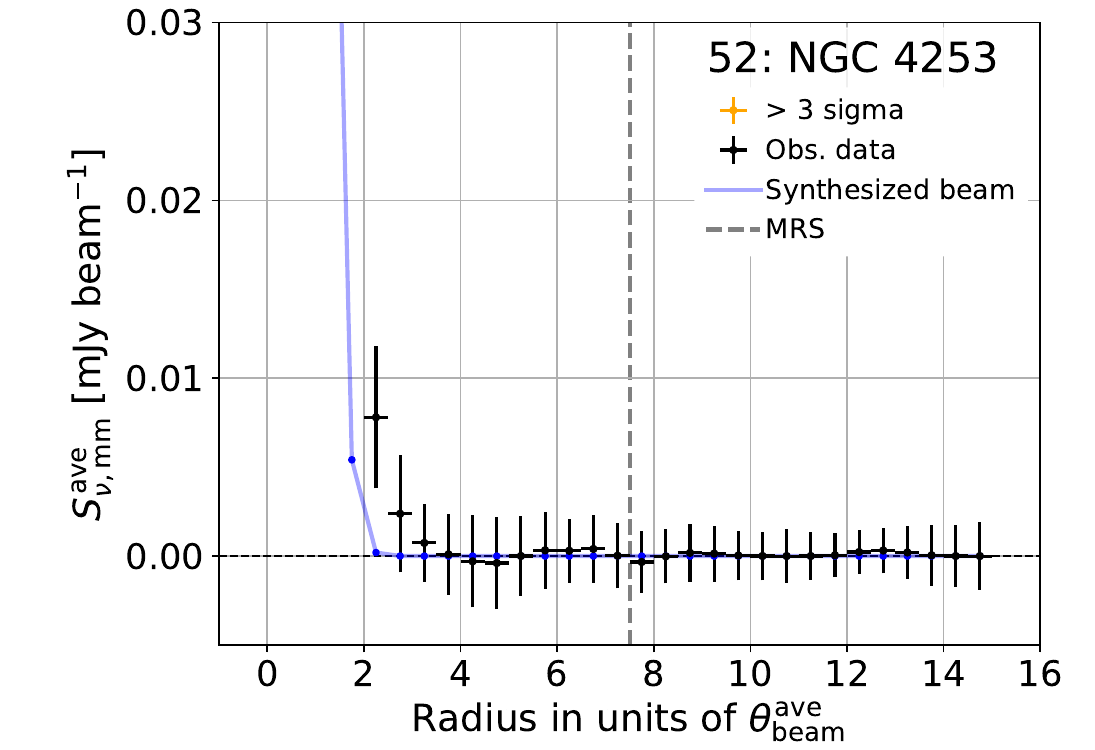}
\includegraphics[width=5.9cm]{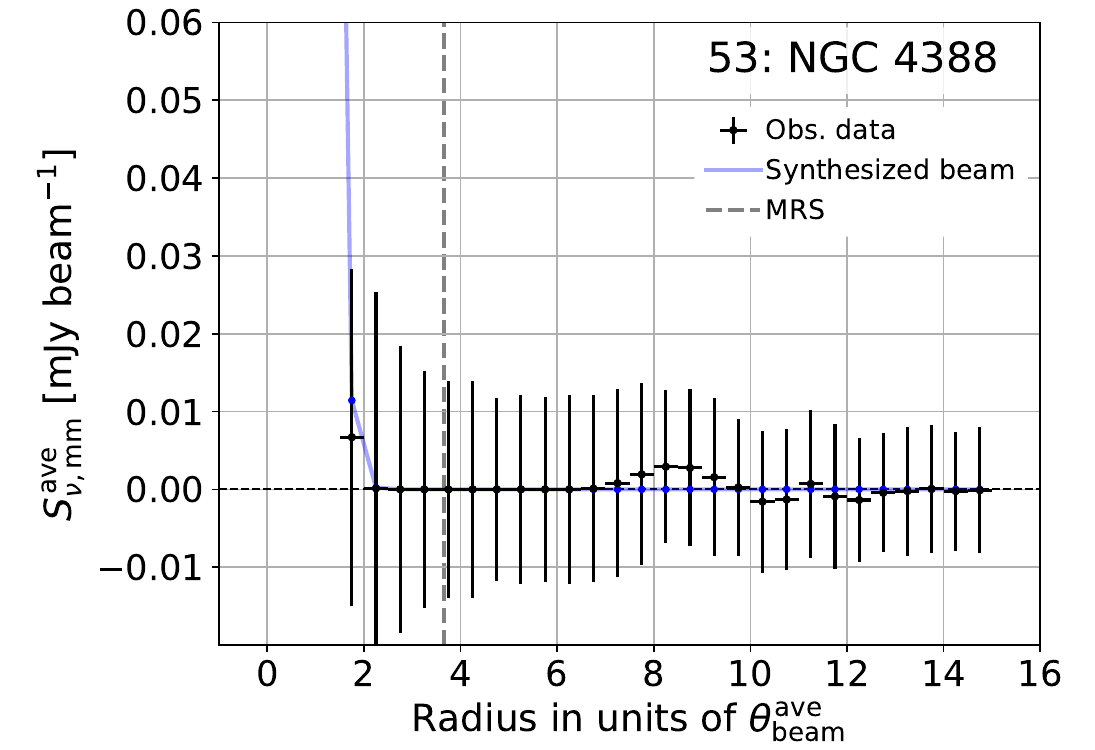}
\includegraphics[width=5.9cm]{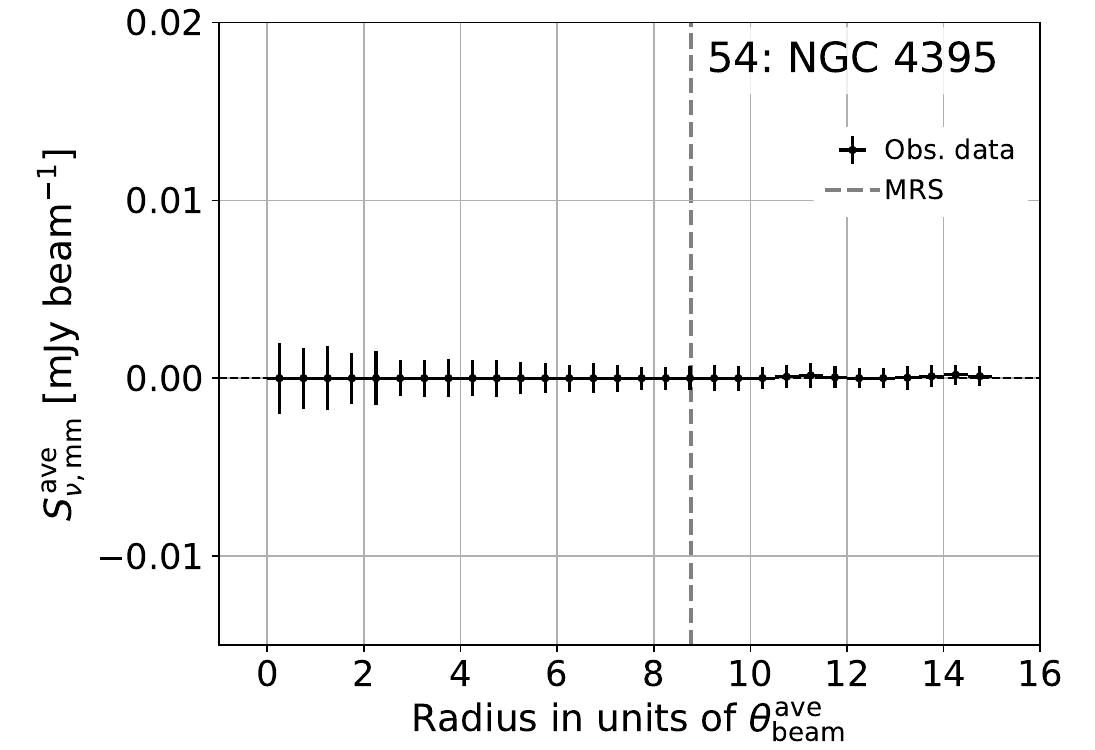}
\includegraphics[width=5.9cm]{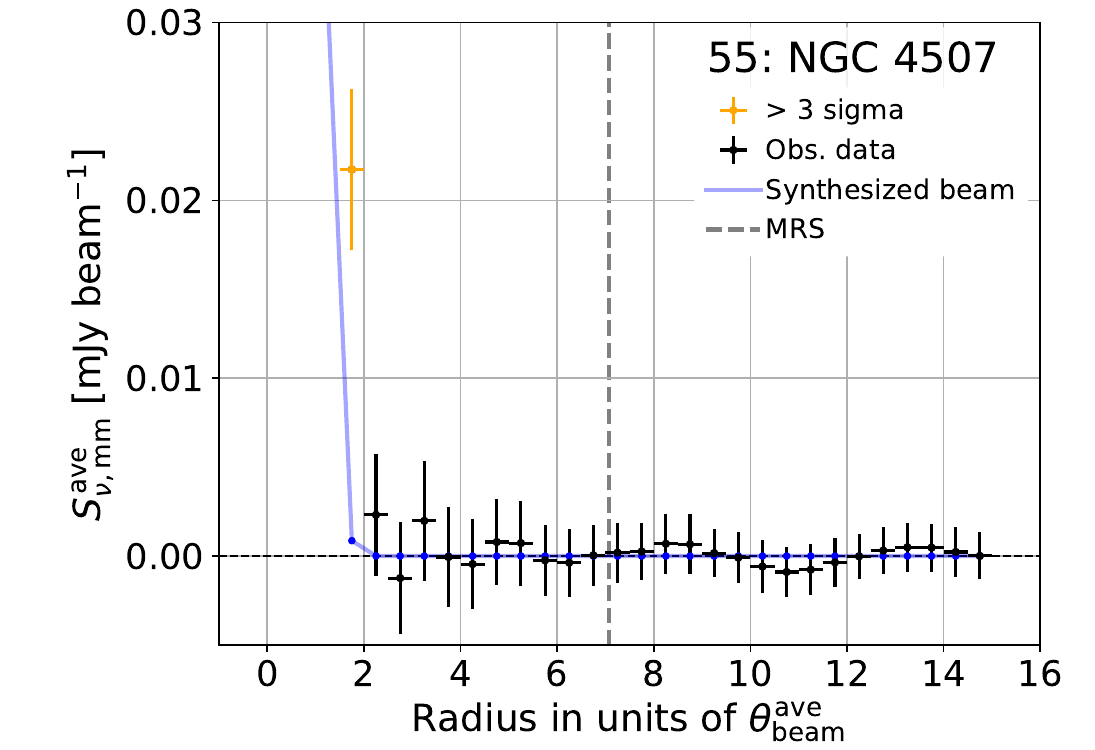}
\includegraphics[width=5.9cm]{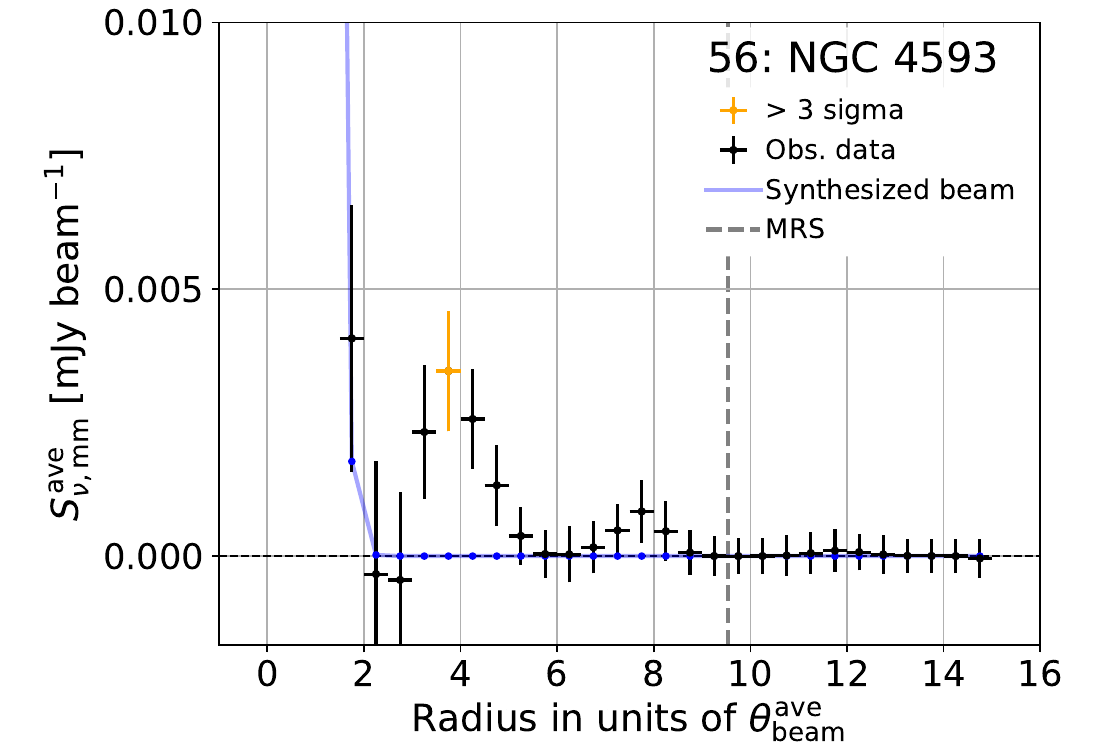}
\includegraphics[width=5.9cm]{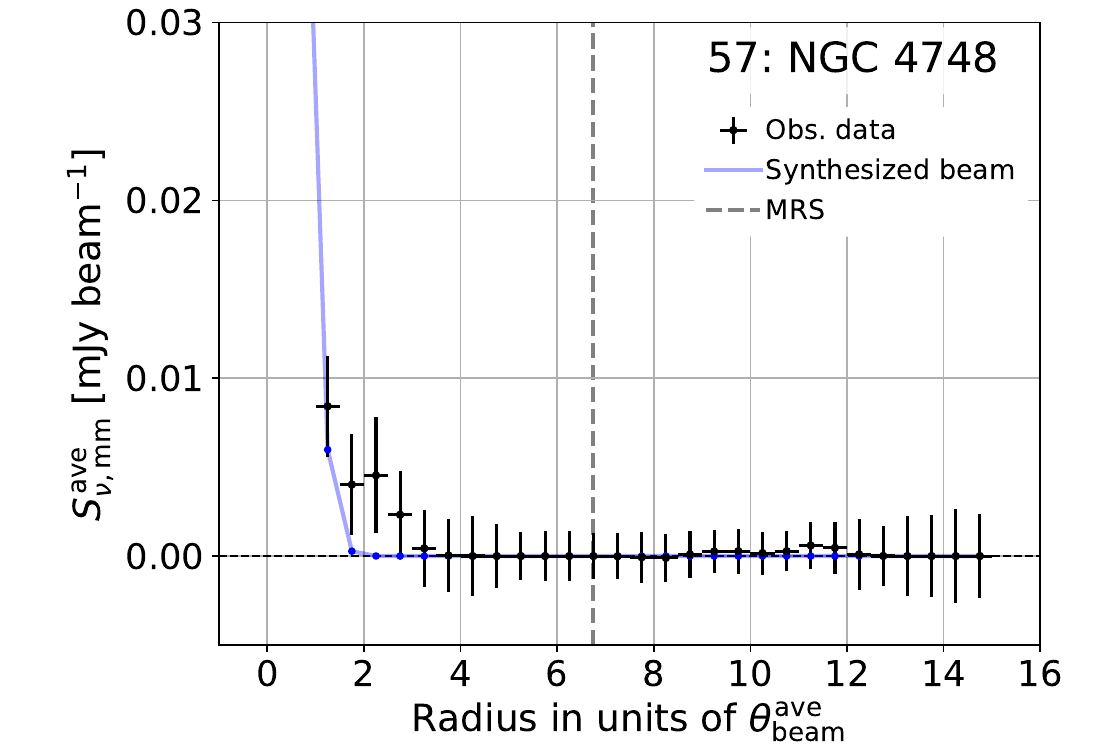}
\includegraphics[width=5.9cm]{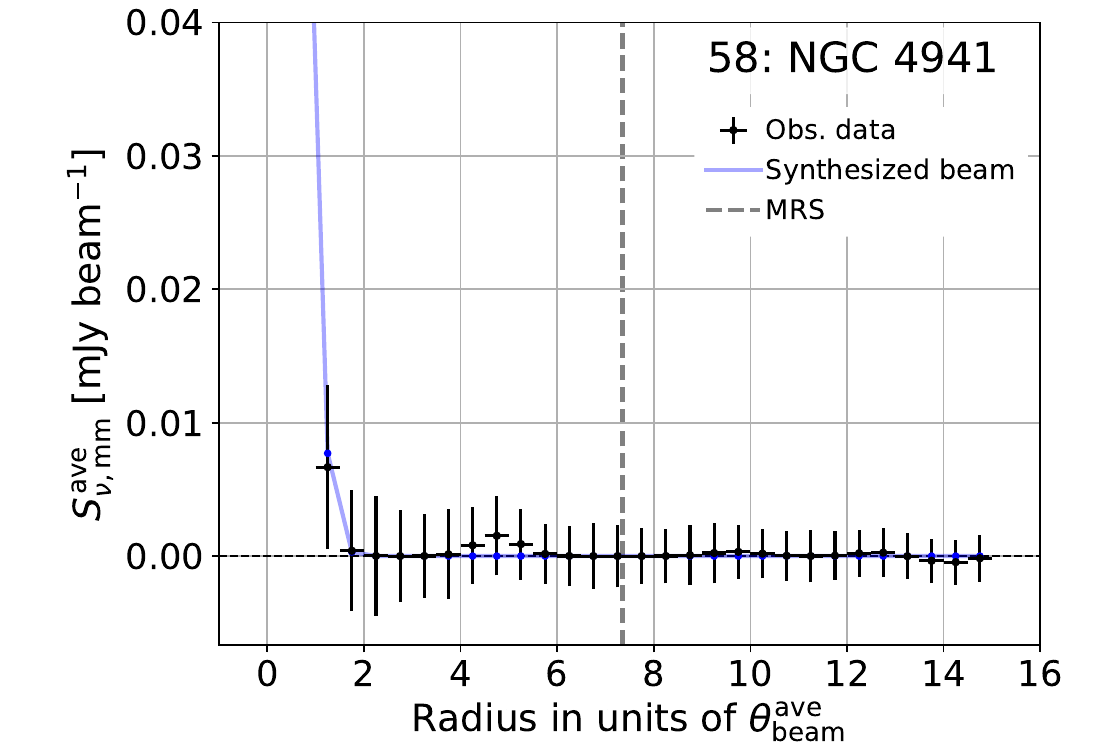}
\includegraphics[width=5.9cm]{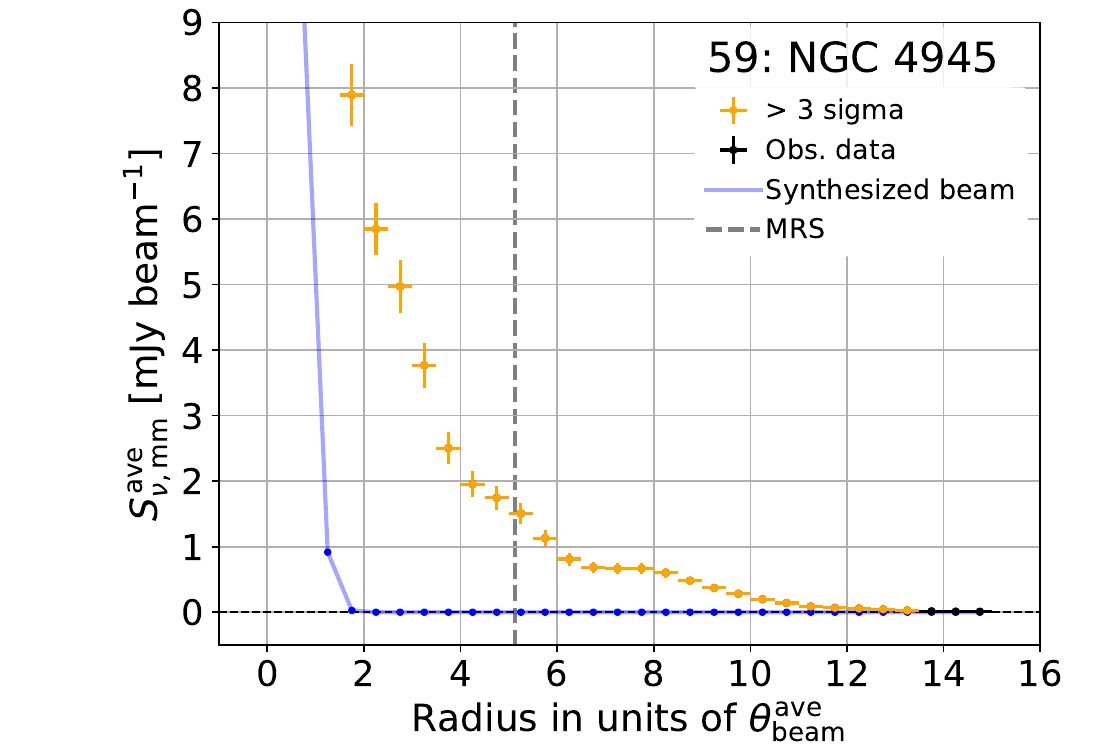}
\includegraphics[width=5.9cm]{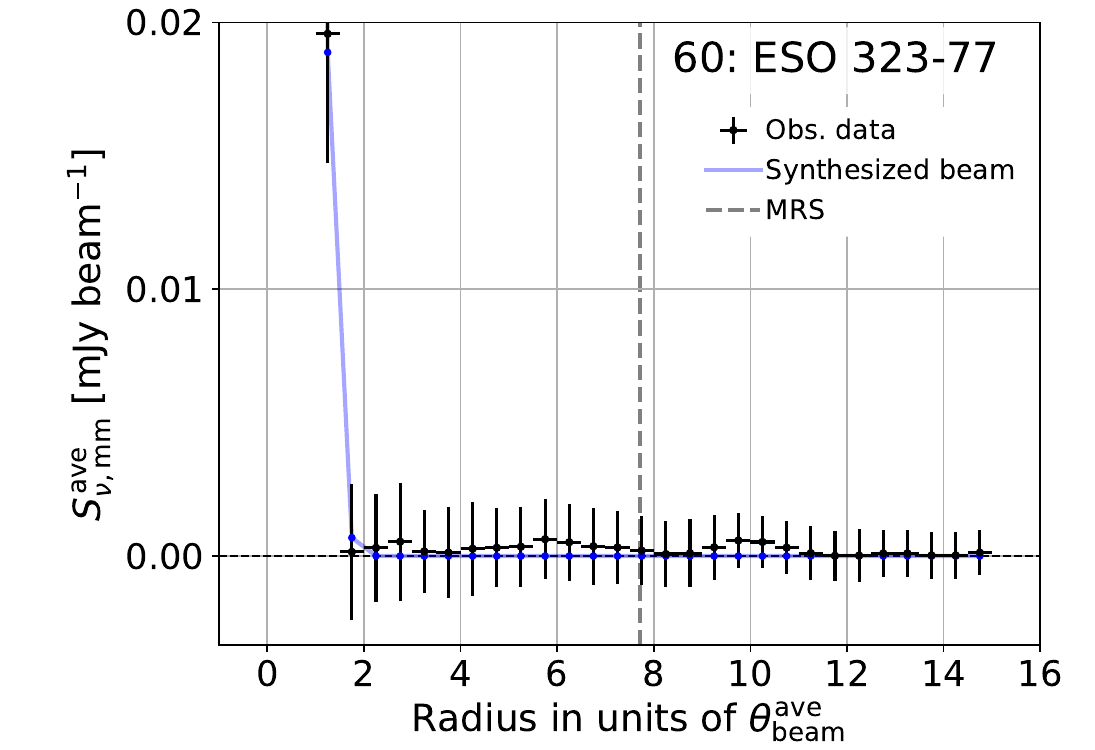}
\caption{Continued. 
    }
\end{figure*}

\addtocounter{figure}{-1}

\begin{figure*}
    \centering    
\includegraphics[width=5.9cm]{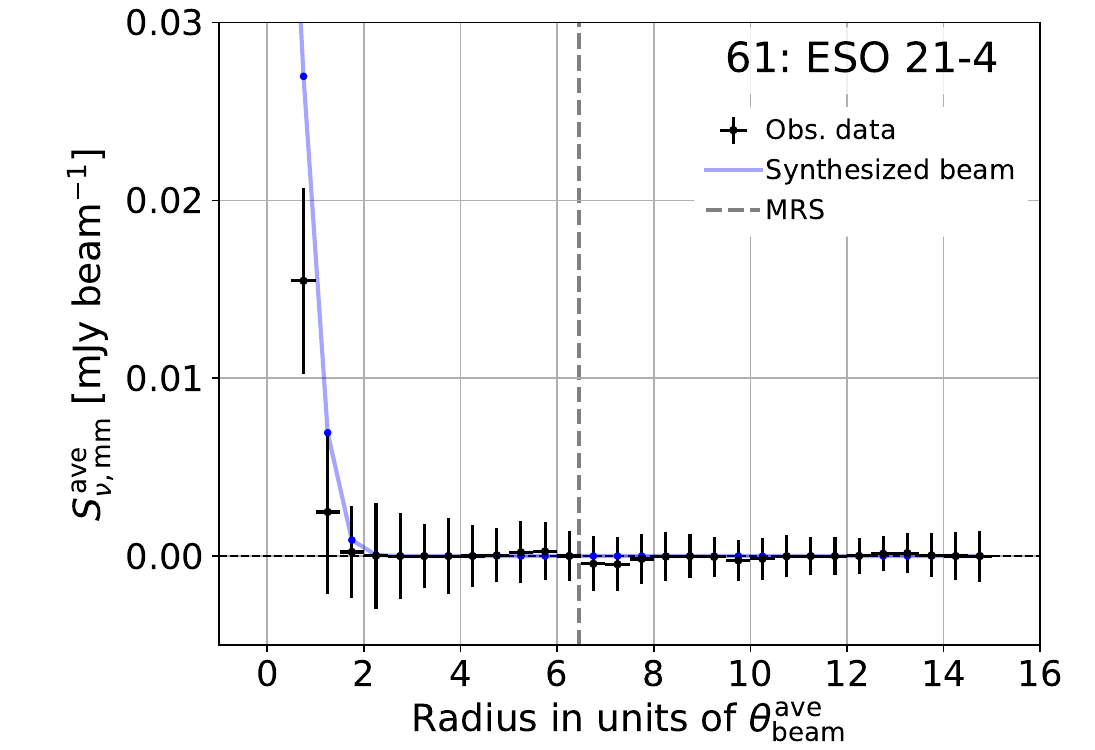}
\includegraphics[width=5.9cm]{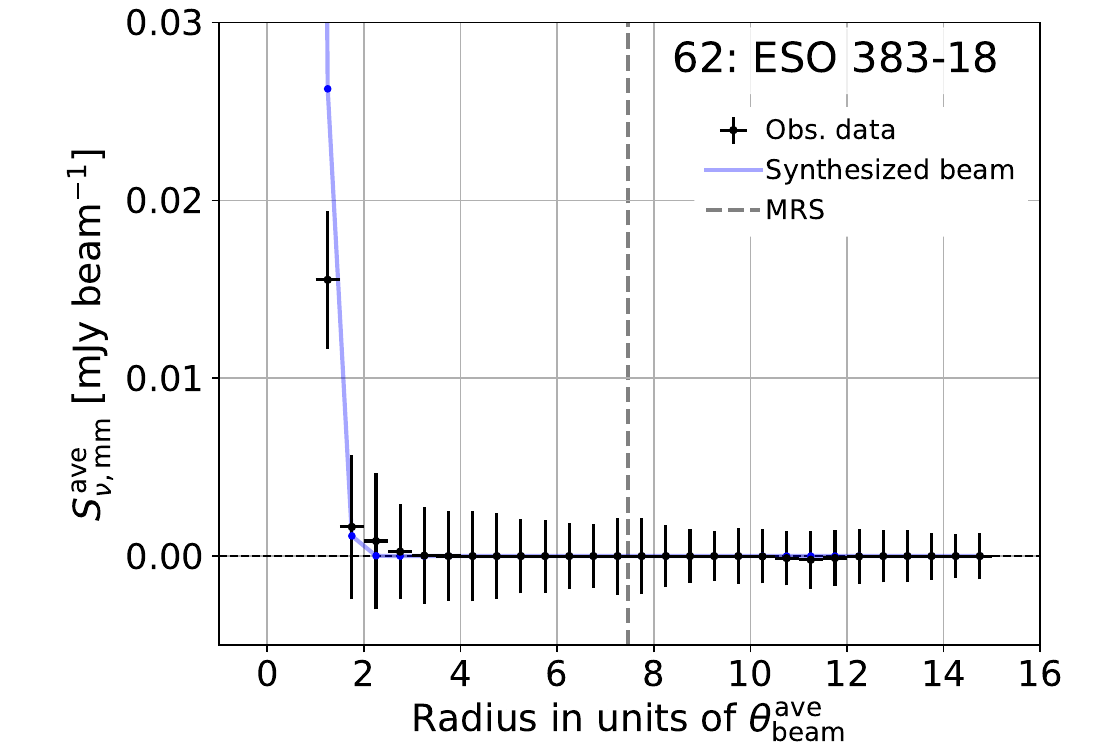}
\includegraphics[width=5.9cm]{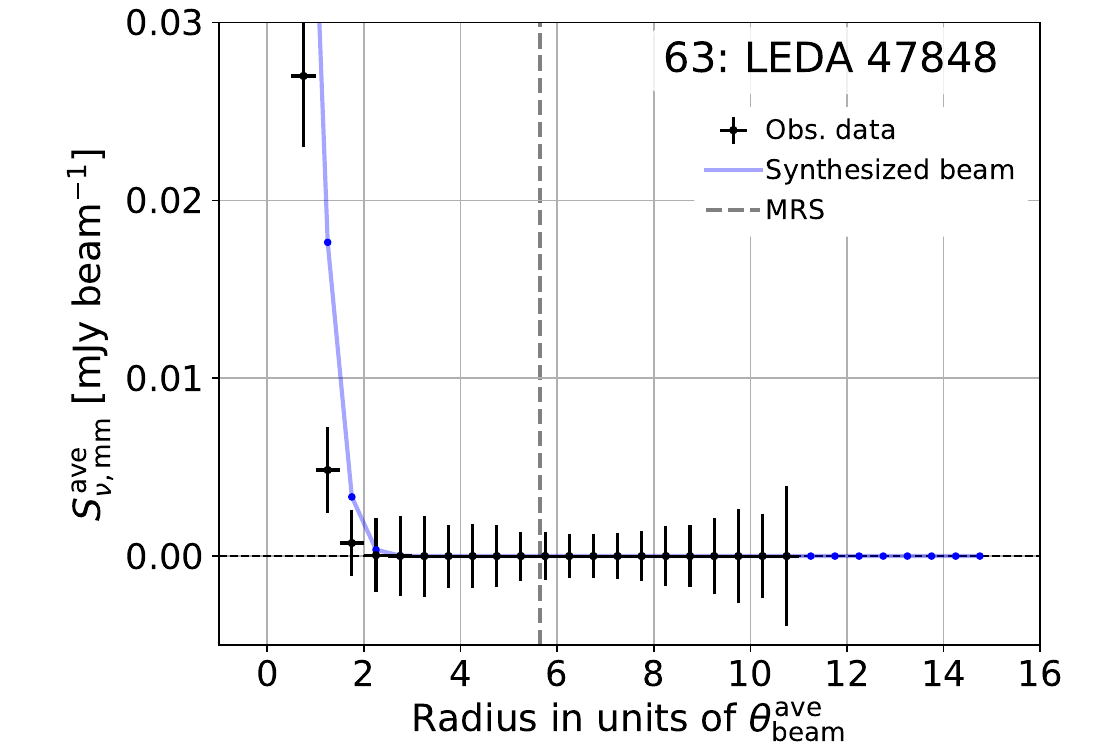}
\includegraphics[width=5.9cm]{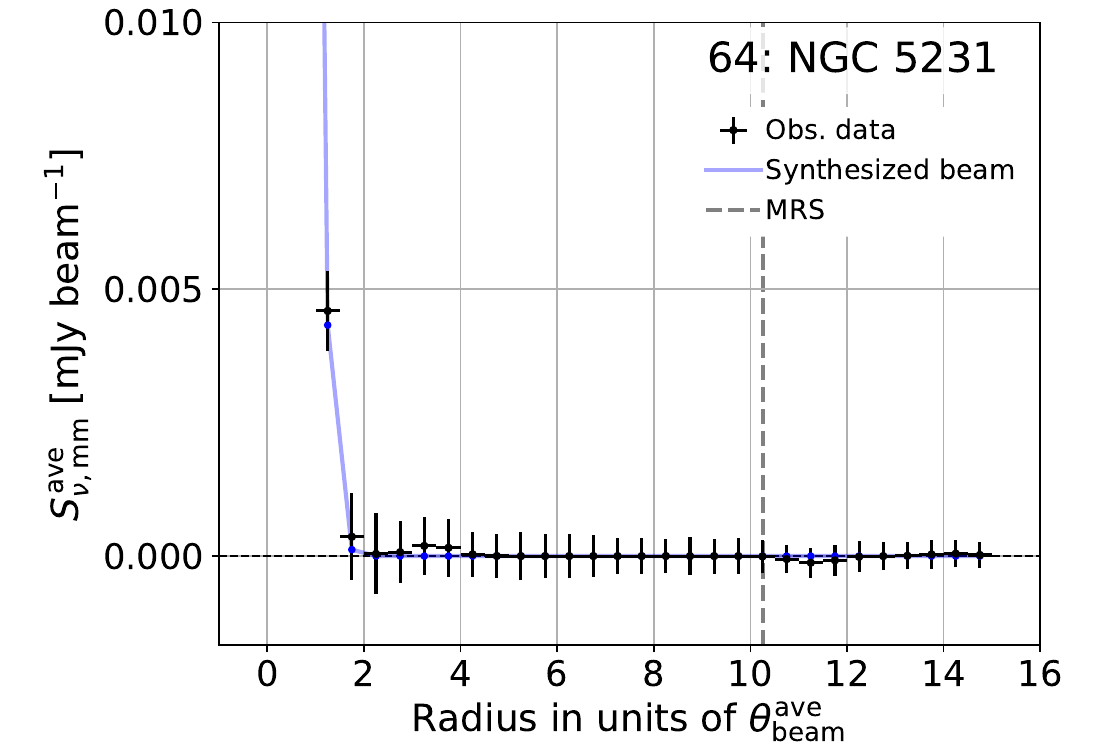}
\includegraphics[width=5.9cm]{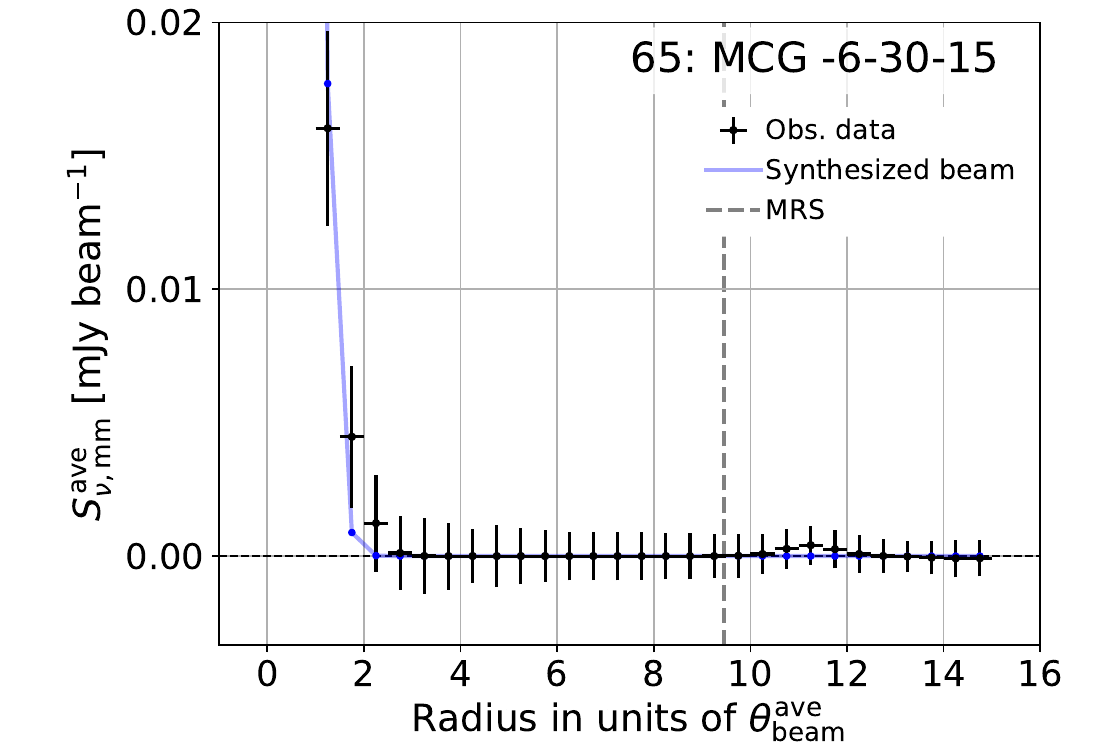}
\includegraphics[width=5.9cm]{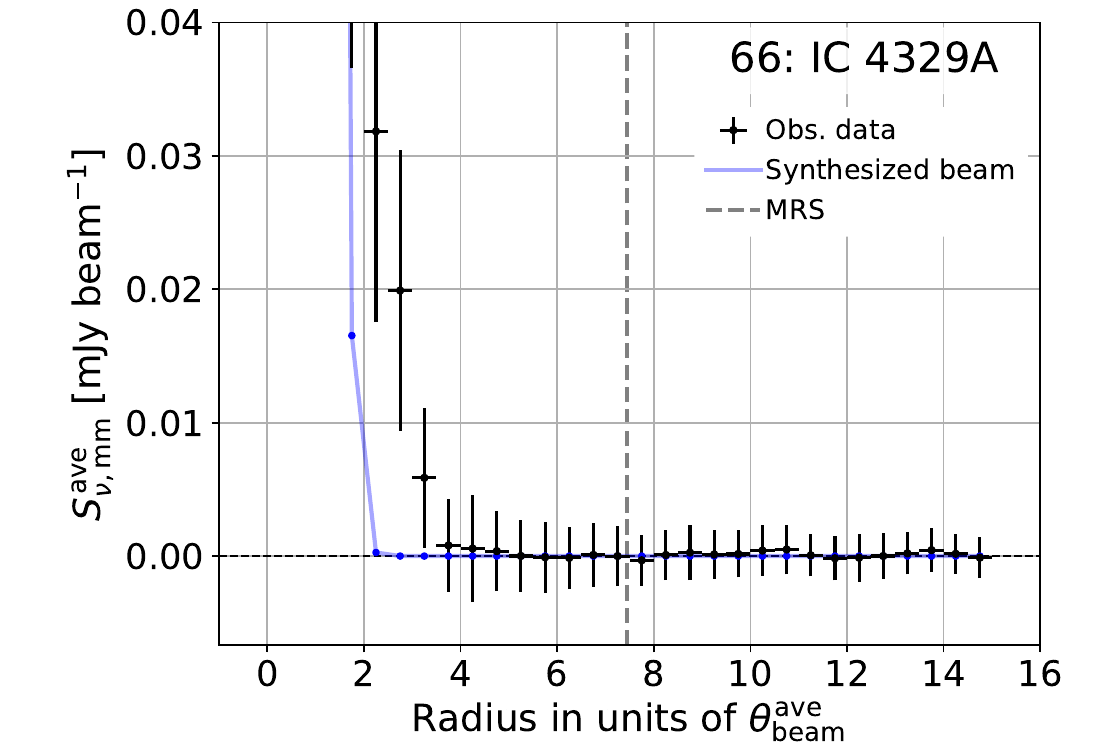}
\includegraphics[width=5.9cm]{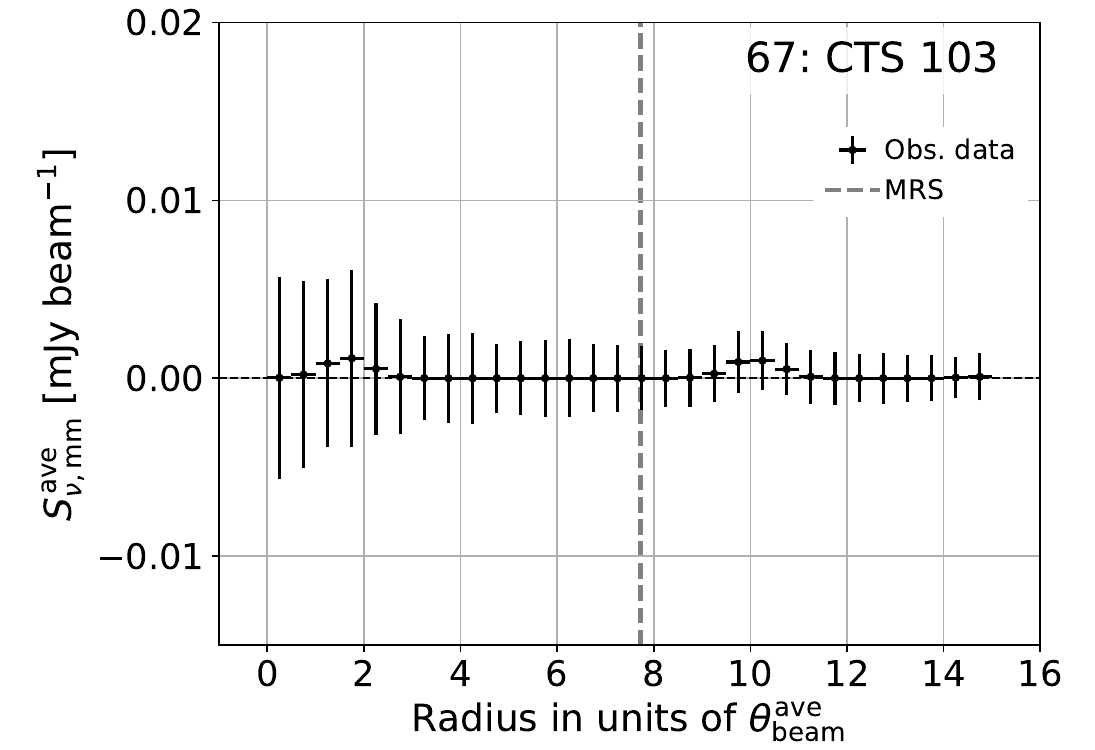}
\includegraphics[width=5.9cm]{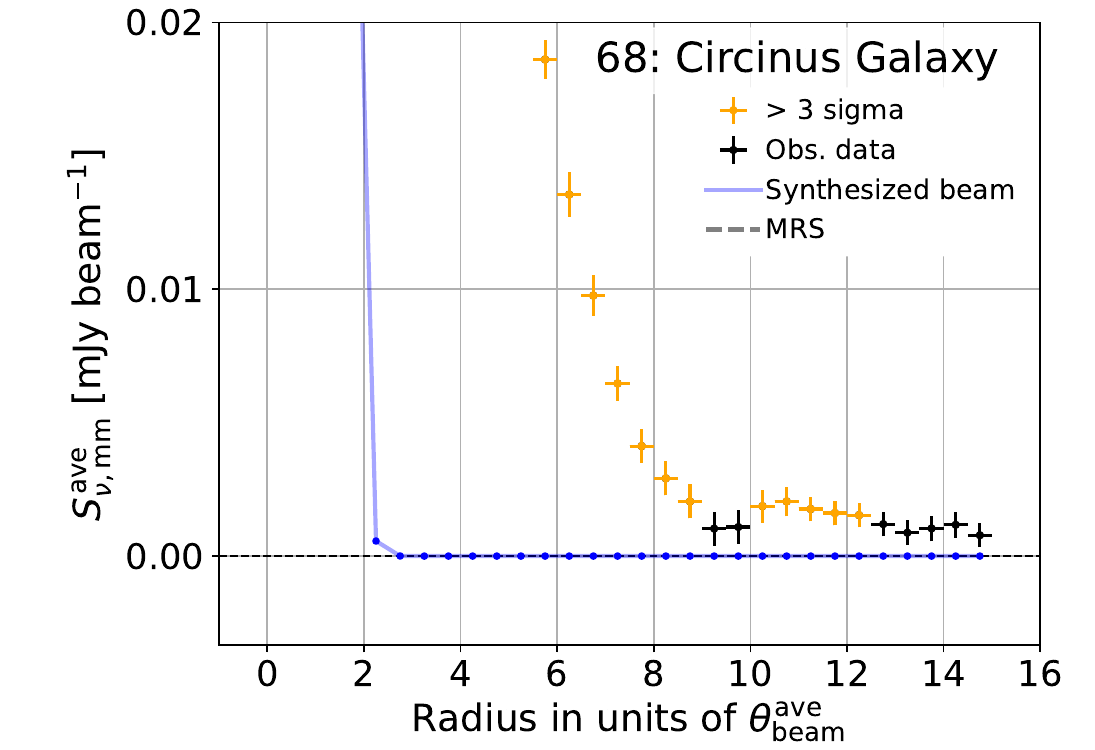}
\includegraphics[width=5.9cm]{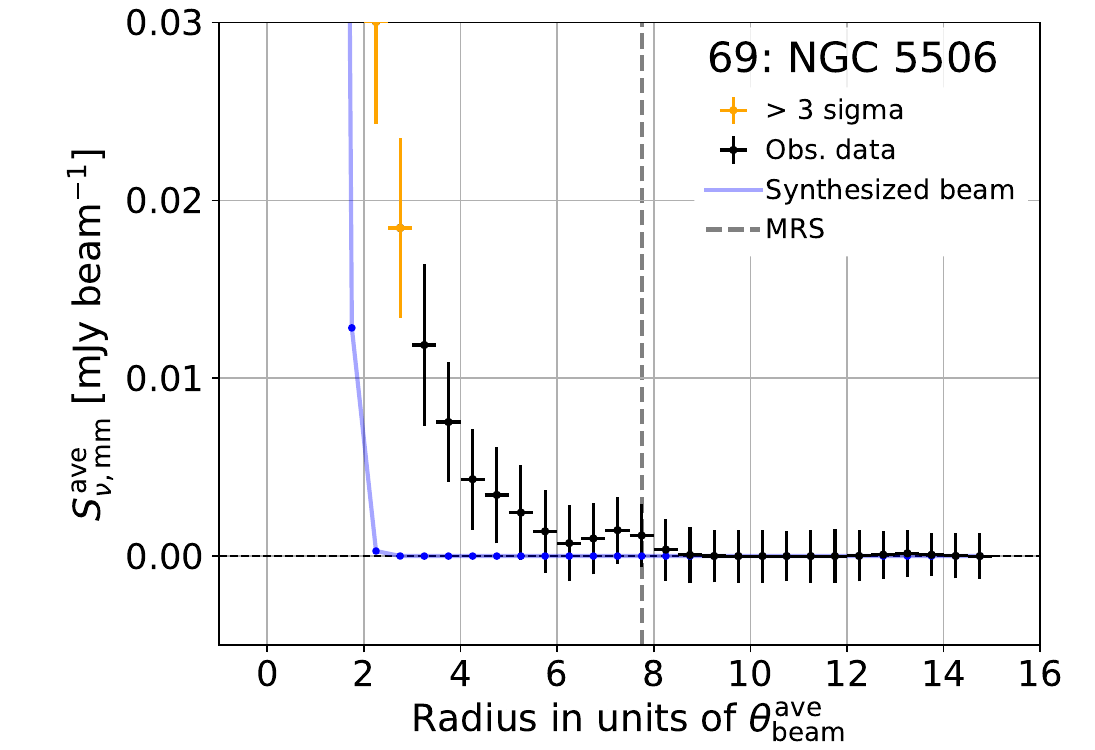}
\includegraphics[width=5.9cm]{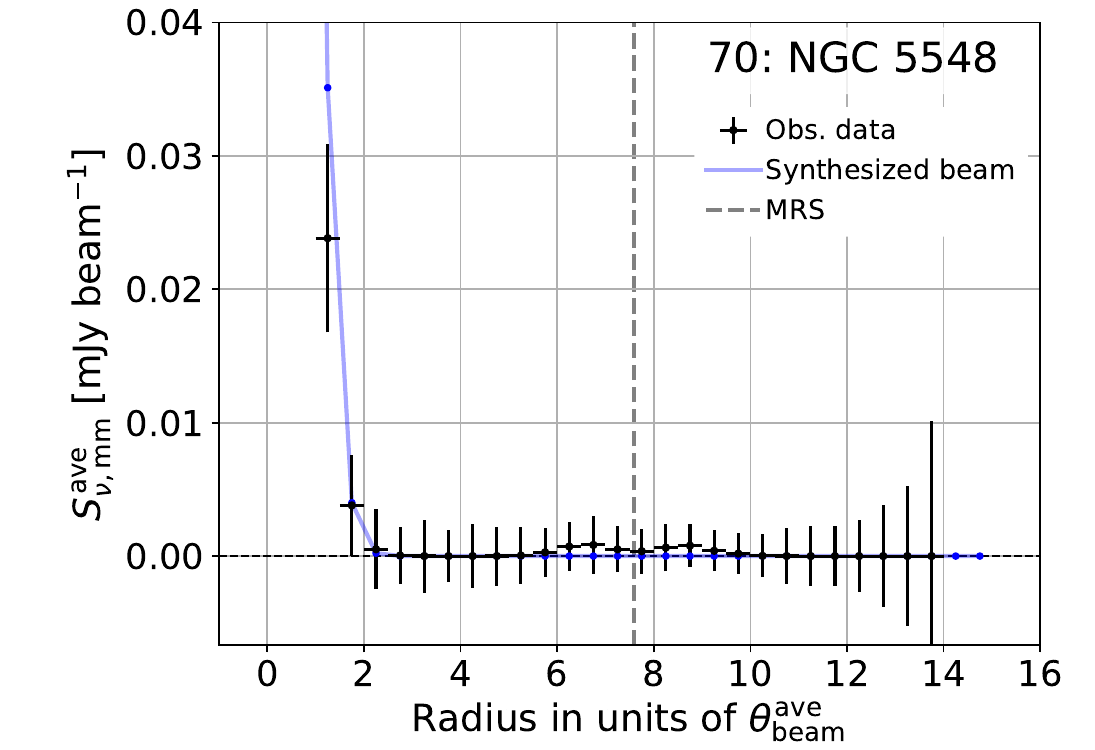}
\includegraphics[width=5.9cm]{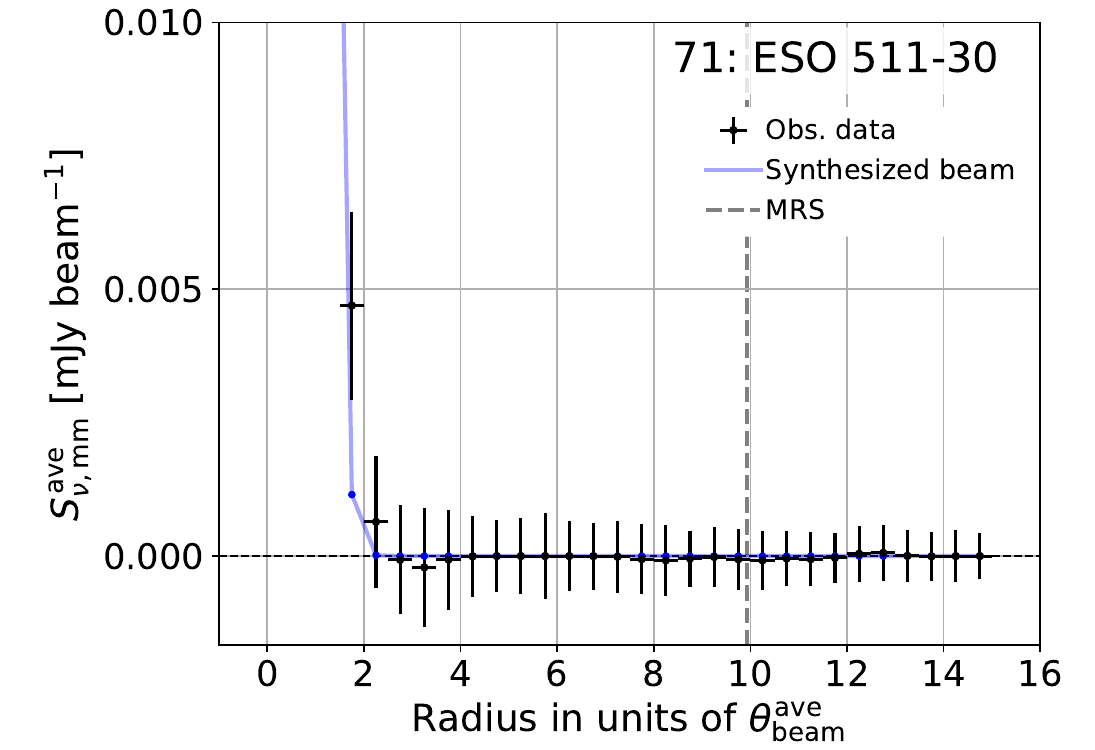}
\includegraphics[width=5.9cm]{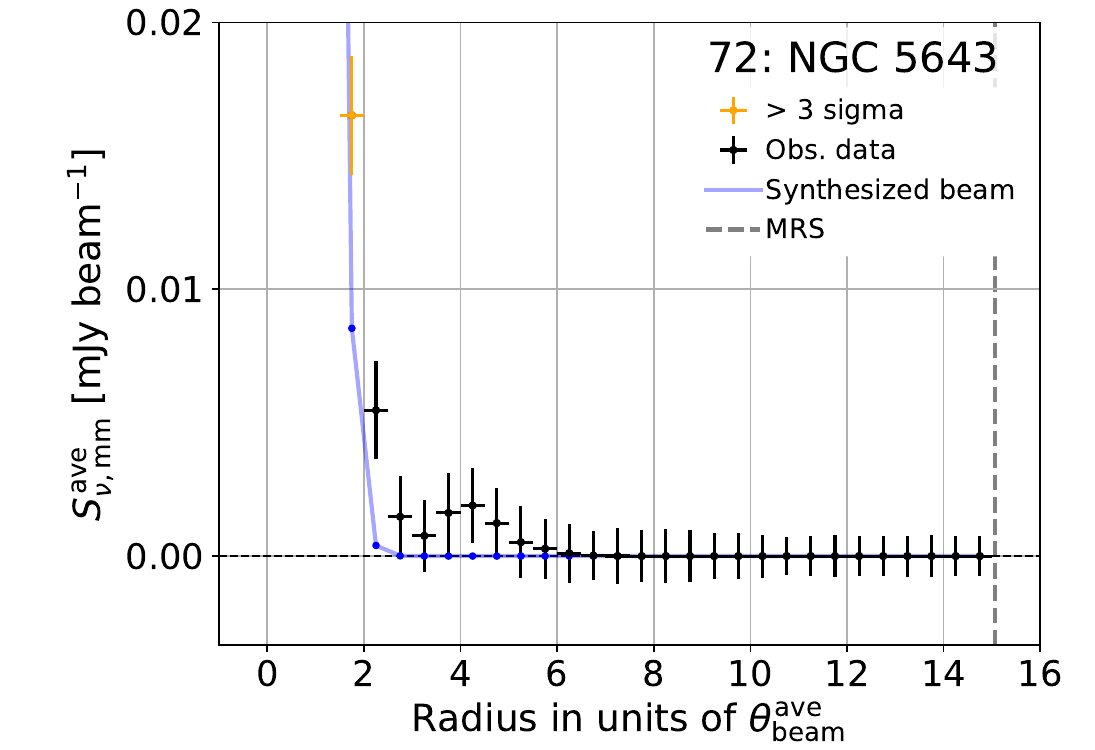}
\includegraphics[width=5.9cm]{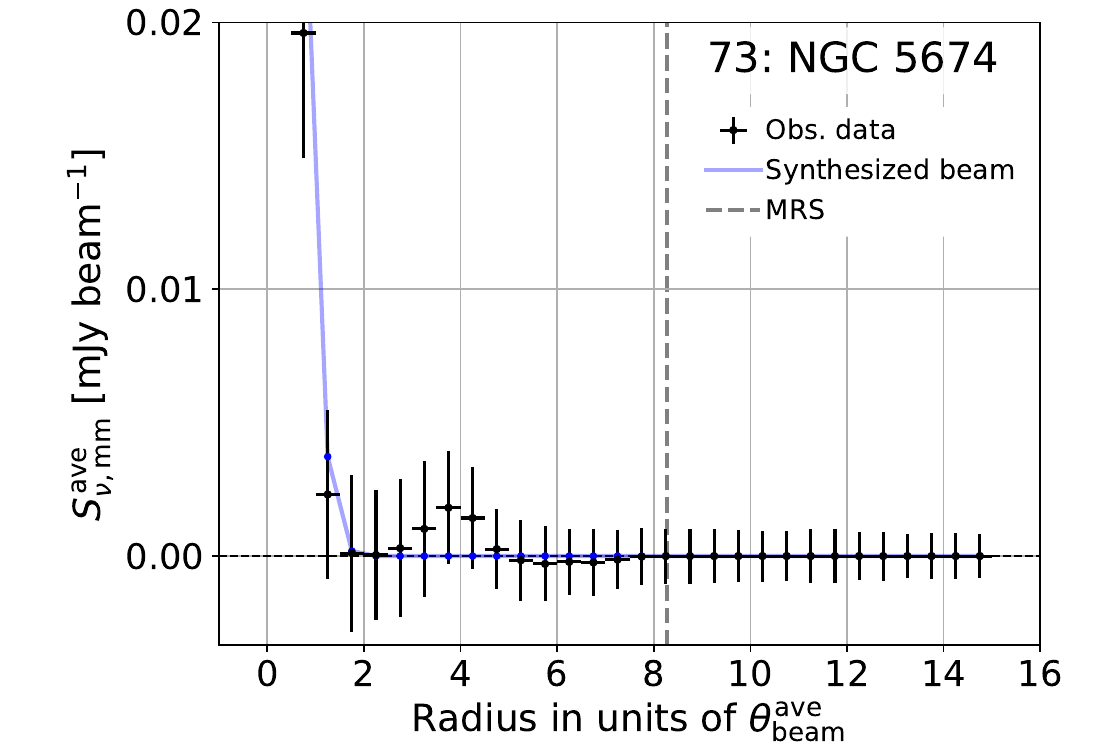}
\includegraphics[width=5.9cm]{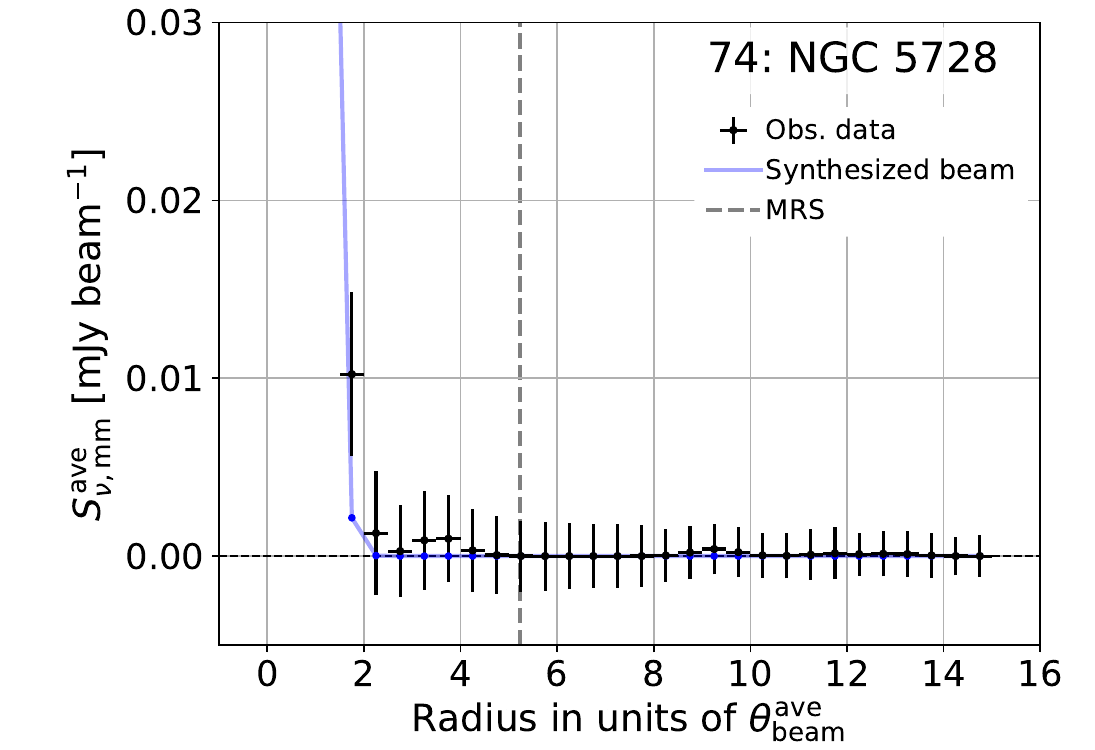}
\includegraphics[width=5.9cm]{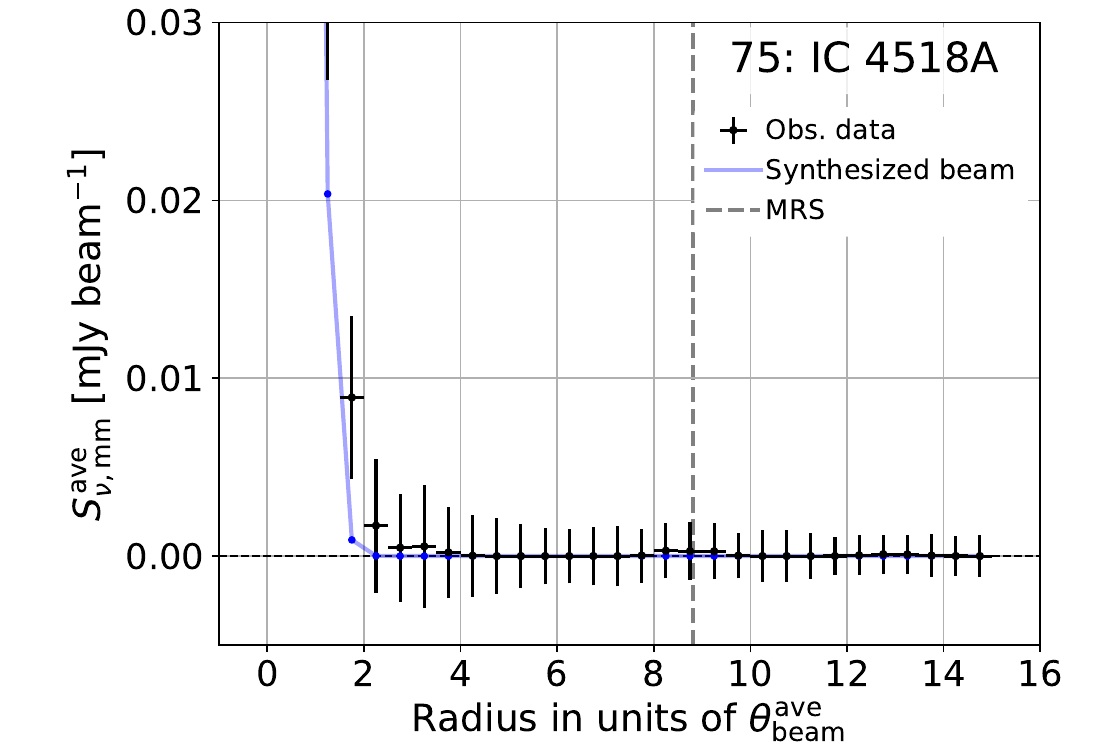}
\caption{Continued. 
    }
\end{figure*}

\addtocounter{figure}{-1}

\begin{figure*}
    \centering    
\includegraphics[width=5.9cm]{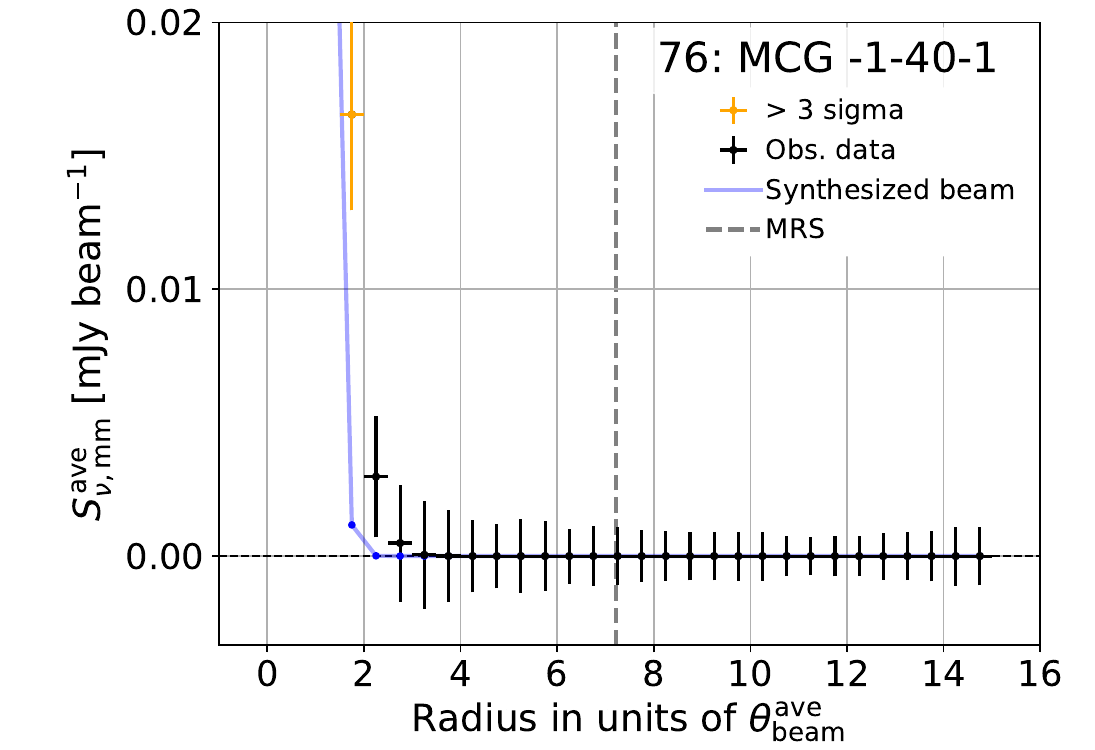}
\includegraphics[width=5.9cm]{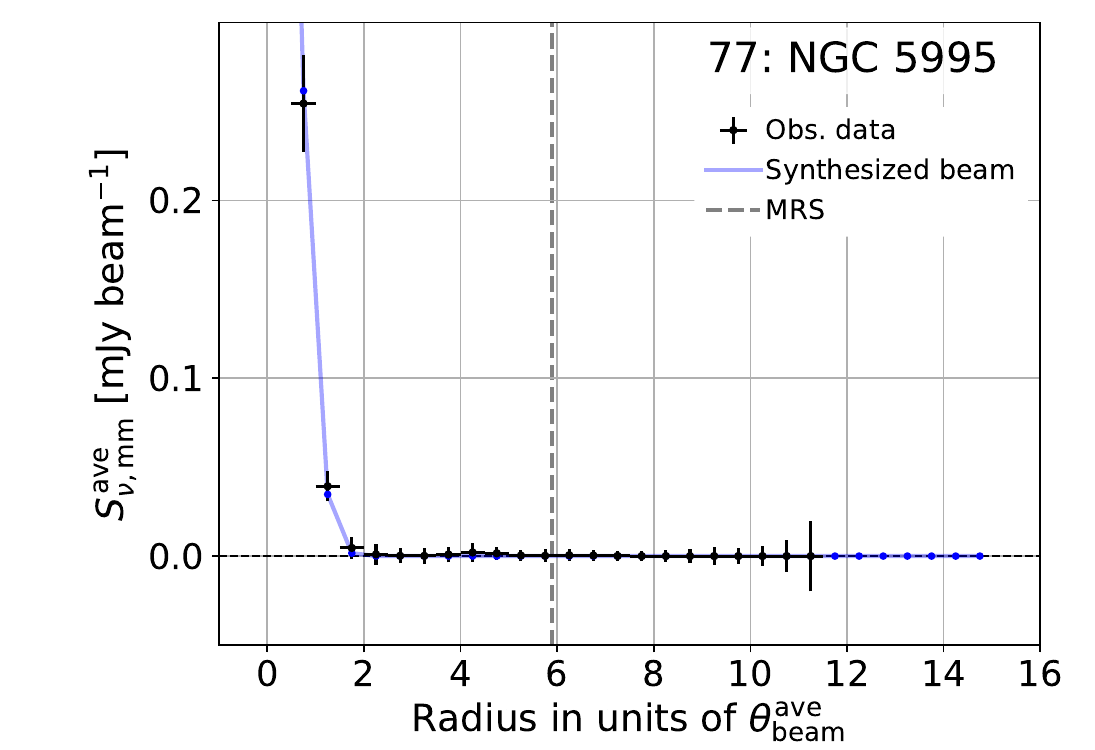}
\includegraphics[width=5.9cm]{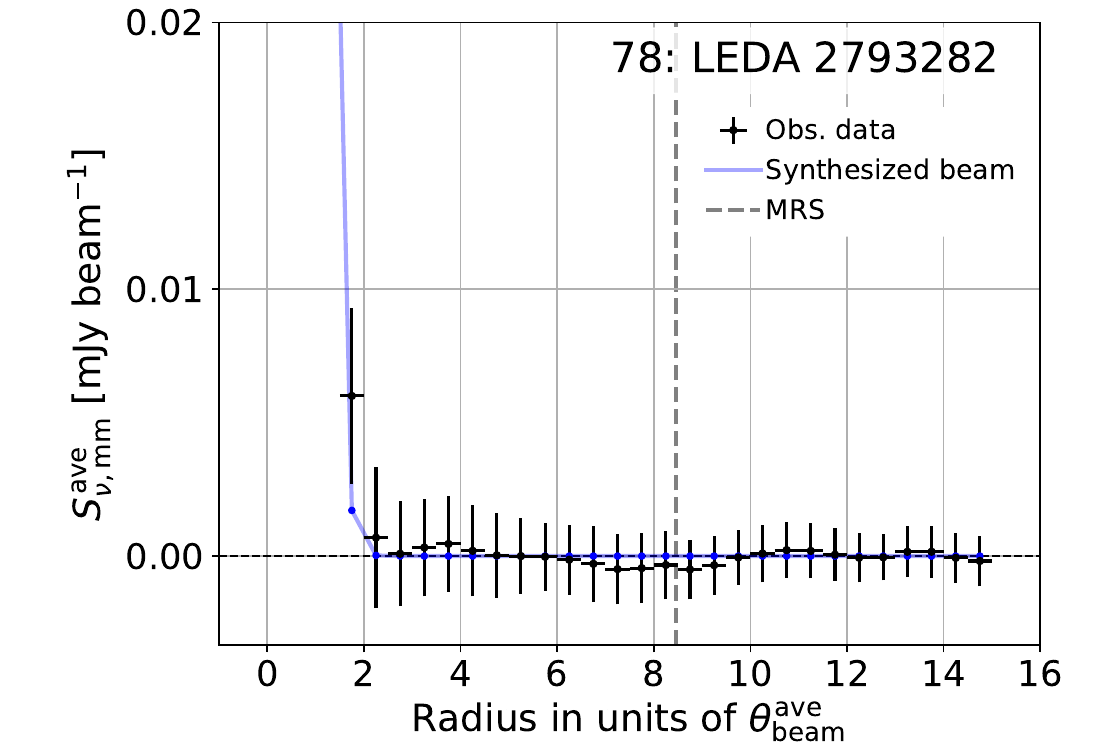}
\includegraphics[width=5.9cm]{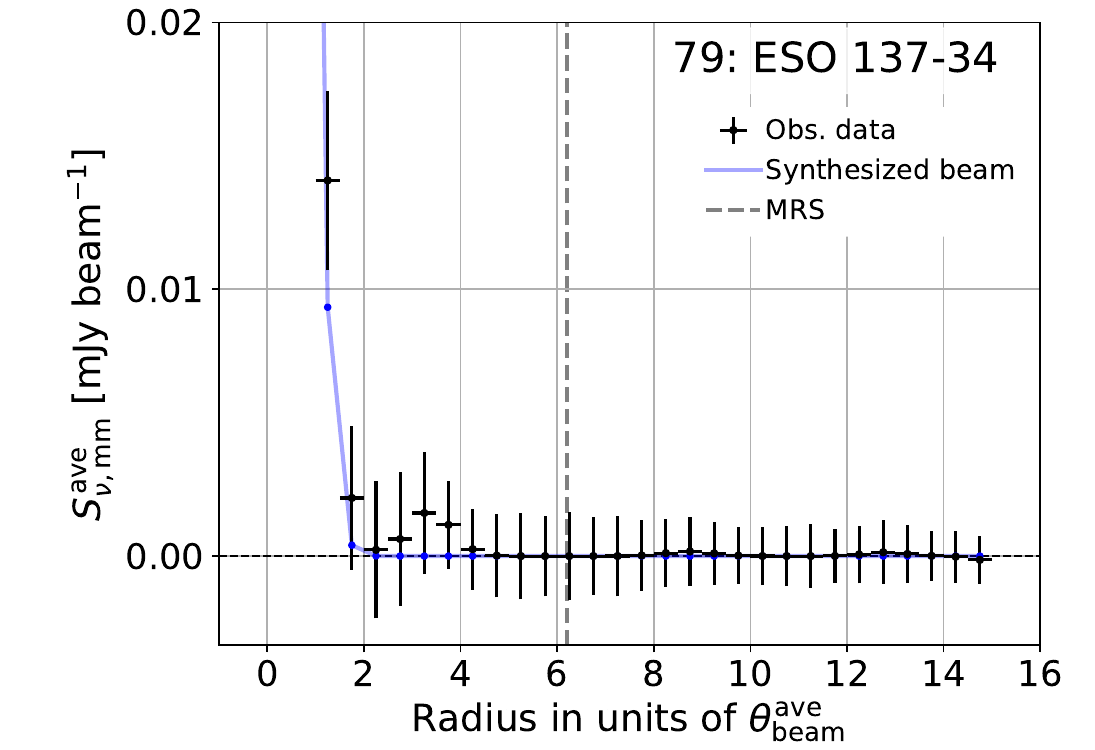}
\includegraphics[width=5.9cm]{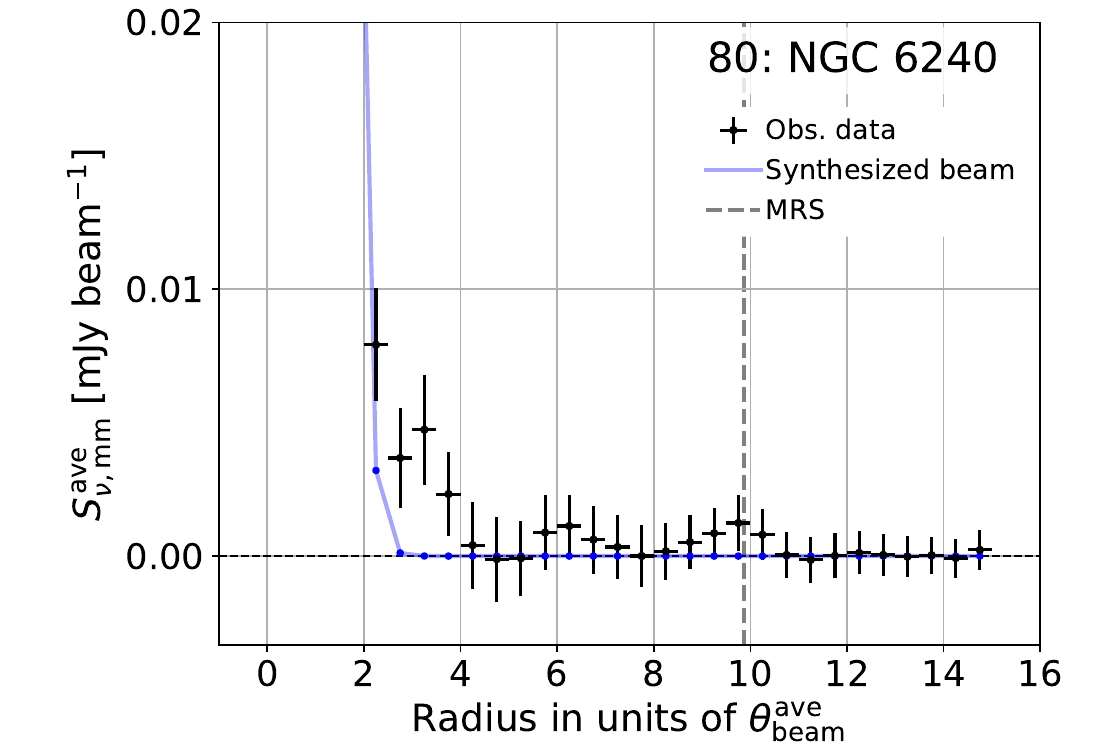}
\includegraphics[width=5.9cm]{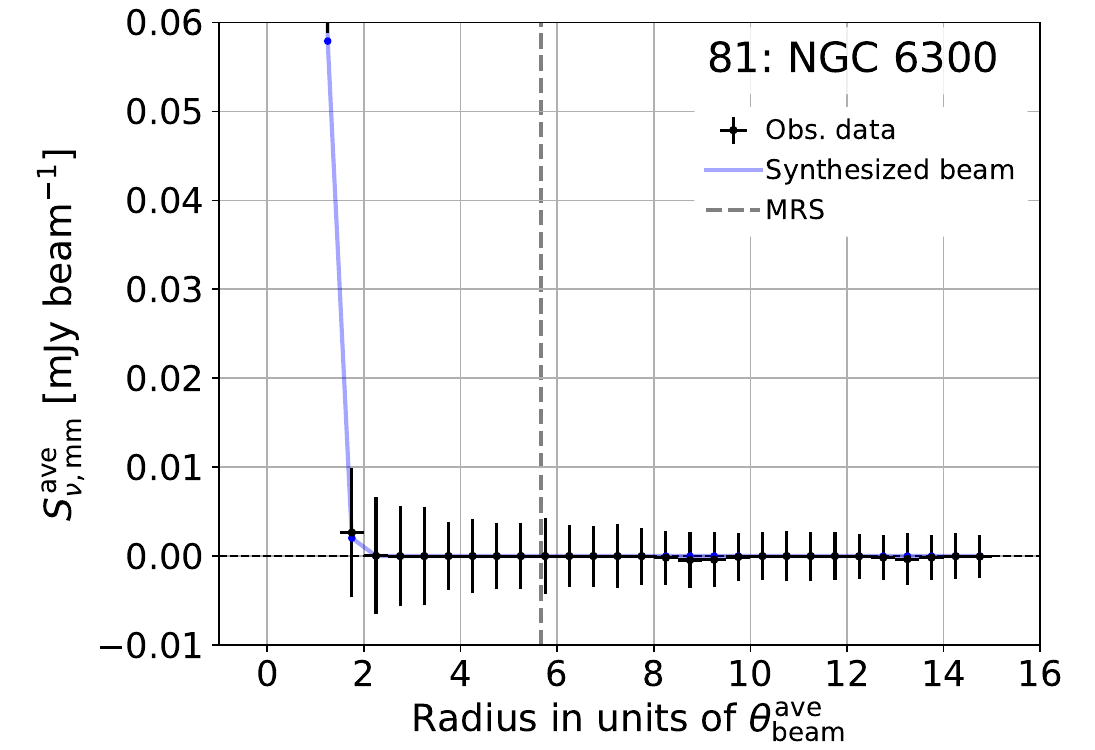}
\includegraphics[width=5.9cm]{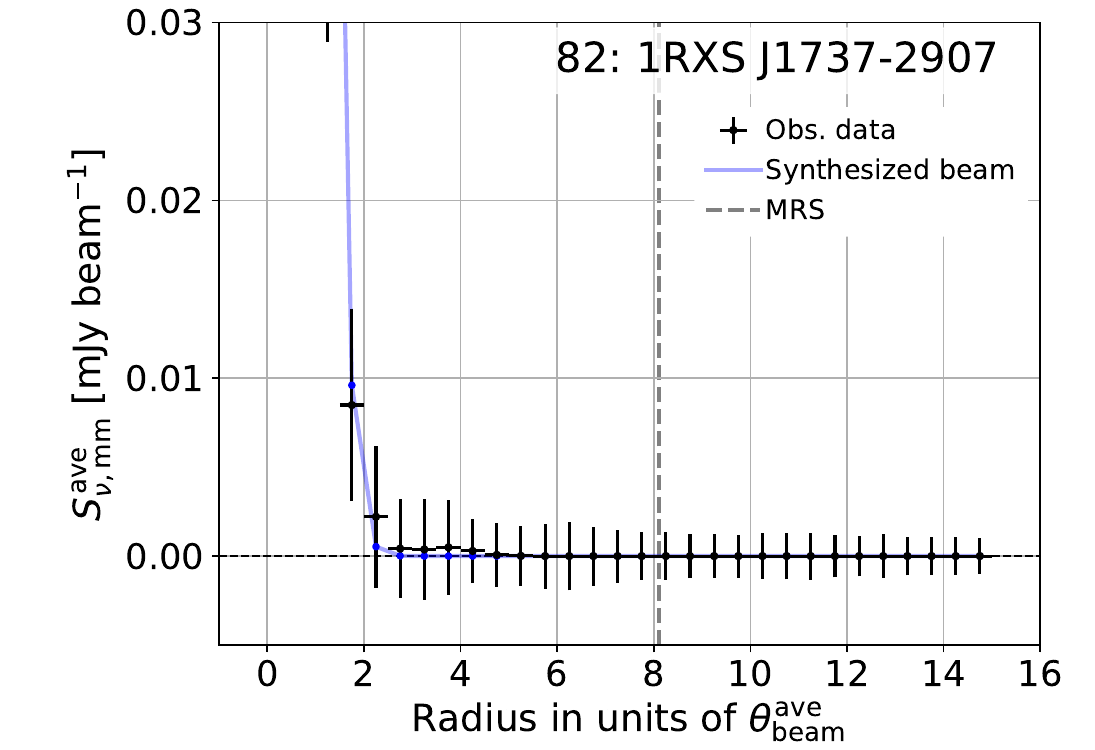}
\includegraphics[width=5.9cm]{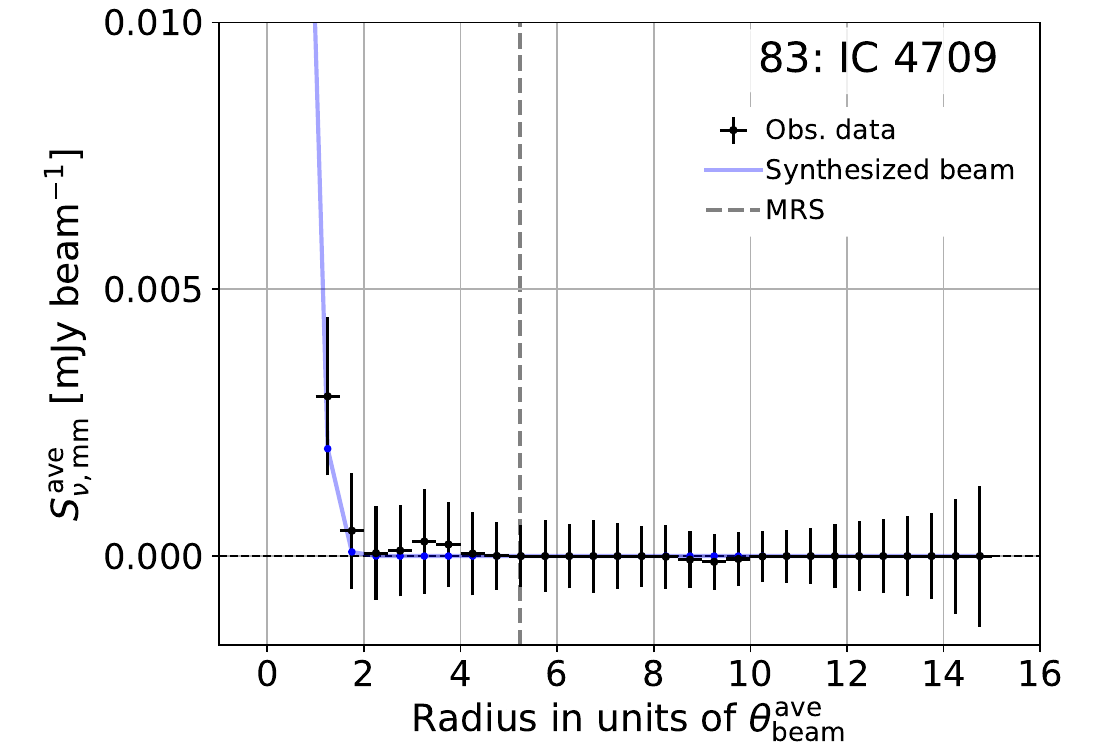}
\includegraphics[width=5.9cm]{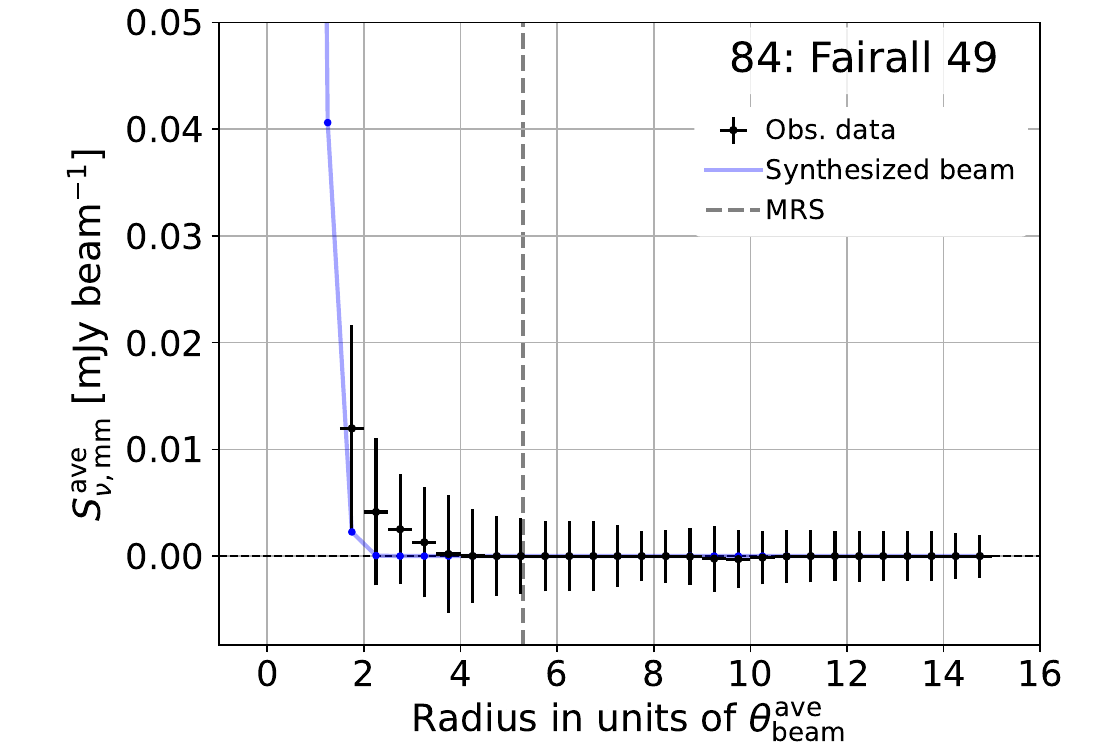}
\includegraphics[width=5.9cm]{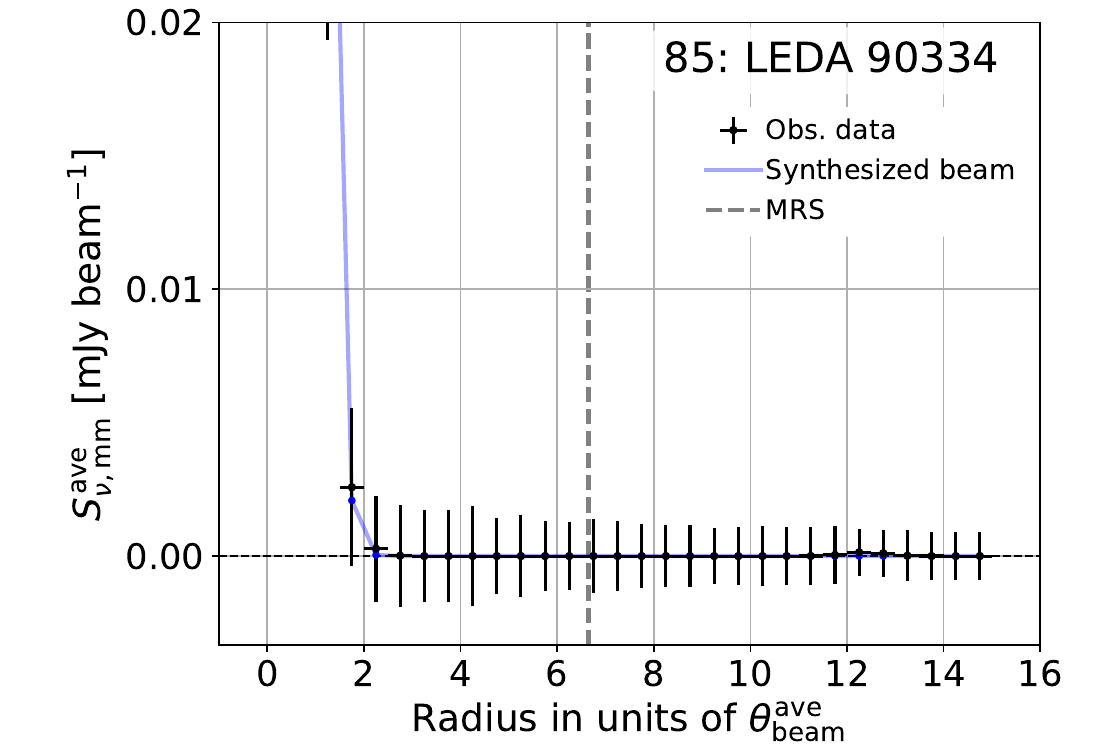}
\includegraphics[width=5.9cm]{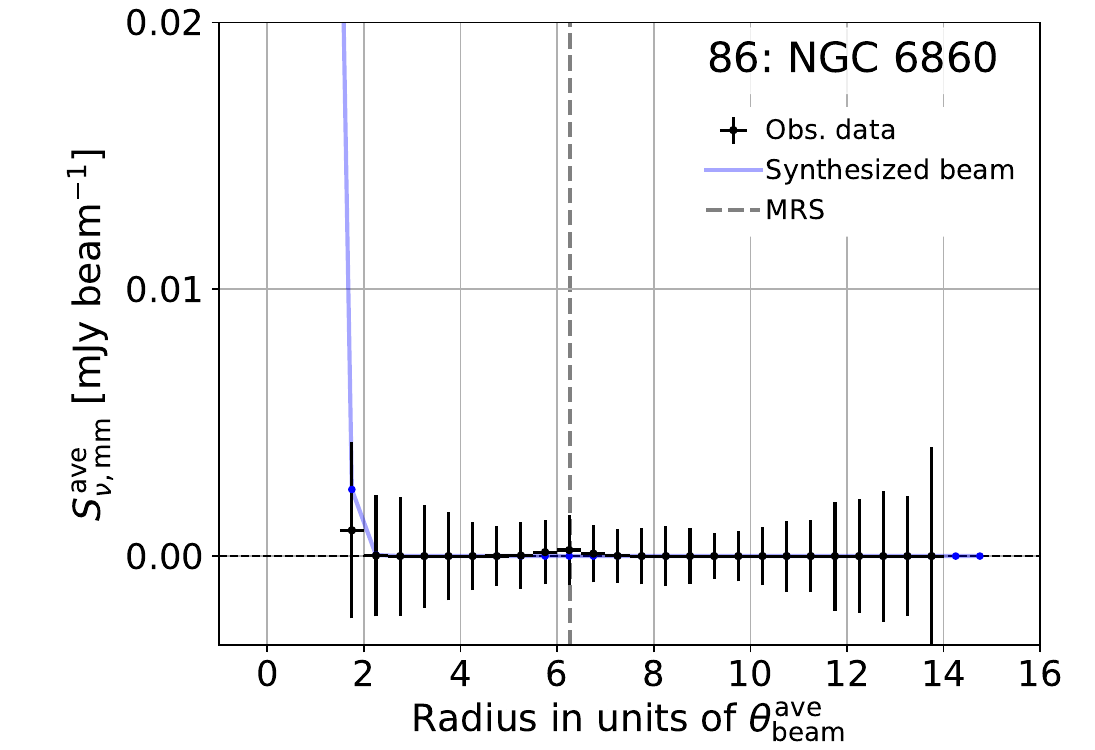}
\includegraphics[width=5.9cm]{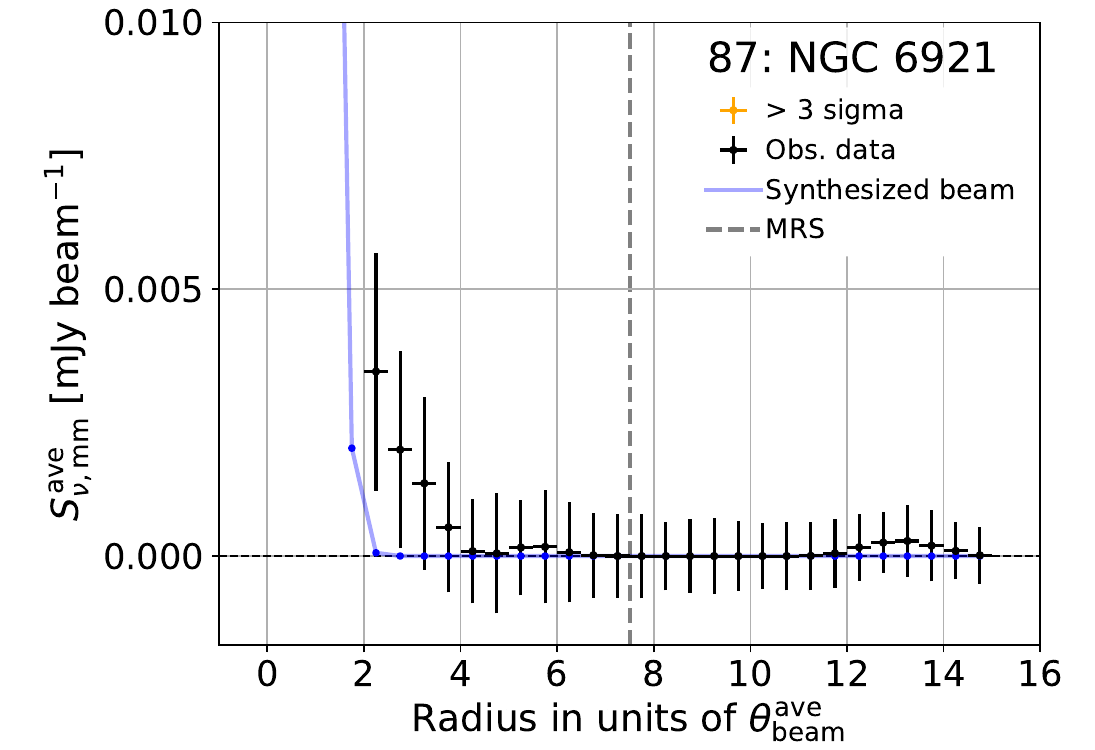}
\includegraphics[width=5.9cm]{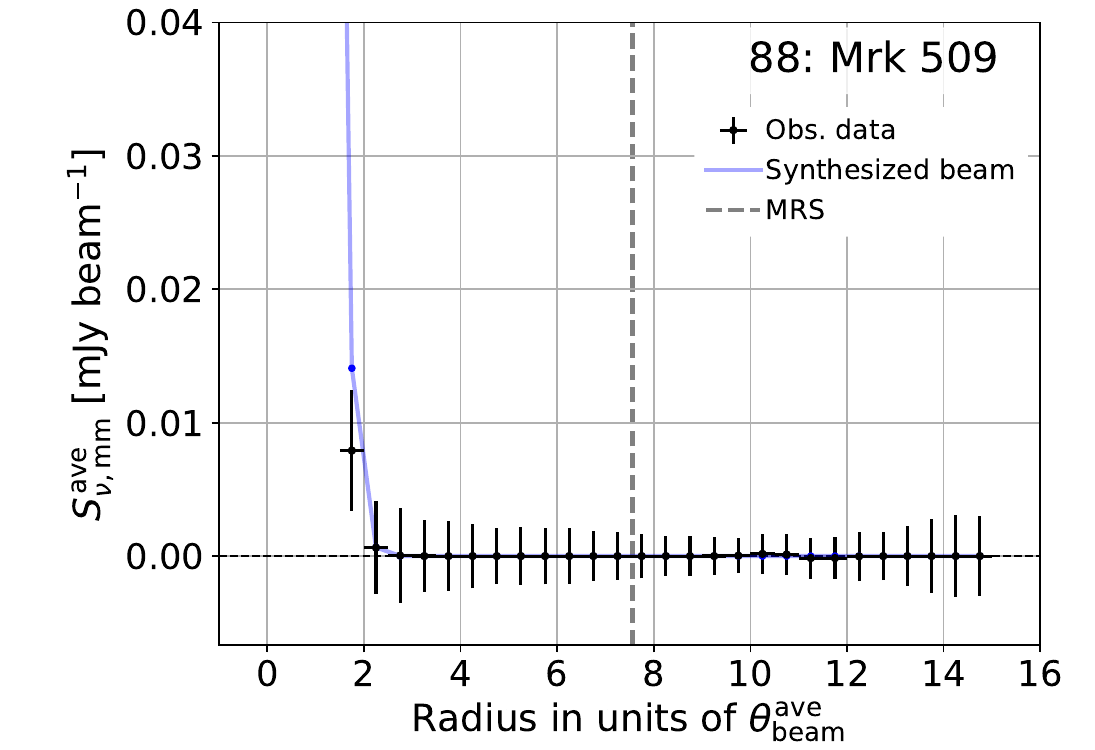}
\includegraphics[width=5.9cm]{089_IC_5063_radpro.pdf}
\includegraphics[width=5.9cm]{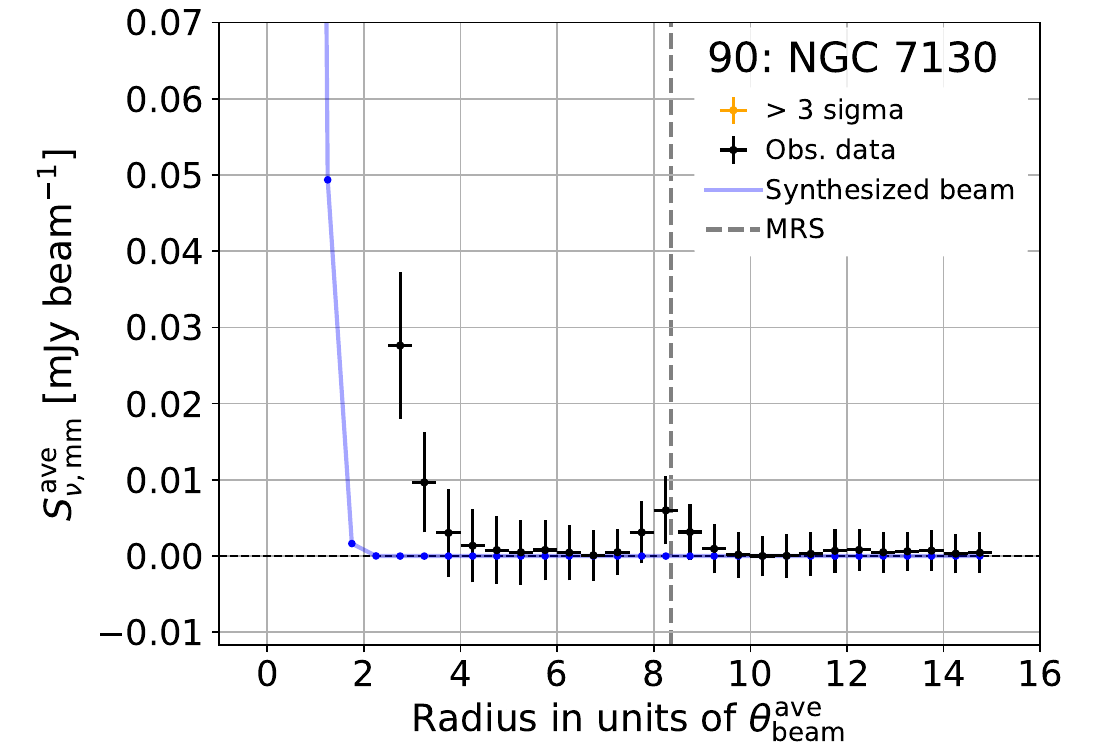}
\caption{Continued. 
    }
\end{figure*}

\addtocounter{figure}{-1}

\begin{figure*}
    \centering    
\includegraphics[width=5.9cm]{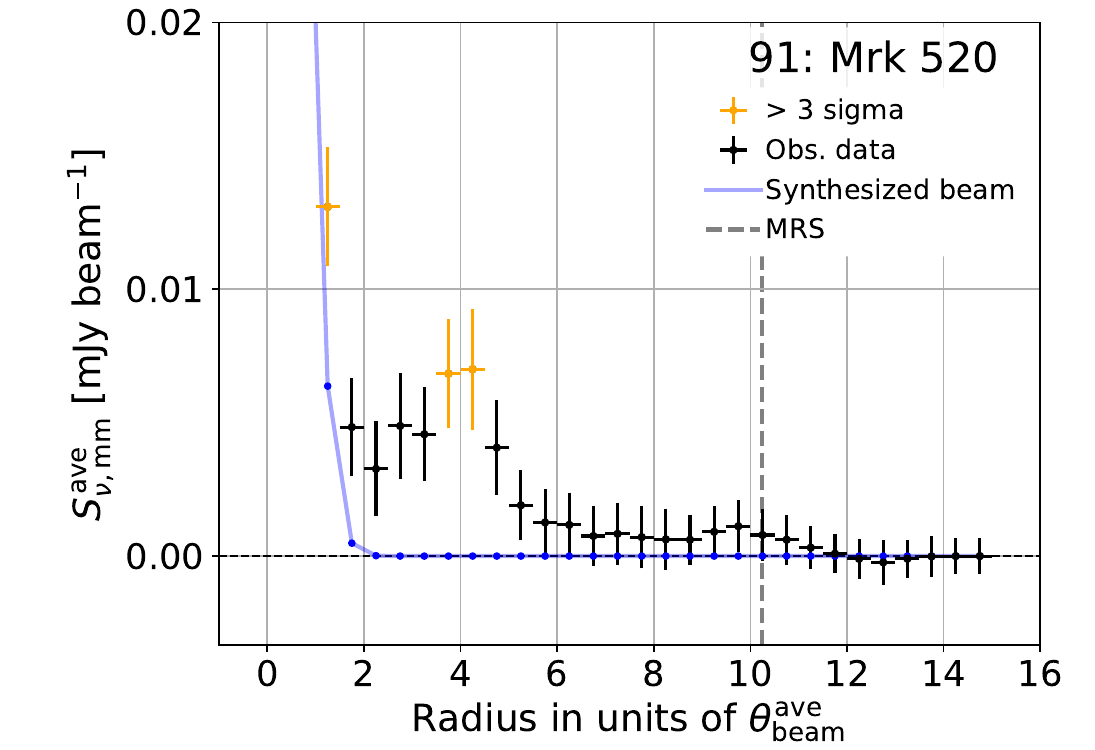}
\includegraphics[width=5.9cm]{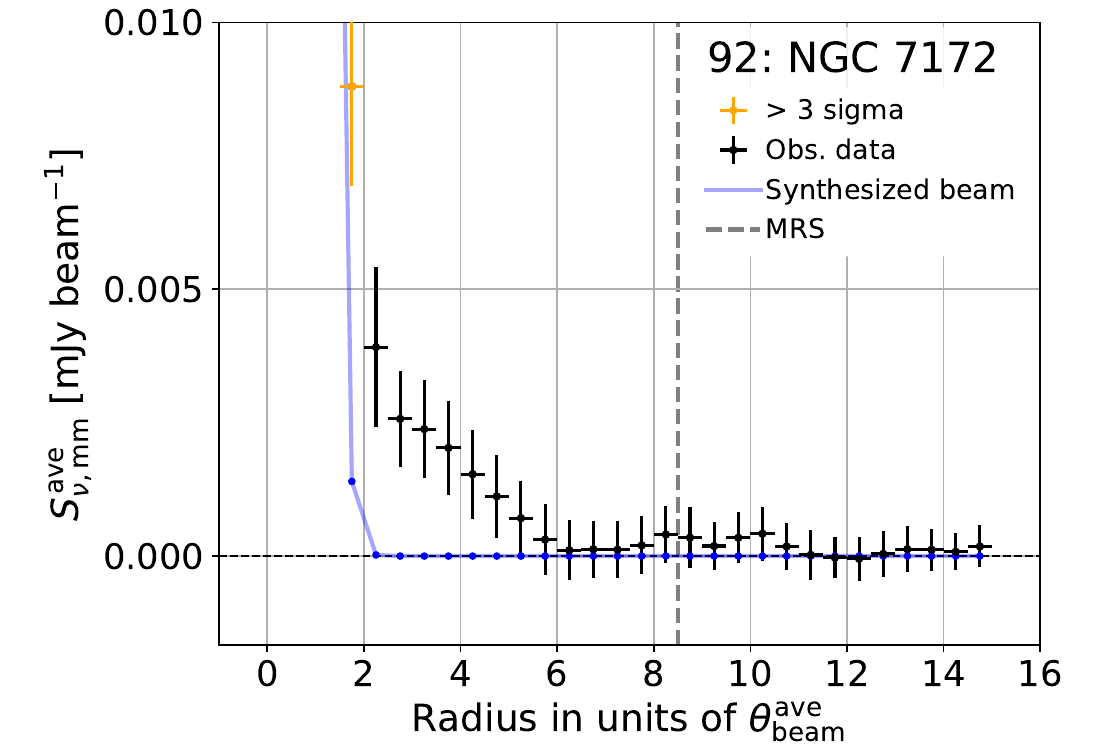}
\includegraphics[width=5.9cm]{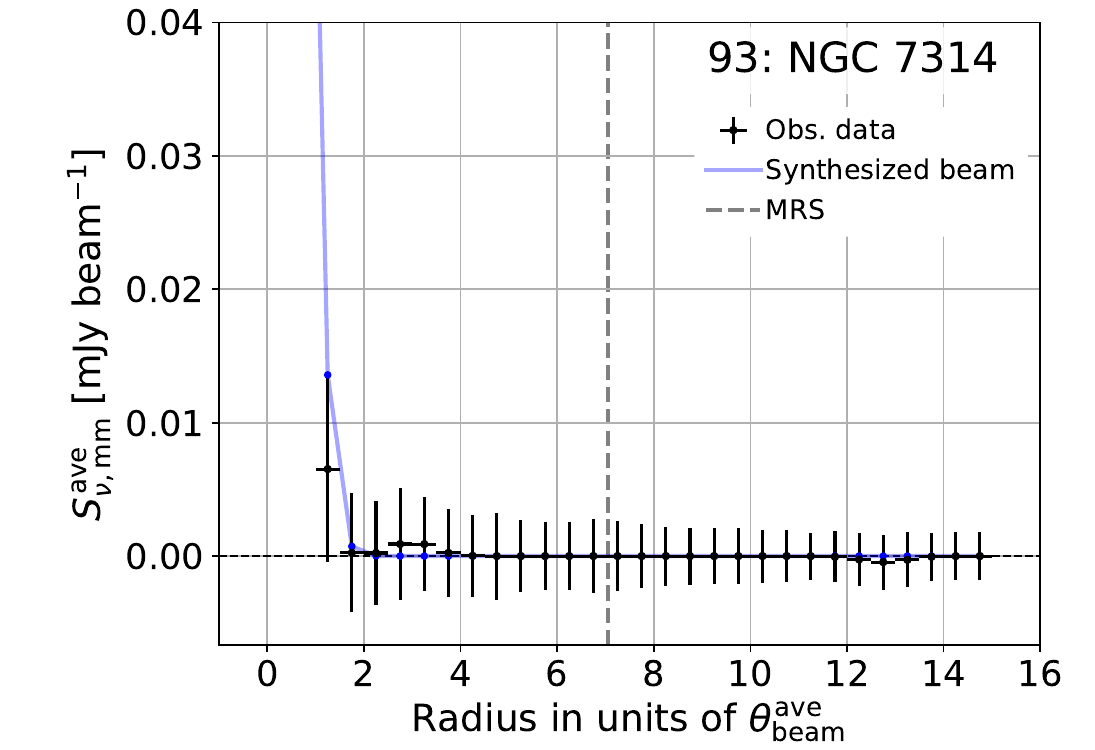}
\includegraphics[width=5.9cm]{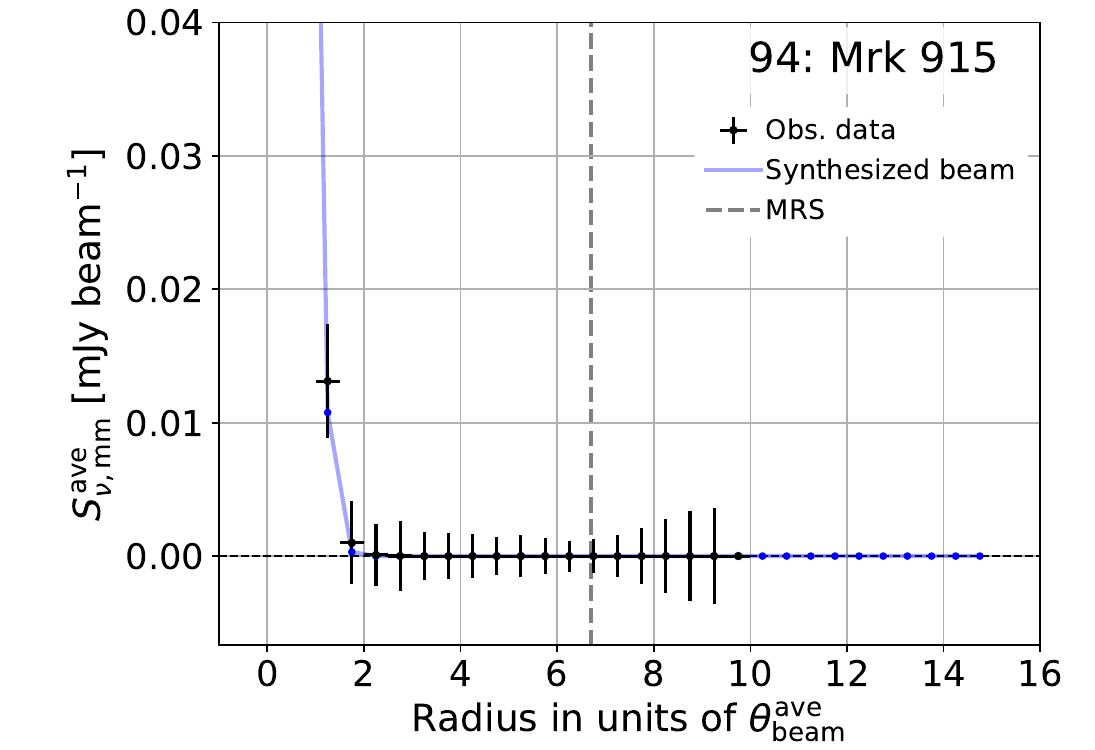}
\includegraphics[width=5.9cm]{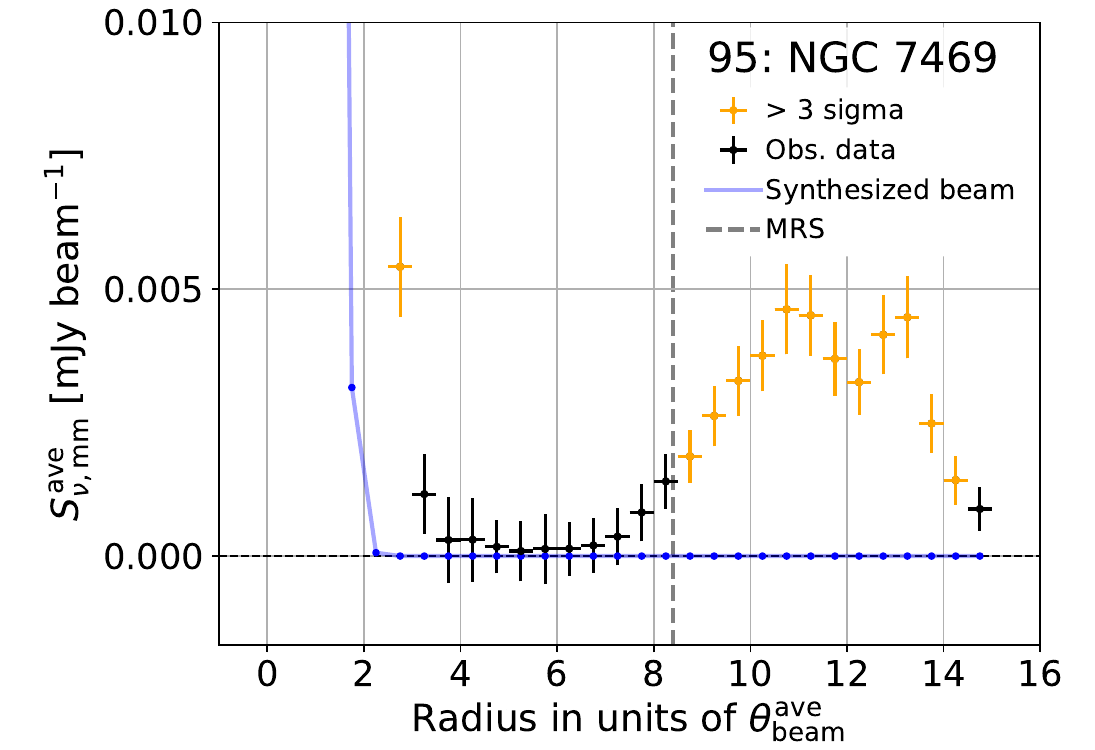}
\includegraphics[width=5.9cm]{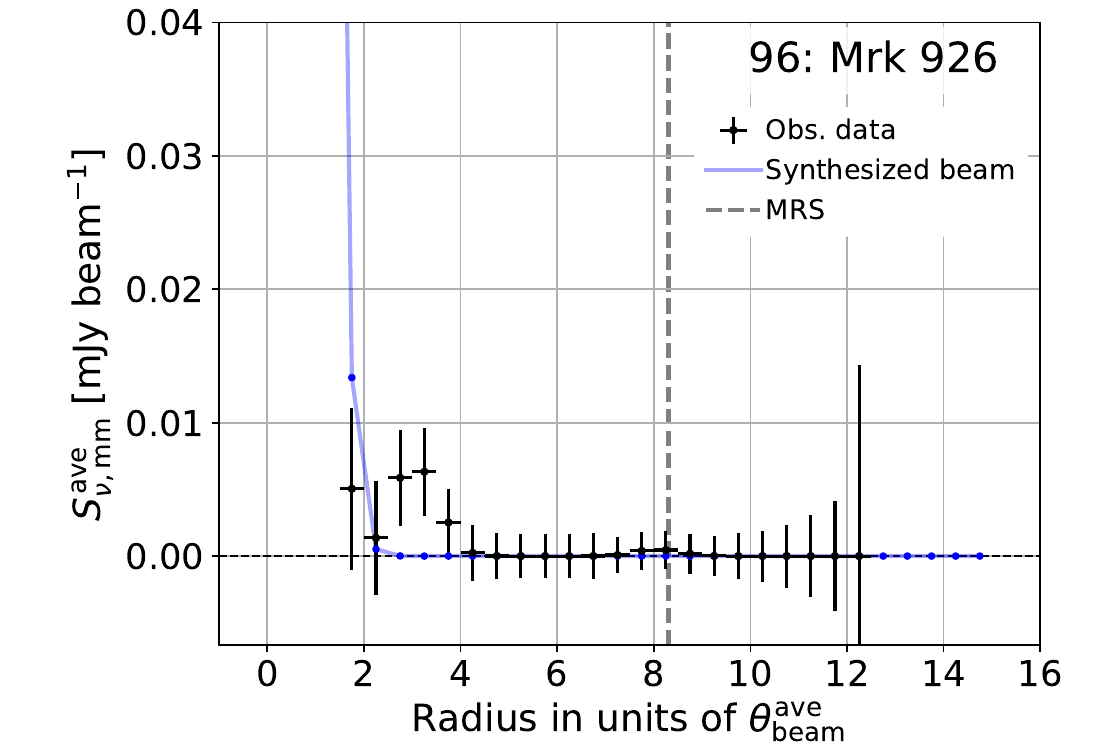}
\includegraphics[width=5.9cm]{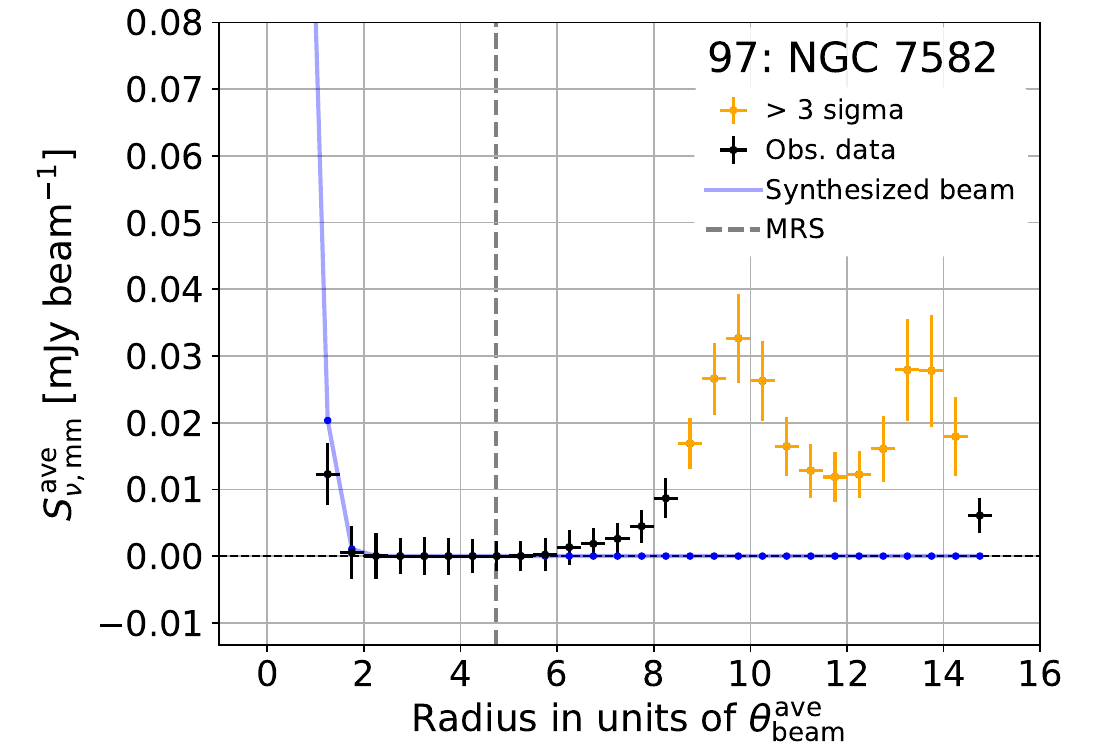}
\includegraphics[width=5.9cm]{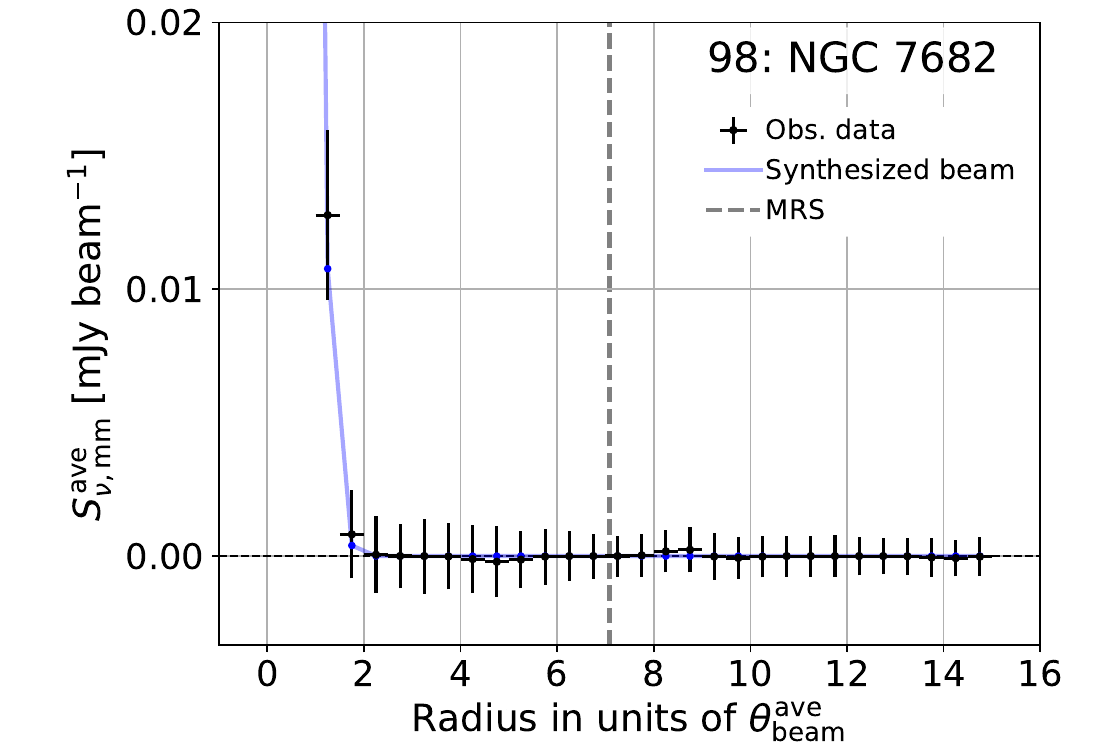}
\caption{Continued. 
    }
\end{figure*}

\clearpage


\section{Notes on individual objects}\label{sec_app:notes}

We present our discussions for 28 individual objects that showed both or one of largely extended or blob-like components, as detailed in Section~\ref{sec:ext}. 
For clarity of the discussions, we show the contour images of the targets in  Figure~\ref{fig:img4notes}. 

\begin{figure*}
    \figurenum{C}
    \centering
    \includegraphics[width=17.7cm]{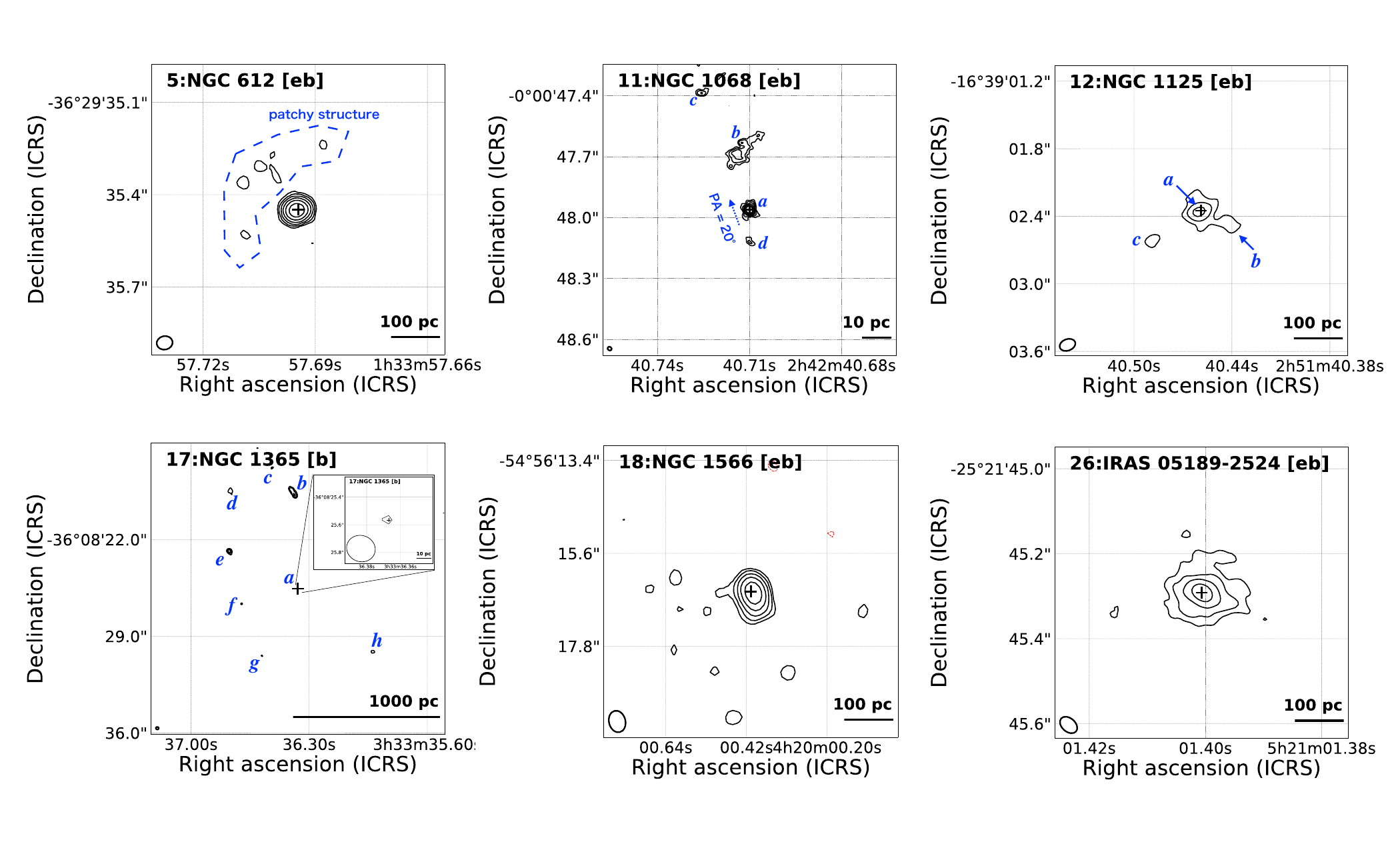} \\ \vspace{-0.7cm}
    \includegraphics[width=17.7cm]{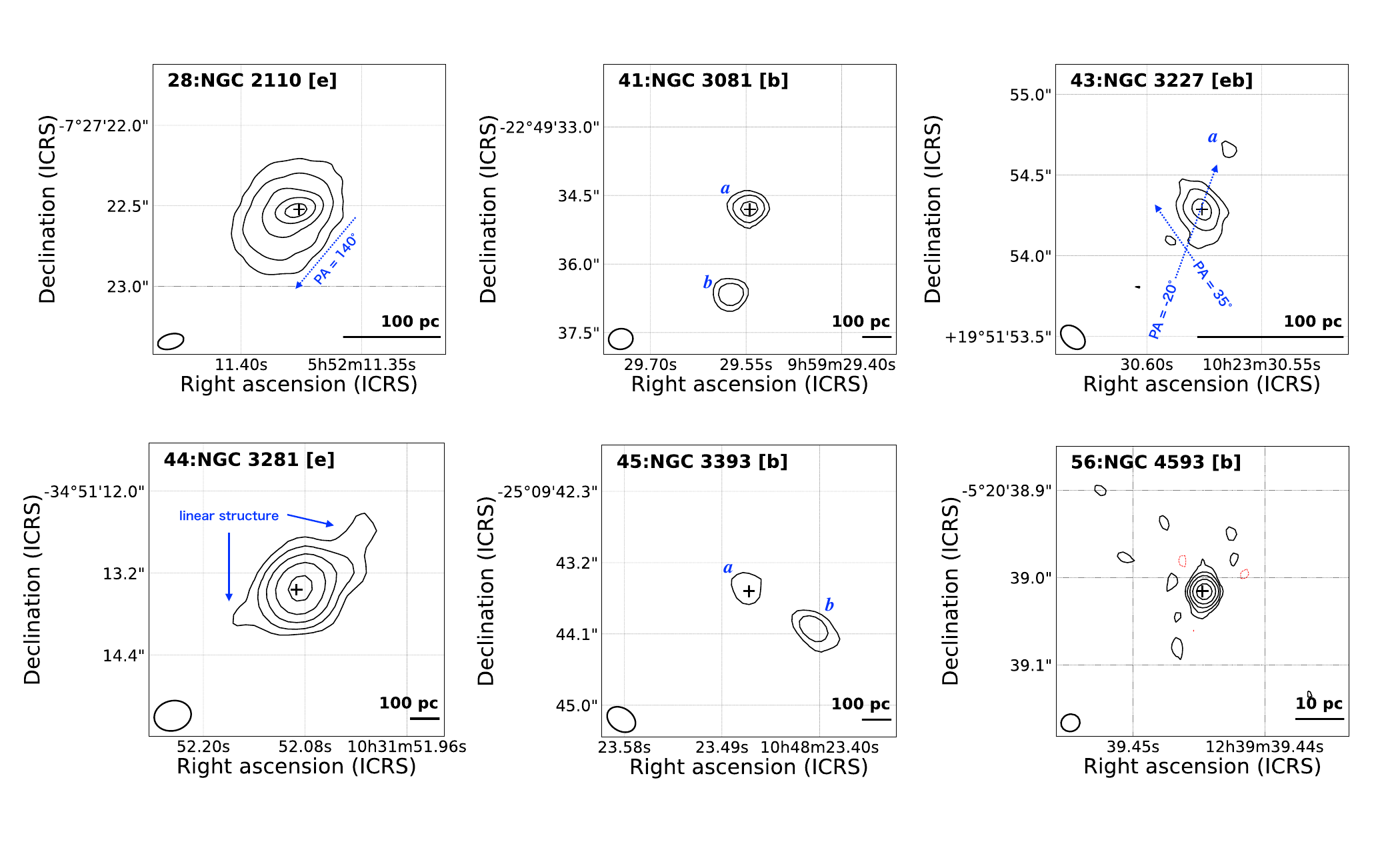}
    \caption{
    ALMA Band-6 images of 28 objects that showed
    both or one of largely extended or blob-like components.
    North is up and east is to the left. 
    Beam sizes are shown by the white ellipses in the left bottom corners. 
    Each central black cross indicates the peak of the nuclear mm-wave emission.
    Black solid contours indicate where flux densities per beam are $5\sigma_{\rm mm}$, $10\sigma_{\rm mm}$, $20\sigma_{\rm mm}$, $40\sigma_{\rm mm}$, 
    $80\sigma_{\rm mm}$ and $160\sigma_{\rm mm}$, while red dotted ones are drawn where flux densities per beam are $-5\sigma_{\rm mm}$. 
    For three objects (NGC 5506, Mrk 520, and NGC 7172), flux densities per beam at 
    $3\sigma_{\rm mm}$ are indicated by black dotted contours. 
    The individual values of $\sigma_{\rm mm}$ are listed in Table~\ref{tab_app:mm_prop}.
    The annotations in blue are used for the discussions in Section~\ref{sec_app:notes}.
    }\label{fig:img4notes}
\end{figure*}    


\begin{figure*}
    \figurenum{C}
    \centering
    \includegraphics[width=17.7cm]{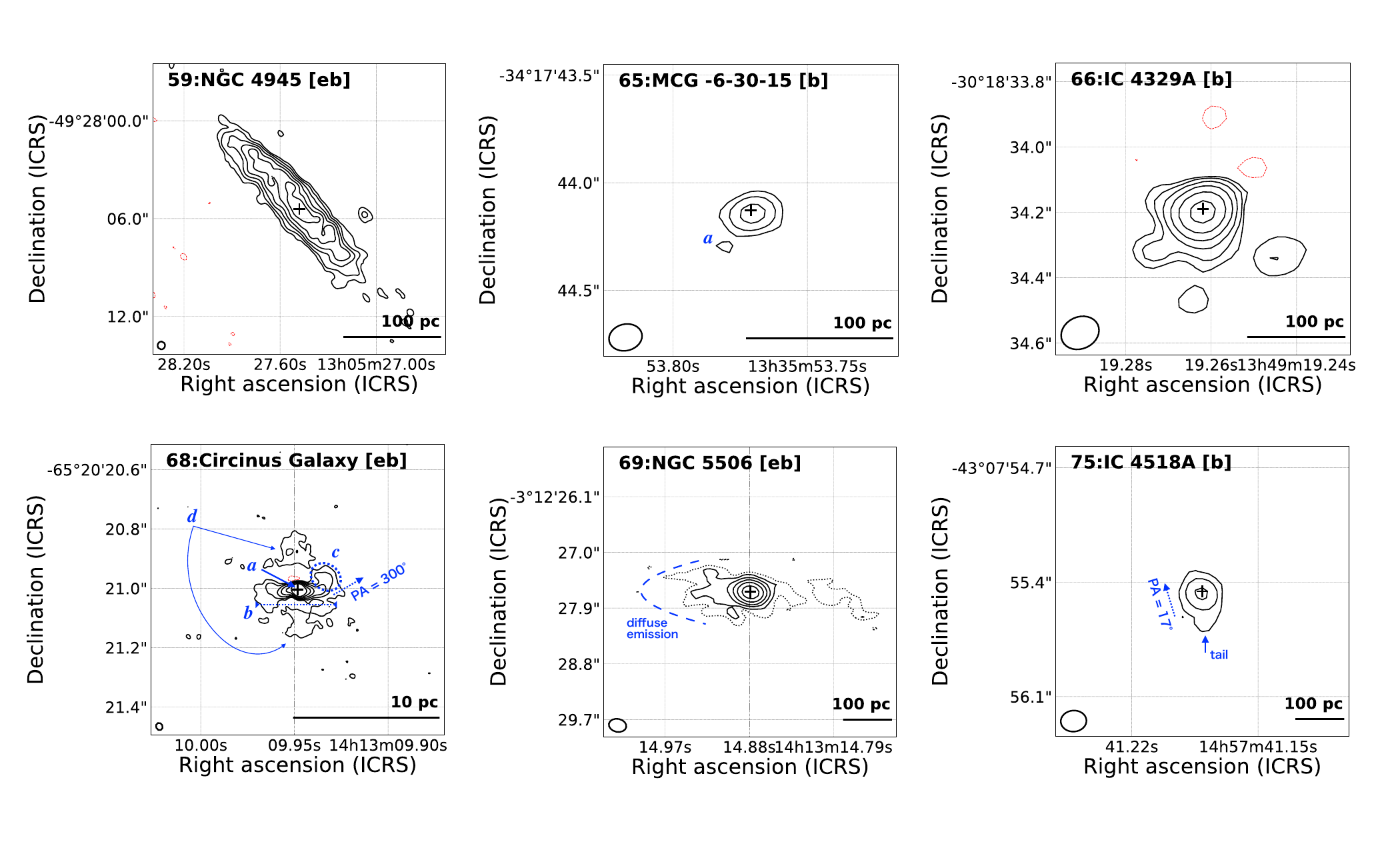} \\ \vspace{-0.7cm}
    \includegraphics[width=17.7cm]{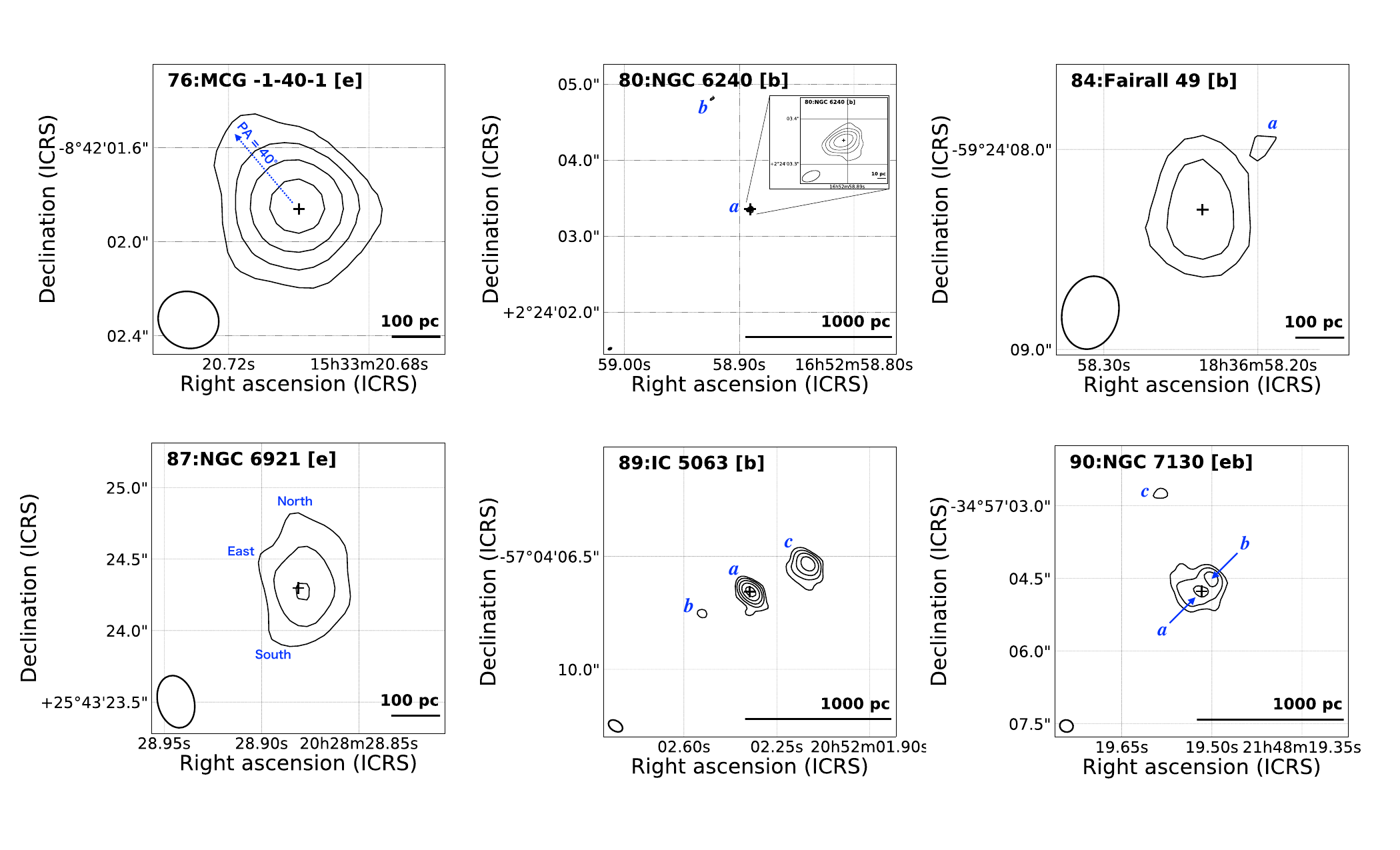}
    \caption{
    Continued.
    }
\end{figure*}

\begin{figure*}
    \figurenum{C}
    \centering
    \includegraphics[width=17.7cm]{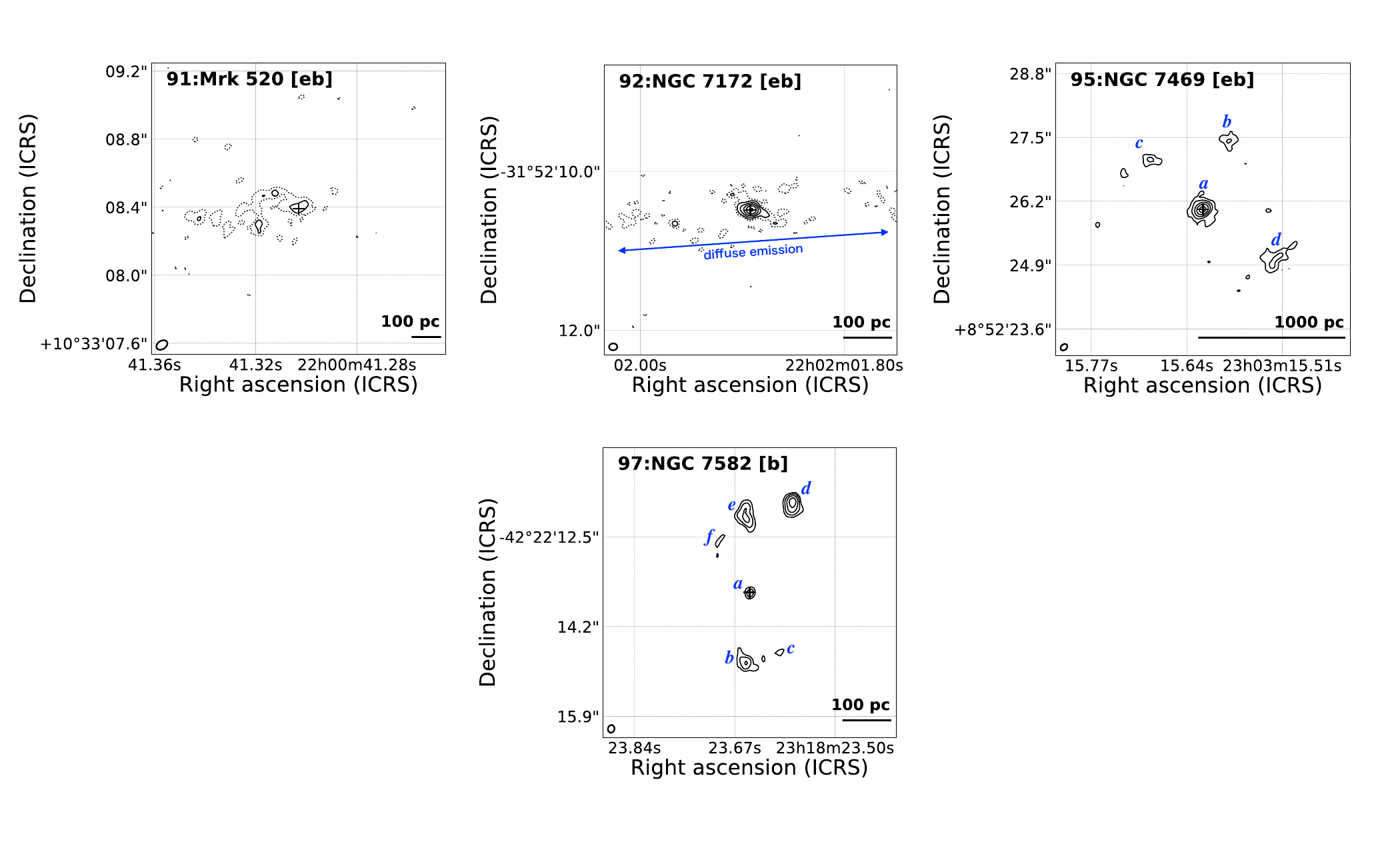} 
    \caption{
    Continued.
    }
\end{figure*}    

\clearpage 

\subsection{05: NGC 612}

The ALMA image of this object was taken at a resolution of 0\farcs05, revealing a patchy structure around the bright nuclear component within a radius of 0\farcs2. 
Based on a VLA observation at 5\,GHz, \cite{Eke78} reported an unresolved component whose diameter size was constrained to be $<$ 0\farcs3. 
This resolution would be almost the highest ever achieved for this object, except for that of the ALMA observation, and thus there seem to be no data that help us to better understand the patchy structure. 

\subsection{11: NGC 1068}

Our 266\,GHz image at a resolution of $\sim$ 0\farcs02 shows diffuse emission around the nucleus (\textit{a} in the corresponding panel of Figure~\ref{fig:img4notes}, hereafter italic characters are used to indicate such components in the same way) and also three other diffuse components $\sim$ 0\farcs3 away at a position angle (PA)\footnote{The PA is defined so that 0\arcdeg\ is the north, and that increases counter-clockwise.} of $\sim$ 10\arcdeg\ (\textit{b}), $\sim$ 0\farcs6 away at a PA of $\sim$ 20\arcdeg\ (\textit{c}), and $\sim$ 0\farcs2 away at a PA of $\sim$ 180\arcdeg\ (\textit{d}). A high-resolution continuum image obtained at almost the same resolution and frequency (i.e., $\approx$ 0\farcs02 and 257\,GHz) was presented by \cite{Imp19}. Their image revealed an X-shaped structure with a PA of $\sim$ 20\arcdeg\ (blue dotted arrow) around the nucleus, and they interpreted this as being due to edge-brightening of a bicone-like ionized region. The X-shaped structure is clearly seen also in our image. \cite{Imp19} measured the mm-wave and radio fluxes within a 0\farcs13 square area with their ALMA data and VLA 43 GHz data, and then derived a spectral index of $\sim 0.1$ for the central emission. 
This is consistent with free-free emission ($\alpha_{\rm mm}$ = 0.1). However, based on an SED covering a wider range of $\sim$ 1--1000\,GHz, \cite{Michi23} presented two other possible scenarios as well: self-absorbed synchrotron emission from an area with a radius of $\approx$ 30 Schwarzschild radii and synchrotron emission from a jet. 

The other diffuse component of \textit{b} appears to be closely located in an area with bright emission found in the radio band ($\sim$ 1--30\,GHz). For example, a 1.4\,GHz component, labeled C in \cite{Gal04}, was found at a resolution $\sim$ 0\farcs01 \cite[see also][]{Mux96,Gal96}. However, it is seen that the diffuse mm-wave emission (\textit{b}) may be closer to the center by $\sim$ 0\farcs04 (i.e., 2\,pc) than the radio emission. Further investigation is crucial to confirm this and, if true, to understand this offset. Similarly, the components \textit{c} and \textit{d} appear to virtually coincide with the radio-bright NE area and S2, reported in \cite{Gal04}. In their paper, it was suggested that the NE emission might originate in an internal shock or denser jet plasma, and the S2 emission may be the result of jet-driven shock. 

In addition to the radio emission, IR emission at 12.5 $\mu$m, for example, was mapped at a resolution of 0\farcs5 within $\sim$ 1\arcsec\ by \cite{Boc00}. 
They revealed a structure that extends from the nucleus in the north and south directions. Apparently, no peculiar MIR structures are associated with the structures \textit{b}, \textit{c}, and \textit{d}. 

Finally, we mention the secondary AGN recently suggested by \cite{Shi21}. The proposed area is $\sim$ 3\arcsec\ away from the nucleus at a PA of $\sim$ 44\arcdeg\ (i.e., it thus lies outside the cutout of the image shown). No counterpart in the mm-wave band is however found. Given the very low X-ray luminosity of $\sim 10^{39}$ erg\,s$^{-1}$ estimated by \cite{Shi21}, the non-detection may be natural. 
Our ALMA data have a sensitivity down to $\sim$ $10^{37}$ erg s$^{-1}$ in the mm-wave band, which
can be converted to a 2--10\,keV luminosity of $\sim 10^{41}$ erg\,s$^{-1}$ according to a relation presented in Paper~I. Thus, the current data would be just insufficient to assess the presence of the suggested secondary AGN. 





\subsection{12: NGC 1125}     

This source was observed at a resolution of $\approx$ 0\farcs1  using ALMA and the image shows central core emission (\textit{a}) and two non-nuclear components to
the southwest (SW) and the southeast (SE), labeled \textit{b} and \textit{c}, respectively.
A moderate resolution ($\approx$ 0\farcs3) image at 8.4\,GHz was obtained with VLA by 
\cite{The00} and revealed a linear structure, which extends at a PA of 120$^\circ$ and has two peaks in the SE and northwest (NW) regions, looking like radio lobes.  
The brightest radio component in the SE direction appears to coincide with the mm-wave blob of \textit{c}. No mm-wave counterpart is however found for the fainter NW radio component. 
This could be because mm-wave emission is also fainter and thus cannot be detected at the achieved sensitivity. 
Regarding the component of \textit{b}, no radio counterpart is found, perhaps due to the lower resolution of the radio image.
Interestingly, the relative strength between the mm-wave and radio emission is highest in the center.
This could suggest that the central and other lobe-like structures (\textit{c} and maybe \textit{b} as well) have different physical origins.

\subsection{17: NGC 1365}     

The image at a resolution of $\approx$ 0\farcs2 
shows seven significant components, labeled \textit{b}, \textit{c}, \textit{d}, \textit{e}, \textit{f}, \textit{g}, and \textit{h} around the central component labeled \textit{a}.
Radio maps at different frequencies were presented in past studies \citep[5\,GHz $\sim$ 15\,GHz;][]{Sand82,Ste99,The00,Sak07}. 
At \textit{a}, there appears to be a radio component according to a 
$\sim$ 1\arcsec\ resolution map created by \cite{Ste99}. 
Furthermore, the five mm-wave components of \textit{b}, \textit{c}, \textit{d}, \textit{e}, and \textit{f} appear to have radio counterparts that have been labeled D, E, G, H, and F, respectively, in previous studies \citep[e.g.,][]{Sand82}. \cite{Sak07} suggested that each of the D, E, G, and H regions possibly hosts a super cluster. 
\cite{Ste99} suggested that the components have moderate to steep spectral indices and are consistent with non-thermal emission. 
Regarding the region F, the interpretation is not so clear, as discussed in \cite{Ste99}. 
It is also currently unclear why the \textit{g} and \textit{h} components have no radio counterparts.

\subsection{18: NGC 1566}

Complex mm-wave structures appear within a radius of $\sim$ 3\arcsec. The central emission is slightly elongated to the north and is surrounded by several blob-like components. 
\cite{Com14} presented a submm-wave ALMA Band-7 image with 
a beam size of 0\farcs5, comparable to our mm-wave data of 0\farcs4. 
The authors argued that non-nuclear submm-wave emission exists and is from dust considering that the submm-wave emission follows a two-arm spiral structure detected in the CO($J$=3--2) line. Also, they estimated that the dust emission is likely to be stronger than synchrotron emission in the submm-wave band.
However, our Band-6 continuum distribution appears to be different from the Band-7 one, suggesting that the Band-6 emission may have a different origin.
Also, we mention the paper by \cite{Mez15}, who presented $\approx$ 0\farcs1 HST optical and VLT near-IR data including $K$-band emission and H$_2$ line emission. Apparently, none of the distributions seems to be spatially similar to the Band-6 structures. Eventually, nothing is found probably relevant to the extended and patchy components.

\subsection{26: IRAS 05189$-$2524}  

This is an ultraluminous IR galaxy and is considered to be in the final stage of a galaxy merger \citep[e.g.,][]{ric17b,Ric21b}. 
Around the central peak emission, diffuse emission 
within a radius of $\sim$ 0\farcs1 is revealed at a resolution of 0\farcs04. 
The encompassing area contains emission of $\sim$ 4 mJy, of which $\approx$ 1.3 mJy beam$^{-1}$ is ascribed to the peak emission. Thus, the contribution of the diffuse component is strong. 
Although there were few images taken at comparable resolutions, to our knowledge, NIR (2.2 $\mu$m and 3.8 $\mu$m) images at resolutions of 0\farcs1--0\farcs2 were obtained by \cite{Ima14nir}, who utilized a camera of Subaru/IRCS in an adaptive-optics mode. The obtained NIR nuclear images seem to be overwhelmed by only a point source. Using shorter wavelength ($\approx$ 1--2 $\mu$m) HST/NICMOS data at similar resolutions of 0\farcs1--0\farcs2, \cite{Sco00} also reported an unresolved component, but additionally mentioned the possible presence of low-level extended emission.
To discuss more about whether the mm-wave diffuse emission has a NIR counterpart, higher-resolution images with better sensitivities are desired.

\subsection{28: NGC 2110}

A core-like component appears to be surrounded by a halo, whose major axis is oriented at a PA of $\sim$ 140\arcdeg. 
In the adjacent cm-wave band, a jet in a north-south direction was clearly
revealed up to 4\arcsec on each side \citep[e.g.,][]{Ulv83}. The difference in angle between the radio jet and the mm-wave halo would suggest that the mm-wave emission would not be dominated by the jet. Instead, ionization cones traced by optical ionization lines (e.g., [O\,{\sc iii}] and [S\,{\sc  ii}]+H$\alpha$) have similar angles to that of the halo \citep[e.g.,][]{Mul94}. Spatially resolved X-ray iron emission that could be due to X-ray irradiation of gas by the AGN also has a similar angle \citep{Kaw20,Kaw21}. 
Thus, the mm-wave halo could be more relevant to the AGN photo-ionization process than the cm-wave jet.

\subsection{41: NGC 3081}

In addition to the central nuclear emission with $\sim$ 0.7 mJy beam$^{-1}$ (\textit{a}), our image at a resolution of $\sim$ 0\farcs5 unveils a comparably bright component (\textit{b}) with $\sim$ 0.5 mJy beam$^{-1}$ $\sim$ 2\arcsec\ away from the nucleus in the south. 
An 8.4\,GHz image was obtained at a similar beam size of $\sim$ 0\farcs4 by \cite{Mun09} using VLA, and shows radio emission ($\sim$ 1 mJy) spatially consistent with \textit{a}.
On the other hand, there is no significant radio emission around the south mm-wave component (\textit{b}). The upper limit of the radio emission, corresponding to $3\times$rms, is estimated to be $\sim$ 0.05 mJy beam$^{-1}$, leading to an upper limit of $< -$0.7 for the spectral index. This is smaller than that for the central emission ($\sim 0.7$). 
As an insight from different-wavelength data, 
\cite{Ma21} spatially revealed ionized gas at resolutions below 0\farcs1. According to 
optical emission line diagnostics with the [N\,{\sc  ii}]/H$\alpha$ and [O\,{\sc iii}]/H$\beta$, AGN-like or LINER-like optical emission line ratios were found around \textit{b}. Thus, the mm-wave emission in \textit{b} may be due to the AGN.

\subsection{43: NGC 3227}

Mm-wave continuum emission in the central 1\arcsec$\times$1\arcsec\ region of the object was discussed in detail by \cite{Alo19}, who used the same ALMA data as we used. Here, we give a brief summary of their discussion, and those who are interested in more details should check \cite{Alo19} and references therein. 

The authors suggested that there exist two mm-wave components with different PAs of $\sim 35$\arcdeg\ and $\sim -20$\arcdeg, indicated in Figure~\ref{fig:img4notes}. 
The centrally concentrated component with a PA $\sim$ 35\arcdeg\ 
may be relevant to an ionization cone and a radio jet. 
By HST observations of [N\,{\sc ii}] and H$\alpha$, an ionization cone was mapped at a similar PA. 
The presence of a jet in the direction, or synchrotron emission, was inferred from a spectral-index map that \cite{Alo19} made by combining the mm-wave data and an additional submm-wave ALMA one. 

The other component at the PA $\sim -20$\arcdeg\ ($a$) is located 0\farcs5 away from the nuclear peak, and the angle almost aligns with the axis of a nuclear disk. Mm-wave-to-submm-wave spectral indices therein were found to be consistent with thermal processes and free-free emission. While considering this fact and other multi-wavelength data (e.g., Br$\gamma$), \cite{Alo19} discussed that SF may contribute to the mm-wave emission. 
In the direction of the mm-wave component, cm-wave radio emission was also revealed by \cite{Mun95} and could originate from supernova remnants in the above context. 

\subsection{44: NGC 3281}
    
A NW-to-SE linear structure, parallel to a PA of $\sim$ 100\arcdeg--130\arcdeg, can be seen. In contrast, \cite{Sch01} reported only unresolved radio (8.46\,GHz) emission with a size upper limit of $<$ 60\,pc based on a spatial resolution of $\sim$ 0\farcs2. This is better than that achieved by the ALMA observation (0\farcs5). 
As a spatially extended signature of the AGN activity, a narrow line region was mapped by using optical [O III] emission in \cite{Sch03}, and this extends toward the NE. Eventually, from the radio and optical observations, we cannot find mm-wave counterparts. 
Instead, it seems that the major axis of the galaxy disk has a similar angle of $\sim$ 130\arcdeg\ \citep{Sch03}. Thus, the linear structure could be associated with gas, or SF processes, distributed in the galaxy disk. 
    
\subsection{45: NGC 3393} 


This object is known to have three radio spots in NE, nuclear, and SW areas \citep[e.g.,][]{Coo00,Kos15,Mak17,Fin18}, and our ALMA image at a resolution of 0\farcs3 shows two significant components: \textit{a} in the nucleus and \textit{b} in the SW direction. Thus, mm-wave emission spatially corresponding to the NE radio spot is undetected. 
At a similar resolution of $\sim$ 0\farcs2, 
\cite{Sch01} estimated the 8.5\,GHz flux densities for the SW, nuclear, and NE components to be $\sim$ 1\,mJy, $\sim$ 1\,mJy, and $\sim$ 8\,mJy, respectively. 
On the other hand, their mm-wave flux densities per beam are $\lesssim$ 0.1\,mJy beam$^{-1}$ ($5\sigma_{\rm mm}$ upper limit), $\sim$ 0.2\,mJy beam$^{-1}$, and $\sim$ 0.4\,mJy beam$^{-1}$, respectively. 
If the SW and NE radio components have similar spectral indices of 0.7--0.9, as inferred from a study of \cite{Mor99}, 
the non-detection of the NE component in the ALMA image is reasonable. In contrast, the nuclear component looks relatively bright compared with the SW one (i.e., a higher ratio of mm-wave and radio flux densities). Indeed, the nuclear emission may have a spectral index of $\sim$ 0.5 according to the 8.5\,GHz and 229\,GHz flux densities. This may suggest different physical origins for the central and other components. 

According to \cite{Mak19}, the radio jet drives the shock, then producing X-ray emission and optical [O~{\sc iii}] emission. Moreover, the authors suggested that the shocked gas has enough energy to evacuate gas from the galaxy. Thus, the mm-wave emission (\textit{b}) may also be tracing such a phenomenon, suggesting that mm-wave observations can be a tool for unveiling such shocked regions. 

We note that \cite{Fab11} suggested the presence of another AGN in the SE region. No significant emission is, however, found around that position in our ALMA image.

\subsection{56: NGC 4593}

In addition to the central core emission, a north-south component appears to extend up to 0\farcs03 on each side, and also there exist blob-like components. 
Data publicly presented to date seem to have achieved resolutions of $\sim$ 0\farcs2 at best 
\citep[e.g.,][]{The00,Sch01,Fis21}. Thus, 
we cannot discuss the origin with the help of other data.

\subsection{59: NGC 4945}  

Bright mm-wave emission is seen across the galaxy disk in the NE-SW direction. The flux density for the disk that we define as a 3\arcsec\ $\times$ 14\arcsec\ rectangular region with a PA of 43$^\circ$ is $\sim$ 600 mJy, and 
its intra-band spectral index is constrained to be $-2.1\pm0.2$.
These are largely different from those at the central peak (\textit{a}) of $S^{\rm peak}_{\nu, \rm mm} \sim$ 40 mJy beam$^{-1}$ and $\alpha_{\rm mm} = -0.4\pm0.2$, indicating that 
the significant component is different between the central peak and the disk region.
An SED in the 5--350\,GHz band was analyzed by \cite{Ben16} with a focus on a central disk region within 12\arcsec\ $\times$ 30\arcsec\ \citep[see also][]{Hen18}, and the 230\,GHz continuum flux of 1.3\,Jy  was basically reproduced by combining dust thermal emission and
free-free emission, resulting in an index of $\sim -3$. 
This index may be expected in our data as well,  but is different from our constraint of $-2.1\pm0.2$. This could be because thermal dust grains are more broadly distributed than free-free emitting photoionized gas, as mentioned by \cite{Ben16}, and the mm-wave slope is flattened in our high-resolution data where dust emission with a steep slope ($\alpha_{\rm mm} = -4 \sim -3$) is relatively missed.




\subsection{65: MCG $-$6$-$30$-$15}

At a PA of $\approx$ 140--150\arcdeg, a blob-like component (\textit{a}) is revealed at 
a resolution of $\sim$ 0\farcs1. 
Radio observations at 8\,GHz and 1.5 GHz were performed \citep{Nag99}, but the resolutions were much 
coarser (a few arcseconds), and detected radio components were not spatially resolved. 
HST images were reported in some works \citep{Fer00,Sch03}, and 
show that the host galaxy and also a NLR traced by the optical [O\,{\sc iii}] emission 
have similar PAs of $-65$\arcdeg, slightly different from the PA of the mm-wave component. 
Thus, the origin could be due neither to phenomena in the main body of the host galaxy nor to the AGN, and currently the origin is unclear. 

\subsection{66: IC 4329A}

In our image at a resolution of 0\farcs1, nuclear emission is prominent, and patchy components are seen.   
Radio (8.4\,GHz) emission was imaged with a beam of 0\farcs3$\times$0\farcs7 by \cite{The00} and significant radio emission with 4.9 mJy beam$^{-1}$ exists around the nuclear mm-wave emission. However, for further discussion, as the radio image is very noisy, better radio images with higher sensitivities and also higher spatial resolutions are desired to be compared with the ALMA one.

\subsection{68: Circinus Galaxy}

Around the bright nuclear core (\textit{a}), extended emission in the east-west direction (\textit{b}) exists on a scale of a few pc. In addition, 
there appear to be two components: one located to the NW (\textit{c}) and the symmetrically distributed structure from north to south (\textit{d}). 
In an area within a radius of 0\farcs35 which includes all the above structures, the flux density is $\sim$ 30 mJy. Since the peak flux density per beam is $\sim$ 15 mJy beam$^{-1}$, an important fraction of the mm-wave emission within the area is resolved at our resolution of $\sim$ 0.5\,pc. 

Submm-wave ALMA Band-7 continuum emission was imaged at a resolution of $\sim$ 0\farcs3 by \cite{Izu18}. 
The resolution is coarser than achieved in our data (i.e., 0\farcs02), thus hampering a visual comparison with our image. However, the authors reported the presence of an elongated structure on a scale of 0\farcs05 ($\sim$ 1\,pc) at a PA of $\sim$ 300$^\circ$ (blue dotted arrow) based on visibility modeling. They suggested that this would trace the polar dust structure found by MIR observations \citep{Tri14}. 
The component \textit{c} appears to coincide with that, thus perhaps indicating the dust emission as the origin of \textit{c}. 
The more prominent east-west emission of \textit{b} could have a different origin, however, given its direction, and currently, it is unclear what the origin is. 
Finally, the origin of the component \textit{d} is also unclear due to the lack of meaningful data available. Each of the north and south components spreads on a scale of 1\arcsec and has a flux density of $\sim$ 0.9 mJy. 
To conclude the physical mechanism for each component, it is important to determine the index, but our ALMA data are not sufficient to constrain it well. 
Thus, additional data at adjacent different wavelengths or those obtained at higher sensitivities are desired.


\subsection{69: NGC 5506}\label{sec:ngc5506}

At a $\approx$ 0\farcs2 resolution, in addition to the bright central emission, diffuse emission is seen, particularly in the east direction, while there may be a west component as well. 
Exceptionally, dotted contours at 3$\sigma_{\rm mm}$, which is formally insignificant but could be real, are presented.
\cite{Ori10} presented three maps at 4.8\,GHz, 8.4\,GHz, and 14.9\,GHz by analyzing VLA data, whose resolutions are $\sim$ 0\farcs4, 0\farcs2, and 0\farcs6, respectively. 
Interestingly, while diffuse halo structures were revealed at the two lowest frequencies within a radius of $\sim$ 2\arcsec, the 14.9\,GHz map shows that a component, extending to the east and west directions by $\sim$ 2\arcsec on each side, is prominent. Its physical origin was not discussed in the paper. The 14.9\,GHz structure looks similar to the mm-wave one at least in the east, although it is slightly tilted counterclockwise by $\sim$20\arcdeg--40\arcdeg. This suggests their association. Furthermore, given that the PAs of the 14.9\,GHz and mm-wave components are similar to that of the galaxy disk, they may be related to gas in the galaxy disk. Probably, a higher inclination angle (i.e., edge-on view) to NGC 5506, where line-of-sight column densities of gas can be higher, would make it easier to identify emission relevant to the gas in the galaxy disk.

\subsection{75: IC 4518A} 

This is a member of a merging system with IC 4518B. 
The prominent core emission, which appears to have a tail-like structure with a PA of 200\arcdeg--210\arcdeg, is seen in our data at a $\approx$ 0\farcs2 resolution. 
To our knowledge, no high-resolution radio images have been published, while \cite{Con96} showed a 1.4\,GHz image at a resolution $\sim$ 10\arcsec. 
As data taken at comparable resolutions, \cite{Asm19} showed 12\,$\mu$m and 18\,$\mu$m images at resolutions of $\sim$ 0\farcs4--0\farcs5.
The author revealed an elongated MIR component with a PA of 17\arcdeg\ (blue dotted arrow), which is almost parallel to the mm-wave tail. 
Given an argument by \cite{Asm19} that such a MIR component is due to dusty wind by an AGN, the mm-wave tail could be also driven by the AGN. 
As a different supportive result for the AGN-driven emission, \cite{Dia10} found extended [S~iv] 10.51 $\mu$m line emission out to $\sim$ 0\farcs8 to the south direction, and interpreted this as a sign of the presence of a narrow line region, consistent with the AGN origin. 


\subsection{76: MCG -1-40-1} 

In addition to the central core-like emission, our high-resolution image created by a beam of 0\farcs26$\times$0\farcs24 revealed a horn-like structure at a PA of $\sim$ 40\arcdeg. However, as no meaningful data taken at comparable resolutions are available,
discussion on the origin is currently difficult. 

\subsection{80: NGC 6240}

This is a well-studied merging system, and two active nuclei have been identified at various wavelengths \cite[e.g.,][]{Kom03,Gal04,Gal04ngc6240,Pac16}. 
According to a recent study by \cite{Kol20}, 
there may be another in-active nucleus to the south of the active nucleus in the south. 
Our image at a 0\farcs03 resolution shows two mm-wave compact components: one at the center (\textit{a}) and the other in the north (\textit{b}), probably associated with the first and second active nuclei \cite[e.g.,][]{Gal04}. On the other hand, insignificant emission is found around the third nucleus (\textit{c}), as already confirmed by \cite{Tre20}, who analyzed the same ALMA data. 
\cite{Med19} suggested that within a radius of 0\farcs077 around each of the two active nuclei, dust thermal emission is dominant at 242\,GHz. This argument was based on the result that the extrapolated mm-wave flux densities from 5\,GHz ones by adopting a spectral index of $\approx 1$ are lower than observed. 
In contrast, our higher-resolution data suggest a spectral index of $\alpha_{\rm mm} \sim$ 1.0$\pm$0.4 for the nucleus (\textit{a}), inconsistent with the dominance of thermal emission. Regarding the north nucleus (\textit{b}), the spectral index is not constrained, as its mm-wave emission is not detected in individual spectral windows. 

We comment on the relation of mm-wave and X-ray luminosities for the two active nuclei. In Paper~I, we added their luminosities together to be consistent with the X-ray flux derived in \cite{ric17b}, but 
we also confirm that the two nuclei follow a mm-wave-to-X-ray regression line 
(Table~1 in Paper~I) within 2 sigma, even if we consider them separately. 
The used X-ray luminosities are taken from \cite{Pac16}, who derived intrinsic 2--10\,keV luminosity for each using NuSTAR, Chandra, XMM-Newton, and BeppoSAX data. 



 

\subsection{84: Fairall 49}

Our ALMA image at a resolution of 0\farcs3 shows an isolated signal at a PA of $-50\sim-40$\arcdeg\ (\textit{a}). 
According to a sub-arcsec MIR observation by \cite{Asm16}, 
an extended MIR component possibly exists, and its PA was constrained to be 42\arcdeg \citep[see also][]{Isb21}. 
The PA does not seem to indicate that the mm-wave signal is associated with the MIR component. 
Instead, the host galaxy has a PA of 120\arcdeg and thus seems aligned to the signal. To furthermore 
discuss whether a host-galaxy phenomenon is indeed relevant, detailed analyses of host-galaxy images are necessary.

\subsection{87: NGC 6921} 

Three horn-like structures in the north, east, and south directions are seen in our mm-wave image constructed with a beam of 0\farcs37$\times$0\farcs25. 
Images at comparable resolutions are not found. Additional observations are desired for discussing the non-nuclear components.

\subsection{89: IC 5063}

Our ALMA image produced at a $\approx$ 0\farcs4 resolution shows three compact components: one at the central AGN position (\textit{a}) and the other two to the SE and NW (\textit{b} and \textit{c}). They are almost linearly aligned at an angle of $\sim$ 115\arcdeg, which is similar to that found for linear structures at frequencies of $\sim$ 10--20\,GHz. 
The mm-wave components were already discussed by \cite{Mor15} with a different ALMA data at $\sim$ 230\,GHz. 
Using 17\,GHz radio data from \cite{Mor07ic5063} together, 
the authors constrained the spectral indices for the SE, NW, and central components to be 0.87$\pm$0.02, 0.93$\pm$0.01, and 0.33$\pm$0.01.
The resolutions of the used 17\,GHz and 230\,GHz data were $\sim$ 0\farcs4--0\farcs5, comparable to ours. 
The indices favor nonthermal processes as the origins of the mm-wave components, and moreover the flatter slope of the central component could suggest that 
its origin is different from those of the other SE and NW ones. 

We moreover note that IC 5063 exhibits extended X-ray emission along the mm-wave and radio structures \citep{Gom17}. Recently, \cite{Tra21} performed a spatially resolved X-ray spectral analysis using Chandra and suggested that extended X-ray emission in the soft X-ray band ($<$ 3\,keV) can be reproduced by either two photo-ionized gas models or a sum of photo-ionized and collisionally ionized gas models. 
Furthermore, hard X-ray Fe\,XXV emission was detected 
at the NW radio structure. On the basis of these facts, the authors 
suggested that the X-ray emission may be caused by the interaction of the radio jet with the ISM. 
Thus, the mm-wave emission seems to also trace the interaction regions. 


\subsection{90: NGC 7130}

Our mm-wave image at a resolution of $\approx$ 0\farcs3 reveals two bright nuclear peaks (\textit{a} and \textit{b}). 
In addition, to the north, a compact component (\textit{c}) is seen. 
\cite{Zha16} studied this object using not only ALMA Band-9 data at a similar resolution of $\approx$ 0\farcs2 which observed CO($J$=6--5), but also other supplementary data at different wavelengths (e.g., optical and radio bands). 
They found a radio core near the SE nucleus (\textit{a}) using VLA 8.4 GHz continuum data provided in \cite{The00}, and also a higher 8.4\,GHz-to-CO($J$=6--5) luminosity ratio than expected for SF regions therein. These support that an AGN is located at the SE position of \textit{a}. We note that \textit{a} is consistent with a peak of optical emission therein  \citep{Gon98uvopt}. 
Following the analysis and suggestion of \cite{Zha16}, we defined the SE nucleus (\textit{a}) as an AGN position. 
On the other hand, \cite{Zha16} did not suggest the presence of an AGN in the NW region (\textit{b}), due to the absence of compact radio emission and a SF-region-like radio-to-CO($J$=6--5) ratio in the area. 
We note that we created a high-resolution Chandra image ($\approx$ 0\farcs1) in the 3--7\,keV band, probably tracing AGN emission \citep{Lev05}, and we did not find clear double nuclei.

Lastly, we mention that the Band-9 emission around the north mm-wave component (\textit{c}) was attributed to dust emission by \cite{Zha16}. 
The peak emission is $\approx$ 10--20 mJy beam$^{-1}$, and if a thermal component with $\alpha_{\rm mm} = -3.5$ is dominant, a predicted flux at 262\,GHz (our mm-wave frequency) is $\sim$ 0.5 mJy beam$^{-1}$. 
This is consistent with our observed value of 0.5\,mJy beam$^{-1}$, suggesting that the Band-6 emission would also be dust emission. 

\subsection{91: Mrk 520}

Our mm-wave image at a high resolution of 0\farcs06, corresponding to $\sim$ 30\,pc for the object, shows that overall the morphology looks diffuse. This can be noticed more clearly by considering fainter emission at $3\sigma_{\rm mm}$ (dotted lines) additionally. 
A 5\,GHz image was presented at a much coarser resolution of $\sim$ 1\farcs4 by \cite{Lon92}, and radio emission has a core-like morphology. This radio emission seems to coincide with the peak of the mm-wave image. However, to confirm whether most of the radio emission is indeed associated with the mm-wave peak or appears like the diffuse mm-wave emission, it is important to obtain a higher-resolution radio image.

\subsection{92: NGC 7172}

In addition to the central core emission, diffuse emission in the east-west direction appears at $3\sigma_{\rm mm}$, although this is formally insignificant. 
The morphology appears to trace the galaxy disk \citep[e.g.,][]{Mal98}. Therefore the diffuse emission would trace gas in the disk, or SF regions. As in NGC 5506 (Section~\ref{sec:ngc5506}), the edge-on view to NGC 7172 would enable us to identify that emission. 
We note that a possible radio counterpart for the core was found with a 0\farcs4 beam in \cite{The00}. 



\subsection{95: NGC 7469} 

Around the strongest central component (\textit{a}), there are three blobs (\textit{b}, \textit{c}, and \textit{d}). These are likely to be encompassed by an annular star-burst ring with radii of $\sim$ 1\farcs5--2\farcs5 found at different wavelengths \cite[e.g.,][]{Soi03,Dia07,Ori10}, and 
are spatially coincident with significant 349.7\,GHz components presented at an $\approx$ 0\farcs50$\times$0\farcs40 resolution by \cite{Izu15}. 
At that 350-GHz band, the three components have flux densities per beam of $\approx$ 1.0--1.7 mJy beam$^{-1}$ \citep{Izu15}. With Band-6 ALMA data, different from ours, \cite{Ima16} also measured their flux densities at $\sim$ 260--266\,GHz with a similar beam size of 0\farcs57$\times$0\farcs52. By correcting for the slight difference in beam size, the spectral slopes between the two bands are estimated to be $\approx -2.6$. This suggests that thermal emission with $\alpha_{\rm mm} \sim -3.5$ may not be the only component. 
We note that, in our mm-wave data, the three components were not detected in a sufficient number of windows to derive the spectral index. 



\subsection{97: NGC 7582}

Our mm-wave image reveals nuclear emission (\textit{a}) and also five surrounding components labeled \textit{b}, \textit{c}, \textit{d}, \textit{e}, and \textit{f}.
Hints on the origins can be obtained by comparing the ALMA image with a MIR [Ne II] emission line image of \cite{Wol06}. 
The MIR image revealed six areas with bright MIR emission, labeled M1, M2, M3, M4, M5, and M6. 
The mm-wave components of \textit{b}, \textit{c}, \textit{d}, \textit{e}, and \textit{f} appear to coincide with 
M1, M2, M3+M4, M5, and M6, respectively. 
The authors described that the six MIR structures may be due to young star clusters embedded in dust and, otherwise, to clusters that are older and obscured by a screen of dust.
However, a recent study by \cite{Ric18ngc7582} suggested that AGN-driven jets/winds would be related to the MIR structures, based on their associations with NIR emission lines (i.e., [Fe II] and Br$\gamma$), while the contributions of young stellar clusters were not completely excluded. 
Thus, the mm-wave components may originate from the AGN jets/winds.

\end{document}

%% file: table4sample.tex
01  &      28  &           J0042.9$-$2332  &                      NGC 235A  &      10.720029  &   $-$23.541049  &       A  &   0.02207  &    96.1  &   0.1(a)  &        1.9 \\ 
02  &      31  &           J0042.9$-$1135  &             MCG $-$2$-$2$-$95  &      10.786582  &   $-$11.601037  &       W  &   0.01938  &    84.3  &   0.1(a)  &          2 \\ 
03  &      58  &           J0111.4$-$3808  &                       NGC 424  &      17.865171  &   $-$38.083467  &       A  &   0.01088  &    51.0  &   0.1(a)  &        1.9 \\ 
04  &      72  &           J0123.8$-$3504  &                      NGC 526A  &      20.976579  &   $-$35.065489  &       A  &   0.01890  &    82.1  &   0.1(a)  &        1.9 \\ 
05  &      84  &           J0134.1$-$3625  &                       NGC 612  &      23.490394  &   $-$36.493180  &       A  &   0.02990  &   131.0  &   0.1(a)  &          2 \\ 
06  &     102  &           J0201.0$-$0648  &                       NGC 788  &      30.276937  &    $-$6.815879  &       A  &   0.01365  &    59.1  &   0.1(a)  &          2 \\ 
07  &    112B  &         J0209.5$-$1010D1  &                       NGC 833  &      32.336852  &   $-$10.133077  &       W  &   0.01355  &    58.6  &   0.1(a)  &          2 \\ 
08  &     112  &         J0209.5$-$1010D2  &                       NGC 835  &      32.352549  &   $-$10.135861  &       W  &   0.01334  &    57.7  &   0.1(a)  &          2 \\ 
09  &     131  &           J0231.6$-$3645  &                       IC 1816  &      37.962408  &   $-$36.672156  &       A  &   0.01665  &    72.2  &   0.1(a)  &          2 \\ 
10  &     134  &           J0234.6$-$0848  &                       NGC 985  &      38.657667  &    $-$8.787805  &       A  &   0.04302  &   190.3  &   0.1(a)  &          1 \\ 
11  &     144  &           J0242.6$+$0000  &                      NGC 1068  &      40.669625  &    $-$0.013323  &       A  &   0.00348  &    14.4  &     2.6  &        1.9 \\ 
12  &     153  &           J0251.6$-$1639  &                      NGC 1125  &      42.918579  &   $-$16.650653  &       A  &   0.01112  &    48.0  &   0.1(a)  &          2 \\ 
13  &     156  &           J0252.7$-$0822  &             MCG $-$2$-$8$-$14  &      43.097500  &    $-$8.510283  &       A  &   0.01674  &    72.6  &   0.1(a)  &          2 \\ 
14  &     159  &           J0256.4$-$3212  &                   ESO 417$-$6  &      44.089754  &   $-$32.185619  &       A  &   0.01679  &    72.8  &   0.1(a)  &          2 \\ 
15  &     163  &           J0304.1$-$0108  &                      NGC 1194  &      45.954617  &    $-$1.103756  &       A  &   0.01357  &    58.7  &   0.1(a)  &          2 \\ 
16  &     182  &           J0331.4$-$0510  &             MCG $-$1$-$9$-$45  &      52.846033  &    $-$5.141687  &       A  &   0.01301  &    56.3  &   0.1(a)  &          2 \\ 
17  &     184  &           J0333.6$-$3607  &                      NGC 1365  &      53.401529  &   $-$36.140435  &       A  &   0.00509  &    19.6  &     1.7  &          1 \\ 
18  &     216  &           J0420.0$-$5457  &                      NGC 1566  &      65.001700  &   $-$54.937918  &       A  &   0.00474  &    17.9  &     2.0  &          1 \\ 
19  &     237  &           J0444.1$+$2813  &                    LEDA 86269  &      71.037558  &      28.216873  &       A  &   0.01053  &    46.1  &   0.1(a)  &          2 \\ 
20  &     242  &           J0451.4$-$0346  &            MCG $-$1$-$13$-$25  &      72.922996  &    $-$3.809381  &       A  &   0.01547  &    67.0  &   0.1(a)  &          1 \\ 
21  &     252  &           J0502.1$+$0332  &                    LEDA 75258  &      75.537700  &       3.530526  &       A  &   0.01589  &    68.9  &   0.1(a)  &          1 \\ 
22  &     260  &           J0508.1$+$1727  &     2MASX J05081967$+$1721483  &      77.082108  &      17.363351  &       A  &   0.01736  &    75.3  &   0.1(a)  &        1.9 \\ 
23  &     261  &           J0510.7$+$1629  &             IRAS 05078$+$1626  &      77.689746  &      16.499493  &       A  &   0.01735  &    75.3  &   0.1(a)  &          1 \\ 
24  &     266  &           J0516.2$-$0009  &                        Ark120  &      79.047550  &    $-$0.149769  &       A  &   0.03257  &   143.0  &   0.1(a)  &          1 \\ 
25  &     269  &           J0501.9$-$3239  &                  ESO 362$-$18  &      79.899233  &   $-$32.657763  &       A  &   0.01246  &    53.9  &   0.1(a)  &          1 \\ 
26  &     272  &           J0521.0$-$2522  &             IRAS 05189$-$2524  &      80.255836  &   $-$25.362581  &       A  &   0.04090  &   180.6  &   0.1(a)  &          2 \\ 
27  &     301  &           J0543.9$-$2749  &                  ESO 424$-$12  &      85.887329  &   $-$27.651423  &       W  &   0.00972  &    41.9  &   0.1(a)  &          1 \\ 
28  &     308  &           J0552.2$-$0727  &                      NGC 2110  &      88.047400  &    $-$7.456256  &       A  &   0.00750  &    34.3  &     6.3  &          2 \\ 
29  &     313  &           J0557.9$-$3822  &                  H 0557$-$385  &      89.508578  &   $-$38.334572  &       A  &   0.03310  &   145.4  &   0.1(a)  &          1 \\ 
30  &     319  &           J0601.9$-$8636  &                     ESO 5$-$4  &      91.422656  &   $-$86.631591  &     CXO  &   0.00616  &    28.2  &     5.6  &          2 \\ 
31  &     330  &           J0623.8$-$3215  &                   ESO 426$-$2  &      95.943412  &   $-$32.216533  &       A  &   0.02244  &    97.7  &   0.1(a)  &          2 \\ 
32  &     404  &           J0804.2$+$0507  &                      Mrk 1210  &     121.024425  &       5.113849  &       A  &   0.01355  &    58.7  &   0.1(a)  &        1.9 \\ 
33  &     416  &           J0823.4$-$0457  &                   Fairall 272  &     125.754558  &    $-$4.934869  &       A  &   0.02224  &    96.9  &   0.1(a)  &          2 \\ 
34  &     423  &           J0838.4$-$3557  &                  Fairall 1146  &     129.628233  &   $-$35.992590  &       A  &   0.03182  &   139.6  &   0.1(a)  &        1.9 \\ 
35  &     453  &           J0920.8$-$0805  &            MCG $-$1$-$24$-$12  &     140.192754  &    $-$8.056100  &       A  &   0.01961  &    85.2  &   0.1(a)  &          2 \\ 
36  &     460  &           J0926.2$+$1244  &                       Mrk 705  &     141.513633  &      12.734366  &       A  &   0.02862  &   125.3  &   0.1(a)  &          1 \\ 
37  &     465  &           J0934.7$-$2156  &                  ESO 565$-$19  &     143.681479  &   $-$21.927726  &       A  &   0.01596  &    69.2  &   0.1(a)  &          2 \\ 
38  &     471  &           J0945.6$-$1420  &                      NGC 2992  &     146.424775  &   $-$14.326272  &       A  &   0.00767  &    38.0  &     7.0  &        1.9 \\ 
39  &     472  &           J0947.6$-$3057  &            MCG $-$5$-$23$-$16  &     146.917242  &   $-$30.948884  &       A  &   0.00841  &    36.2  &   0.1(a)  &        1.9 \\ 
40  &     475  &           J0951.9$-$0649  &                      NGC 3035  &     147.979300  &    $-$6.822918  &       A  &   0.01441  &    62.4  &   0.1(a)  &        1.9 \\ 
41  &     480  &           J0959.5$-$2248  &                      NGC 3081  &     149.873100  &   $-$22.826334  &       A  &   0.00807  &      32  &      12  &          2 \\ 
42  &     486  &           J1005.9$-$2305  &                  ESO 499$-$41  &     151.480797  &   $-$23.057012  &     CXO  &   0.01294  &    56.0  &   0.1(a)  &          1 \\ 
43  &     497  &           J1023.5$+$1952  &                      NGC 3227  &     155.877400  &      19.865080  &       A  &   0.00328  &    22.9  &     3.2  &          1 \\ 
44  &     502  &           J1031.7$-$3451  &                      NGC 3281  &     157.967042  &   $-$34.853734  &       A  &   0.01113  &    48.1  &   0.1(a)  &          2 \\ 
45  &     518  &           J1048.4$-$2511  &                      NGC 3393  &     162.097771  &   $-$25.162100  &       A  &   0.01294  &    56.0  &   0.1(a)  &          2 \\ 
46  &     520  &          J1051.2$-$1704A  &                      NGC 3431  &     162.812671  &   $-$17.008094  &       A  &   0.01744  &    75.7  &   0.1(a)  &          2 \\ 
47  &     558  &           J1139.0$-$3743  &                      NGC 3783  &     174.757129  &   $-$37.738614  &       A  &   0.00896  &      38  &      14  &          1 \\ 
48  &     576  &           J1152.1$-$1122  &                 PG 1149$-$110  &     178.014804  &   $-$11.373347  &       A  &   0.04858  &   215.7  &   0.1(a)  &          1 \\ 
49  &     583  &           J1201.2$-$0341  &                      Mrk 1310  &     180.309800  &    $-$3.678049  &       A  &   0.01946  &    84.6  &   0.1(a)  &          1 \\ 
50  &     586  &           J1204.5$+$2019  &                      NGC 4074  &     181.123617  &      20.316280  &       A  &   0.02274  &    99.1  &   0.1(a)  &          2 \\ 
51  &     607  &           J1217.3$+$0714  &                      NGC 4235  &     184.291171  &       7.191571  &       A  &   0.00793  &    26.6  &     3.7  &          1 \\ 
52  &     608  &           J1218.5$+$2952  &                      NGC 4253  &     184.610500  &      29.812926  &       A  &   0.01292  &    55.9  &   0.1(a)  &          1 \\ 
53  &     615  &           J1225.8$+$1240  &                      NGC 4388  &     186.444917  &      12.662158  &       A  &   0.00834  &    18.1  &     3.6  &          2 \\ 
54  &     616  &           J1202.5$+$3332  &                      NGC 4395  &     186.453600  &      33.546920  &     CXO  &   0.00111  &     4.8  &     0.5  &          1 \\ 
55  &     626  &           J1235.6$-$3954  &                      NGC 4507  &     188.902646  &   $-$39.909354  &       A  &   0.01169  &    50.5  &   0.1(a)  &          2 \\ 
56  &     631  &           J1239.6$-$0519  &                      NGC 4593  &     189.914353  &    $-$5.344171  &       A  &   0.00832  &    37.2  &     6.8  &          1 \\ 
57  &     641  &           J1252.3$-$1323  &                      NGC 4748  &     193.051958  &   $-$13.414787  &       A  &   0.01414  &    61.2  &   0.1(a)  &          1 \\ 
58  &     653  &           J1304.3$-$0532  &                      NGC 4941  &     196.054596  &    $-$5.551558  &       A  &   0.00388  &    20.4  &     3.8  &          2 \\ 
59  &     655  &           J1305.4$-$4928  &                      NGC 4945  &     196.364500  &   $-$49.468169  &       A  &   0.00226  &     3.5  &     0.3  &          2 \\ 
60  &     657  &          J1306.4$-$4025A  &                  ESO 323$-$77  &     196.608829  &   $-$40.414618  &       A  &   0.01563  &    67.7  &   0.1(a)  &          1 \\ 
61  &     676  &           J1332.0$-$7754  &                    ESO 21$-$4  &     203.169458  &   $-$77.844624  &       A  &   0.00963  &    41.6  &   0.1(a)  &          2 \\ 
62  &     677  &           J1333.5$-$3401  &                  ESO 383$-$18  &     203.358758  &   $-$34.014800  &       A  &   0.01287  &    55.7  &   0.1(a)  &          2 \\ 
63  &     678  &         J1334.8$-$2328D2  &                    LEDA 47848  &     203.665067  &   $-$23.446548  &       A  &   0.03424  &   150.5  &   0.1(a)  &        1.9 \\ 
64  &     679  &           J1336.0$+$0304  &                      NGC 5231  &     203.951012  &       2.998910  &       A  &   0.02156  &    93.9  &   0.1(a)  &          2 \\ 
65  &     680  &           J1335.8$-$3416  &            MCG $-$6$-$30$-$15  &     203.974046  &   $-$34.295591  &       A  &   0.00706  &    30.4  &   0.1(a)  &          1 \\ 
66  &     694  &           J1349.3$-$3018  &                      IC 4329A  &     207.330258  &   $-$30.309497  &       A  &   0.01598  &    69.3  &   0.1(a)  &          1 \\ 
67  &     696  &           J1351.5$-$1814  &                       CTS 103  &     207.872949  &   $-$18.229518  &       W  &   0.01209  &    52.2  &   0.1(a)  &          1 \\ 
68  &     711  &           J1412.9$-$6522  &               Circinus Galaxy  &     213.291451  &   $-$65.339168  &       A  &   0.00149  &     4.2  &     0.3  &          2 \\ 
69  &     712  &           J1413.2$-$0312  &                      NGC 5506  &     213.311996  &    $-$3.207677  &       A  &   0.00598  &    26.4  &     3.7  &        1.9 \\ 
70  &     717  &           J1417.9$+$2507  &                      NGC 5548  &     214.498071  &      25.136826  &       A  &   0.01669  &    72.4  &   0.1(a)  &          1 \\ 
71  &     719  &           J1419.0$-$2639  &                  ESO 511$-$30  &     214.843358  &   $-$26.644753  &       A  &   0.02294  &   100.0  &   0.1(a)  &          1 \\ 
72  &     731  &           J1432.8$-$4412  &                      NGC 5643  &     218.169588  &   $-$44.174425  &       A  &   0.00404  &    12.7  &     1.0  &          2 \\ 
73  &     733  &           J1433.9$+$0528  &                      NGC 5674  &     218.467833  &       5.458366  &       A  &   0.02477  &   108.1  &   0.1(a)  &        1.9 \\ 
74  &     739  &           J1442.5$-$1715  &                      NGC 5728  &     220.599488  &   $-$17.253064  &       A  &   0.01032  &    37.5  &   0.1(a)  &        1.9 \\ 
75  &     751  &           J1457.8$-$4308  &                      IC 4518A  &     224.421587  &   $-$43.132073  &       A  &   0.01605  &    69.6  &   0.1(a)  &          2 \\ 
76  &     772  &           J1533.2$-$0836  &             MCG $-$1$-$40$-$1  &     233.336250  &    $-$8.700517  &       A  &   0.02302  &   100.3  &   0.1(a)  &        1.9 \\ 
77  &     783  &           J1548.5$-$1344  &                      NGC 5995  &     237.103948  &   $-$13.757565  &       A  &   0.02481  &   108.3  &   0.1(a)  &          1 \\ 
78  &     795  &           J1613.2$-$6043  &                  LEDA 2793282  &     242.964121  &   $-$60.631939  &       A  &   0.01540  &    66.8  &   0.1(a)  &          1 \\ 
79  &     823  &           J1635.0$-$5804  &                  ESO 137$-$34  &     248.808317  &   $-$58.079943  &       A  &   0.00876  &    34.1  &     6.2  &          2 \\ 
80  &     841  &           J1652.9$+$0223  &                      NGC 6240  &     253.245377  &       2.400931  &       A  &   0.02472  &   107.9  &   0.1(a)  &        1.9 \\ 
81  &     875  &           J1717.1$-$6249  &                      NGC 6300  &     259.248108  &   $-$62.820553  &       A  &   0.00313  &    13.2  &     2.6  &          2 \\ 
82  &     896  &           J1737.5$-$2908  &       1RXS J173728.0$-$290759  &     264.368258  &   $-$29.133917  &       A  &   0.02176  &    94.8  &   0.1(a)  &        1.9 \\ 
83  &     970  &           J1824.3$-$5624  &                       IC 4709  &     276.080758  &   $-$56.369095  &       W  &   0.01689  &    73.3  &   0.1(a)  &          2 \\ 
84  &     986  &           J1836.9$-$5924  &                    Fairall 49  &     279.242650  &   $-$59.402306  &       A  &   0.01959  &    85.2  &   0.1(a)  &        1.9 \\ 
85  &    1042  &           J1937.5$-$0613  &                    LEDA 90334  &     294.387612  &    $-$6.218000  &       A  &   0.01036  &    44.7  &   0.1(a)  &          1 \\ 
86  &    1064  &           J2009.0$-$6103  &                      NGC 6860  &     302.195333  &   $-$61.099947  &       A  &   0.01468  &    63.6  &   0.1(a)  &        1.9 \\ 
87  &    1077  &         J2028.5$+$2543D1  &                      NGC 6921  &     307.120338  &      25.723416  &       A  &   0.01406  &    60.9  &   0.1(a)  &          2 \\ 
88  &    1090  &           J2044.2$-$1045  &                       Mrk 509  &     311.040621  &   $-$10.723543  &       A  &   0.03470  &   152.5  &   0.1(a)  &          1 \\ 
89  &    1092  &           J2052.0$-$5704  &                       IC 5063  &     313.009800  &   $-$57.068776  &       A  &   0.01127  &    45.9  &   0.1(a)  &          2 \\ 
90  &    1127  &           J2148.3$-$3454  &                      NGC 7130  &     327.081321  &   $-$34.951324  &       A  &   0.01583  &    68.6  &   0.1(a)  &        1.9 \\ 
91  &    1133  &           J2200.9$+$1032  &                       Mrk 520  &     330.172096  &      10.552331  &       A  &   0.02749  &   120.2  &   0.1(a)  &        1.9 \\ 
92  &    1135  &           J2201.9$-$3152  &                      NGC 7172  &     330.507883  &   $-$31.869580  &       A  &   0.00851  &      34  &      12  &          2 \\ 
93  &    1157  &           J2235.9$-$2602  &                      NGC 7314  &     338.942492  &   $-$26.050438  &       A  &   0.00461  &    16.8  &     3.4  &        1.9 \\ 
94  &    1161  &           J2236.7$-$1233  &                       Mrk 915  &     339.193767  &   $-$12.545231  &       A  &   0.02394  &   104.4  &   0.1(a)  &        1.9 \\ 
95  &    1182  &           J2303.3$+$0852  &                      NGC 7469  &     345.815075  &       8.873899  &       A  &   0.01600  &    69.4  &   0.1(a)  &          1 \\ 
96  &    1183  &           J2304.8$-$0843  &                       Mrk 926  &     346.181158  &    $-$8.685728  &       A  &   0.04774  &   211.9  &   0.1(a)  &          1 \\ 
97  &    1188  &           J2318.4$-$4223  &                      NGC 7582  &     349.598521  &   $-$42.370432  &       A  &   0.00525  &    22.5  &     4.5  &          2 \\ 
98  &    1198  &           J2328.9$+$0328  &                      NGC 7682  &     352.266333  &       3.533292  &       A  &   0.01706  &    74.0  &   0.1(a)  &          2 \\ 

%% file: table4almadata.tex
 01 &  2018.1.00538.S &    A001/X133d/X1fce &       ... &   3.4 &    24.9 &    4990 &  2018-11-04 &    0.16 &       0.13 &         75 &         61 \\ 
 02 &  2018.1.00538.S &    A001/X133d/X1fbe &       ... &   4.0 &    24.9 &    2026 &  2018-10-26 &    0.33 &       0.24 &        130 &         99 \\ 
 03 &  2017.1.00236.S &     A001/X1296/X91b &       ... &   1.6 &    24.9 &    5685 &  2017-12-02 &   0.099 &      0.076 &         24 &         18 \\ 
 04 &  2018.1.00538.S &    A001/X133d/X1fe8 &       ... &   3.8 &    24.8 &    1845 &  2018-10-22 &    0.25 &       0.24 &        100 &         94 \\ 
 05 &  2017.1.00904.S &     A001/X1296/X5bd &       ... &  0.80 &    25.1 &     435 &  2017-11-19 &   0.052 &      0.044 &         33 &         27 \\ 
 06 &  2018.1.00538.S &    A001/X133d/X1ff0 &       ... &   4.5 &    26.3 &     514 &  2018-11-02 &    0.34 &       0.26 &         97 &         73 \\ 
 07 &  2018.1.00657.S &    A001/X133d/X21f5 &    mosaic &   6.0 &   168.8 &      78 &  2019-04-27 &    0.48 &       0.42 &        130 &        120 \\ 
 08 &  2018.1.00657.S &    A001/X133d/X21f5 &    mosaic &   6.0 &   168.8 &      78 &  2019-04-27 &    0.48 &       0.42 &        130 &        110 \\ 
 09 &  2018.1.00538.S &    A001/X133d/X1fa4 &       ... &   4.2 &    26.3 &    1119 &  2018-10-26 &    0.31 &       0.28 &        100 &         99 \\ 
 10 &  2016.1.01140.S &      A001/X87a/X78a &       ... &   1.5 &    26.2 &    1089 &  2017-08-18 &    0.11 &      0.089 &        100 &         82 \\ 
 11 &  2018.1.00037.S &     A001/X133d/Xb82 &       ... &  0.30 &    21.9 &   19813 &  2019-06-06 &   0.021 &      0.018 &        1.5 &        1.3 \\ 
 12 &  2017.1.00236.S &     A001/X1296/X917 &       ... &   2.1 &    24.9 &    4052 &  2017-12-09 &    0.14 &       0.10 &         33 &         23 \\ 
 13 &  2018.1.00538.S &    A001/X133d/X1fc2 &       ... &   3.5 &    24.8 &    1149 &  2018-10-09 &    0.26 &       0.25 &         92 &         88 \\ 
 14 &  2018.1.00538.S &    A001/X133d/X1f8c &       ... &   3.9 &    26.4 &    1179 &  2018-10-22 &    0.27 &       0.26 &         95 &         90 \\ 
 15 &  2016.1.01553.S &      A001/X87d/X5c7 &       ... &   2.7 &    25.4 &    1300 &  2016-10-24 &    0.26 &       0.19 &         73 &         54 \\ 
 16 &  2018.1.00538.S &    A001/X133d/X1fb0 &       ... &   4.3 &    26.3 &     454 &  2018-11-04 &    0.30 &       0.28 &         82 &         75 \\ 
 17 &  2013.1.01161.S &      A001/X12f/X321 &    mosaic &   1.9 &    92.0 &     100 &  2015-08-08 &    0.21 &       0.19 &         19 &         18 \\ 
 18 &  2012.1.00474.S &   A002/X5d7935/X283 &       ... &   4.1 &    24.7 &    3810 &  2014-06-29 &    0.52 &       0.39 &         45 &         33 \\ 
 19 &  2017.1.01439.S &     A001/X1296/X654 &       ... &   2.9 &    24.8 &    1361 &  2017-12-31 &    0.22 &       0.15 &         49 &         33 \\ 
 20 &  2018.1.00538.S &    A001/X133d/X1fb4 &       ... &   6.8 &    26.3 &     816 &  2018-10-27 &    0.32 &       0.30 &        100 &         96 \\ 
 20 &  2018.1.00538.S &    A001/X133d/X1fb6 &       ... &   ... &     ... &     212 &  2018-12-27 &     ... &        ... &        ... &        ... \\ 
 21 &  2018.1.00538.S &    A001/X133d/X1f70 &       ... &   4.9 &    26.3 &     998 &  2018-10-26 &    0.41 &       0.25 &        130 &         83 \\ 
 22 &  2018.1.00538.S &    A001/X133d/X1f7c &       ... &   4.2 &    26.4 &    4627 &  2018-11-03 &    0.17 &       0.13 &         60 &         48 \\ 
 23 &  2018.1.00576.S &     A001/X134b/X176 &       ... &   1.8 &    25.0 &     490 &  2019-08-12 &    0.13 &       0.11 &         47 &         39 \\ 
 24 &  2017.1.01439.S &     A001/X1296/X62e &       ... &   2.9 &    25.2 &    1482 &  2017-12-17 &    0.23 &       0.12 &        160 &         85 \\ 
 25 &  2018.1.00538.S &    A001/X133d/X1f84 &       ... &   4.5 &    26.2 &     363 &  2018-11-03 &    0.35 &       0.28 &         92 &         72 \\ 
 26 &  2016.1.01385.S &      A001/X87c/X704 &       ... &  0.60 &    23.5 &    4173 &  2017-09-21 &   0.046 &      0.032 &         40 &         28 \\ 
 27 &  2019.1.01742.S &     A001/X1465/X2be &       ... &   4.2 &    24.8 &     696 &  2019-10-03 &    0.28 &       0.23 &         57 &         45 \\ 
 28 &  2016.1.00839.S &      A001/X87d/X304 &       ... &   5.1 &    25.4 &    1421 &  2017-07-30 &    0.16 &      0.092 &         27 &         15 \\ 
 28 &  2016.1.00839.S &      A001/X87d/X306 &       ... &   ... &     ... &     424 &  2018-03-22 &     ... &        ... &        ... &        ... \\ 
 29 &  2016.1.01385.S &      A001/X87c/X6f8 &       ... &  0.60 &    23.3 &    4318 &  2017-09-20 &   0.039 &      0.030 &         27 &         21 \\ 
 30 &  2013.1.00623.S &      A001/X145/X199 &       ... &  13.6 &    24.6 &    1300 &  2015-05-03 &     1.7 &       0.97 &        230 &        130 \\ 
 31 &  2018.1.00538.S &    A001/X133d/X1f90 &       ... &   3.8 &    25.0 &    3175 &  2018-10-03 &    0.26 &       0.24 &        120 &        110 \\ 
 32 &  2017.1.01439.S &     A001/X1296/X658 &       ... &   4.6 &    24.8 &    1240 &  2018-03-11 &    0.38 &       0.32 &        100 &         91 \\ 
 33 &  2018.1.00538.S &    A001/X133d/X1fa0 &       ... &   3.3 &    24.9 &    3629 &  2018-10-03 &    0.20 &       0.16 &         94 &         75 \\ 
 34 &  2017.1.00904.S &     A001/X1296/X5b3 &       ... &   3.9 &    25.1 &     454 &  2018-09-20 &    0.26 &       0.24 &        170 &        160 \\ 
 35 &  2018.1.00538.S &    A001/X133d/X1fba &       ... &   3.8 &    24.9 &    4596 &  2018-10-16 &    0.28 &       0.22 &        110 &         89 \\ 
 36 &  2018.1.00006.S &      A001/X133d/X78 &       ... &  10.0 &    25.0 &     998 &  2019-03-17 &     1.1 &       0.88 &        670 &        530 \\ 
 37 &  2018.1.00538.S &    A001/X133d/X1f9c &       ... &   4.1 &    26.3 &    1996 &  2018-10-23 &    0.30 &       0.25 &        100 &         82 \\ 
 38 &  2017.1.00236.S &     A001/X1296/X8ef &    mosaic &   2.6 &    37.8 &     862 &  2017-12-31 &    0.17 &       0.12 &         32 &         22 \\ 
 39 &  2019.1.01742.S &     A001/X1465/X2c4 &       ... &   4.4 &    24.7 &     696 &  2019-10-03 &    0.38 &       0.29 &         67 &         51 \\ 
 40 &  2018.1.00538.S &    A001/X133d/X1fd2 &       ... &   6.1 &    26.3 &    1270 &  2018-10-25 &    0.31 &       0.27 &         94 &         82 \\ 
 40 &  2018.1.00538.S &    A001/X133d/X1fd4 &       ... &   ... &     ... &     181 &  2018-12-21 &     ... &        ... &        ... &        ... \\ 
 41 &  2015.1.00086.S &      A001/X2fe/X665 &       ... &   5.2 &    24.6 &    1724 &  2016-05-02 &    0.53 &       0.45 &         83 &         70 \\ 
 42 &  2018.1.00538.S &    A001/X133d/X1f94 &       ... &   4.1 &    26.3 &     393 &  2018-11-03 &    0.30 &       0.25 &         80 &         67 \\ 
 43 &  2016.1.00254.S &      A001/X87a/X489 &       ... &   5.6 &    25.3 &    2570 &  2017-09-02 &    0.17 &       0.12 &         19 &         13 \\ 
 43 &  2016.1.00254.S &      A001/X87a/X48b &       ... &   ... &     ... &     786 &  2017-04-29 &     ... &        ... &        ... &        ... \\ 
 44 &  2018.1.00211.S &     A001/X133d/X1d1 &       ... &   6.5 &    26.2 &    1754 &  2019-04-16 &    0.54 &       0.44 &        120 &        100 \\ 
 45 &  2016.1.01553.S &      A001/X87d/X5ca &       ... &   4.6 &    25.4 &    1240 &  2017-05-02 &    0.39 &       0.28 &        100 &         76 \\ 
 46 &  2018.1.00538.S &    A001/X133d/X1fd8 &       ... &   3.7 &    24.8 &    1331 &  2018-10-25 &    0.26 &       0.23 &         96 &         85 \\ 
 47 &  2018.1.00576.S &     A001/X134b/X173 &       ... &   1.7 &    25.0 &     423 &  2019-08-13 &    0.11 &       0.10 &         21 &         18 \\ 
 48 &  2015.1.00872.S &       A001/X5a4/Xb4 &       ... &   4.6 &    25.7 &     242 &  2016-06-09 &    0.22 &       0.19 &        220 &        190 \\ 
 49 &  2018.1.00538.S &    A001/X133d/X1fc6 &       ... &   4.0 &    24.9 &    4415 &  2018-10-20 &    0.27 &       0.24 &        110 &         97 \\ 
 50 &  2018.1.00538.S &    A001/X133d/X1f78 &       ... &   3.2 &    25.0 &    4657 &  2019-09-26 &    0.18 &       0.14 &         88 &         66 \\ 
 51 &  2019.1.01742.S &     A001/X1465/X2ca &       ... &   5.5 &    24.7 &     756 &  2019-10-08 &    0.51 &       0.43 &         65 &         55 \\ 
 52 &  2017.1.01439.S &     A001/X1296/X636 &       ... &   6.4 &    24.8 &    1693 &  2018-09-09 &    0.53 &       0.34 &        140 &         92 \\ 
 53 &  2012.1.00139.S &    A002/X5a9a13/Xed &       ... &   3.8 &    22.1 &      65 &  2014-05-26 &    0.59 &       0.46 &         52 &         40 \\ 
 54 &  2015.1.00597.S &       A001/X2c9/X1e &       ... &  0.70 &    24.6 &    3157 &  2015-11-22 &   0.053 &      0.030 &        1.2 &       0.69 \\ 
 55 &  2018.1.00538.S &    A001/X133d/X1fdc &       ... &   4.0 &    26.2 &     302 &  2018-11-13 &    0.29 &       0.28 &         70 &         67 \\ 
 56 &  2018.1.00978.S &    A001/X133d/X279e &       ... &  0.40 &    24.8 &   11757 &  2019-06-23 &   0.022 &      0.020 &        4.0 &        3.6 \\ 
 57 &  2017.1.01439.S &     A001/X1296/X63c &       ... &   5.8 &    24.8 &     816 &  2018-03-23 &    0.48 &       0.39 &        140 &        110 \\ 
 58 &  2017.1.00236.S &     A001/X1296/X8cb &    mosaic &   6.2 &    63.6 &     150 &  2018-03-22 &    0.46 &       0.39 &         45 &         38 \\ 
 59 &  2016.1.01279.S &       A001/X888/X7a &    mosaic &   4.8 &    39.6 &    1719 &  2016-12-01 &    0.49 &       0.45 &        8.2 &        7.6 \\ 
 60 &  2018.1.00538.S &    A001/X133d/X1f80 &       ... &   4.1 &    26.3 &     877 &  2018-10-27 &    0.28 &       0.25 &         93 &         81 \\ 
 61 &  2019.1.01742.S &     A001/X1465/X2b6 &       ... &   7.0 &    24.8 &     816 &  2019-10-08 &    0.75 &       0.39 &        150 &         79 \\ 
 62 &  2018.1.00538.S &    A001/X133d/X1f88 &       ... &   4.3 &    26.2 &     393 &  2018-11-19 &    0.32 &       0.26 &         85 &         70 \\ 
 63 &  2013.1.00525.S &       A001/X13f/Xcd &       ... &   3.0 &    25.5 &    1421 &  2015-08-26 &    0.41 &       0.17 &        290 &        120 \\ 
 64 &  2018.1.00538.S &    A001/X133d/X1fe0 &       ... &   3.2 &    24.9 &    5080 &  2018-11-13 &    0.16 &       0.15 &         72 &         69 \\ 
 65 &  2017.1.00236.S &     A001/X1296/X8f3 &    mosaic &   2.6 &    37.8 &     862 &  2017-12-28 &    0.15 &       0.12 &         22 &         18 \\ 
 66 &  2018.1.00576.S &     A001/X134b/X170 &       ... &   1.6 &    25.0 &     423 &  2019-08-13 &    0.12 &      0.097 &         39 &         32 \\ 
 67 &  2018.1.00538.S &    A001/X133d/X1f74 &       ... &   5.2 &    26.2 &     333 &  2018-11-28 &    0.38 &       0.29 &         97 &         74 \\ 
 68 &  2018.1.00581.S &    A001/X133d/X1047 &       ... &   1.0 &    22.5 &    5407 &  2019-06-06 &   0.026 &      0.022 &       0.53 &       0.45 \\ 
 68 &  2018.1.00581.S &    A001/X133d/X1049 &       ... &   ... &     ... &         &  2019-08-15 &     ... &        ... &        ... &        ... \\ 
 69 &  2017.1.00236.S &     A001/X1296/X8e7 &    mosaic &   3.8 &    47.3 &     330 &  2018-01-15 &    0.28 &       0.21 &         36 &         27 \\ 
 70 &  2017.1.01439.S &     A001/X1296/X640 &       ... &   6.4 &    24.9 &    1089 &  2018-09-09 &    0.56 &       0.32 &        190 &        110 \\ 
 71 &  2018.1.00538.S &    A001/X133d/X1f98 &       ... &   2.7 &    24.9 &    3992 &  2018-11-12 &    0.15 &       0.12 &         71 &         60 \\ 
 72 &  2016.1.00254.S &      A001/X87a/X491 &       ... &   5.2 &    25.3 &    2147 &  2017-07-18 &    0.22 &       0.13 &         13 &        8.2 \\ 
 72 &  2016.1.00254.S &      A001/X87a/X493 &       ... &   ... &     ... &     665 &  2016-12-24 &     ... &        ... &        ... &        ... \\ 
 73 &  2017.1.01439.S &     A001/X1296/X65c &       ... &   2.5 &    25.0 &    2510 &  2018-01-04 &    0.17 &       0.13 &         89 &         70 \\ 
 74 &  2015.1.00086.S &      A001/X2fe/X671 &       ... &   5.0 &    24.6 &    1724 &  2016-05-15 &    0.51 &       0.45 &         92 &         81 \\ 
 75 &  2017.1.00255.S &    A001/X1284/X1395 &       ... &   2.5 &    24.8 &     998 &  2017-12-23 &    0.16 &       0.13 &         52 &         43 \\ 
 76 &  2017.1.01439.S &     A001/X1296/X660 &       ... &   3.6 &    24.9 &    1845 &  2018-09-18 &    0.26 &       0.24 &        120 &        110 \\ 
 77 &  2017.1.00904.S &     A001/X1296/X5cf &       ... &   3.1 &    25.0 &     308 &  2019-07-17 &    0.29 &       0.24 &        150 &        120 \\ 
 77 &  2017.1.00904.S &     A001/X1296/X5d1 &       ... &   ... &     ... &     302 &  2018-09-17 &     ... &        ... &        ... &        ... \\ 
 78 &  2018.1.00538.S &    A001/X133d/X1ff4 &       ... &   5.2 &    26.3 &     968 &  2018-11-24 &    0.34 &       0.28 &        100 &         91 \\ 
 79 &  2019.1.01742.S &     A001/X1465/X2ba &       ... &   5.5 &    24.7 &     786 &  2019-10-07 &    0.49 &       0.40 &         80 &         66 \\ 
 80 &  2015.1.00370.S &      A001/X2d8/X395 &       ... &  0.60 &    24.9 &    2159 &  2017-09-29 &   0.042 &      0.022 &         21 &         11 \\ 
 81 &  2017.1.00886.L &     A001/X12a3/X4ba &    mosaic &   9.2 &   227.1 &      73 &  2018-05-03 &    0.86 &       0.76 &         55 &         48 \\ 
 82 &  2018.1.00576.S &     A001/X134b/X16a &       ... &   1.9 &    25.0 &     423 &  2019-08-14 &    0.16 &      0.088 &         71 &         40 \\ 
 83 &  2018.1.00538.S &    A001/X133d/X1fa8 &       ... &   4.0 &    26.3 &    1240 &  2018-11-23 &    0.41 &       0.35 &        140 &        120 \\ 
 84 &  2017.1.00904.S &     A001/X1296/X5ab &       ... &   3.4 &    24.8 &     308 &  2017-11-23 &    0.37 &       0.28 &        150 &        110 \\ 
 84 &  2017.1.00904.S &     A001/X1296/X5ad &       ... &   ... &     ... &     302 &  2018-09-17 &     ... &        ... &        ... &        ... \\ 
 85 &  2017.1.01439.S &     A001/X1296/X644 &       ... &   6.1 &    24.8 &    1633 &  2018-03-29 &    0.53 &       0.40 &        110 &         86 \\ 
 86 &  2018.1.00211.S &     A001/X133d/X1d5 &       ... &   6.1 &    24.8 &    2026 &  2019-04-18 &    0.54 &       0.44 &        160 &        130 \\ 
 87 &  2017.1.00598.S &    A001/X1288/X10f3 &       ... &   4.6 &    22.2 &    3810 &  2018-09-14 &    0.37 &       0.25 &        100 &         74 \\ 
 88 &  2018.1.00576.S &     A001/X134b/X17c &       ... &   2.7 &    25.0 &     423 &  2018-10-12 &    0.23 &       0.14 &        170 &        100 \\ 
 89 &  2016.1.01279.S &       A001/X888/X76 &    mosaic &   3.9 &    39.1 &    1956 &  2016-11-10 &    0.48 &       0.31 &        100 &         69 \\ 
 90 &  2017.1.00598.S &    A001/X1288/X10f7 &       ... &   4.4 &    22.2 &     605 &  2018-09-09 &    0.28 &       0.25 &         92 &         83 \\ 
 91 &  2017.1.01439.S &     A001/X1296/X650 &       ... &   1.2 &    25.1 &    1306 &  2017-11-21 &   0.070 &      0.049 &         40 &         28 \\ 
 92 &  2018.1.00248.S &     A001/X133d/X8e7 &       ... &   1.6 &    24.6 &    6592 &  2019-08-23 &    0.10 &      0.086 &         16 &         14 \\ 
 93 &  2017.1.00236.S &     A001/X1296/X8d3 &    mosaic &   4.3 &    50.7 &     111 &  2018-09-13 &    0.35 &       0.27 &         28 &         21 \\ 
 94 &  2018.1.00211.S &     A001/X133d/X1cd &       ... &   5.8 &    25.0 &     968 &  2019-04-21 &    0.44 &       0.42 &        220 &        210 \\ 
 95 &  2017.1.00395.S &     A001/X1284/Xd7e &    mosaic &   2.3 &    38.0 &    1800 &  2017-12-12 &    0.16 &       0.12 &         52 &         40 \\ 
 96 &  2018.1.00576.S &     A001/X134b/X17f &       ... &   2.8 &    25.0 &     847 &  2018-10-12 &    0.21 &       0.13 &        210 &        130 \\ 
 97 &  2016.1.00254.S &      A001/X87a/X499 &       ... &   1.3 &    25.3 &    2087 &  2016-10-12 &    0.15 &       0.12 &         16 &         13 \\ 
 98 &  2018.1.00538.S &    A001/X133d/X1fec &       ... &   4.0 &    24.8 &    1179 &  2018-10-25 &    0.30 &       0.26 &        100 &         94 \\ 

%% file: table4alma.tex
 01 &  4/4 &  233.6/233.6 &  6.2/7.2 &    0.44 &  0.01 &    39 & ...   & $-$14.99 &   39.06 &    0.05 &     1.1 &     0.8 &   ... &   ... \\ 
 02 &  0/4 &  234.3/234.3 &  6.1/7.2 &    0.05 &  0.02 &   2.8 & $<$   & $-$15.51 &   38.42 &     ... &     ... &     ... &   ... &   ... \\ 
 03 &  4/4 &  233.9/233.7 &  4.8/7.6 &    1.19 &  0.01 &   114 & ...   & $-$14.56 &   38.93 &    0.04 &     0.6 &     0.3 &    eb &     1 \\ 
 04 &  4/4 &  234.3/234.3 &  5.9/7.2 &    4.02 &  0.02 &   258 & ...   & $-$14.03 &   39.88 &    0.04 &     0.7 &     0.2 &   ... &   ... \\ 
 05 &  4/4 &  232.2/232.4 &  6.3/7.8 &   22.38 &  0.07 &   300 & ...   & $-$13.30 &   41.02 &    0.04 &     0.3 &     0.2 &    eb &     4 \\ 
 06 &  4/4 &  221.7/221.7 &  6.1/7.2 &    0.36 &  0.03 &    12 & ...   & $-$15.09 &   38.53 &    0.06 &       0 &       3 &   ... &   ... \\ 
 07 &  0/4 &  233.5/233.5 &  6.5/7.4 &    0.33 &  0.09 &   3.6 & $<$   & $-$14.72 &   38.89 &     ... &     ... &     ... &   ... &   ... \\ 
 08 &  0/4 &  233.5/233.5 &  6.2/7.4 &    0.18 &  0.09 &   2.0 & $<$   & $-$14.85 &   38.75 &     ... &     ... &     ... &   ... &   ... \\ 
 09 &  4/4 &  221.0/221.0 &  6.2/7.2 &    0.67 &  0.02 &    33 & ...   & $-$14.83 &   38.97 &    0.05 &     0.7 &     0.9 &   ... &   ... \\ 
 10 &  4/4 &  222.0/222.0 &  5.6/7.9 &    1.22 &  0.02 &    54 & ...   & $-$14.57 &   40.07 &    0.04 &     0.7 &     0.5 &   ... &   ... \\ 
 11 &  3/3 &  266.3/266.4 &  3.4/5.4 &    8.09 &  0.01 &   734 & ...   & $-$13.63 &   38.76 &    0.05 &     ... &     ... &    eb &    12 \\ 
 12 &  4/4 &  234.1/233.9 &  5.5/7.6 &    0.33 &  0.01 &    26 & ...   & $-$15.10 &   38.34 &    0.05 &       2 &       1 &    eb &     6 \\ 
 13 &  4/4 &  234.8/234.8 &  6.0/7.3 &    0.33 &  0.02 &    16 & ...   & $-$15.12 &   38.68 &    0.06 &       0 &       2 &   ... &   ... \\ 
 14 &  4/4 &  220.9/220.9 &  6.3/7.2 &    0.77 &  0.02 &    42 & ...   & $-$14.78 &   39.02 &    0.05 &     1.7 &     0.8 &   ... &   ... \\ 
 15 &  4/4 &  228.8/229.0 &  4.0/4.9 &    1.53 &  0.02 &    63 & ...   & $-$14.46 &   39.15 &    0.05 &     ... &     ... &   ... &   ... \\ 
 16 &  4/4 &  221.8/221.8 &  6.2/7.2 &    0.63 &  0.04 &    18 & ...   & $-$14.87 &   38.71 &    0.05 &       3 &       2 &   ... &   ... \\ 
 17 &  0/4 &  237.3/237.3 &  6.2/7.0 &    0.44 &  0.08 &   5.4 & ...   & $-$14.99 &   37.67 &    0.09 &     ... &     ... &     b &   ... \\ 
 18 &  4/4 &  235.5/235.4 &  6.3/7.6 &    3.58 &  0.03 &   140 & ...   & $-$14.09 &   38.50 &    0.04 &   $-$0.1 &     0.3 &    eb &     1 \\ 
 19 &  4/4 &  235.0/235.0 &  6.0/7.4 &    0.84 &  0.02 &    37 & ...   & $-$14.73 &   38.68 &    0.05 &   $-$0.4 &     0.8 &   ... &   ... \\ 
 20 &  4/4 &  221.3/221.3 &  6.0/7.2 &    6.19 &  0.03 &   246 & ...   & $-$13.87 &   39.86 &    0.04 &     1.2 &     0.2 &   ... &   ... \\ 
 21 &  4/4 &  221.3/221.3 &  6.8/7.2 &    0.55 &  0.02 &    26 & ...   & $-$14.91 &   38.84 &    0.05 &       1 &       1 &   ... &   ... \\ 
 22 &  4/4 &  220.8/220.8 &  5.9/7.2 &    0.63 &  0.01 &    53 & ...   & $-$14.85 &   38.98 &    0.04 &     0.3 &     0.6 &   ... &   ... \\ 
 23 &  4/4 &  233.0/233.0 &  5.5/7.9 &    0.90 &  0.03 &    31 & ...   & $-$14.69 &   39.15 &    0.05 &     0.3 &     0.9 &     e &     1 \\ 
 24 &  4/4 &  231.3/231.3 &  6.1/7.4 &    0.97 &  0.03 &    34 & ...   & $-$14.65 &   39.73 &    0.05 &     0.6 &     0.9 &     b &   ... \\ 
 25 &  4/4 &  221.9/221.9 &  6.2/7.3 &    0.53 &  0.04 &    12 & ...   & $-$14.91 &   38.63 &    0.06 &     $-$1 &       2 &   ... &   ... \\ 
 26 &  4/4 &  248.0/248.0 &  5.5/7.2 &    1.33 &  0.02 &    66 & ...   & $-$14.61 &   39.99 &    0.05 &   $-$1.4 &     0.5 &    eb &     7 \\ 
 27 &  0/4 &  235.2/235.2 &  5.8/7.8 &    0.09 &  0.03 &   3.0 & $<$   & $-$15.26 &   38.06 &     ... &     ... &     ... &   ... &   ... \\ 
 28 &  1/1 &  228.8/228.8 &  1.0/1.9 &    17.9 &   0.2 &   113 & ...   & $-$13.39 &   39.76 &    0.04 &     ... &     ... &     e &     6 \\ 
 29 &  4/4 &  249.8/249.8 &  7.2/7.2 &    2.58 &  0.01 &   196 & ...   & $-$14.17 &   40.23 &    0.05 &     1.4 &     0.3 &     b &   ... \\ 
 30 &  0/4 &  237.2/237.2 &  5.5/7.9 &     0.5 &   0.2 &   3.1 & $<$   & $-$14.53 &   38.45 &     ... &     ... &     ... &   ... &   ... \\ 
 31 &  4/4 &  232.8/232.8 &  6.1/7.0 &    0.18 &  0.01 &    16 & ...   & $-$15.40 &   38.66 &    0.06 &     $-$0 &       2 &   ... &   ... \\ 
 32 &  4/4 &  235.0/235.0 &  6.0/7.0 &    2.97 &  0.03 &    93 & ...   & $-$14.17 &   39.44 &    0.04 &   $-$0.0 &     0.4 &   ... &   ... \\ 
 33 &  4/4 &  233.6/233.6 &  5.9/7.2 &    0.27 &  0.01 &    24 & ...   & $-$15.23 &   38.82 &    0.05 &     $-$1 &       1 &   ... &   ... \\ 
 34 &  4/4 &  231.8/231.9 &  5.7/7.8 &    0.45 &  0.03 &    15 & ...   & $-$14.96 &   39.41 &    0.06 &       2 &       2 &   ... &   ... \\ 
 35 &  4/4 &  234.2/234.2 &  6.1/7.2 &    0.85 &  0.01 &    68 & ...   & $-$14.72 &   39.22 &    0.04 &     0.1 &     0.5 &     e &     1 \\ 
 36 &  4/4 &  232.5/232.5 &  5.9/7.5 &    0.60 &  0.03 &    21 & ...   & $-$14.89 &   39.38 &    0.05 &     $-$1 &       1 &   ... &   ... \\ 
 37 &  4/4 &  221.1/221.1 &  6.0/7.2 &    0.76 &  0.02 &    41 & ...   & $-$14.77 &   38.99 &    0.05 &     0.5 &     0.8 &   ... &   ... \\ 
 38 &  4/4 &  234.9/234.8 &  5.3/7.6 &    2.99 &  0.02 &   136 & ...   & $-$14.15 &   39.09 &    0.04 &     1.6 &     0.3 &     e &     1 \\ 
 39 &  4/4 &  235.6/235.6 &  5.5/7.8 &    5.52 &  0.04 &   134 & ...   & $-$13.90 &   39.30 &    0.04 &     0.3 &     0.3 &   ... &   ... \\ 
 40 &  4/4 &  221.5/221.5 &  6.2/7.2 &    0.28 &  0.02 &    15 & ...   & $-$15.20 &   38.47 &    0.06 &     $-$0 &       2 &   ... &   ... \\ 
 41 &  4/4 &  236.8/236.9 &  6.0/7.7 &    0.67 &  0.02 &    28 & ...   & $-$14.80 &   38.31 &    0.05 &       1 &       1 &     b &   ... \\ 
 42 &  0/4 &  221.8/221.8 &  6.1/7.2 &    0.17 &  0.04 &   4.1 & $<$   & $-$15.06 &   38.51 &     ... &     ... &     ... &   ... &   ... \\ 
 43 &  2/2 &  230.4/230.4 &  2.5/3.3 &    0.81 &  0.03 &    28 & ...   & $-$14.73 &   38.07 &    0.05 &     ... &     ... &    eb &    13 \\ 
 44 &  4/4 &  222.7/222.7 &  4.9/7.4 &    2.24 &  0.02 &   100 & ...   & $-$14.29 &   39.15 &    0.04 &     0.1 &     0.4 &     e &     4 \\ 
 45 &  3/4 &  229.2/229.2 &  4.4/4.9 &    0.25 &  0.03 &  10.0 & ...   & $-$15.25 &   38.32 &    0.06 &     ... &     ... &     b &   ... \\ 
 46 &  0/4 &  234.7/234.7 &  5.9/7.3 &    0.12 &  0.02 &   7.0 & ...   & $-$15.54 &   38.30 &    0.08 &     ... &     ... &   ... &   ... \\ 
 47 &  4/4 &  233.0/233.0 &  6.3/7.9 &    1.16 &  0.03 &    40 & ...   & $-$14.57 &   38.68 &    0.05 &     0.7 &     0.7 &   ... &   ... \\ 
 48 &  1/4 &  226.5/226.5 &  5.2/7.9 &    0.45 &  0.07 &   6.8 & ...   & $-$15.00 &   39.75 &    0.08 &     ... &     ... &   ... &   ... \\ 
 49 &  4/4 &  234.2/234.2 &  6.2/7.2 &    0.22 &  0.01 &    18 & ...   & $-$15.30 &   38.63 &    0.06 &       0 &       2 &   ... &   ... \\ 
 50 &  0/4 &  233.1/233.1 &  6.0/7.3 &    0.09 &  0.01 &   7.3 & ...   & $-$15.67 &   38.40 &    0.08 &     ... &     ... &   ... &   ... \\ 
 51 &  4/4 &  235.8/235.8 &  5.5/7.8 &    0.83 &  0.04 &    21 & ...   & $-$14.74 &   38.19 &    0.05 &     $-$1 &       1 &   ... &   ... \\ 
 52 &  4/4 &  234.5/234.5 &  5.4/6.5 &    0.85 &  0.04 &    24 & ...   & $-$14.74 &   38.83 &    0.05 &     $-$2 &       1 &     e &     2 \\ 
 53 &  4/4 &  263.0/263.0 &  6.3/7.9 &     5.8 &   0.2 &    32 & ...   & $-$13.82 &   38.78 &    0.08 &       1 &       1 &   ... &   ... \\ 
 54 &  1/8 &  237.1/237.1 &  6.5/7.7 &    0.12 &  0.03 &   4.7 & $<$   & $-$15.23 &   36.20 &     ... &     ... &     ... &   ... &   ... \\ 
 55 &  4/4 &  222.1/222.1 &  6.1/7.2 &    1.39 &  0.04 &    39 & ...   & $-$14.48 &   39.01 &    0.05 &   $-$2.5 &     0.8 &     e &     2 \\ 
 56 &  4/4 &  234.6/234.5 &  5.7/7.7 &   1.131 &  0.008 &   137 & ...   & $-$14.57 &   38.65 &    0.04 &     1.7 &     0.3 &     b &     1 \\ 
 57 &  2/4 &  234.9/234.9 &  5.9/7.0 &    0.25 &  0.03 &  10.0 & ...   & $-$15.24 &   38.41 &    0.06 &     ... &     ... &   ... &   ... \\ 
 58 &  1/4 &  235.8/235.6 &  5.5/7.6 &    0.45 &  0.06 &   7.2 & ...   & $-$14.98 &   37.72 &    0.08 &     ... &     ... &   ... &   ... \\ 
 59 &  8/8 &  222.9/222.9 &  5.8/7.2 &   37.01 &  0.09 &   415 & ...   & $-$13.07 &   38.09 &    0.04 &   $-$0.4 &     0.2 &    eb &    26 \\ 
 60 &  4/4 &  221.2/221.2 &  5.4/7.2 &    0.58 &  0.03 &    23 & ...   & $-$14.88 &   38.86 &    0.05 &     $-$0 &       1 &   ... &   ... \\ 
 61 &  0/4 &  235.2/235.2 &  4.8/7.8 &    0.19 &  0.04 &   5.2 & ...   & $-$15.36 &   37.95 &    0.10 &     ... &     ... &   ... &   ... \\ 
 62 &  4/4 &  221.9/221.9 &  5.4/7.2 &    0.91 &  0.04 &    23 & ...   & $-$14.67 &   38.90 &    0.05 &     $-$1 &       1 &   ... &   ... \\ 
 63 &  0/4 &  228.6/228.6 &  6.2/7.0 &    0.23 &  0.03 &   7.0 & ...   & $-$15.28 &   39.16 &    0.08 &     ... &     ... &   ... &   ... \\ 
 64 &  4/4 &  233.7/233.7 &  6.0/7.2 &    0.23 &  0.01 &    20 & ...   & $-$15.22 &   38.81 &    0.05 &       5 &       2 &   ... &   ... \\ 
 65 &  4/4 &  234.8/234.7 &  5.4/7.6 &    0.55 &  0.02 &    27 & ...   & $-$14.89 &   38.15 &    0.05 &       1 &       1 &     b &   ... \\ 
 66 &  4/4 &  233.0/233.0 &  6.2/7.9 &    9.19 &  0.04 &   215 & ...   & $-$13.68 &   40.08 &    0.04 &   $-$0.0 &     0.2 &     b &   ... \\ 
 67 &  0/4 &  222.1/222.1 &  6.4/7.3 &    0.13 &  0.03 &   3.7 & $<$   & $-$15.16 &   38.36 &     ... &     ... &     ... &   ... &   ... \\ 
 68 &  4/4 &  258.8/258.8 &  6.3/7.3 &   14.60 &  0.01 &  1147 & ...   & $-$13.46 &   37.86 &    0.04 &     0.2 &     0.2 &    eb &    20 \\ 
 69 &  4/4 &  235.2/235.1 &  5.4/7.6 &    6.24 &  0.04 &   142 & ...   & $-$13.83 &   39.09 &    0.04 &     1.2 &     0.3 &    eb &     3 \\ 
 70 &  4/4 &  234.0/234.0 &  6.1/7.4 &    0.53 &  0.03 &    15 & ...   & $-$14.95 &   38.85 &    0.06 &     $-$1 &       2 &   ... &   ... \\ 
 71 &  4/4 &  233.4/233.4 &  6.1/7.2 &    0.82 &  0.01 &    63 & ...   & $-$14.70 &   39.37 &    0.04 &     1.8 &     0.6 &   ... &   ... \\ 
 72 &  2/2 &  230.4/230.3 &  2.7/3.3 &    1.11 &  0.03 &    40 & ...   & $-$14.59 &   37.69 &    0.04 &     ... &     ... &     e &     1 \\ 
 73 &  1/3 &  233.0/233.0 &  4.2/5.3 &    0.19 &  0.03 &   6.8 & ...   & $-$15.36 &   38.78 &    0.08 &     ... &     ... &   ... &   ... \\ 
 74 &  4/4 &  236.4/236.5 &  5.5/7.7 &    2.08 &  0.03 &    66 & ...   & $-$14.34 &   38.89 &    0.04 &   $-$0.8 &     0.5 &   ... &   ... \\ 
 75 &  4/4 &  234.8/234.8 &  5.4/7.4 &    0.64 &  0.03 &    23 & ...   & $-$14.82 &   38.94 &    0.05 &       1 &       1 &     b &   ... \\ 
 76 &  4/4 &  233.6/233.6 &  5.4/7.2 &    1.57 &  0.03 &    61 & ...   & $-$14.43 &   39.65 &    0.04 &     1.1 &     0.5 &     e &     3 \\ 
 77 &  4/4 &  232.9/233.3 &  5.1/7.8 &    1.06 &  0.05 &    21 & ...   & $-$14.60 &   39.55 &    0.05 &       2 &       2 &   ... &   ... \\ 
 78 &  4/4 &  221.3/221.3 &  6.1/7.2 &    1.19 &  0.03 &    45 & ...   & $-$14.57 &   39.16 &    0.05 &     0.1 &     0.7 &   ... &   ... \\ 
 79 &  3/4 &  235.4/235.4 &  5.2/7.8 &    0.32 &  0.03 &    12 & ...   & $-$15.13 &   38.01 &    0.06 &     ... &     ... &   ... &   ... \\ 
 80 &  3/3 &  235.4/234.7 &  3.9/5.6 &    2.40 &  0.03 &    81 & ...   & $-$14.21 &   39.94 &    0.04 &     1.0 &     0.5 &     b &   ... \\ 
 81 &  4/4 &  224.1/224.2 &  5.6/6.6 &     2.5 &   0.1 &    25 & ...   & $-$14.27 &   38.05 &    0.05 &       2 &       1 &   ... &   ... \\ 
 82 &  4/4 &  233.0/233.0 &  6.3/7.9 &    1.10 &  0.04 &    31 & ...   & $-$14.59 &   39.44 &    0.05 &     0.7 &     0.9 &   ... &   ... \\ 
 83 &  0/4 &  221.0/221.0 &  5.8/7.2 &    0.11 &  0.01 &   7.2 & ...   & $-$15.60 &   38.20 &    0.08 &     ... &     ... &   ... &   ... \\ 
 84 &  4/4 &  234.4/234.6 &  5.7/7.8 &    1.05 &  0.05 &    19 & ...   & $-$14.61 &   39.33 &    0.06 &       1 &       2 &     b &   ... \\ 
 85 &  4/4 &  235.0/235.0 &  5.9/7.2 &    0.98 &  0.03 &    37 & ...   & $-$14.63 &   38.75 &    0.05 &     1.5 &     0.8 &   ... &   ... \\ 
 86 &  4/4 &  235.0/235.1 &  5.8/7.7 &    1.60 &  0.02 &    68 & ...   & $-$14.44 &   39.24 &    0.04 &   $-$0.0 &     0.5 &   ... &   ... \\ 
 87 &  2/2 &  262.8/262.8 &  3.4/3.7 &    0.43 &  0.02 &    21 & ...   & $-$14.97 &   38.68 &    0.09 &     ... &     ... &     e &     3 \\ 
 88 &  3/3 &  232.0/232.0 &  4.7/6.0 &    1.97 &  0.05 &    38 & ...   & $-$14.37 &   40.08 &    0.05 &   $-$0.4 &     0.9 &   ... &   ... \\ 
 89 &  8/8 &  220.9/220.9 &  5.8/7.2 &    6.74 &  0.06 &   114 & ...   & $-$13.83 &   39.57 &    0.04 &     1.0 &     0.3 &     b &     5 \\ 
 90 &  2/2 &  262.3/262.3 &  1.9/3.7 &    1.96 &  0.07 &    26 & ...   & $-$14.32 &   39.44 &    0.09 &     ... &     ... &    eb &     4 \\ 
 91 &  3/4 &  231.8/231.8 &  5.1/7.0 &    0.16 &  0.02 &   6.9 & ...   & $-$15.44 &   38.80 &    0.08 &     ... &     ... &    eb &     3 \\ 
 92 &  4/4 &  236.6/236.6 &  6.0/7.5 &   0.786 &  0.009 &    90 & ...   & $-$14.74 &   38.40 &    0.04 &     0.5 &     0.4 &    eb &     1 \\ 
 93 &  2/4 &  235.5/235.3 &  5.4/7.6 &    0.50 &  0.05 &   9.9 & ...   & $-$14.93 &   37.60 &    0.06 &     ... &     ... &   ... &   ... \\ 
 94 &  4/4 &  233.1/233.1 &  5.6/7.7 &    0.53 &  0.02 &    23 & ...   & $-$14.88 &   39.24 &    0.05 &       3 &       1 &   ... &   ... \\ 
 95 &  8/8 &  234.3/234.3 &  5.5/7.2 &    1.44 &  0.01 &   116 & ...   & $-$14.49 &   39.27 &    0.04 &   $-$0.2 &     0.3 &    eb &    16 \\ 
 96 &  3/3 &  232.0/232.0 &  4.7/6.0 &    2.19 &  0.04 &    61 & ...   & $-$14.29 &   40.44 &    0.05 &     1.0 &     0.6 &   ... &   ... \\ 
 97 &  2/2 &  230.3/230.1 &  2.2/3.3 &    0.59 &  0.03 &    17 & ...   & $-$14.86 &   37.92 &    0.05 &     ... &     ... &     b &    12 \\ 
 98 &  4/4 &  234.8/234.8 &  6.2/7.3 &    0.36 &  0.02 &    20 & ...   & $-$15.06 &   38.76 &    0.06 &       2 &       2 &   ... &   ... \\ 

%% file: BASS_AGNs_Milli_Prop_arXiv.bbl
\begin{thebibliography}{}
\expandafter\ifx\csname natexlab\endcsname\relax\def\natexlab#1{#1}\fi
\providecommand{\url}[1]{\href{#1}{#1}}
\providecommand{\dodoi}[1]{doi:~\href{http://doi.org/#1}{\nolinkurl{#1}}}
\providecommand{\doeprint}[1]{\href{http://ascl.net/#1}{\nolinkurl{http://ascl.net/#1}}}
\providecommand{\doarXiv}[1]{\href{https://arxiv.org/abs/#1}{\nolinkurl{https://arxiv.org/abs/#1}}}

\bibitem[{{Alonso-Herrero} {et~al.}(2019){Alonso-Herrero},
  {Garc{\'\i}a-Burillo}, {Pereira-Santaella}, {Davies}, {Combes},
  {Vestergaard}, {Raimundo}, {Bunker}, {D{\'\i}az-Santos}, {Gandhi},
  {Garc{\'\i}a-Bernete}, {Hicks}, {H{\"o}nig}, {Hunt}, {Imanishi}, {Izumi},
  {Levenson}, {Maciejewski}, {Packham}, {Ramos Almeida}, {Ricci}, {Rigopoulou},
  {Roche}, {Rosario}, {Schartmann}, {Usero}, \& {Ward}}]{Alo19}
{Alonso-Herrero}, A., {Garc{\'\i}a-Burillo}, S., {Pereira-Santaella}, M.,
  {et~al.} 2019, \aap, 628, A65, \dodoi{10.1051/0004-6361/201935431}

\bibitem[{{Asmus}(2019)}]{Asm19}
{Asmus}, D. 2019, \mnras, 489, 2177, \dodoi{10.1093/mnras/stz2289}

\bibitem[{{Asmus} {et~al.}(2016){Asmus}, {H{\"o}nig}, \& {Gandhi}}]{Asm16}
{Asmus}, D., {H{\"o}nig}, S.~F., \& {Gandhi}, P. 2016, \apj, 822, 109,
  \dodoi{10.3847/0004-637X/822/2/109}

\bibitem[{{Balokovi{\'c}} {et~al.}(2020){Balokovi{\'c}}, {Harrison},
  {Madejski}, {Comastri}, {Ricci}, {Annuar}, {Ballantyne}, {Boorman}, {Brandt},
  {Brightman}, {Gandhi}, {Kamraj}, {Koss}, {Marchesi}, {Marinucci}, {Masini},
  {Matt}, {Stern}, \& {Urry}}]{Bal20}
{Balokovi{\'c}}, M., {Harrison}, F.~A., {Madejski}, G., {et~al.} 2020, \apj,
  905, 41, \dodoi{10.3847/1538-4357/abc342}

\bibitem[{{Baumgartner} {et~al.}(2013){Baumgartner}, {Tueller}, {Markwardt},
  {Skinner}, {Barthelmy}, {Mushotzky}, {Evans}, \& {Gehrels}}]{bau13}
{Baumgartner}, W.~H., {Tueller}, J., {Markwardt}, C.~B., {et~al.} 2013, \apjs,
  207, 19, \dodoi{10.1088/0067-0049/207/2/19}

\bibitem[{{Behar} {et~al.}(2015){Behar}, {Baldi}, {Laor}, {Horesh}, {Stevens},
  \& {Tzioumis}}]{Beh15}
{Behar}, E., {Baldi}, R.~D., {Laor}, A., {et~al.} 2015, \mnras, 451, 517,
  \dodoi{10.1093/mnras/stv988}

\bibitem[{{Behar} {et~al.}(2018){Behar}, {Vogel}, {Baldi}, {Smith}, \&
  {Mushotzky}}]{Beh18}
{Behar}, E., {Vogel}, S., {Baldi}, R.~D., {Smith}, K.~L., \& {Mushotzky}, R.~F.
  2018, \mnras, 478, 399, \dodoi{10.1093/mnras/sty850}

\bibitem[{{Bendo} {et~al.}(2016){Bendo}, {Henkel}, {D'Cruze}, {Dickinson},
  {Fuller}, \& {Karim}}]{Ben16}
{Bendo}, G.~J., {Henkel}, C., {D'Cruze}, M.~J., {et~al.} 2016, \mnras, 463,
  252, \dodoi{10.1093/mnras/stw1659}

\bibitem[{{Bock} {et~al.}(2000){Bock}, {Neugebauer}, {Matthews}, {Soifer},
  {Becklin}, {Ressler}, {Marsh}, {Werner}, {Egami}, \& {Blandford}}]{Boc00}
{Bock}, J.~J., {Neugebauer}, G., {Matthews}, K., {et~al.} 2000, \aj, 120, 2904,
  \dodoi{10.1086/316871}

\bibitem[{{Burlon} {et~al.}(2011){Burlon}, {Ajello}, {Greiner}, {Comastri},
  {Merloni}, \& {Gehrels}}]{Bur11}
{Burlon}, D., {Ajello}, M., {Greiner}, J., {et~al.} 2011, \apj, 728, 58,
  \dodoi{10.1088/0004-637X/728/1/58}

\bibitem[{{Chiaraluce} {et~al.}(2020){Chiaraluce}, {Panessa}, {Bruni}, {Baldi},
  {Behar}, {Vagnetti}, {Tombesi}, \& {McHardy}}]{Chi20}
{Chiaraluce}, E., {Panessa}, F., {Bruni}, G., {et~al.} 2020, \mnras, 495, 3943,
  \dodoi{10.1093/mnras/staa1393}

\bibitem[{{Combes} {et~al.}(2014){Combes}, {Garc{\'\i}a-Burillo}, {Casasola},
  {Hunt}, {Krips}, {Baker}, {Boone}, {Eckart}, {Marquez}, {Neri}, {Schinnerer},
  \& {Tacconi}}]{Com14}
{Combes}, F., {Garc{\'\i}a-Burillo}, S., {Casasola}, V., {et~al.} 2014, \aap,
  565, A97, \dodoi{10.1051/0004-6361/201423433}

\bibitem[{{Condon}(1992)}]{Con92}
{Condon}, J.~J. 1992, \araa, 30, 575,
  \dodoi{10.1146/annurev.aa.30.090192.003043}

\bibitem[{{Condon} {et~al.}(1996){Condon}, {Helou}, {Sanders}, \&
  {Soifer}}]{Con96}
{Condon}, J.~J., {Helou}, G., {Sanders}, D.~B., \& {Soifer}, B.~T. 1996, \apjs,
  103, 81, \dodoi{10.1086/192270}

\bibitem[{{Cooke} {et~al.}(2000){Cooke}, {Baldwin}, {Ferland}, {Netzer}, \&
  {Wilson}}]{Coo00}
{Cooke}, A.~J., {Baldwin}, J.~A., {Ferland}, G.~J., {Netzer}, H., \& {Wilson},
  A.~S. 2000, \apjs, 129, 517, \dodoi{10.1086/313422}

\bibitem[{{Courtois} {et~al.}(2017){Courtois}, {Tully}, {Hoffman},
  {Pomar{\`e}de}, {Graziani}, \& {Dupuy}}]{Cou17}
{Courtois}, H.~M., {Tully}, R.~B., {Hoffman}, Y., {et~al.} 2017, \apjl, 847,
  L6, \dodoi{10.3847/2041-8213/aa88b2}

\bibitem[{{D{\'\i}az-Santos} {et~al.}(2010){D{\'\i}az-Santos},
  {Alonso-Herrero}, {Colina}, {Packham}, {Levenson}, {Pereira-Santaella},
  {Roche}, \& {Telesco}}]{Dia10}
{D{\'\i}az-Santos}, T., {Alonso-Herrero}, A., {Colina}, L., {et~al.} 2010,
  \apj, 711, 328, \dodoi{10.1088/0004-637X/711/1/328}

\bibitem[{{D{\'\i}az-Santos} {et~al.}(2007){D{\'\i}az-Santos},
  {Alonso-Herrero}, {Colina}, {Ryder}, \& {Knapen}}]{Dia07}
{D{\'\i}az-Santos}, T., {Alonso-Herrero}, A., {Colina}, L., {Ryder}, S.~D., \&
  {Knapen}, J.~H. 2007, \apj, 661, 149, \dodoi{10.1086/513089}

\bibitem[{{Ekers} {et~al.}(1978){Ekers}, {Goss}, {Kotanyi}, \&
  {Skellern}}]{Eke78}
{Ekers}, R.~D., {Goss}, W.~M., {Kotanyi}, C.~G., \& {Skellern}, D.~J. 1978,
  \aap, 69, L21

\bibitem[{{Elvis} {et~al.}(1994){Elvis}, {Wilkes}, {McDowell}, {Green},
  {Bechtold}, {Willner}, {Oey}, {Polomski}, \& {Cutri}}]{Elv94}
{Elvis}, M., {Wilkes}, B.~J., {McDowell}, J.~C., {et~al.} 1994, \apjs, 95, 1,
  \dodoi{10.1086/192093}

\bibitem[{{Fabbiano} {et~al.}(2011){Fabbiano}, {Wang}, {Elvis}, \&
  {Risaliti}}]{Fab11}
{Fabbiano}, G., {Wang}, J., {Elvis}, M., \& {Risaliti}, G. 2011, \nat, 477,
  431, \dodoi{10.1038/nature10364}

\bibitem[{{Ferrarese} \& {Merritt}(2000)}]{Fer00}
{Ferrarese}, L., \& {Merritt}, D. 2000, \apjl, 539, L9, \dodoi{10.1086/312838}

\bibitem[{{Finlez} {et~al.}(2018){Finlez}, {Nagar}, {Storchi-Bergmann},
  {Schnorr-M{\"u}ller}, {Riffel}, {Lena}, {Mundell}, \& {Elvis}}]{Fin18}
{Finlez}, C., {Nagar}, N.~M., {Storchi-Bergmann}, T., {et~al.} 2018, \mnras,
  479, 3892, \dodoi{10.1093/mnras/sty1555}

\bibitem[{{Fischer} {et~al.}(2021){Fischer}, {Secrest}, {Johnson}, {Dorland},
  {Cigan}, {Fernandez}, {Hunt}, {Koss}, {Schmitt}, \& {Zacharias}}]{Fis21}
{Fischer}, T.~C., {Secrest}, N.~J., {Johnson}, M.~C., {et~al.} 2021, \apj, 906,
  88, \dodoi{10.3847/1538-4357/abca3c}

\bibitem[{{Francis} {et~al.}(2020){Francis}, {Johnstone}, {Herczeg}, {Hunter},
  \& {Harsono}}]{Fra20}
{Francis}, L., {Johnstone}, D., {Herczeg}, G., {Hunter}, T.~R., \& {Harsono},
  D. 2020, \aj, 160, 270, \dodoi{10.3847/1538-3881/abbe1a}

\bibitem[{{Gallimore} {et~al.}(2004){Gallimore}, {Baum}, \& {O'Dea}}]{Gal04}
{Gallimore}, J.~F., {Baum}, S.~A., \& {O'Dea}, C.~P. 2004, \apj, 613, 794,
  \dodoi{10.1086/423167}

\bibitem[{{Gallimore} {et~al.}(1996){Gallimore}, {Baum}, {O'Dea}, \&
  {Pedlar}}]{Gal96}
{Gallimore}, J.~F., {Baum}, S.~A., {O'Dea}, C.~P., \& {Pedlar}, A. 1996, \apj,
  458, 136, \dodoi{10.1086/176798}

\bibitem[{{Gallimore} \& {Beswick}(2004)}]{Gal04ngc6240}
{Gallimore}, J.~F., \& {Beswick}, R. 2004, \aj, 127, 239,
  \dodoi{10.1086/379959}

\bibitem[{{Gandhi} {et~al.}(2009){Gandhi}, {Horst}, {Smette}, {H{\"o}nig},
  {Comastri}, {Gilli}, {Vignali}, \& {Duschl}}]{Gan09}
{Gandhi}, P., {Horst}, H., {Smette}, A., {et~al.} 2009, \aap, 502, 457,
  \dodoi{10.1051/0004-6361/200811368}

\bibitem[{{G{\'o}mez-Guijarro} {et~al.}(2017){G{\'o}mez-Guijarro},
  {Gonz{\'a}lez-Mart{\'\i}n}, {Ramos Almeida}, {Rodr{\'\i}guez-Espinosa}, \&
  {Gallego}}]{Gom17}
{G{\'o}mez-Guijarro}, C., {Gonz{\'a}lez-Mart{\'\i}n}, O., {Ramos Almeida}, C.,
  {Rodr{\'\i}guez-Espinosa}, J.~M., \& {Gallego}, J. 2017, \mnras, 469, 2720,
  \dodoi{10.1093/mnras/stx1037}

\bibitem[{{Gonz{\'a}lez Delgado} {et~al.}(1998){Gonz{\'a}lez Delgado},
  {Heckman}, {Leitherer}, {Meurer}, {Krolik}, {Wilson}, {Kinney}, \&
  {Koratkar}}]{Gon98uvopt}
{Gonz{\'a}lez Delgado}, R.~M., {Heckman}, T., {Leitherer}, C., {et~al.} 1998,
  \apj, 505, 174, \dodoi{10.1086/306154}

\bibitem[{{Henkel} {et~al.}(2018){Henkel}, {M{\"u}hle}, {Bendo}, {J{\'o}zsa},
  {Gong}, {Viti}, {Aalto}, {Combes}, {Garc{\'\i}a-Burillo}, {Hunt}, {Mangum},
  {Mart{\'\i}n}, {Muller}, {Ott}, {van der Werf}, {Malawi}, {Ismail},
  {Alkhuja}, {Asiri}, {Aladro}, {Alves}, {Ao}, {Baan}, {Costagliola}, {Fuller},
  {Greene}, {Impellizzeri}, {Kamali}, {Klessen}, {Mauersberger}, {Tang},
  {Tristram}, {Wang}, \& {Zhang}}]{Hen18}
{Henkel}, C., {M{\"u}hle}, S., {Bendo}, G., {et~al.} 2018, \aap, 615, A155,
  \dodoi{10.1051/0004-6361/201732174}

\bibitem[{{Hickox} \& {Alexander}(2018)}]{Hic18}
{Hickox}, R.~C., \& {Alexander}, D.~M. 2018, \araa, 56, 625,
  \dodoi{10.1146/annurev-astro-081817-051803}

\bibitem[{{Ho}(2008)}]{Ho08}
{Ho}, L.~C. 2008, \araa, 46, 475,
  \dodoi{10.1146/annurev.astro.45.051806.110546}

\bibitem[{{Imanishi} {et~al.}(2016){Imanishi}, {Nakanishi}, \& {Izumi}}]{Ima16}
{Imanishi}, M., {Nakanishi}, K., \& {Izumi}, T. 2016, \aj, 152, 218,
  \dodoi{10.3847/0004-6256/152/6/218}

\bibitem[{{Imanishi} \& {Saito}(2014)}]{Ima14nir}
{Imanishi}, M., \& {Saito}, Y. 2014, \apj, 780, 106,
  \dodoi{10.1088/0004-637X/780/1/106}

\bibitem[{{Impellizzeri} {et~al.}(2019){Impellizzeri}, {Gallimore}, {Baum},
  {Elitzur}, {Davies}, {Lutz}, {Maiolino}, {Marconi}, {Nikutta}, {O'Dea}, \&
  {Sani}}]{Imp19}
{Impellizzeri}, C.~M.~V., {Gallimore}, J.~F., {Baum}, S.~A., {et~al.} 2019,
  \apjl, 884, L28, \dodoi{10.3847/2041-8213/ab3c64}

\bibitem[{{Inoue} \& {Doi}(2018)}]{Ino18}
{Inoue}, Y., \& {Doi}, A. 2018, \apj, 869, 114,
  \dodoi{10.3847/1538-4357/aaeb95}

\bibitem[{{Inoue} {et~al.}(2020){Inoue}, {Khangulyan}, \& {Doi}}]{Ino20}
{Inoue}, Y., {Khangulyan}, D., \& {Doi}, A. 2020, \apjl, 891, L33,
  \dodoi{10.3847/2041-8213/ab7661}

\bibitem[{{Isbell} {et~al.}(2021){Isbell}, {Burtscher}, {Asmus}, {Pott},
  {Couzy}, {Stalevski}, {G{\'a}mez Rosas}, \& {Meisenheimer}}]{Isb21}
{Isbell}, J.~W., {Burtscher}, L., {Asmus}, D., {et~al.} 2021, \apj, 910, 104,
  \dodoi{10.3847/1538-4357/abdfd3}

\bibitem[{{Izumi} {et~al.}(2018){Izumi}, {Wada}, {Fukushige}, {Hamamura}, \&
  {Kohno}}]{Izu18}
{Izumi}, T., {Wada}, K., {Fukushige}, R., {Hamamura}, S., \& {Kohno}, K. 2018,
  \apj, 867, 48, \dodoi{10.3847/1538-4357/aae20b}

\bibitem[{{Izumi} {et~al.}(2015){Izumi}, {Kohno}, {Aalto}, {Doi}, {Espada},
  {Fathi}, {Harada}, {Hatsukade}, {Hattori}, {Hsieh}, {Ikarashi}, {Imanishi},
  {Iono}, {Ishizuki}, {Krips}, {Mart{\'\i}n}, {Matsushita}, {Meier}, {Nagai},
  {Nakai}, {Nakajima}, {Nakanishi}, {Nomura}, {Regan}, {Schinnerer}, {Sheth},
  {Takano}, {Tamura}, {Terashima}, {Tosaki}, {Turner}, {Umehata}, \&
  {Wiklind}}]{Izu15}
{Izumi}, T., {Kohno}, K., {Aalto}, S., {et~al.} 2015, \apj, 811, 39,
  \dodoi{10.1088/0004-637X/811/1/39}

\bibitem[{{Kawamuro} {et~al.}(2020){Kawamuro}, {Izumi}, {Onishi}, {Imanishi},
  {Nguyen}, \& {Baba}}]{Kaw20}
{Kawamuro}, T., {Izumi}, T., {Onishi}, K., {et~al.} 2020, \apj, 895, 135,
  \dodoi{10.3847/1538-4357/ab8b62}

\bibitem[{{Kawamuro} {et~al.}(2021){Kawamuro}, {Ricci}, {Izumi}, {Imanishi},
  {Baba}, {Nguyen}, \& {Onishi}}]{Kaw21}
{Kawamuro}, T., {Ricci}, C., {Izumi}, T., {et~al.} 2021, arXiv e-prints,
  arXiv:2109.09742.
\newblock \doarXiv{2109.09742}

\bibitem[{{Kawamuro} {et~al.}(2016{\natexlab{a}}){Kawamuro}, {Ueda}, {Tazaki},
  {Ricci}, \& {Terashima}}]{Kaw16b}
{Kawamuro}, T., {Ueda}, Y., {Tazaki}, F., {Ricci}, C., \& {Terashima}, Y.
  2016{\natexlab{a}}, \apjs, 225, 14, \dodoi{10.3847/0067-0049/225/1/14}

\bibitem[{{Kawamuro} {et~al.}(2013){Kawamuro}, {Ueda}, {Tazaki}, \&
  {Terashima}}]{Kaw13}
{Kawamuro}, T., {Ueda}, Y., {Tazaki}, F., \& {Terashima}, Y. 2013, \apj, 770,
  157, \dodoi{10.1088/0004-637X/770/2/157}

\bibitem[{{Kawamuro} {et~al.}(2016{\natexlab{b}}){Kawamuro}, {Ueda}, {Tazaki},
  {Terashima}, \& {Mushotzky}}]{Kaw16c}
{Kawamuro}, T., {Ueda}, Y., {Tazaki}, F., {Terashima}, Y., \& {Mushotzky}, R.
  2016{\natexlab{b}}, \apj, 831, 37, \dodoi{10.3847/0004-637X/831/1/37}

\bibitem[{{Kawamuro} {et~al.}(2022){Kawamuro}, {Ricci}, {Imanishi},
  {Mushotzky}, {Izumi}, {Ricci}, {Bauer}, {Koss}, {Trakhtenbrot}, {Ichikawa},
  {Rojas}, {Smith}, {Shimizu}, {Oh}, {den Brok}, {Baba}, {Balokovi{\'c}},
  {Chang}, {Kakkad}, {Pfeifle}, {Privon}, {Temple}, {Ueda}, {Harrison},
  {Powell}, {Stern}, {Urry}, \& {Sanders}}]{Kaw22}
{Kawamuro}, T., {Ricci}, C., {Imanishi}, M., {et~al.} 2022, arXiv e-prints,
  arXiv:2208.03880.
\newblock \doarXiv{2208.03880}

\bibitem[{{Kollatschny} {et~al.}(2020){Kollatschny}, {Weilbacher}, {Ochmann},
  {Chelouche}, {Monreal-Ibero}, {Bacon}, \& {Contini}}]{Kol20}
{Kollatschny}, W., {Weilbacher}, P.~M., {Ochmann}, M.~W., {et~al.} 2020, \aap,
  633, A79, \dodoi{10.1051/0004-6361/201936540}

\bibitem[{{Komossa} {et~al.}(2003){Komossa}, {Burwitz}, {Hasinger}, {Predehl},
  {Kaastra}, \& {Ikebe}}]{Kom03}
{Komossa}, S., {Burwitz}, V., {Hasinger}, G., {et~al.} 2003, \apjl, 582, L15,
  \dodoi{10.1086/346145}

\bibitem[{{Koss} {et~al.}(2015){Koss}, {Romero-Ca{\~n}izales}, {Baronchelli},
  {Teng}, {Balokovi{\'c}}, {Puccetti}, {Bauer}, {Ar{\'e}valo}, {Assef},
  {Ballantyne}, {Brandt}, {Brightman}, {Comastri}, {Gandhi}, {Harrison}, {Luo},
  {Schawinski}, {Stern}, \& {Treister}}]{Kos15}
{Koss}, M.~J., {Romero-Ca{\~n}izales}, C., {Baronchelli}, L., {et~al.} 2015,
  \apj, 807, 149, \dodoi{10.1088/0004-637X/807/2/149}

\bibitem[{{Koss} {et~al.}(2018){Koss}, {Blecha}, {Bernhard}, {Hung}, {Lu},
  {Trakhtenbrot}, {Treister}, {Weigel}, {Sartori}, {Mushotzky}, {Schawinski},
  {Ricci}, {Veilleux}, \& {Sanders}}]{Kos18}
{Koss}, M.~J., {Blecha}, L., {Bernhard}, P., {et~al.} 2018, \nat, 563, 214,
  \dodoi{10.1038/s41586-018-0652-7}

\bibitem[{{Koss} {et~al.}(2022{\natexlab{a}}){Koss}, {Ricci}, {Trakhtenbrot},
  {Oh}, {den Brok}, {Mej{\'\i}a-Restrepo}, {Stern}, {Privon}, {Treister},
  {Powell}, {Mushotzky}, {Bauer}, {Ananna}, {Balokovi{\'c}}, {B{\"a}r},
  {Becker}, {Bessiere}, {Burtscher}, {Caglar}, {Congiu}, {Evans}, {Harrison},
  {Heida}, {Ichikawa}, {Kamraj}, {Lamperti}, {Pacucci}, {Ricci}, {Riffel},
  {Rojas}, {Schawinski}, {Temple}, {Urry}, {Veilleux}, \&
  {Williams}}]{Kos22_catalog}
{Koss}, M.~J., {Ricci}, C., {Trakhtenbrot}, B., {et~al.} 2022{\natexlab{a}},
  \apjs, 261, 2, \dodoi{10.3847/1538-4365/ac6c05}

\bibitem[{{Koss} {et~al.}(2022{\natexlab{b}}){Koss}, {Trakhtenbrot}, {Ricci},
  {Bauer}, {Treister}, {Mushotzky}, {Urry}, {Ananna}, {Balokovi{\'c}}, {den
  Brok}, {Cenko}, {Harrison}, {Ichikawa}, {Lamperti}, {Lein},
  {Mej{\'\i}a-Restrepo}, {Oh}, {Pacucci}, {Pfeifle}, {Powell}, {Privon},
  {Ricci}, {Salvato}, {Schawinski}, {Shimizu}, {Smith}, \&
  {Stern}}]{Kos22_overview}
{Koss}, M.~J., {Trakhtenbrot}, B., {Ricci}, C., {et~al.} 2022{\natexlab{b}},
  \apjs, 261, 1, \dodoi{10.3847/1538-4365/ac6c8f}

\bibitem[{{Laor} \& {Behar}(2008)}]{Lao08}
{Laor}, A., \& {Behar}, E. 2008, \mnras, 390, 847,
  \dodoi{10.1111/j.1365-2966.2008.13806.x}

\bibitem[{{Levenson} {et~al.}(2005){Levenson}, {Weaver}, {Heckman}, {Awaki}, \&
  {Terashima}}]{Lev05}
{Levenson}, N.~A., {Weaver}, K.~A., {Heckman}, T.~M., {Awaki}, H., \&
  {Terashima}, Y. 2005, \apj, 618, 167, \dodoi{10.1086/425913}

\bibitem[{{Lonsdale} {et~al.}(1992){Lonsdale}, {Lonsdale}, \& {Smith}}]{Lon92}
{Lonsdale}, C.~J., {Lonsdale}, C.~J., \& {Smith}, H.~E. 1992, \apj, 391, 629,
  \dodoi{10.1086/171377}

\bibitem[{{Ma} {et~al.}(2021){Ma}, {Maksym}, {Fabbiano}, {Elvis},
  {Storchi-Bergmann}, {Karovska}, {Wang}, \& {Travascio}}]{Ma21}
{Ma}, J., {Maksym}, W.~P., {Fabbiano}, G., {et~al.} 2021, \apj, 908, 155,
  \dodoi{10.3847/1538-4357/abcfc1}

\bibitem[{{Mainzer} {et~al.}(2011){Mainzer}, {Bauer}, {Grav}, {Masiero},
  {Cutri}, {Dailey}, {Eisenhardt}, {McMillan}, {Wright}, {Walker}, {Jedicke},
  {Spahr}, {Tholen}, {Alles}, {Beck}, {Brandenburg}, {Conrow}, {Evans},
  {Fowler}, {Jarrett}, {Marsh}, {Masci}, {McCallon}, {Wheelock}, {Wittman},
  {Wyatt}, {DeBaun}, {Elliott}, {Elsbury}, {Gautier}, {Gomillion}, {Leisawitz},
  {Maleszewski}, {Micheli}, \& {Wilkins}}]{Mai11}
{Mainzer}, A., {Bauer}, J., {Grav}, T., {et~al.} 2011, \apj, 731, 53,
  \dodoi{10.1088/0004-637X/731/1/53}

\bibitem[{{Makarov} {et~al.}(2014){Makarov}, {Prugniel}, {Terekhova},
  {Courtois}, \& {Vauglin}}]{Mak14}
{Makarov}, D., {Prugniel}, P., {Terekhova}, N., {Courtois}, H., \& {Vauglin},
  I. 2014, \aap, 570, A13, \dodoi{10.1051/0004-6361/201423496}

\bibitem[{{Maksym} {et~al.}(2017){Maksym}, {Fabbiano}, {Elvis}, {Karovska},
  {Paggi}, {Raymond}, {Wang}, \& {Storchi-Bergmann}}]{Mak17}
{Maksym}, W.~P., {Fabbiano}, G., {Elvis}, M., {et~al.} 2017, \apj, 844, 69,
  \dodoi{10.3847/1538-4357/aa78a4}

\bibitem[{{Maksym} {et~al.}(2019){Maksym}, {Fabbiano}, {Elvis}, {Karovska},
  {Paggi}, {Raymond}, {Wang}, {Storchi-Bergmann}, \& {Risaliti}}]{Mak19}
---. 2019, \apj, 872, 94, \dodoi{10.3847/1538-4357/aaf4f5}

\bibitem[{{Malkan} {et~al.}(1998){Malkan}, {Gorjian}, \& {Tam}}]{Mal98}
{Malkan}, M.~A., {Gorjian}, V., \& {Tam}, R. 1998, \apjs, 117, 25,
  \dodoi{10.1086/313110}

\bibitem[{{Massaro} {et~al.}(2009){Massaro}, {Giommi}, {Leto}, {Marchegiani},
  {Maselli}, {Perri}, {Piranomonte}, \& {Sclavi}}]{Mas09}
{Massaro}, E., {Giommi}, P., {Leto}, C., {et~al.} 2009, \aap, 495, 691,
  \dodoi{10.1051/0004-6361:200810161}

\bibitem[{{McMullin} {et~al.}(2007){McMullin}, {Waters}, {Schiebel}, {Young},
  \& {Golap}}]{McM07}
{McMullin}, J.~P., {Waters}, B., {Schiebel}, D., {Young}, W., \& {Golap}, K.
  2007, Astronomical Society of the Pacific Conference Series, Vol. 376, {CASA
  Architecture and Applications}, ed. R.~A. {Shaw}, F.~{Hill}, \& D.~J. {Bell},
  127

\bibitem[{{Medling} {et~al.}(2019){Medling}, {Privon}, {Barcos-Mu{\~n}oz},
  {Treister}, {Cicone}, {Messias}, {Sanders}, {Scoville}, {U}, {Armus},
  {Bauer}, {Chang}, {Comerford}, {Evans}, {Max}, {M{\"u}ller-S{\'a}nchez},
  {Nagar}, \& {Sheth}}]{Med19}
{Medling}, A.~M., {Privon}, G.~C., {Barcos-Mu{\~n}oz}, L., {et~al.} 2019,
  \apjl, 885, L21, \dodoi{10.3847/2041-8213/ab4db7}

\bibitem[{{Mezcua} {et~al.}(2015){Mezcua}, {Prieto}, {Fern{\'a}ndez-Ontiveros},
  {Tristram}, {Neumayer}, \& {Kotilainen}}]{Mez15}
{Mezcua}, M., {Prieto}, M.~A., {Fern{\'a}ndez-Ontiveros}, J.~A., {et~al.} 2015,
  \mnras, 452, 4128, \dodoi{10.1093/mnras/stv1408}

\bibitem[{{Michiyama} {et~al.}(2023){Michiyama}, {Inoue}, \& {Doi}}]{Michi23}
{Michiyama}, T., {Inoue}, Y., \& {Doi}, A. 2023, arXiv e-prints,
  arXiv:2306.15950, \dodoi{10.48550/arXiv.2306.15950}

\bibitem[{{Morganti} {et~al.}(2007){Morganti}, {Holt}, {Saripalli},
  {Oosterloo}, \& {Tadhunter}}]{Mor07ic5063}
{Morganti}, R., {Holt}, J., {Saripalli}, L., {Oosterloo}, T.~A., \&
  {Tadhunter}, C.~N. 2007, \aap, 476, 735, \dodoi{10.1051/0004-6361:20077888}

\bibitem[{{Morganti} {et~al.}(2015){Morganti}, {Oosterloo}, {Oonk},
  {Frieswijk}, \& {Tadhunter}}]{Mor15}
{Morganti}, R., {Oosterloo}, T., {Oonk}, J.~B.~R., {Frieswijk}, W., \&
  {Tadhunter}, C. 2015, \aap, 580, A1, \dodoi{10.1051/0004-6361/201525860}

\bibitem[{{Morganti} {et~al.}(1999){Morganti}, {Tsvetanov}, {Gallimore}, \&
  {Allen}}]{Mor99}
{Morganti}, R., {Tsvetanov}, Z.~I., {Gallimore}, J., \& {Allen}, M.~G. 1999,
  \aaps, 137, 457, \dodoi{10.1051/aas:1999258}

\bibitem[{{Mulchaey} {et~al.}(1994){Mulchaey}, {Wilson}, {Bower}, {Heckman},
  {Krolik}, \& {Miley}}]{Mul94}
{Mulchaey}, J.~S., {Wilson}, A.~S., {Bower}, G.~A., {et~al.} 1994, \apj, 433,
  625, \dodoi{10.1086/174671}

\bibitem[{{Mundell} {et~al.}(2009){Mundell}, {Ferruit}, {Nagar}, \&
  {Wilson}}]{Mun09}
{Mundell}, C.~G., {Ferruit}, P., {Nagar}, N., \& {Wilson}, A.~S. 2009, \apj,
  703, 802, \dodoi{10.1088/0004-637X/703/1/802}

\bibitem[{{Mundell} {et~al.}(1995){Mundell}, {Holloway}, {Pedlar}, {Meaburn},
  {Kukula}, \& {Axon}}]{Mun95}
{Mundell}, C.~G., {Holloway}, A.~J., {Pedlar}, A., {et~al.} 1995, \mnras, 275,
  67, \dodoi{10.1093/mnras/275.1.67}

\bibitem[{{Muxlow} {et~al.}(1996){Muxlow}, {Pedlar}, {Holloway}, {Gallimore},
  \& {Antonucci}}]{Mux96}
{Muxlow}, T.~W.~B., {Pedlar}, A., {Holloway}, A.~J., {Gallimore}, J.~F., \&
  {Antonucci}, R.~R.~J. 1996, \mnras, 278, 854, \dodoi{10.1093/mnras/278.3.854}

\bibitem[{{Nagar} {et~al.}(1999){Nagar}, {Wilson}, {Mulchaey}, \&
  {Gallimore}}]{Nag99}
{Nagar}, N.~M., {Wilson}, A.~S., {Mulchaey}, J.~S., \& {Gallimore}, J.~F. 1999,
  \apjs, 120, 209, \dodoi{10.1086/313183}

\bibitem[{{Novak} {et~al.}(2017){Novak}, {Smol{\v{c}}i{\'c}}, {Delhaize},
  {Delvecchio}, {Zamorani}, {Baran}, {Bondi}, {Capak}, {Carilli}, {Ciliegi},
  {Civano}, {Ilbert}, {Karim}, {Laigle}, {Le F{\`e}vre}, {Marchesi},
  {McCracken}, {Miettinen}, {Salvato}, {Sargent}, {Schinnerer}, \&
  {Tasca}}]{Nov17}
{Novak}, M., {Smol{\v{c}}i{\'c}}, V., {Delhaize}, J., {et~al.} 2017, \aap, 602,
  A5, \dodoi{10.1051/0004-6361/201629436}

\bibitem[{{Orienti} \& {Prieto}(2010)}]{Ori10}
{Orienti}, M., \& {Prieto}, M.~A. 2010, \mnras, 401, 2599,
  \dodoi{10.1111/j.1365-2966.2009.15837.x}

\bibitem[{{Paliya} {et~al.}(2019){Paliya}, {Koss}, {Trakhtenbrot}, {Ricci},
  {Oh}, {Ajello}, {Stern}, {Powell}, {Urry}, {Harrison}, {Lamperti},
  {Mushotzky}, {Marcotulli}, {Mej{\'\i}a-Restrepo}, \& {Hartmann}}]{pal19}
{Paliya}, V.~S., {Koss}, M., {Trakhtenbrot}, B., {et~al.} 2019, \apj, 881, 154,
  \dodoi{10.3847/1538-4357/ab2f8b}

\bibitem[{{Panessa} {et~al.}(2019){Panessa}, {Baldi}, {Laor}, {Padovani},
  {Behar}, \& {McHardy}}]{Pan19}
{Panessa}, F., {Baldi}, R.~D., {Laor}, A., {et~al.} 2019, Nature Astronomy, 3,
  387, \dodoi{10.1038/s41550-019-0765-4}

\bibitem[{{Puccetti} {et~al.}(2016){Puccetti}, {Comastri}, {Bauer}, {Brandt},
  {Fiore}, {Harrison}, {Luo}, {Stern}, {Urry}, {Alexander}, {Annuar},
  {Ar{\'e}valo}, {Balokovi{\'c}}, {Boggs}, {Brightman}, {Christensen}, {Craig},
  {Gandhi}, {Hailey}, {Koss}, {La Massa}, {Marinucci}, {Ricci}, {Walton},
  {Zappacosta}, \& {Zhang}}]{Pac16}
{Puccetti}, S., {Comastri}, A., {Bauer}, F.~E., {et~al.} 2016, \aap, 585, A157,
  \dodoi{10.1051/0004-6361/201527189}

\bibitem[{{Ricci} {et~al.}(2015){Ricci}, {Ueda}, {Koss}, {Trakhtenbrot},
  {Bauer}, \& {Gandhi}}]{ric15}
{Ricci}, C., {Ueda}, Y., {Koss}, M.~J., {et~al.} 2015, \apjl, 815, L13,
  \dodoi{10.1088/2041-8205/815/1/L13}

\bibitem[{{Ricci} {et~al.}(2017{\natexlab{a}}){Ricci}, {Trakhtenbrot}, {Koss},
  {Ueda}, {Schawinski}, {Oh}, {Lamperti}, {Mushotzky}, {Treister}, {Ho},
  {Weigel}, {Bauer}, {Paltani}, {Fabian}, {Xie}, \& {Gehrels}}]{Ric17nat}
{Ricci}, C., {Trakhtenbrot}, B., {Koss}, M.~J., {et~al.} 2017{\natexlab{a}},
  \nat, 549, 488, \dodoi{10.1038/nature23906}

\bibitem[{{Ricci} {et~al.}(2017{\natexlab{b}}){Ricci}, {Trakhtenbrot}, {Koss},
  {Ueda}, {Del Vecchio}, {Treister}, {Schawinski}, {Paltani}, {Oh}, {Lamperti},
  {Berney}, {Gandhi}, {Ichikawa}, {Bauer}, {Ho}, {Asmus}, {Beckmann}, {Soldi},
  {Balokovi{\'c}}, {Gehrels}, \& {Markwardt}}]{ric17c}
---. 2017{\natexlab{b}}, \apjs, 233, 17, \dodoi{10.3847/1538-4365/aa96ad}

\bibitem[{{Ricci} {et~al.}(2017{\natexlab{c}}){Ricci}, {Bauer}, {Treister},
  {Schawinski}, {Privon}, {Blecha}, {Arevalo}, {Armus}, {Harrison}, {Ho},
  {Iwasawa}, {Sanders}, \& {Stern}}]{ric17b}
{Ricci}, C., {Bauer}, F.~E., {Treister}, E., {et~al.} 2017{\natexlab{c}},
  \mnras, 468, 1273, \dodoi{10.1093/mnras/stx173}

\bibitem[{{Ricci} {et~al.}(2021){Ricci}, {Privon}, {Pfeifle}, {Armus},
  {Iwasawa}, {Torres-Alb{\`a}}, {Satyapal}, {Bauer}, {Treister}, {Ho}, {Aalto},
  {Ar{\'e}valo}, {Barcos-Mu{\~n}oz}, {Charmandaris}, {Diaz-Santos}, {Evans},
  {Gao}, {Inami}, {Koss}, {Lansbury}, {Linden}, {Medling}, {Sanders}, {Song},
  {Stern}, {U}, {Ueda}, \& {Yamada}}]{Ric21b}
{Ricci}, C., {Privon}, G.~C., {Pfeifle}, R.~W., {et~al.} 2021, \mnras, 506,
  5935, \dodoi{10.1093/mnras/stab2052}

\bibitem[{{Ricci} {et~al.}(2018){Ricci}, {Steiner}, {May}, {Garcia-Rissmann},
  \& {Menezes}}]{Ric18ngc7582}
{Ricci}, T.~V., {Steiner}, J.~E., {May}, D., {Garcia-Rissmann}, A., \&
  {Menezes}, R.~B. 2018, \mnras, 473, 5334, \dodoi{10.1093/mnras/stx2746}

\bibitem[{{Sakamoto} {et~al.}(2007){Sakamoto}, {Ho}, {Mao}, {Matsushita}, \&
  {Peck}}]{Sak07}
{Sakamoto}, K., {Ho}, P. T.~P., {Mao}, R.-Q., {Matsushita}, S., \& {Peck},
  A.~B. 2007, \apj, 654, 782, \dodoi{10.1086/509775}

\bibitem[{{Sandqvist} {et~al.}(1982){Sandqvist}, {Jorsater}, \&
  {Lindblad}}]{Sand82}
{Sandqvist}, A., {Jorsater}, S., \& {Lindblad}, P.~O. 1982, \aap, 110, 336

\bibitem[{{Schmitt} {et~al.}(2003){Schmitt}, {Donley}, {Antonucci},
  {Hutchings}, \& {Kinney}}]{Sch03}
{Schmitt}, H.~R., {Donley}, J.~L., {Antonucci}, R.~R.~J., {Hutchings}, J.~B.,
  \& {Kinney}, A.~L. 2003, \apjs, 148, 327, \dodoi{10.1086/377440}

\bibitem[{{Schmitt} {et~al.}(2001){Schmitt}, {Ulvestad}, {Antonucci}, \&
  {Kinney}}]{Sch01}
{Schmitt}, H.~R., {Ulvestad}, J.~S., {Antonucci}, R.~R.~J., \& {Kinney}, A.~L.
  2001, \apjs, 132, 199, \dodoi{10.1086/318957}

\bibitem[{{Scoville} {et~al.}(2000){Scoville}, {Evans}, {Thompson}, {Rieke},
  {Hines}, {Low}, {Dinshaw}, {Surace}, \& {Armus}}]{Sco00}
{Scoville}, N.~Z., {Evans}, A.~S., {Thompson}, R., {et~al.} 2000, \aj, 119,
  991, \dodoi{10.1086/301248}

\bibitem[{{She} {et~al.}(2017){She}, {Ho}, \& {Feng}}]{She17}
{She}, R., {Ho}, L.~C., \& {Feng}, H. 2017, \apj, 835, 223,
  \dodoi{10.3847/1538-4357/835/2/223}

\bibitem[{{Shin} {et~al.}(2021){Shin}, {Woo}, {Kim}, \& {Wang}}]{Shi21}
{Shin}, J., {Woo}, J.-H., {Kim}, M., \& {Wang}, J. 2021, \apj, 908, 81,
  \dodoi{10.3847/1538-4357/abd779}

\bibitem[{{Soifer} {et~al.}(2003){Soifer}, {Bock}, {Marsh}, {Neugebauer},
  {Matthews}, {Egami}, \& {Armus}}]{Soi03}
{Soifer}, B.~T., {Bock}, J.~J., {Marsh}, K., {et~al.} 2003, \aj, 126, 143,
  \dodoi{10.1086/375647}

\bibitem[{{Stevens} {et~al.}(1999){Stevens}, {Forbes}, \& {Norris}}]{Ste99}
{Stevens}, I.~R., {Forbes}, D.~A., \& {Norris}, R.~P. 1999, \mnras, 306, 479,
  \dodoi{10.1046/j.1365-8711.1999.02543.x}

\bibitem[{{Tanimoto} {et~al.}(2018){Tanimoto}, {Ueda}, {Kawamuro}, {Ricci},
  {Awaki}, \& {Terashima}}]{Tan18}
{Tanimoto}, A., {Ueda}, Y., {Kawamuro}, T., {et~al.} 2018, \apj, 853, 146,
  \dodoi{10.3847/1538-4357/aaa47c}

\bibitem[{{Thean} {et~al.}(2000){Thean}, {Pedlar}, {Kukula}, {Baum}, \&
  {O'Dea}}]{The00}
{Thean}, A., {Pedlar}, A., {Kukula}, M.~J., {Baum}, S.~A., \& {O'Dea}, C.~P.
  2000, \mnras, 314, 573, \dodoi{10.1046/j.1365-8711.2000.03401.x}

\bibitem[{{Travascio} {et~al.}(2021){Travascio}, {Fabbiano}, {Paggi}, {Elvis},
  {Maksym}, {Morganti}, {Oosterloo}, \& {Fiore}}]{Tra21}
{Travascio}, A., {Fabbiano}, G., {Paggi}, A., {et~al.} 2021, \apj, 921, 129,
  \dodoi{10.3847/1538-4357/ac18c7}

\bibitem[{{Treister} {et~al.}(2020){Treister}, {Messias}, {Privon}, {Nagar},
  {Medling}, {U}, {Bauer}, {Cicone}, {Mu{\~n}oz}, {Evans}, {Muller-Sanchez},
  {Comerford}, {Armus}, {Chang}, {Koss}, {Venturi}, {Schawinski}, {Casey},
  {Urry}, {Sanders}, {Scoville}, \& {Sheth}}]{Tre20}
{Treister}, E., {Messias}, H., {Privon}, G.~C., {et~al.} 2020, \apj, 890, 149,
  \dodoi{10.3847/1538-4357/ab6b28}

\bibitem[{{Tristram} {et~al.}(2014){Tristram}, {Burtscher}, {Jaffe},
  {Meisenheimer}, {H{\"o}nig}, {Kishimoto}, {Schartmann}, \& {Weigelt}}]{Tri14}
{Tristram}, K. R.~W., {Burtscher}, L., {Jaffe}, W., {et~al.} 2014, \aap, 563,
  A82, \dodoi{10.1051/0004-6361/201322698}

\bibitem[{{Tully} {et~al.}(2009){Tully}, {Rizzi}, {Shaya}, {Courtois},
  {Makarov}, \& {Jacobs}}]{Tul09}
{Tully}, R.~B., {Rizzi}, L., {Shaya}, E.~J., {et~al.} 2009, \aj, 138, 323,
  \dodoi{10.1088/0004-6256/138/2/323}

\bibitem[{{Ulvestad} \& {Wilson}(1983)}]{Ulv83}
{Ulvestad}, J.~S., \& {Wilson}, A.~S. 1983, \apjl, 264, L7,
  \dodoi{10.1086/183935}

\bibitem[{{Wold} \& {Galliano}(2006)}]{Wol06}
{Wold}, M., \& {Galliano}, E. 2006, \mnras, 369, L47,
  \dodoi{10.1111/j.1745-3933.2006.00171.x}

\bibitem[{{Wright} {et~al.}(2010){Wright}, {Eisenhardt}, {Mainzer}, {Ressler},
  {Cutri}, {Jarrett}, {Kirkpatrick}, {Padgett}, {McMillan}, {Skrutskie},
  {Stanford}, {Cohen}, {Walker}, {Mather}, {Leisawitz}, {Gautier}, {McLean},
  {Benford}, {Lonsdale}, {Blain}, {Mendez}, {Irace}, {Duval}, {Liu}, {Royer},
  {Heinrichsen}, {Howard}, {Shannon}, {Kendall}, {Walsh}, {Larsen}, {Cardon},
  {Schick}, {Schwalm}, {Abid}, {Fabinsky}, {Naes}, \& {Tsai}}]{Wri00}
{Wright}, E.~L., {Eisenhardt}, P. R.~M., {Mainzer}, A.~K., {et~al.} 2010, \aj,
  140, 1868, \dodoi{10.1088/0004-6256/140/6/1868}

\bibitem[{{Zhao} {et~al.}(2016){Zhao}, {Lu}, {Xu}, {Gao}, {Barcos-Mun{\~o}z},
  {D{\'\i}az-Santos}, {Appleton}, {Charmandaris}, {Armus}, {van der Werf},
  {Evans}, {Cao}, {Inami}, \& {Murphy}}]{Zha16}
{Zhao}, Y., {Lu}, N., {Xu}, C.~K., {et~al.} 2016, \apj, 820, 118,
  \dodoi{10.3847/0004-637X/820/2/118}

\end{thebibliography}
